\documentclass[useAMS,usenatbib]{mn2e}

\usepackage{color}
\usepackage{colortbl}
\usepackage{multirow}
\usepackage{dcolumn}
\usepackage{amsmath}
\usepackage{amssymb}
\usepackage{graphicx}
\usepackage{epsfig}
\usepackage{changebar}
\usepackage{natbib}
\usepackage{multirow}
\usepackage{subfigure}
\usepackage{txfonts}
\usepackage{relsize}
\usepackage{afterpage}

\usepackage{aalongtable}
\usepackage{longtable,lscape}
\usepackage[figuresright]{rotating}

\newcolumntype{d}[1]{D{.}{\cdot}{#1}}

\newcolumntype{.}{D{.}{.}{-1}}

\newcommand{\lsun}{L$_\odot$}

\renewcommand{\footnoterule}{ }

\newcommand{\tastar}{$T_{\mathrm{A}}^{*}$}
\newcommand{\nh}{NH$_3$}

\newcommand{\Tk}{$T_{\mathrm{k}}$}
\newcommand{\Tex}{$T_{\mathrm{ex}}$}

\newcommand{\Vlsr}{$V_{\mathrm{LSR}}$}

\title[Dust and Gas in Star-Forming Environments]{The Correlation of Dust and Gas Emission in Star-Forming Environments}
\author[L. K. Morgan et al.]{L. K. Morgan$^1$\thanks{E-mail:
lkm@astro.livjm.ac.uk}, T. J. T. Moore$^1$, D. J. Eden$^1$, J. Hatchell$^2$, J. S. Urquhart$^3$\\
$^1$Astrophysics Research Institute, Liverpool John Moores University, Twelve Quays House, Egerton Wharf, Birkenhead CH41 1LD \\
$^2$School of Physics, University of Exeter, Stocker Road, Exeter EX4 4QL\\
$^3$Max-Planck-Institut f\"{u}r Radioastronomie, Auf dem H\"{u}gel 69, Bonn, Germany}
\begin{document}

\date{Accepted 2014 February 23. Received 2014 February 17; in original form 2012 December 20}

\pagerange{\pageref{firstpage}--\pageref{lastpage}} \pubyear{2014}

\maketitle

\label{firstpage}

\begin{abstract}
  We present ammonia maps of portions of the W3 and Perseus molecular clouds in order to compare gas emission with submillimetre continuum thermal emission which are commonly used to trace the same mass component in star-forming regions, often under the assumption of LTE.
  
  The Perseus and W3 star-forming regions are found to have significantly different physical characteristics consistent with the difference in size scales traced by our observations. Accounting for the distance of the W3 region does not fully reconcile these differences, suggesting that there may be an underlying difference in the structure of the two regions.

  Peak positions of submillimetre and ammonia emission do not correlate strongly. Also, the extent of diffuse emission is only moderately matched between ammonia and thermal emission. Source sizes measured from our observations are consistent between regions, although there is a noticeable difference between the submillimetre source sizes with sources in Perseus being significantly smaller than those in W3.

  Fractional abundances of ammonia are determined for our sources which indicate a dip in the measured ammonia abundance at the positions of peak submillimetre column density.
  
  Virial ratios are determined which show that our sources are generally bound in both regions, although there is considerable scatter in both samples. We conclude that sources in Perseus are bound on smaller scales than in W3 in a way that may reflect their previous identification as low- and high-mass, respectively.

  Our results indicate that assumptions of local thermal equilibrium and/or the coupling of the dust and gas phases in star-forming regions may not be as robust as commonly assumed.  
\end{abstract}


\section{Introduction}
\label{sec:Introduction}
  In the study of star formation, the estimation of masses, temperatures and the internal gas velocity dispersions of protostellar and pre-stellar cores is fundamental to understanding the physical state of the core and hence to placing constraints on the processes of star formation and early stellar evolution.  The submillimetre continuum emission from cool dust is regarded as the most reliable tracer of mass in cores, since its radiative transfer is simple and its abundance uncertainties relatively low. However, deriving the dust temperatures required for mass calculations from continuum measurements is not always straightforward, usually because of a lack of far-IR data with sufficient spatial resolution and sensitivity to match that available in the (sub)millimetre and mid-IR wavebands. In addition to this, acquiring kinematic information to determine, e.g. the virial state of a core, requires spectroscopy of molecular emission lines that trace the denser core gas.
  
  It is often assumed that the ammonia inversion transitions, with critical densities of a few $\times$10$^3$ cm$^{-3}$ \citep{Maret2009}, can be used to trace much the same mass component as the submillimetre continuum emission (e.g.\citealp{Urquhart2011,Dunham2011}) and, since the gas and dust should be in thermal equilibrium at densities typical of star-forming regions (10$^5$ cm$^{-3}$ e.g. \citealt{Keto2008}), that both gas temperatures and line widths from \nh\ can be taken as representative of the dust-traced cores, allowing the calculation of core masses and virial ratios. This assumption is often necessary in order to perform virial analyses. However, there is evidence in the literature of the position of ammonia peaks failing to match the position of peaks of submillimetre emission (e.g. \citealp{Zhou1991}) and suggestions that \nh\ emission does not trace the densest gas in submillimetre cores \citep{Johnstone2010,Friesen2009}. Ammonia emission observed at the peak of submillimetre emission has also been noted to vary from measurements made at different positions within star-forming cores \citep{Chira2013}.

  Several Green Bank Telescope (GBT) studies have been made of the \nh\ emission from protostellar cores in a number of Galactic star-forming regions and have been used to derive the important physical parameters of velocity dispersion and temperature in the dense gas of the core. One of these \citep{AllsoppThesis} observed beam-spaced, 3 $\times$ 3 position grids centred on $\sim$60 cores detected in the W3 star-forming giant molecular cloud in the 850 \micron\ continuum by Moore et al. (2007). A surprising outcome of this study is that the cores appear to be poorly defined in \nh\ on scales of one arcmin (0.6 pc at the distance of W3), compared to the submillimetre continuum, in which they are compact with radii $<$ 30\arcsec. In contrast, no significant differences were found between grid centre and edge positions in the kinetic temperature, optical depth, column density, or velocity dispersion derived from \nh\ \citep{AllsoppThesis}. The simplest explanation for this is that the \nh\  emission is tracing a significantly larger and less dense gas component than the submillimetre continuum.  Another major study, by \citet{Rosolowsky2008}, observed \nh\ towards a selection of submillimetre-detected cores (among other positions described in the paper) in the Perseus (mainly low-mass) star-forming region. Only a single point per core was observed and so no spatial information is available in \nh, but a comparison between the two studies shows some significant differences between the \nh-traced cores in the two star-forming regions. For example, the Perseus cores have generally lower gas kinetic temperatures, higher \nh\ optical depths and column densities than in the high-mass star-forming regions of W3.

Through observation of fully-sampled, extended maps of ammonia cores associated with submillimetre emission this paper explores the likelihood that \nh\ and submillimetre continuum emission trace different structures in terms of extent and their spatial distribution. A determination of the typical size of cores in the Perseus and W3 regions shows significant systematic differences between the regions. These differences potentially mean that the two emission mechanisms are not tracing the same mass component. This result casts doubt on the use of \nh -derived gas temperatures to calculate clump/core masses from submillimetre continuum fluxes and \nh\ velocity dispersions in combination with continuum masses in calculating virial ratios.

\section{Observations}
\label{sec:Observations}
\subsection{Observational set up}
Fully-sampled maps in the \nh\ (1,1) and (2,2) rotational inversion transition lines were made using the newly-commissioned K-band focal plane array (KFPA) in shared-risk time on the 100-metre Green Bank Telescope (GBT), operated by the National Radio Astronomy Observatory\footnote{The National Radio Astronomy Observatory is a facility of the National Science Foundation operated under cooperative agreement by Associated Universities, Inc.}.

Observations were made over seven sessions from the 16$^{\mathrm{th}}$ of December 2010 to the 28$^{\mathrm{th}}$ of March 2011. Sky subtraction was attained through off-source measurements and flux calibration on the \tastar\ scale was achieved through switching observations of a noise diode. Comparison with previous observations using the dual-feed K-band receiver indicate that absolute flux calibration is accurate to $\lesssim$20\%. 

Atmospheric opacity values (at zenith) were determined from local weather models, collated in the archive maintained by R.Maddalena\footnote{http://www.gb.nrao.edu/$\sim$rmaddale/Weather/}. 
A spectral bandpass of 50 MHz was used, incorporating both the \nh\ (1,1) and (2,2) rotational inversion transitions at $\sim$23.69 GHz and $\sim$23.72 GHz respectively.

 Each map was completed using the `Daisy' pattern, consisting of a petal-shaped scan trajectory with continuous sampling. The oscillation period of the daisy scan is calculated for each map so that full sampling is achieved within a radius of 3.5\arcmin. Outside of this radius the integration time per map pixel is radially dependent, resulting in reduced signal-to-noise ratios towards the map edges. This is compensated for in the reduction process via weighting by effective integration time per pixel. The fiducial radius of each of our maps is then 3.5\arcmin, with some coverage (incompletely sampled) beyond this radius due to the size of the receiver array. Where necessary, some maps were mosaicked together so that complete, fully-sampled coverage of all desired regions could be obtained. Sampling rates were set at 1 or 2 seconds per sample, dependent upon the source and observing conditions, where the oscillation period of the daisy scan was defined with four samples per beam.
  Typical (median) pointing offsets were $\sim$4\arcsec\ and weather conditions were stable during all seven observing sessions. On two occasions (the 13$^{\mathrm{th}}$ of January and the 7$^{\mathrm{th}}$ of March 2011) snow in the GBT dish was a concern with regard to observing efficiency. Through calibrator observations and comparison with previous observations gain values were adjusted where appropriate and we have achieved an absolute flux calibration accuracy of better than 20\%.

  Maps were centred on individual submillimetre continuum sources selected from the catalogues of \citet{Hatchell2005,Hatchell2007} for Perseus and \citet{Moore2007} for W3. Targets were chosen partly at random but also selected to include both clustered regions, so that single maps cover several neighbouring objects, and isolated sources. The degree to which the resulting samples are representative of the populations of both regions in examined in Section \ref{Sec:Source_Associations}. Thirteen maps were completed in Perseus and 15 in the W3 GMC, with some overlaps. The pointing centres, observation dates and total integration times of these maps are listed in Table \ref{tbl:observations}. 
  
\begin{table}
\begin{center}
\caption{Pointing centres and observation dates for each mapped region.}
\label{tbl:observations}
\begin{minipage}{\linewidth}
\begin{center}
\begin{tabular}{lcccc}
\hline
\hline
{}		& \multicolumn{2}{c}{}	& {}				& {Total}\\
{}		& \multicolumn{2}{c}{Pointing Centre Position}	& {Observation}				& {Integration}\\
{Region}	& {R.A.(J2000)}	& {Dec.(J2000)}			& {Date}				& {Time (m)}\\
\hline
Perseus & & & & \\
\hline
Map 01  	& 03:25:48.8	& +30:42:24			& 20$^{\mathrm{th}}$ Dec 2010	& 21 \\
Map 02  	& 03:27:40.0	& +30:12:13			& 7$^{\mathrm{th}}$ Mar 2011  	& 21 \\
Map 03  	& 03:28:32.2	& +31:11:09			& 6$^{\mathrm{th}}$ Jan 2011	& 41 \\
Map 04  	& 03:28:42.6	& +31:06:13			& 21$^{\mathrm{st}}$ Dec 2010	& 21 \\
Map 05  	& 03:29:03.4	& +31:14:58			& 20$^{\mathrm{th}}$ Dec 2010	& 41 \\
Map 06  	& 03:29:10.3	& +31:13:35			& 21$^{\mathrm{st}}$ Dec 2010	& 21 \\
Map 07  	& 03:29:18.5	& +31:25:13			& 6$^{\mathrm{th}}$ Jan 2011	& 32 \\
Map 08  	& 03:32:17.5	& +30:49:49			& 21$^{\mathrm{st}}$ Dec 2010	& 21 \\
Map 09  	& 03:33:13.3	& +31:19:51			& 28$^{\mathrm{th}}$ Mar 2011 	& 41 \\
Map 10  	& 03:33:15.1	& +31:07:04			& 28$^{\mathrm{th}}$ Mar 2011 	& 41 \\
Map 11  	& 03:41:46.0	& +31:57:22			& 5$^{\mathrm{th}}$ Jan 2011	& 41 \\
Map 12  	& 03:43:57.8	& +32:04:06			& 7$^{\mathrm{th}}$ Mar 2011  	& 41 \\
Map 13  	& 03:44:48.8	& +32:00:29			& 7$^{\mathrm{th}}$ Mar 2011  	& 41 \\
W3 & & & & \\
\hline
Map 01  	& 02:19:53.1	& +61:01:55			& 16$^{\mathrm{th}}$ Dec 2010	& 41 \\
Map 02  	& 02:20:41.0	& +61:09:42			& 14$^{\mathrm{th}}$ Jan 2011	& 41 \\
Map 03  	& 02:21:04.0	& +61:06:01			& 14$^{\mathrm{th}}$ Jan 2011	& 41 \\
Map 04  	& 02:21:06.0	& +61:27:28			& 20$^{\mathrm{th}}$ Dec 2010	& 41 \\
Map 05  	& 02:21:41.0	& +61:05:44			& 14$^{\mathrm{th}}$ Jan 2011	& 41 \\
Map 06  	& 02:22:23.0	& +61:06:12			& 14$^{\mathrm{th}}$ Jan 2011	& 41 \\
Map 07  	& 02:23:28.6	& +61:12:04			& 21$^{\mathrm{st}}$ Dec 2010	& 41 \\
Map 08  	& 02:25:31.2	& +62:06:20			& 16$^{\mathrm{th}}$ Dec 2010	& 82 \\
Map 09  	& 02:25:37.7	& +61:13:51			& 19$^{\mathrm{th}}$ Dec 2010	& 41 \\
Map 10  	& 02:26:40.0	& +62:07:00			& 29$^{\mathrm{th}}$ Mar 2011 	& 41 \\
Map 11  	& 02:26:44.6	& +61:29:44			& 20$^{\mathrm{th}}$ Dec 2010	& 41 \\
Map 12  	& 02:27:17.8	& +61:57:14			& 21$^{\mathrm{st}}$ Dec 2010	& 41 \\
Map 13  	& 02:28:00.0	& +61:24:00			& 29$^{\mathrm{th}}$ Mar 2011 	& 41 \\
Map 14  	& 02:28:12.3	& +61:29:40			& 19$^{\mathrm{th}}$ Dec 2010	& 41 \\
Map 15  	& 02:28:26.2	& +61:32:14			& 20$^{\mathrm{th}}$ Dec 2010	& 41 \\
\hline
\end{tabular}\\
\end{center}
\end{minipage}
\end{center}
\end{table}

\subsection{Data Reduction and LTE Analysis}
\label{sec:DR}
Individual feeds, i.e. time series data from each of the 7 receivers, were reduced by running data through the GBT pipeline which calibrates the data and reconstructs spatial image cubes using reference scans and performing sky subtraction as appropriate\footnote{https://safe.nrao.edu/wiki/pub/Kbandfpa/KfpaReduction/kfpaDataReduceGuide-11Dec01.pdf}. The output cubes were produced with an angular pixel size of 6\arcsec\ (in comparison to the GBT beam size of 30\arcsec). The GBT pipeline uses Parseltongue\footnote{http://www.radionet-eu.org/rnwiki/ParselTongue} scripts to control AIPS tasks which produce images of the combined spectral data and removes a second order spectral baseline. The mosaicking and weighting of overlapping mapping observations was performed as part of the GBT pipeline reductions.

 The determination of optical depth, rotational temperature, kinetic temperature and excitation temperature from ammonia (1,1) and (2,2) spectra, assuming LTE, has been described in detail by \citet{Ho1983} and applied to observational data by numerous authors (e.g. \citealt{Morgan2010,Friesen2009,Rosolowsky2008}). The exact processes used here are described in detail by \citet{Morgan2012}. We fit the hyperfine structure of the ammonia (1,1) and (2,2) lines using our own developed IDL\footnote{http://www.exelisvis.com/ProductsServices/IDL.aspx} routines in conjunction with the IDL astronomy library routines \citep{Landsman1993}. The results of the fitting allowed us to produce maps of kinetic and excitation temperature, linewidth, optical depth and column density. It should be noted here that our deduction of ammonia column density makes use of the assumption of local thermodynamic equilibrium (LTE), i.e. that the excitation temperature (\Tex) of a source is equivalent to its kinetic temperature (\Tk). The fact that derived values of \Tex\ are $not$ approximately equal to values of \Tk\ reflects the fact that \Tex\ is highly dependent upon the filling factor of the source in question as it is derived from a single transition. This has ramifications for observations of clumps which may be expected to form multiple cores as well as when making comparisons of regions which might be expected to be structurally different (i.e. filaments vs. cores). This effect is particularly relevant to a study such as this one; the distances to Perseus and W3 are 260 pc and 2 kpc respectively, leading to a significant difference in the physical scale which is being sampled in our observations. The filling factor of each source has been calculated from the assumption of LTE (see \citealt{Morgan2010}).
 
There are few ammonia observations of the W3 region on the scale of our coverage, though there are several of the area around W3(OH) \citep{Wilson1993,Zeng1984} as well as more complete coverage of W3 Main by \citet{Tieftrunk1998}.\citet{Rosolowsky2008} performed a single-pointing survey of ammonia towards selected positions in the Perseus molecular cloud. A comparison of the derived physical parameters found by those authors, at positions for which we have matching data, shows consistency between the measurements at a level reflecting measurement errors of 10-20\% (see Section \ref{sec:Discussion}).

\section{Results}
\label{sec:Results}
\begin{figure*}
\begin{center}
\includegraphics*[width=0.9\textwidth]{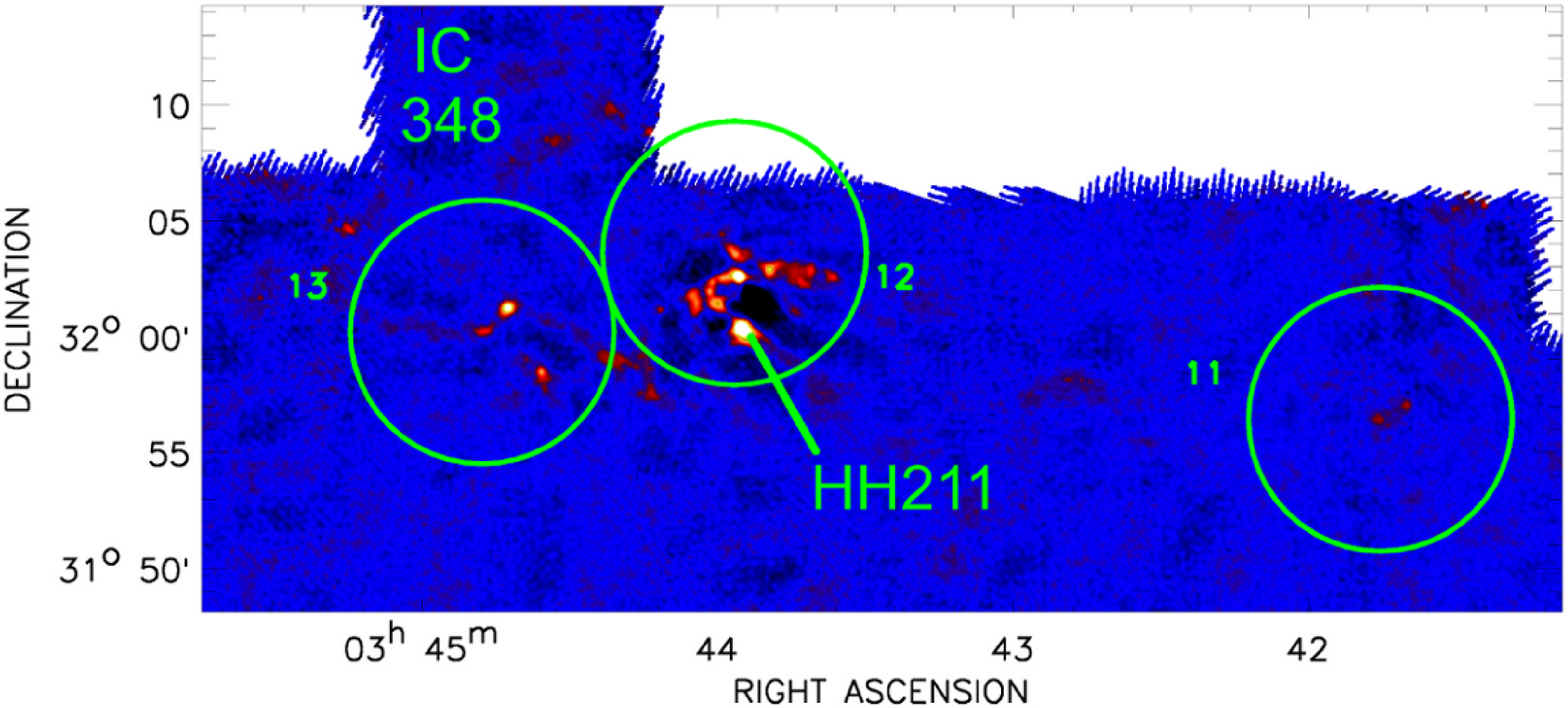}\\
\includegraphics*[width=0.9\textwidth]{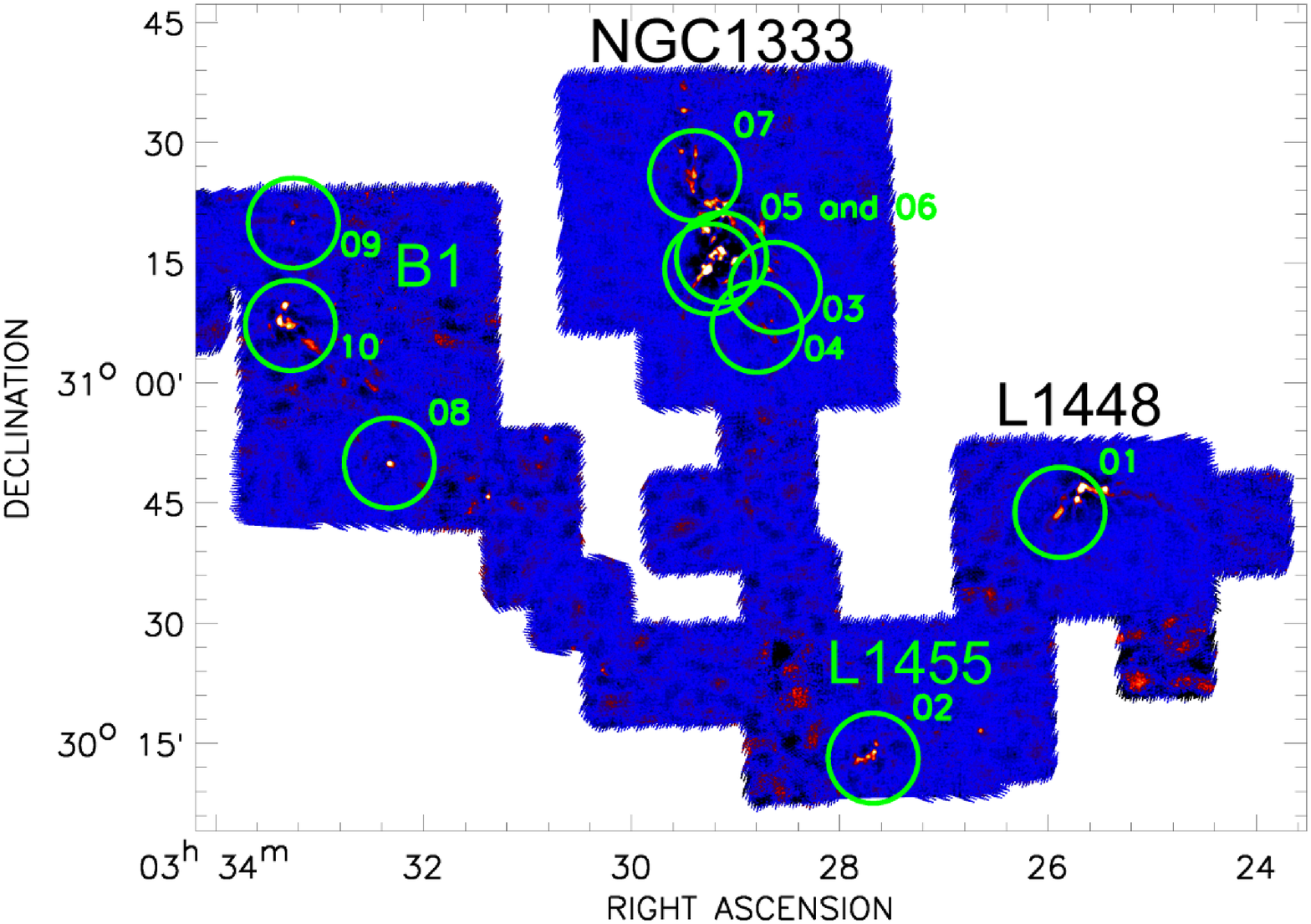}
\caption{Extracts from the 850 \micron\ submillimetre SCUBA map of the Persus region from \citet{Hatchell2005} overlaid with circles showing the coverage of each observed \nh\ inversion line map, including the under-sampled regions outside of a $\sim$3.5\arcmin\ radius. The eastern portion of the Perseus molecular cloud complex is shown in the top panel with the western portion shown in the bottom panel. Some well-known regions are labelled.}
\label{fig:Perseus_Coverage_Map}
\end{center}
\end{figure*}

\begin{figure*}
\begin{center}
\includegraphics*[width=1.5\textwidth]{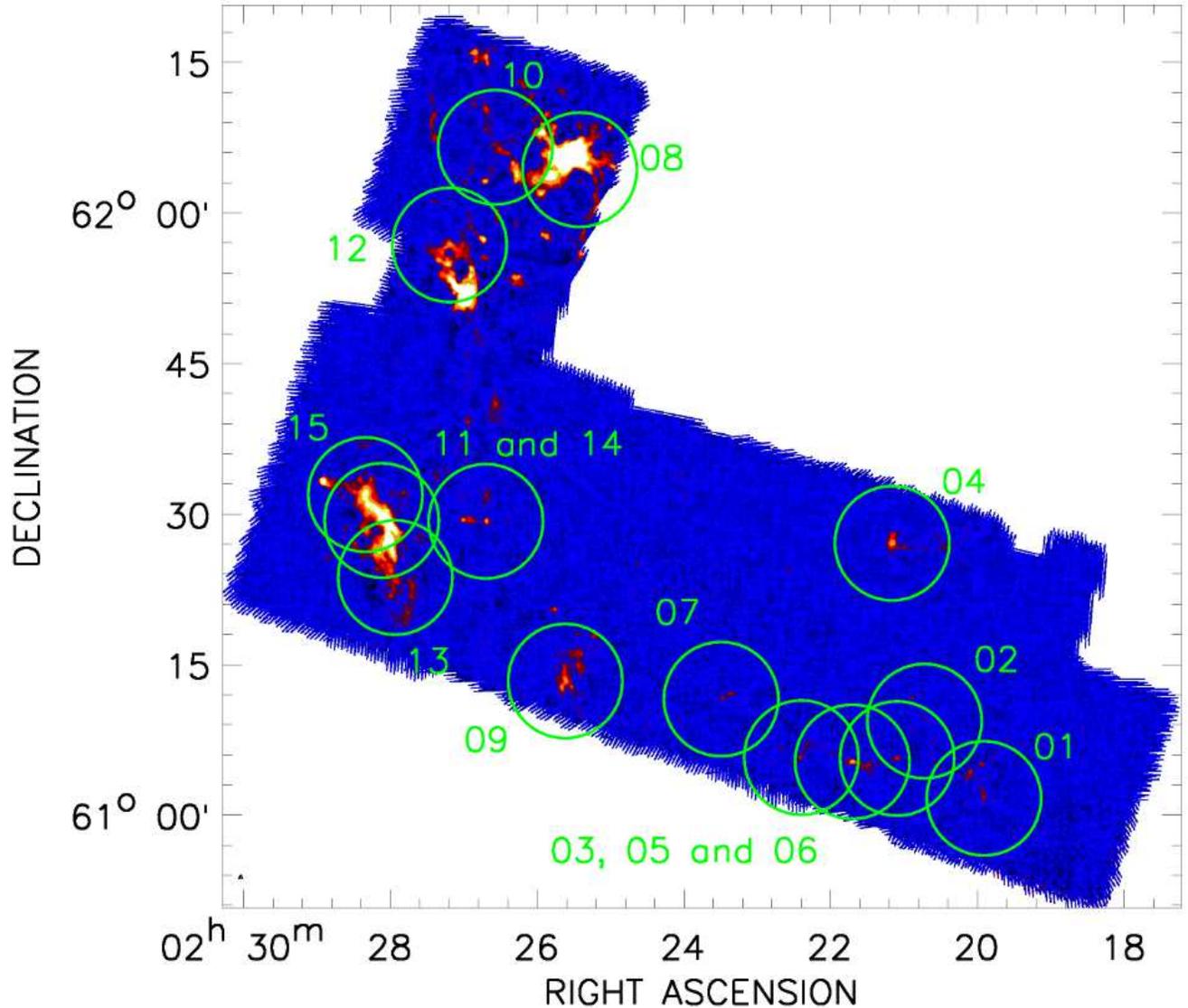}
\caption{The 850 \micron\ submillimetre SCUBA map of the W3 region from \citet{Moore2007} overlaid with circles showing the coverage of each observed \nh\ inversion line map, including the under-sampled regions outside of a $\sim$3.5\arcmin\ radius}.
\label{fig:W3_Coverage_Map}
\end{center}
\end{figure*}

Figures \ref{fig:Perseus_Coverage_Map} and \ref{fig:W3_Coverage_Map} show the submillimetre maps of \citet{Hatchell2005} and \citet{Moore2007} covering the Perseus and W3 star-forming regions respectively. Overlaid on the images are circles depicting the coverage of the GBT-KFPA ammonia maps presented here. Individual \nh\ (1,1) integrated intensity images are presented in Figure \ref{fig:Int_Maps}. Noise values vary slightly between the Perseus and W3 regions, reflecting the lower integration times for some of our Perseus data. Individual r.m.s. values are presented with each map, but the Perseus data have r.m.s. off-source integrated intensity values in the range 0.36 - 0.86 K km/s with a mean of 0.49 K km/s. For W3 the range is 0.18 - 0.47 K km/s with a mean of 0.28 K km/s. As mentioned in Section \ref{sec:Observations}, there is a radial variation in r.m.s. in our maps outside of a 3.5\arcmin radius, due to the inconsistent sampling of the 'daisy' pattern observing mode. This radial variation in noise due to the inconsistent sampling of the 'daisy' pattern observing mode is particularly obvious in Perseus map 11.

The ammonia maps sample the majority of the Perseus and W3 regions associated with submillimetre emission. Although the sampling can be seen to be less than complete, the brightest, most extended sources in the submillimetre are covered by our observations. Our observations of Perseus cover at least some portion of each of the well-known regions identified as B5, IC348, B1, NGC1333, L1455 and L1448. The High Density Layer (HDL) identified by \citet{Lada1978} on the eastern edge of the W3 GMC is covered approximately equally with the more diffuse material in the southwest portion of the cloud. The W3 ammonia maps cover 1178 arcmin$^2$ split approximately equally (51-49\%) between the HDL and the more diffuse region. Despite the equivalence in area between the two regions, more sources are found in the HDL. Our maps cover 192 of the submillimetre sources from the catalogue of \citet{Moore2007}, of these, 127 are within the HDL and 65 are to the southwest of the W3 region.

The integrated intensity maps presented in Figure \ref{fig:Int_Maps} show a mixture of condensed cores (e.g. Perseus Map 08) and extended structure (e.g. W3 Map 13) with multiple examples of apparent filamentary morphology (e.g. Perseus Map 06, W3 Map 14). A large proportion show multiple cores contained within apparently connected extended structures (e.g. Perseus Map 12).

\begin{figure*}
\begin{center} 
\includegraphics*[width=0.4\textwidth]{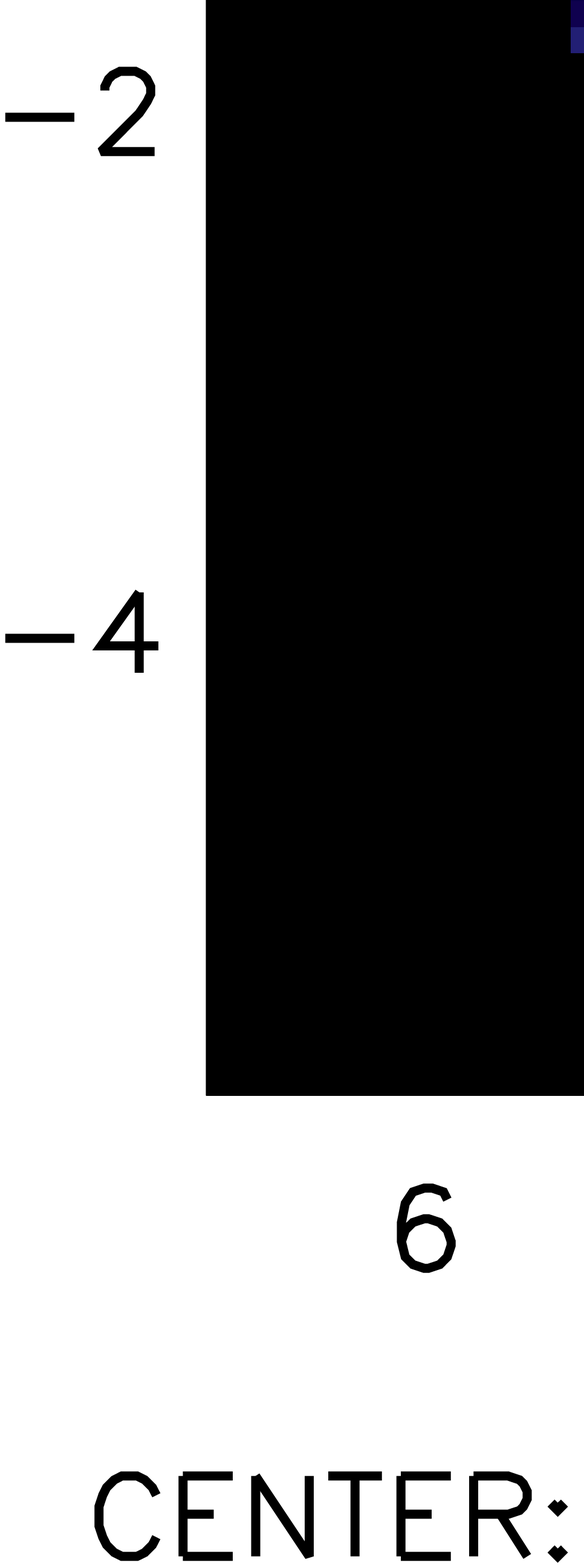}
\includegraphics*[width=0.4\textwidth]{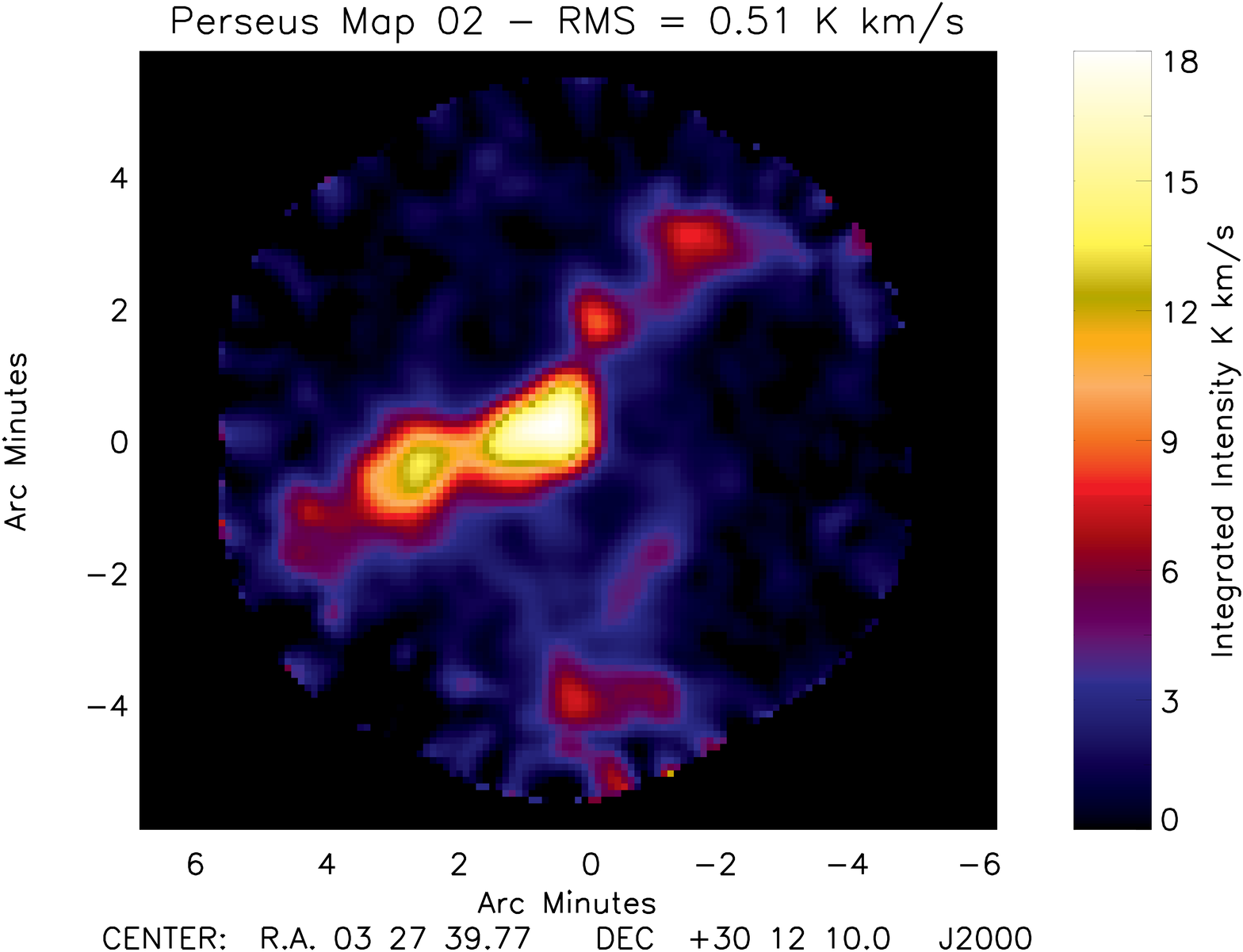}\\
\includegraphics*[width=0.4\textwidth]{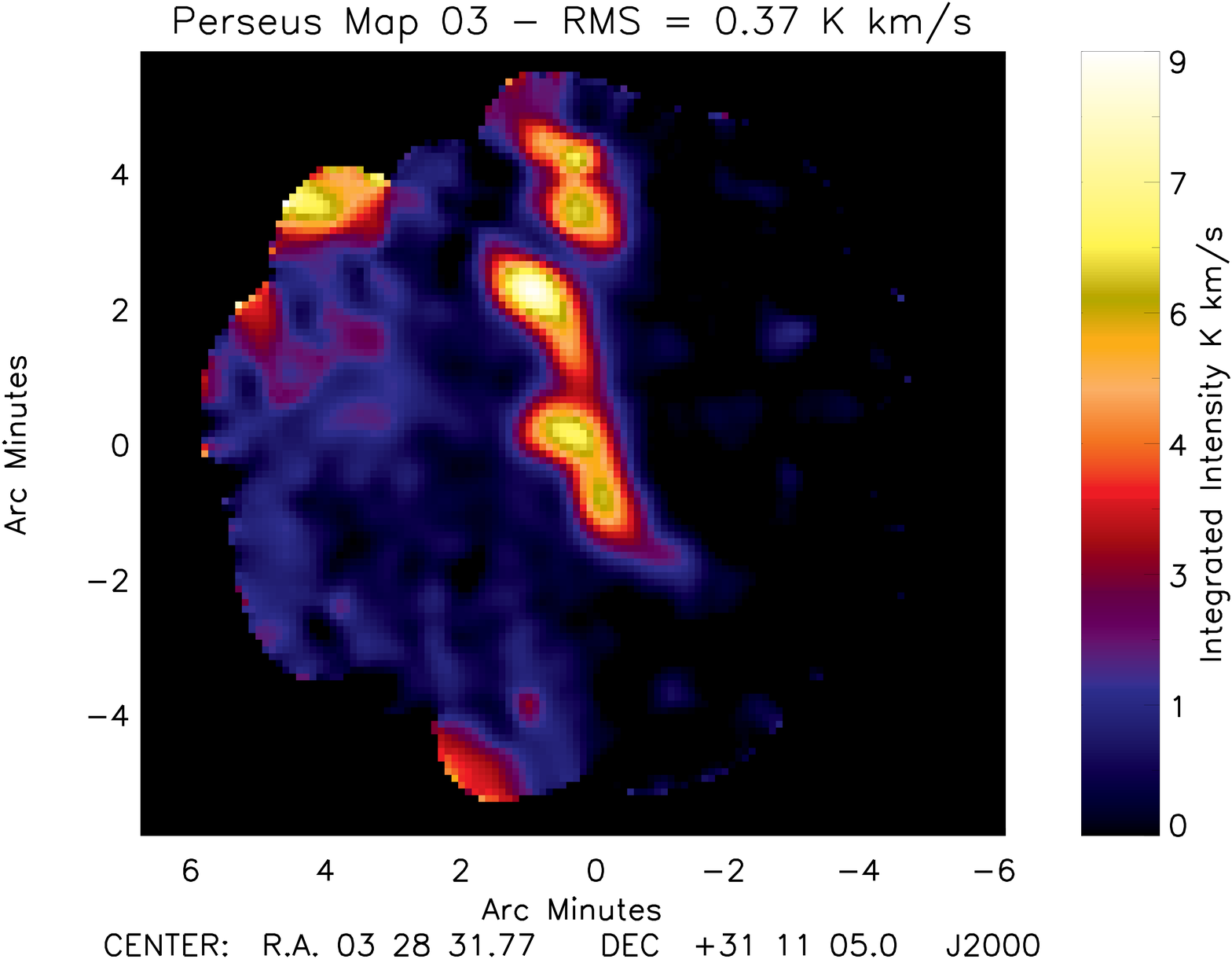}
\includegraphics*[width=0.4\textwidth]{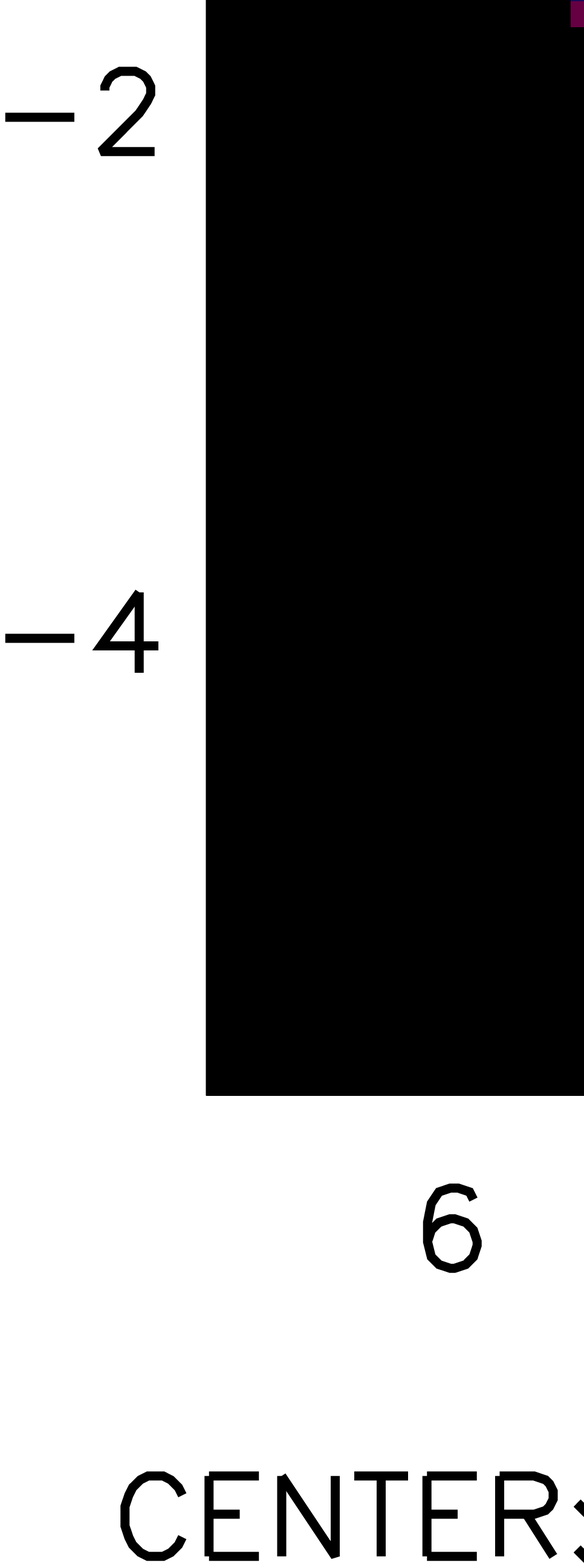}\\
\includegraphics*[width=0.4\textwidth]{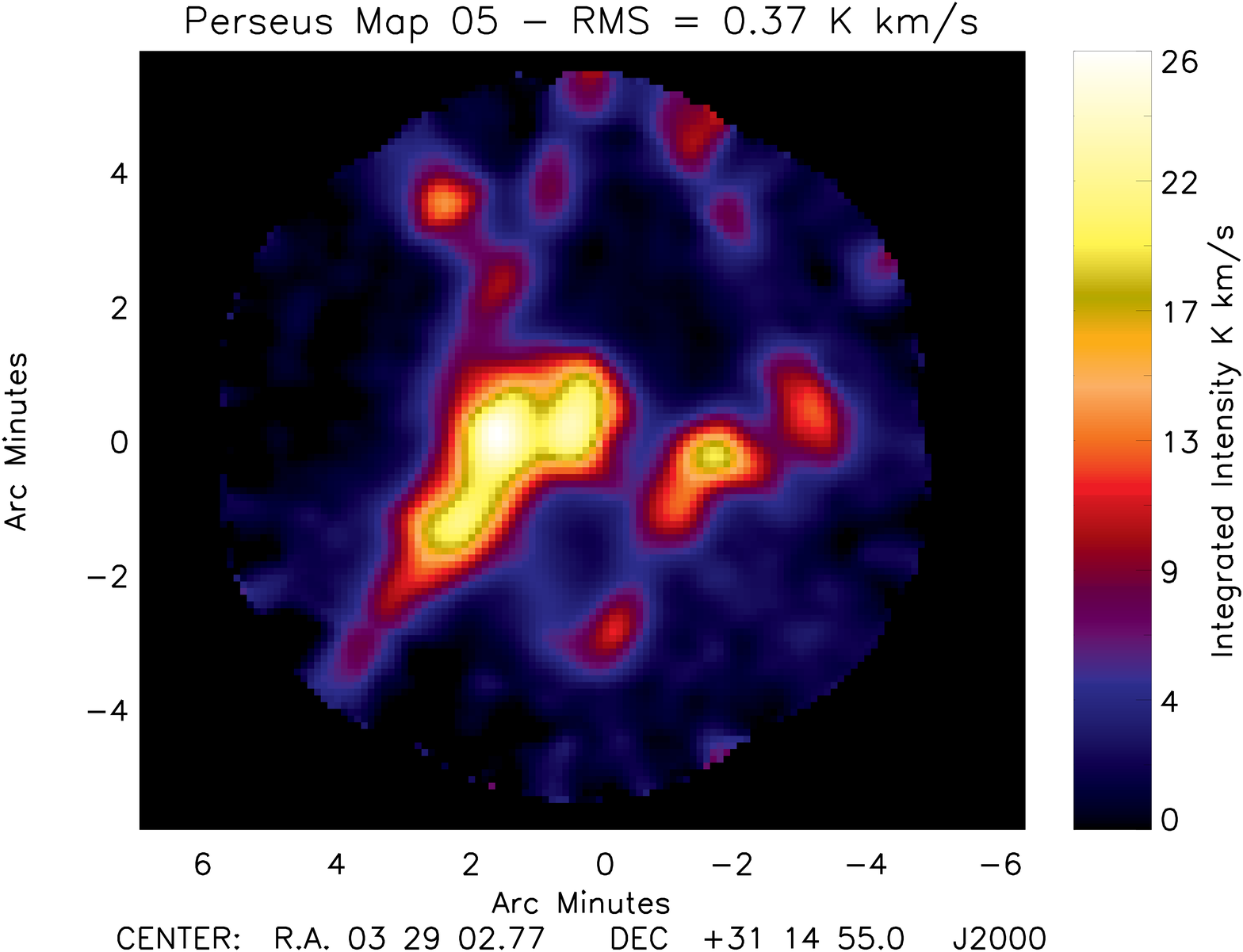}
\includegraphics*[width=0.4\textwidth]{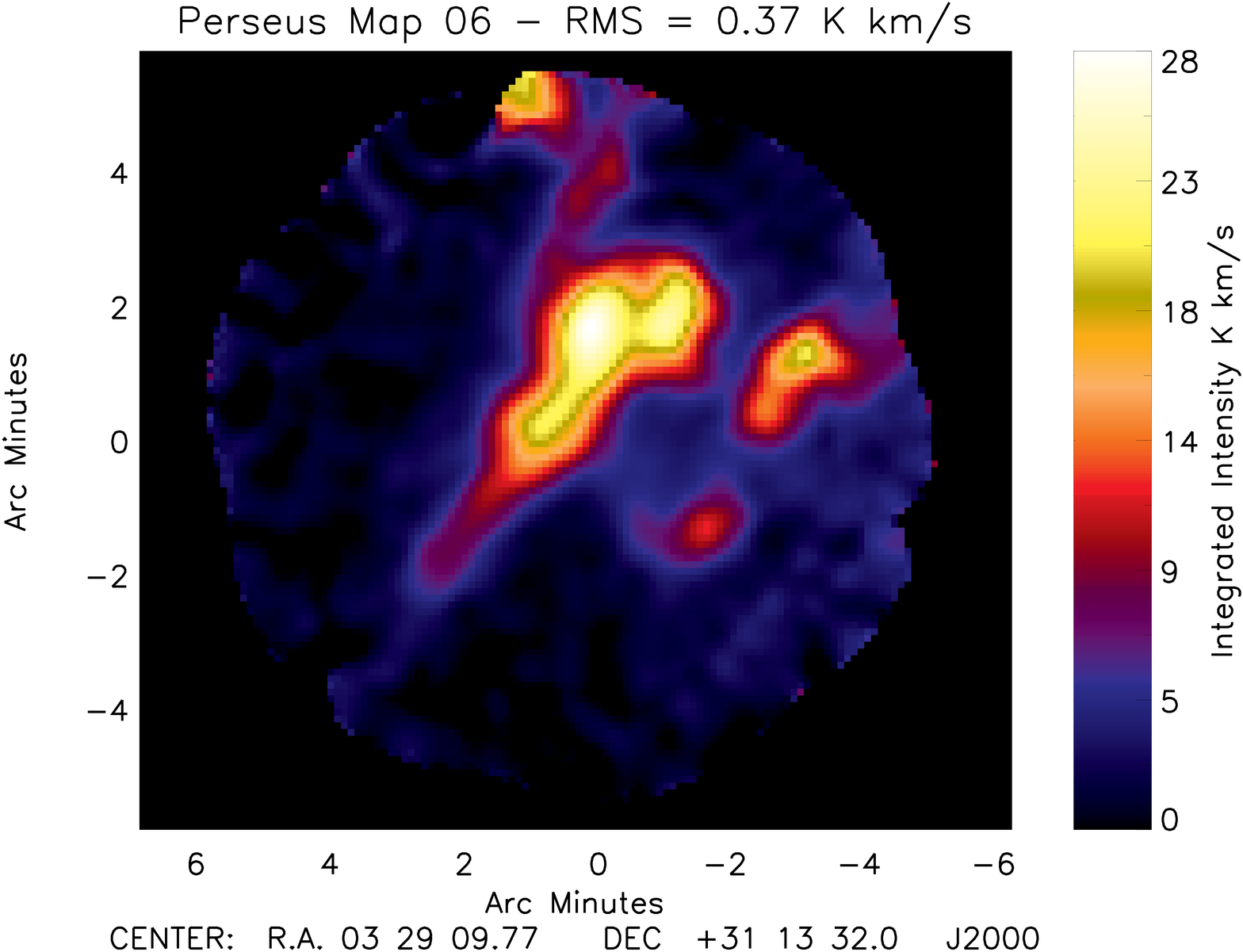}\\
\includegraphics*[width=0.4\textwidth]{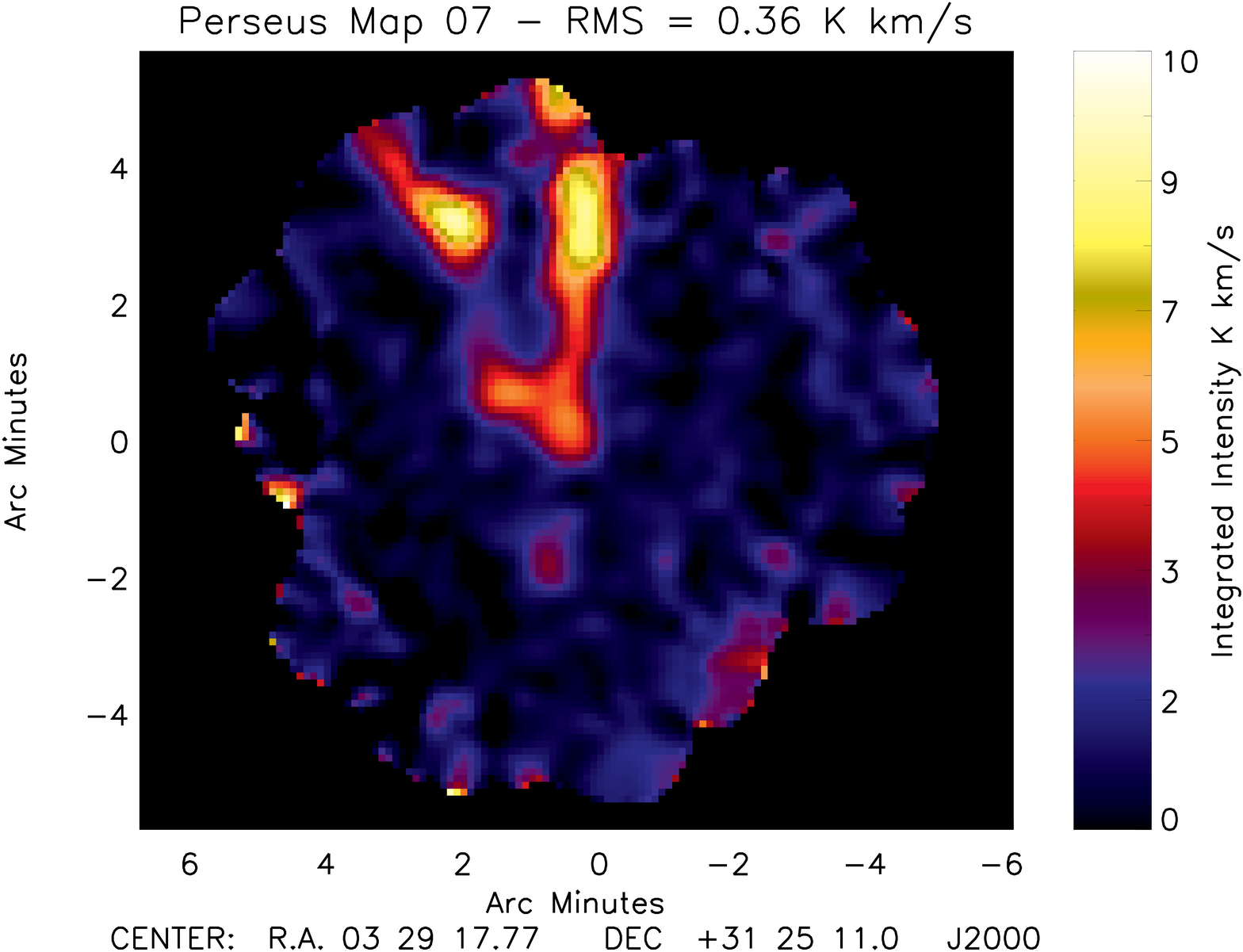}
\includegraphics*[width=0.4\textwidth]{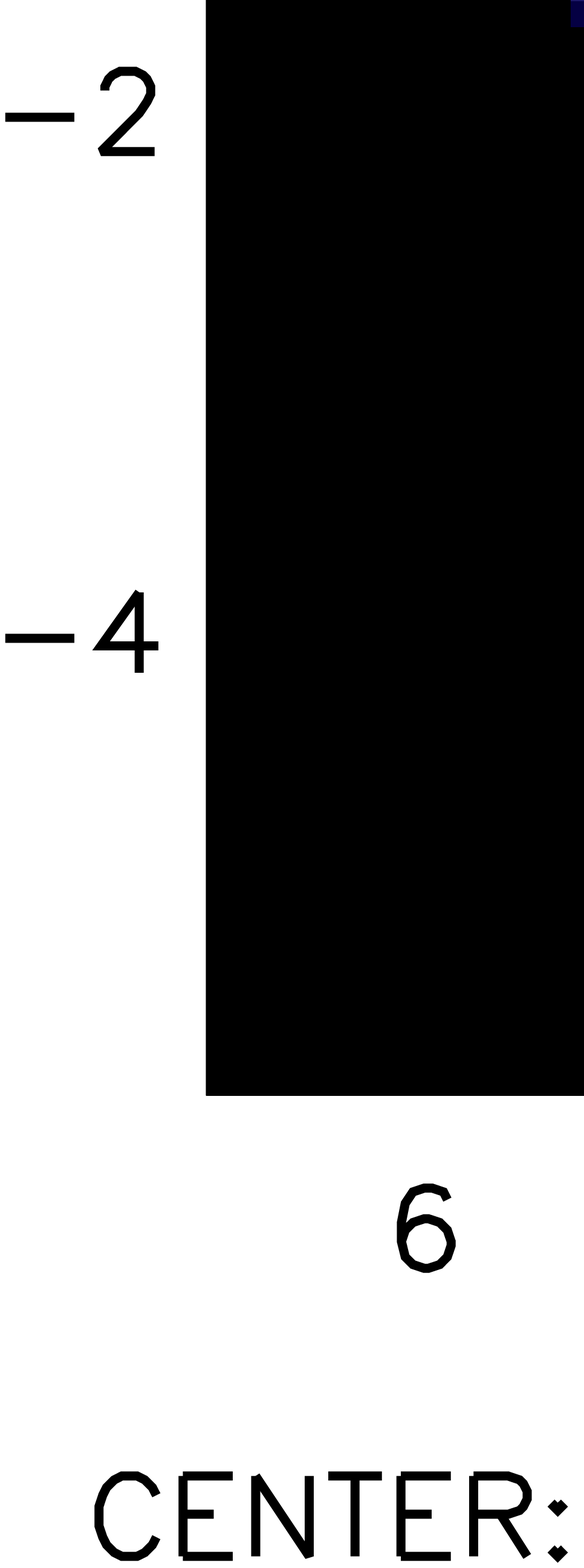}\\
\includegraphics*[width=0.4\textwidth]{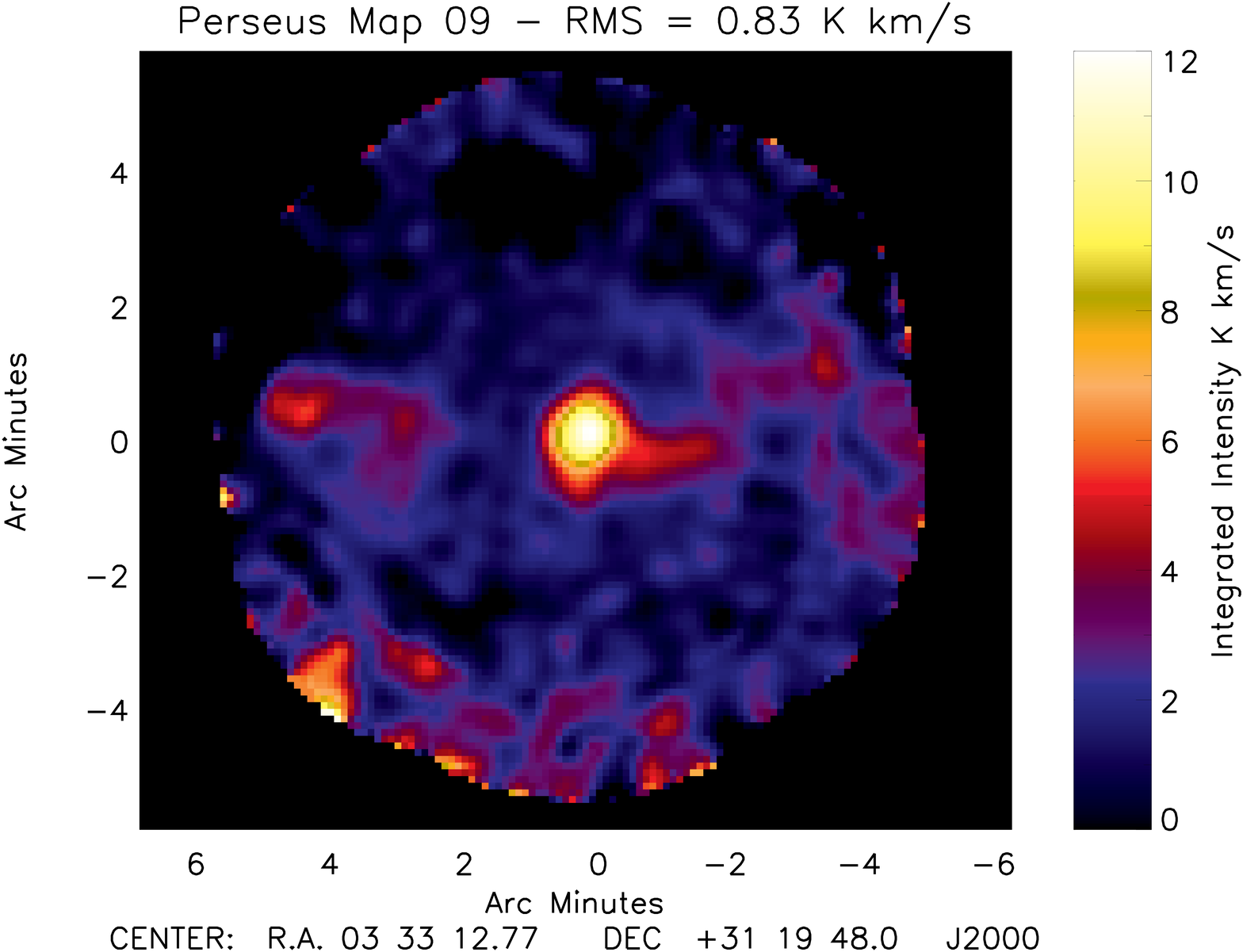}
\includegraphics*[width=0.4\textwidth]{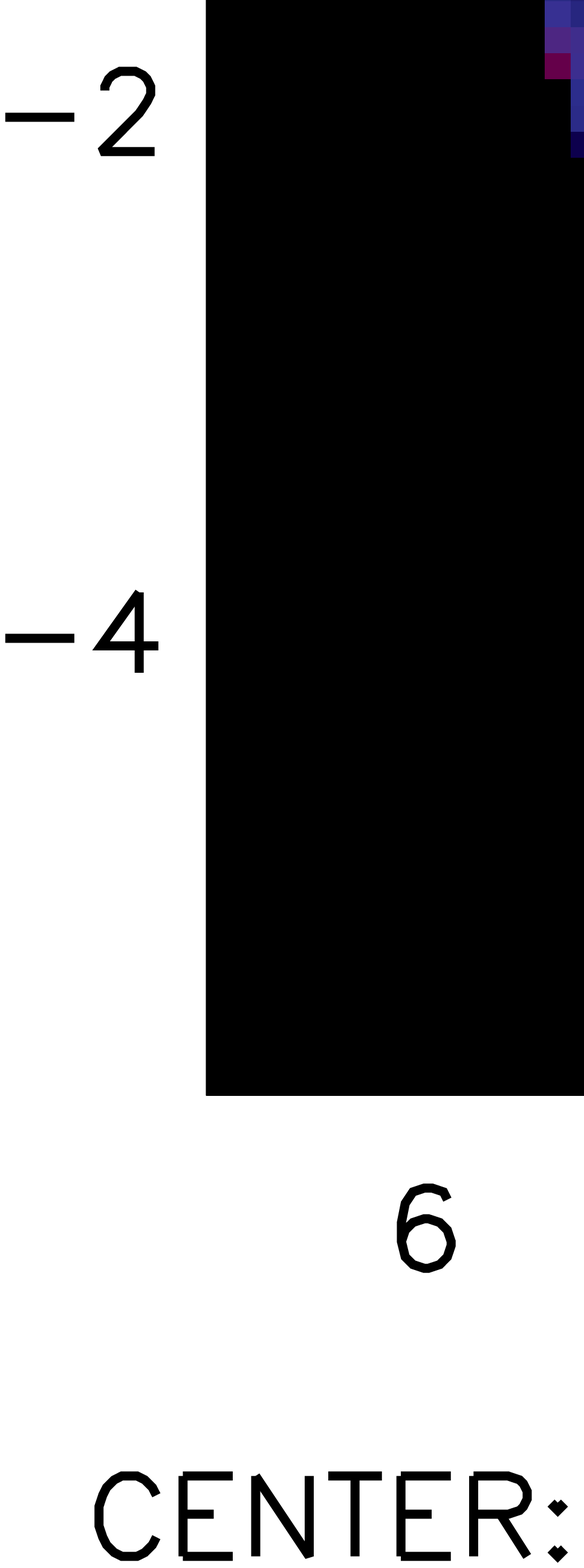}\\
\caption{The Perseus and W3 integrated intensity maps.}
\label{fig:Int_Maps}
\end{center}
\end{figure*}

\addtocounter{figure}{-1}

\begin{figure*}
\begin{center} 
\includegraphics*[width=0.4\textwidth]{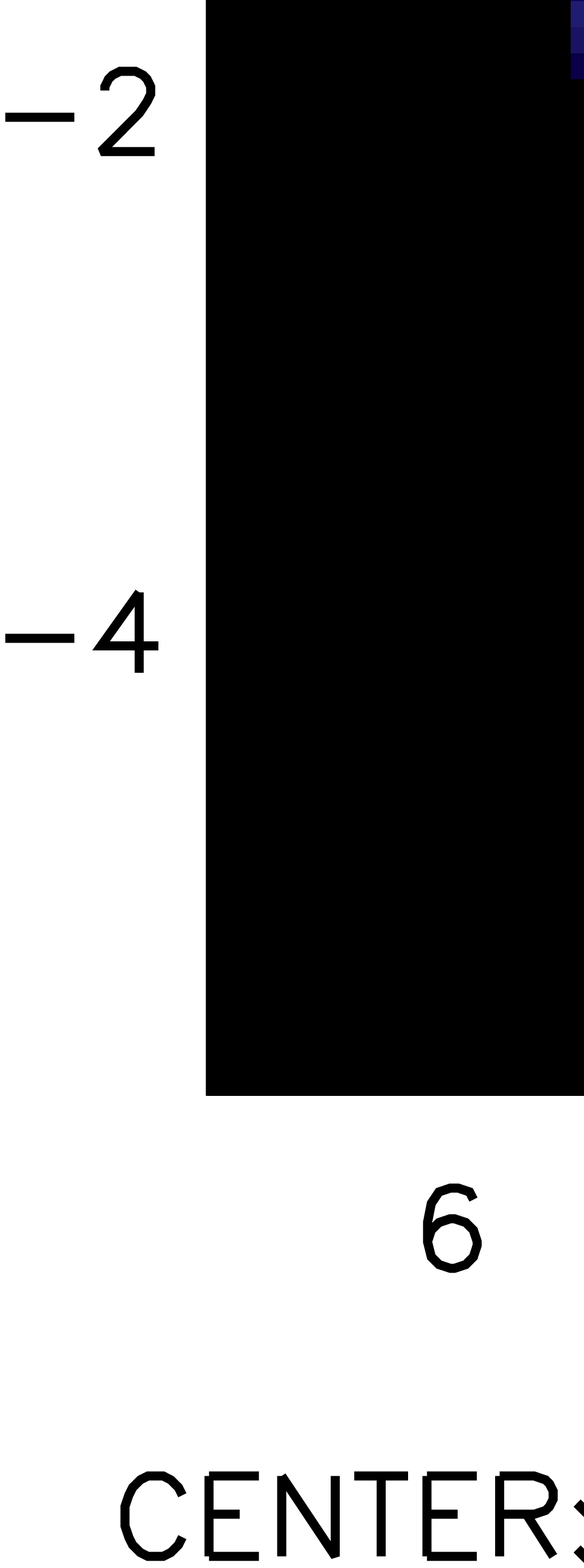}
\includegraphics*[width=0.4\textwidth]{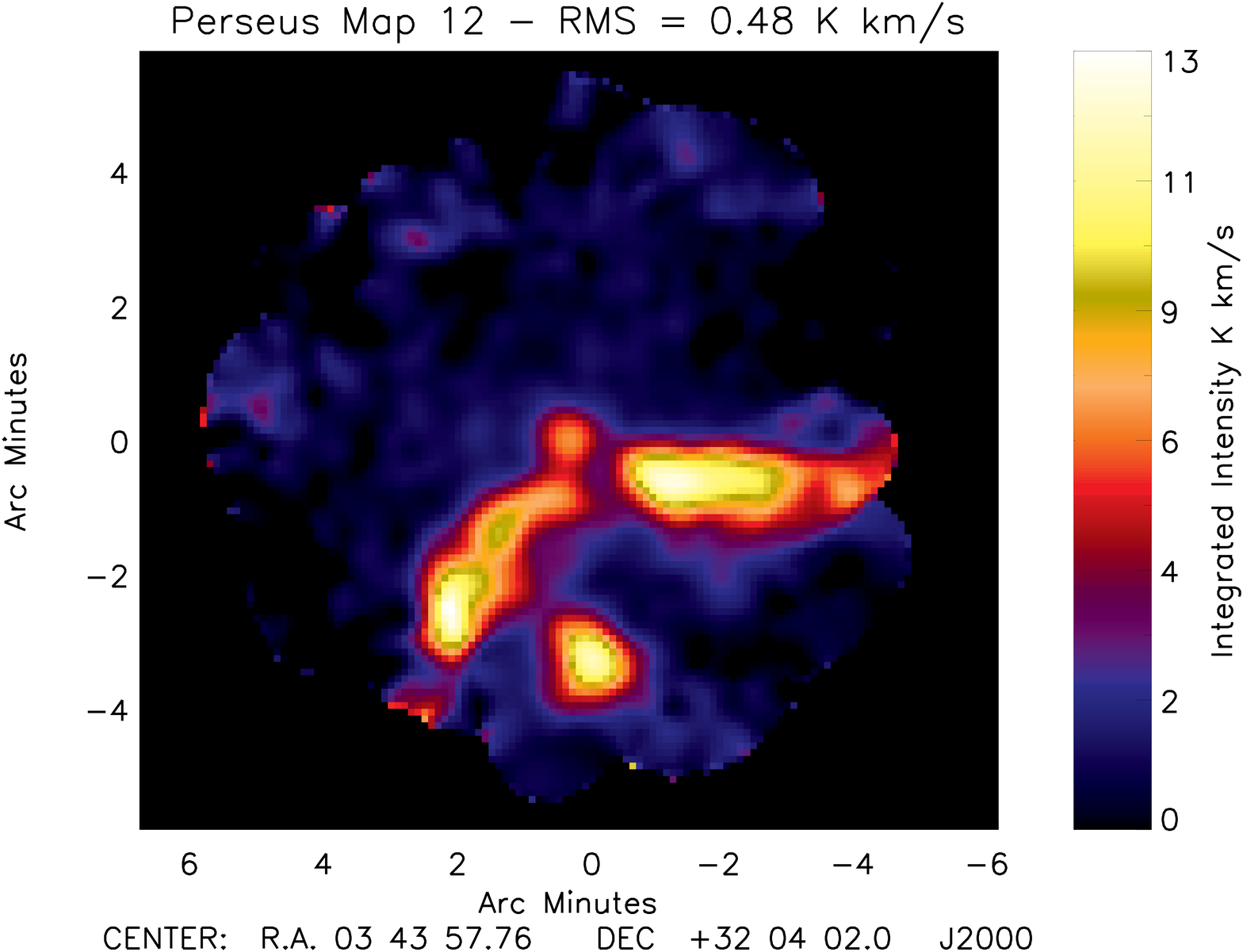}\\
\includegraphics*[width=0.4\textwidth]{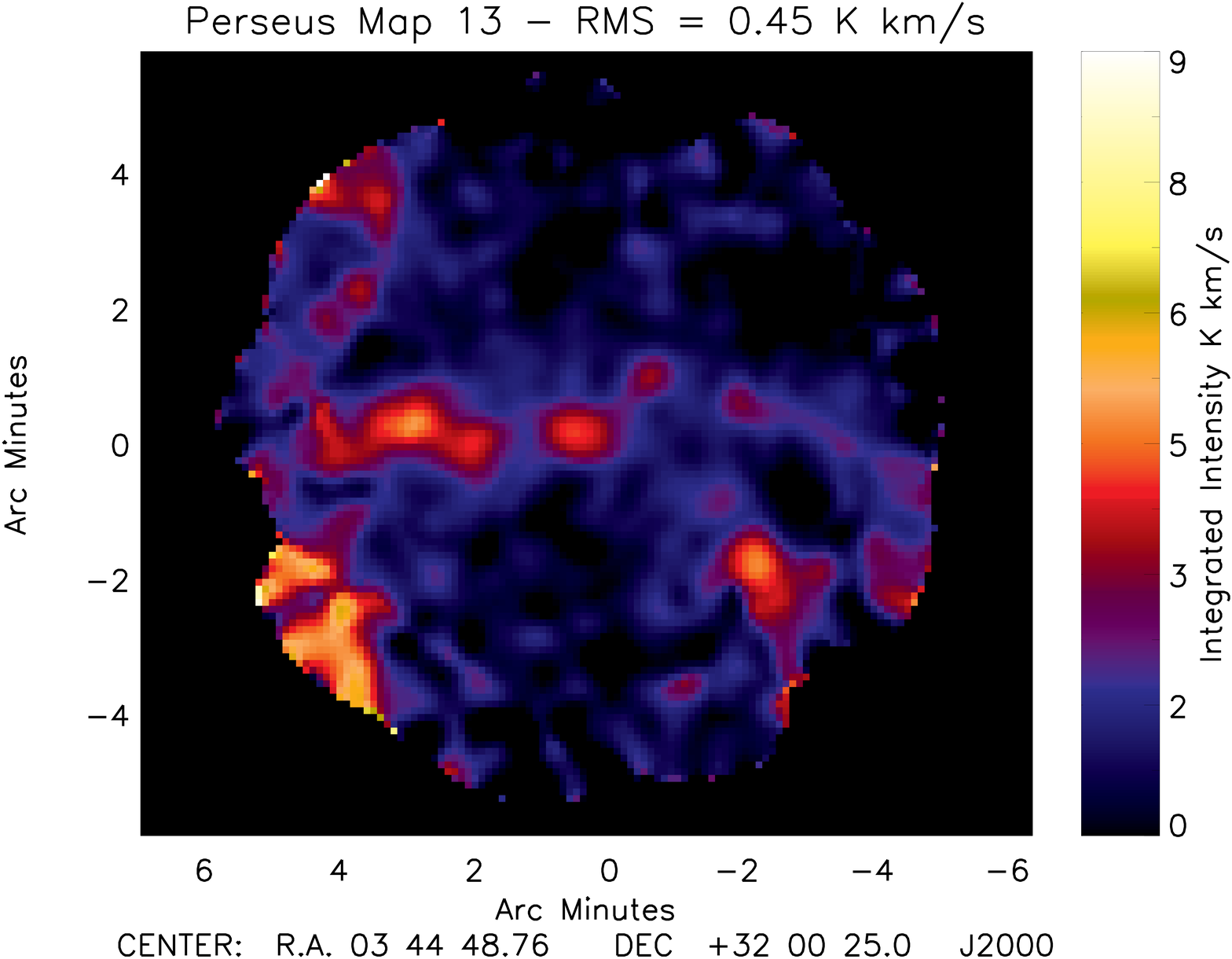}
\includegraphics*[width=0.4\textwidth]{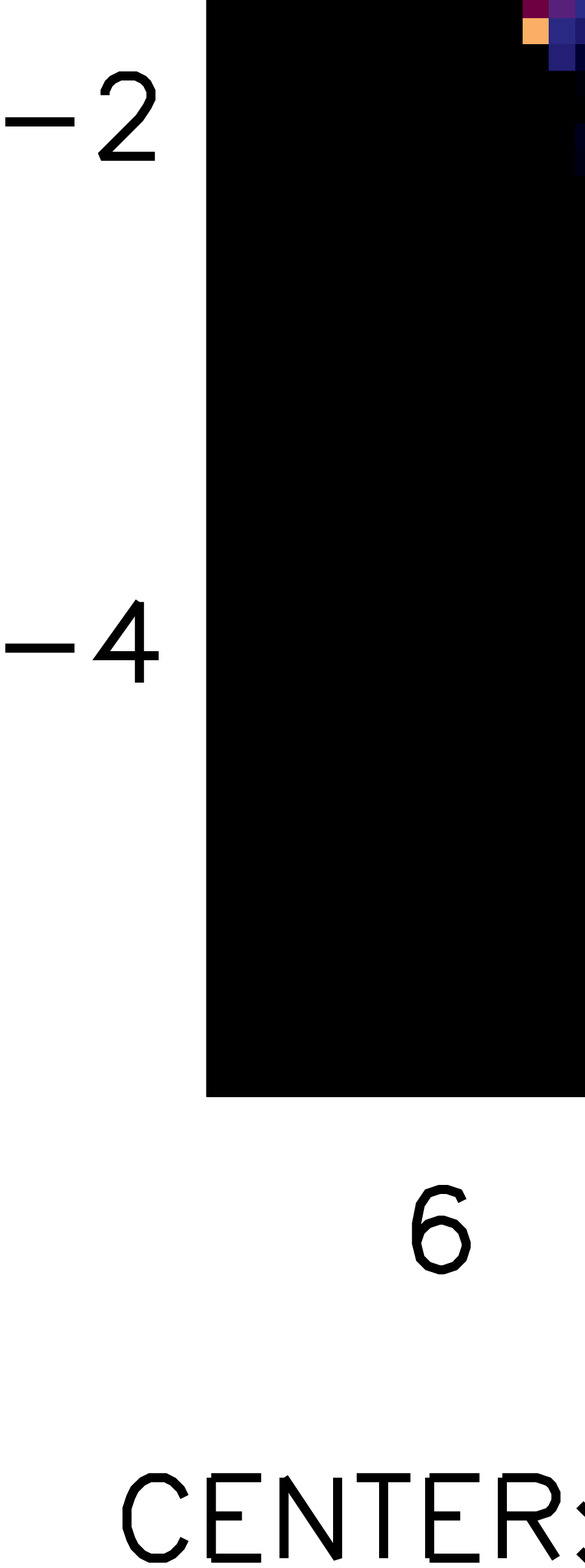}\\
\includegraphics*[width=0.4\textwidth]{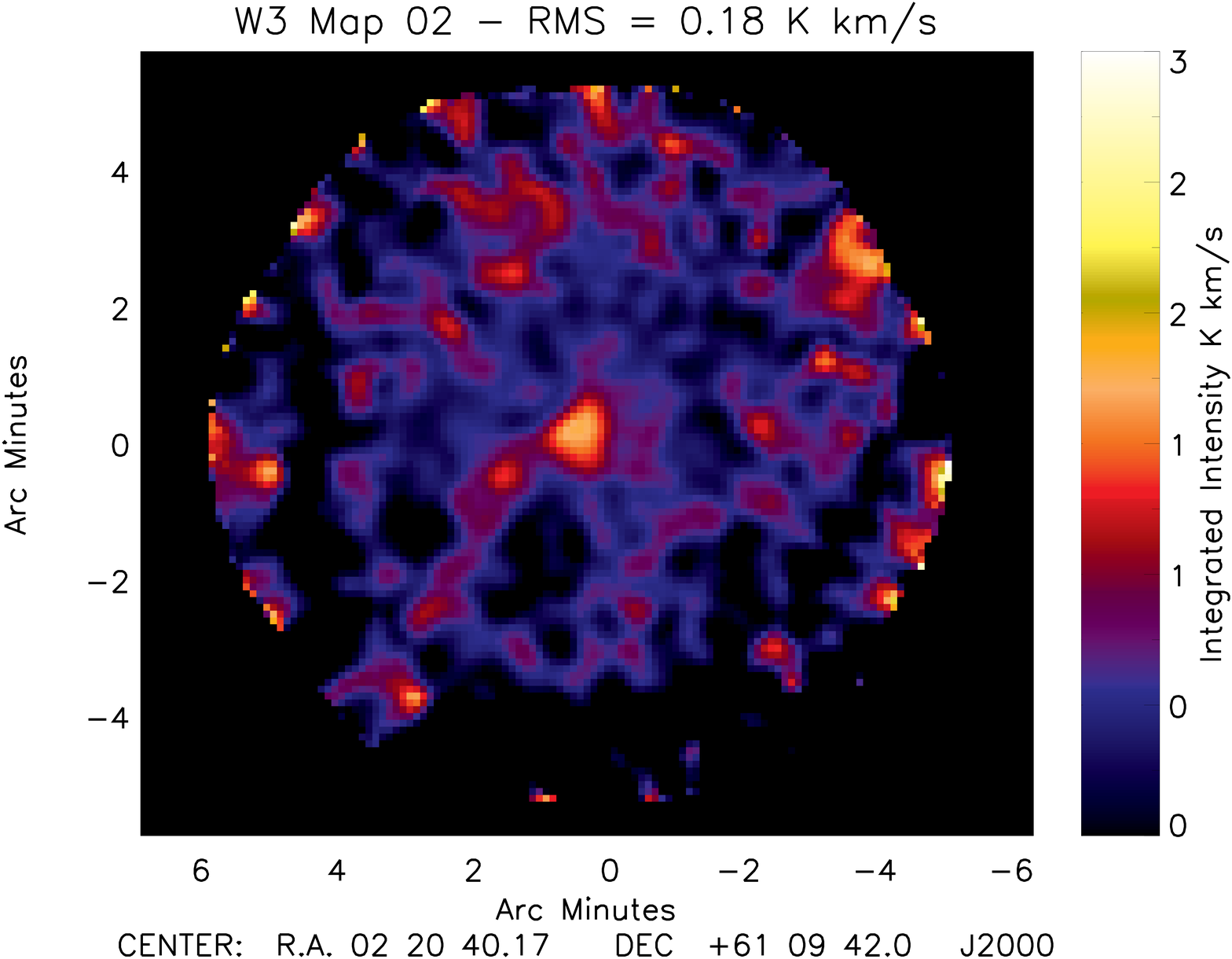}
\includegraphics*[width=0.4\textwidth]{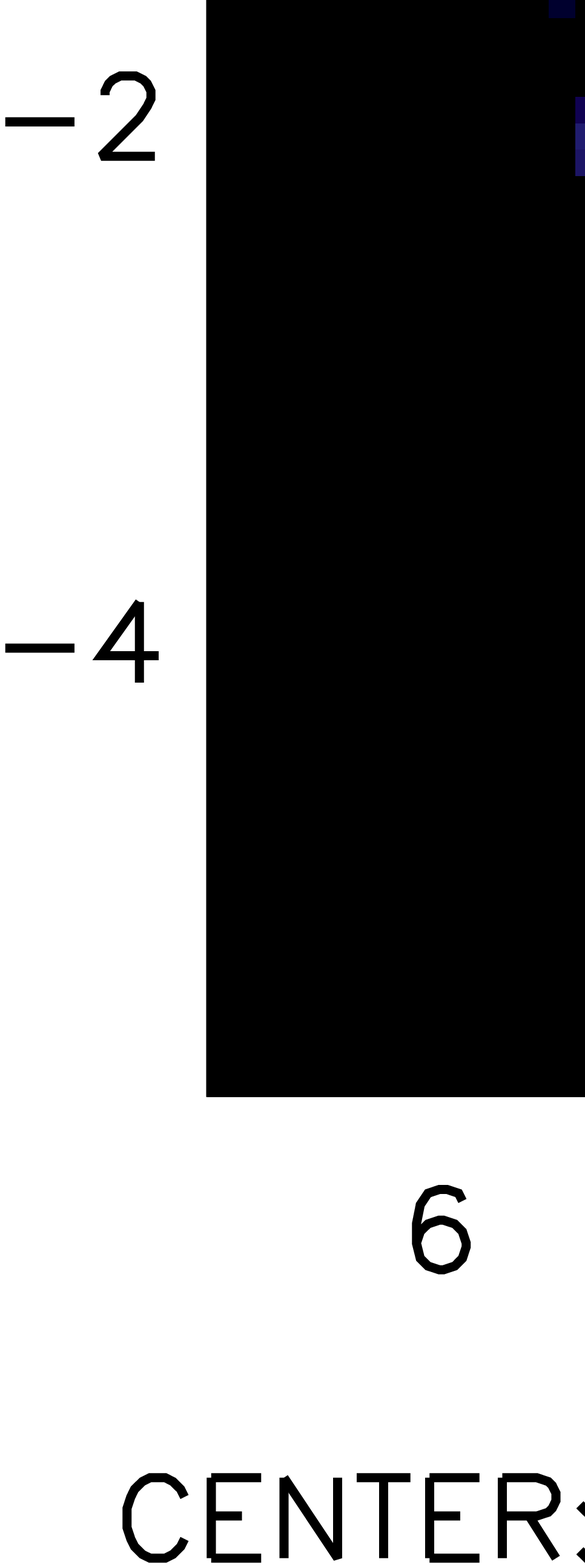}\\
\includegraphics*[width=0.4\textwidth]{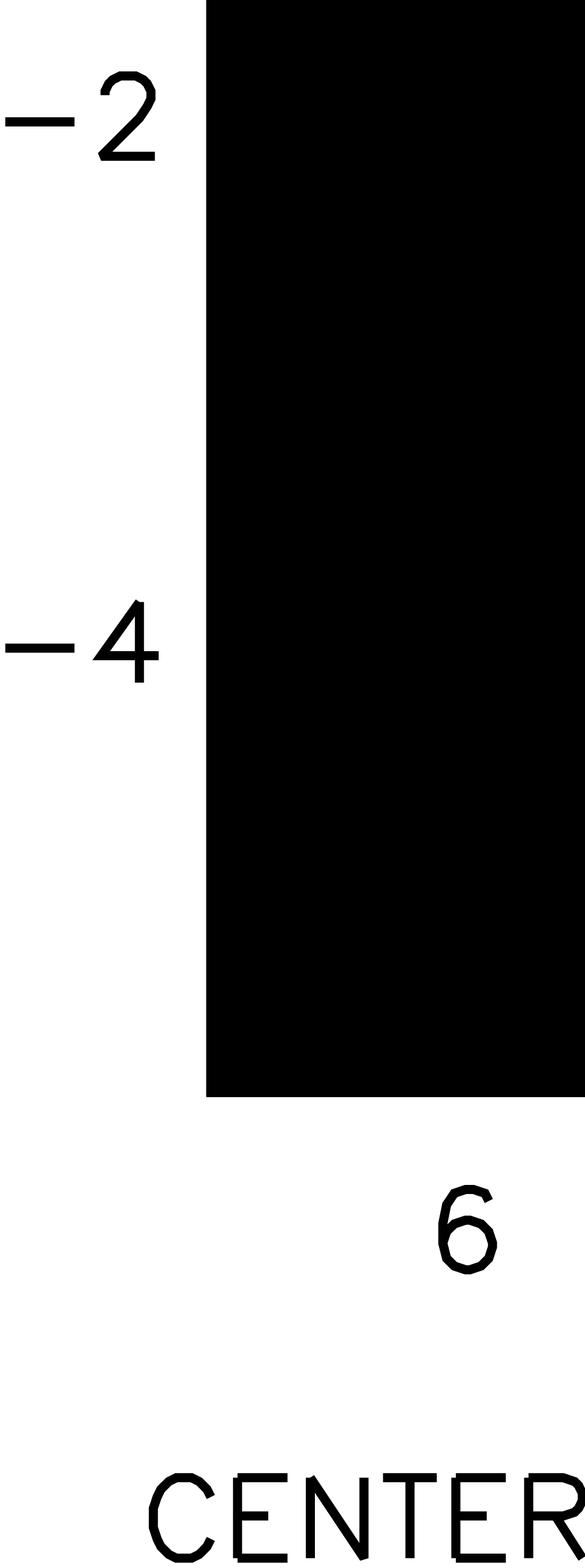}
\includegraphics*[width=0.4\textwidth]{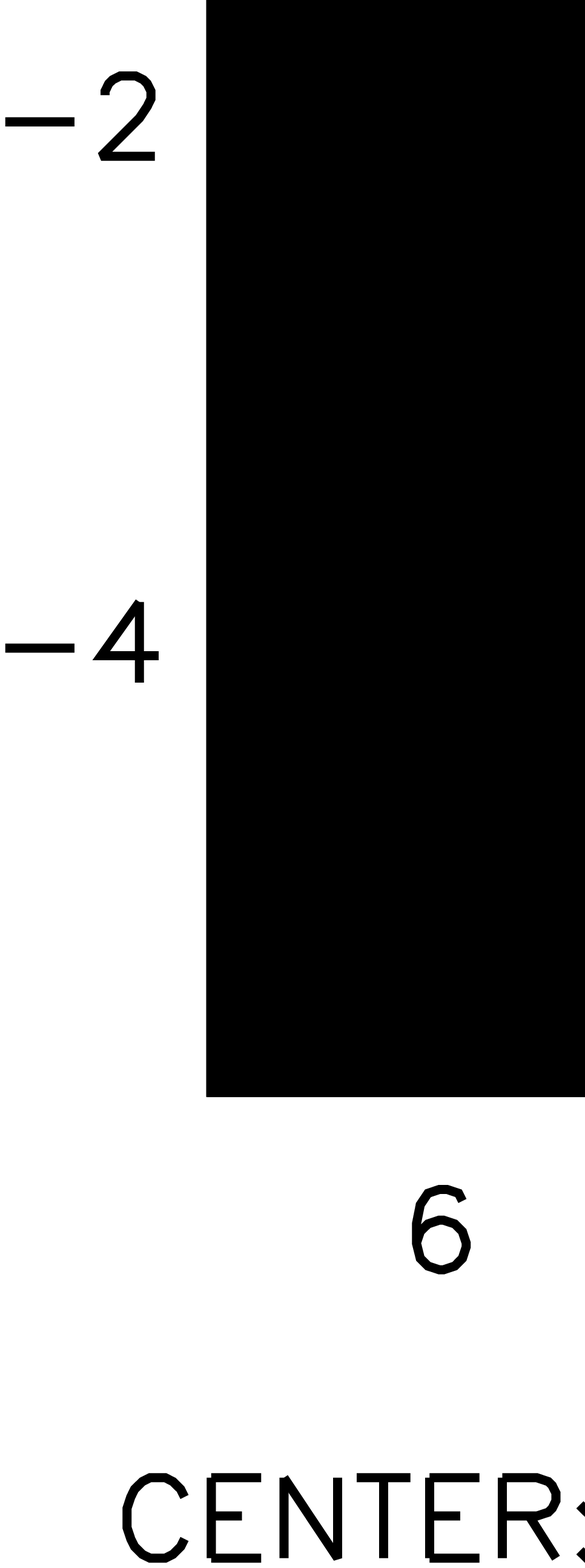}\\
\includegraphics*[width=0.4\textwidth]{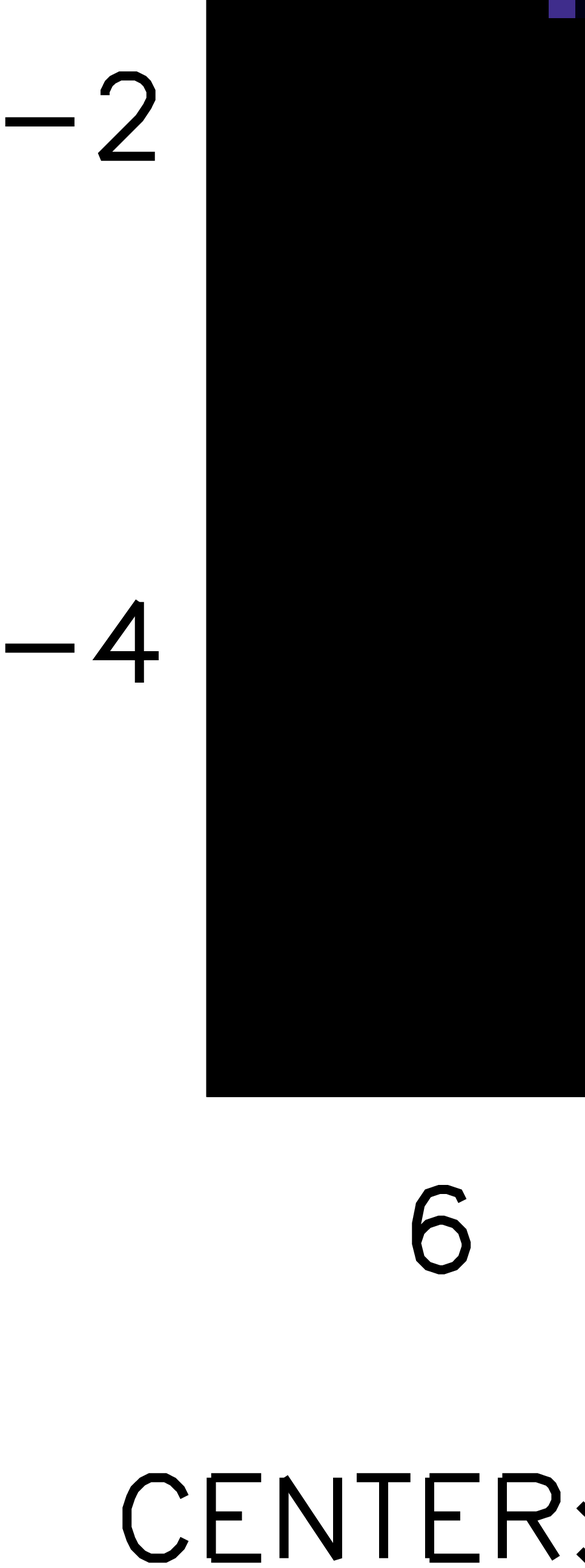}
\includegraphics*[width=0.4\textwidth]{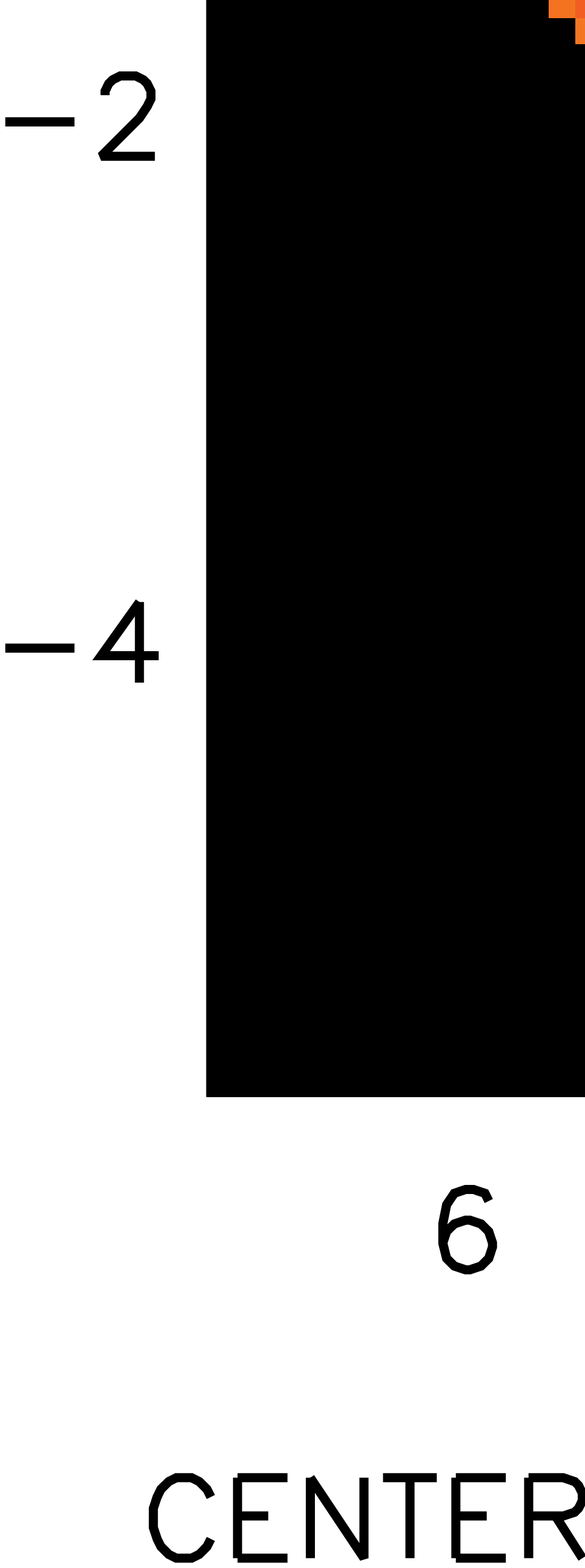}\\
\caption{\textbf{cont.} The Perseus and W3 integrated intensity maps.}
\end{center}
\end{figure*}

\addtocounter{figure}{-1}

\begin{figure*}
\begin{center} 
\includegraphics*[width=0.4\textwidth]{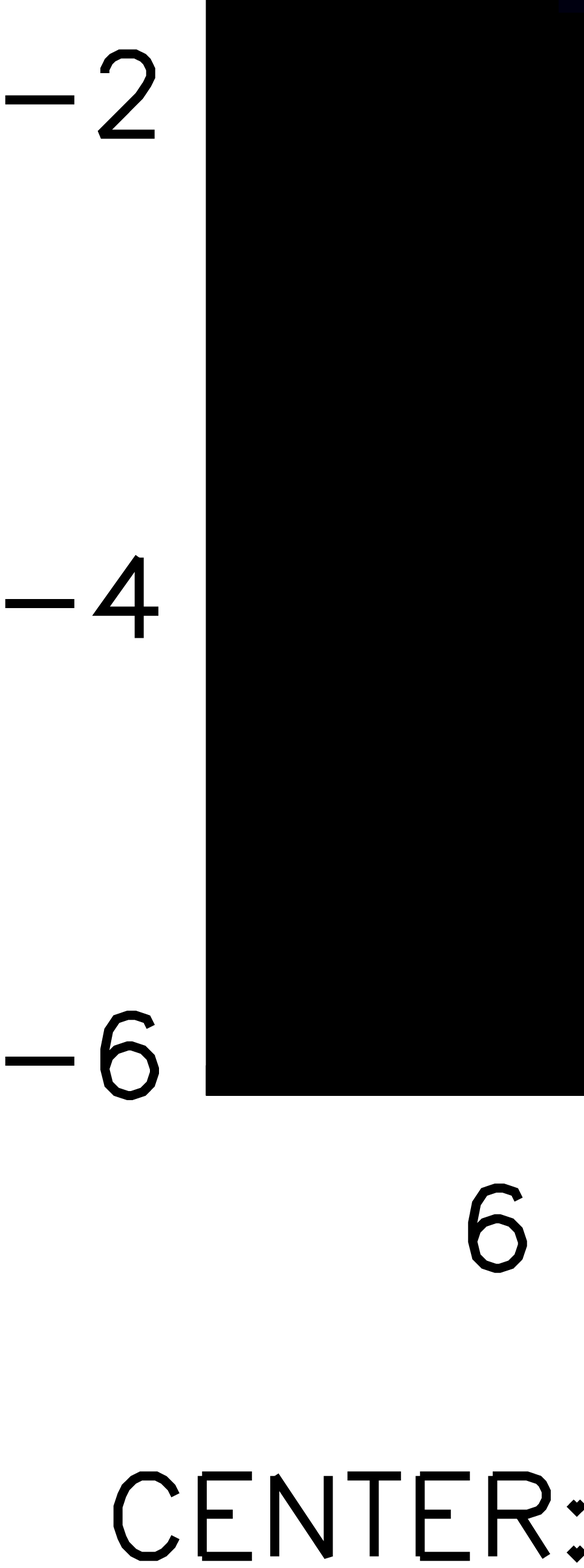}
\includegraphics*[width=0.4\textwidth]{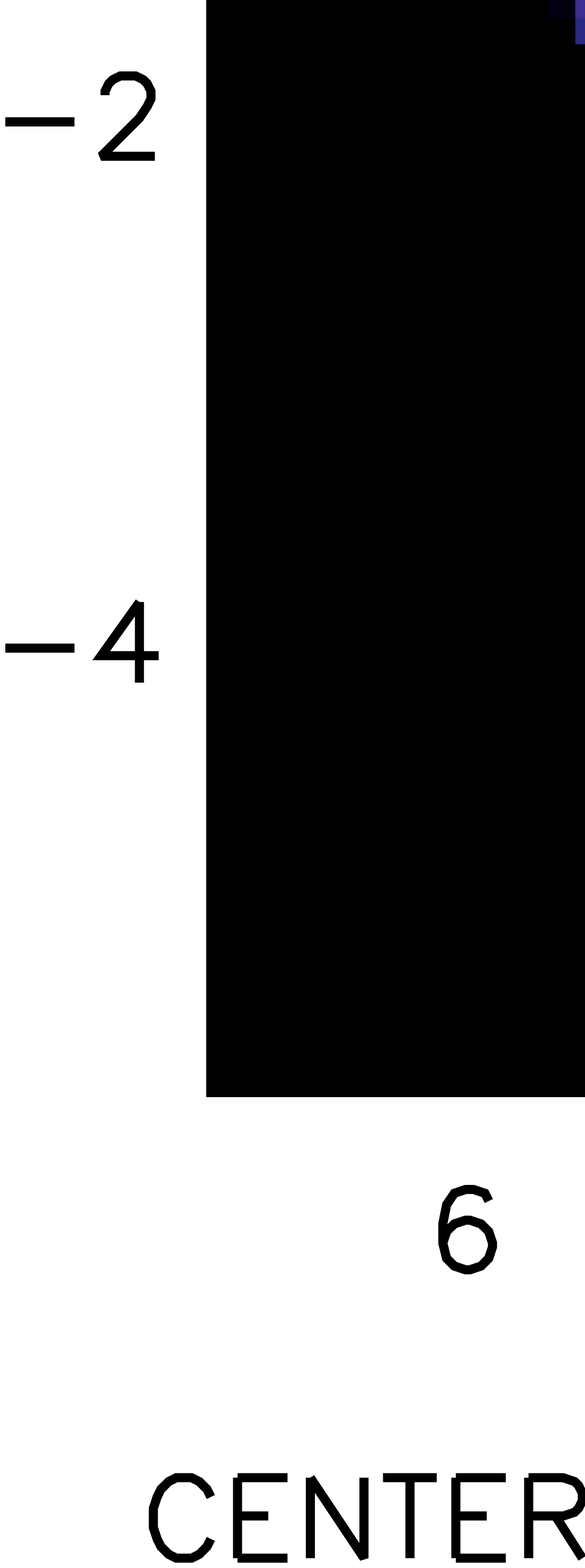}\\
\includegraphics*[width=0.4\textwidth]{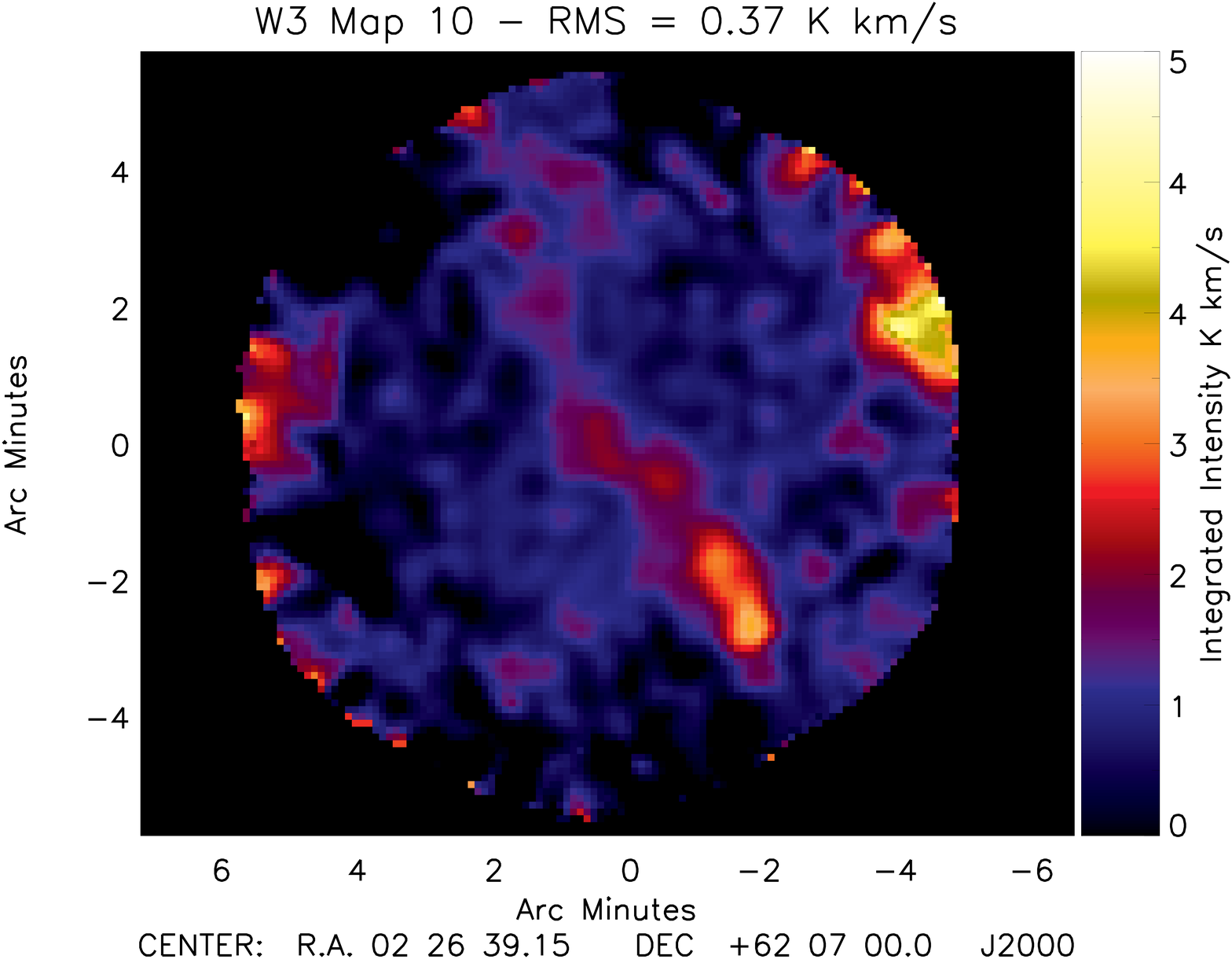}
\includegraphics*[width=0.4\textwidth]{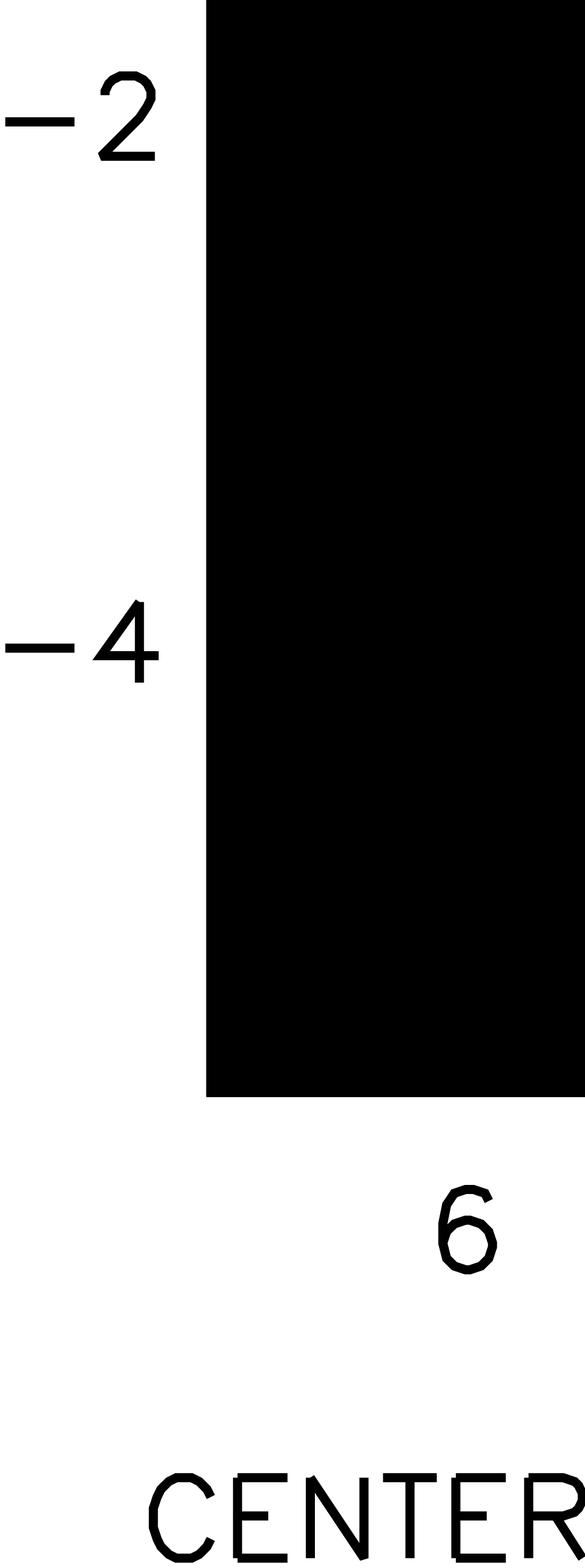}\\
\includegraphics*[width=0.4\textwidth]{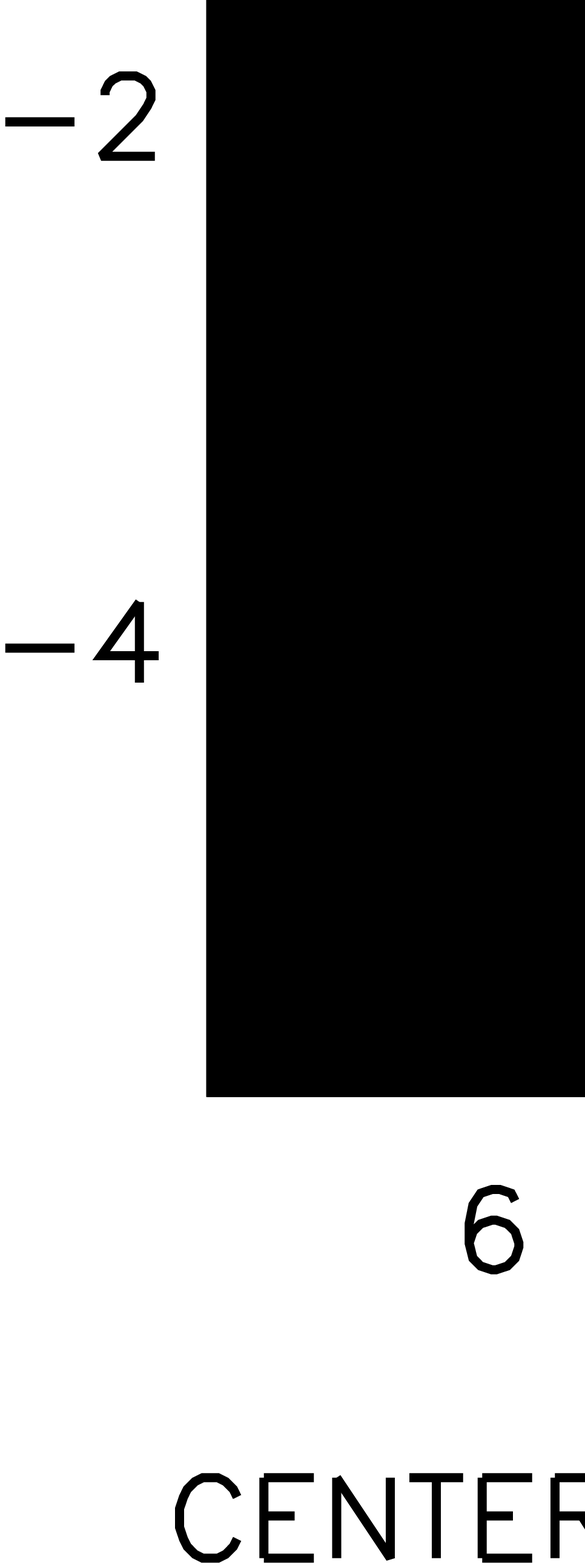}
\includegraphics*[width=0.4\textwidth]{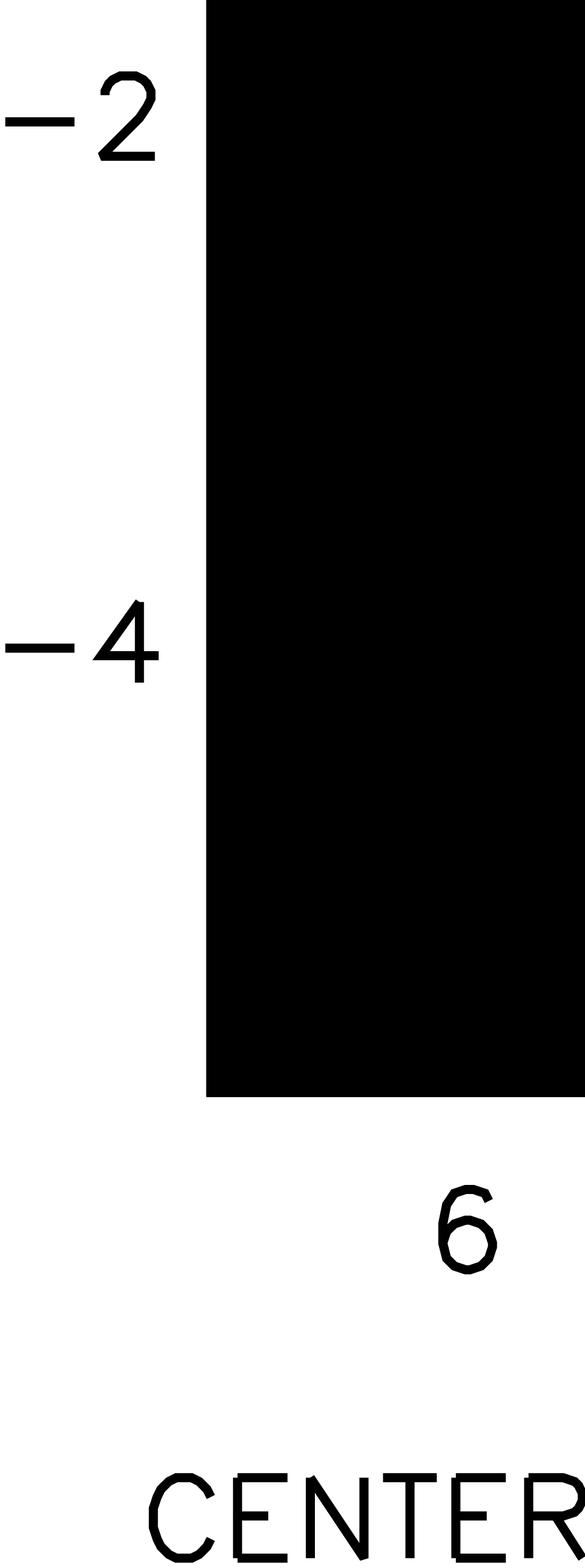}\\
\includegraphics*[width=0.4\textwidth]{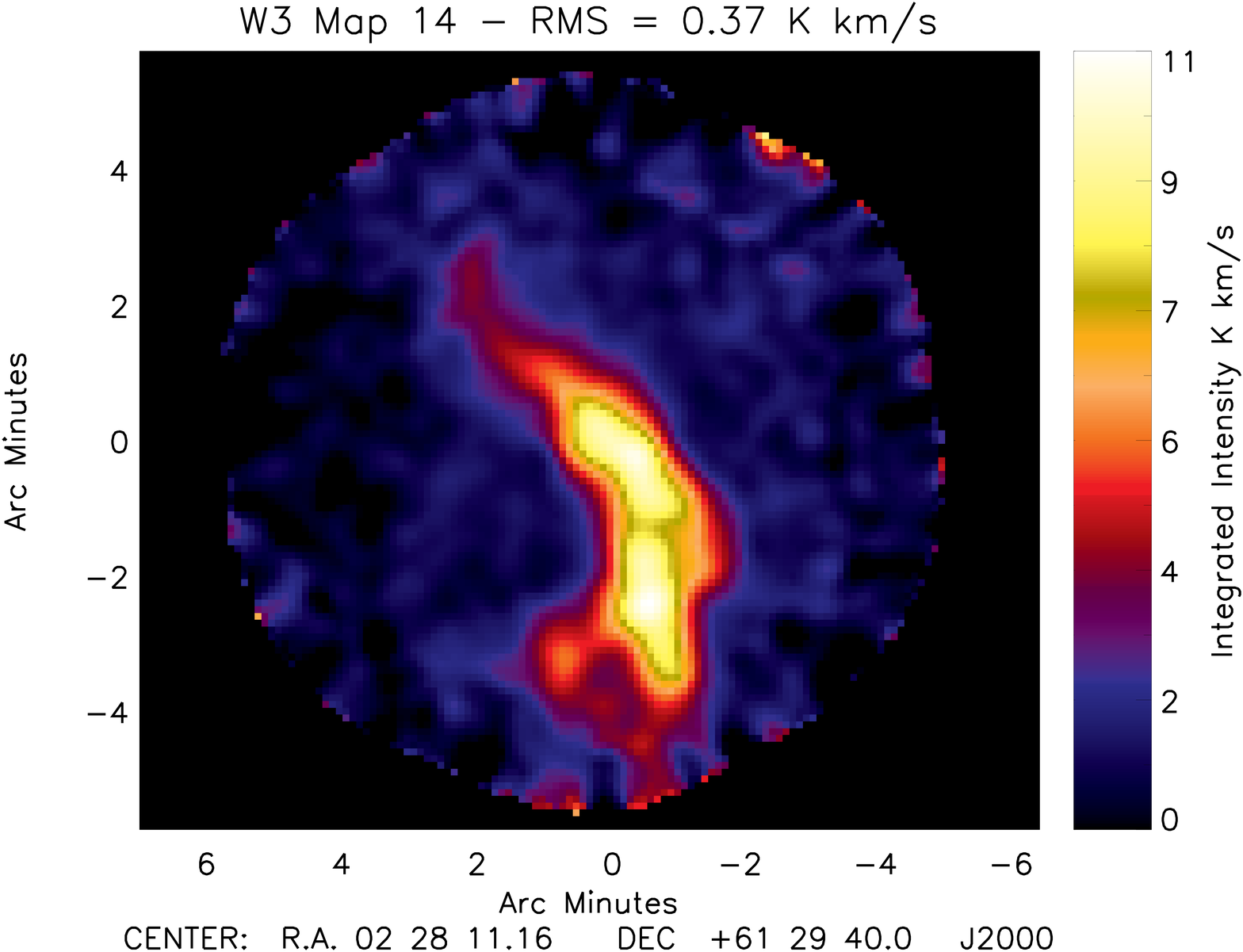}
\includegraphics*[width=0.4\textwidth]{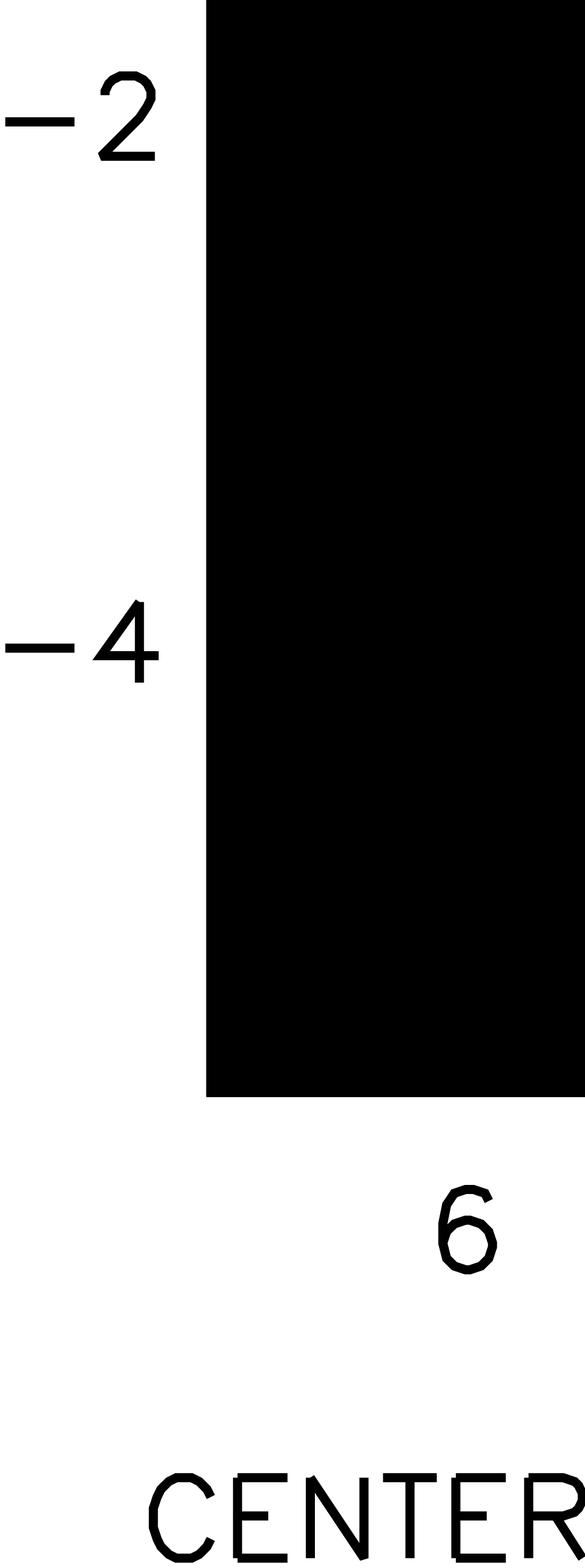}\\
\caption{\textbf{cont.} The Perseus and W3 integrated intensity maps.}
\end{center}
\end{figure*}

\subsection{Source Extraction}
In order to locate individual \nh\ sources, the Starlink-CUPID\footnote{http://docs.jach.hawaii.edu/star/sun255.htx/sun255.html} implementation of the $Clumpfind$ \citep{Williams1994} package was run on integrated emission maps produced from the reduced spectral cubes. The reasons for running the two-dimensional implementation of $Clumpfind$ rather than a full three-dimensional analysis on our spectral cubes are two-fold; firstly, some results show that the three-dimensional implementation of $Clumpfind$ may underestimate the mass of clumps in star formation regions \citep{Smith2008} and is not as robust as the two-dimensional version \citep{Pineda2009}, particularly in cases of limiting resolution (c.f. \citet{Ward2012}, who find that the three-dimensional extraction is the most accurate implementation when recovering simulated clumps). Secondly, we found that no clump velocity separations exceeded the maximum linewidth found in the relevant map from which the clumps were extracted. Thus, the separations of any multiple spectral components along the line-of-sight, are smaller than or approximately equal to typical clump linewidths and so the need for a three-dimensional extraction is somewhat negligible. \citet{Rosolowsky2008} performed a single-pointing survey of suspected cores in Perseus. Of the 193 sources in their catalogue, only six showed unambiguous evidence of multiple components with a mean velocity separation of 0.63 km/s. The mean observed linewidth of sources in our sample of Perseus cores is 0.64 km/s (see Table \ref{tbl:Props_Comparison}) and so we are unlikely to be able to separate any multiple components in this region, though they are likely present.

  $Clumpfind$ was run using a lowest contour level of three times the r.m.s. off-source noise ($\sigma$) measurements within the completely sampled portion of each individual map and successive contours spaced by 1$\sigma$. The minimum pixel count required to define a source was left at the default 7 (6\arcsec) pixels, although preliminary tests showed that results were insensitive to an increase in this parameter to 20 pixels, the size of the beam. Where individual maps overlap, sources in the overlap region were extracted from the mosaicked image rather than the individual Daisy scans. Each extracted source was examined in order to identify and discard any sources unduly affected by the higher noise values at the edges of our maps.

\subsection{Source Properties}
  With the above search parameters, $Clumpfind$ detected 84 discrete ammonia sources in Perseus and 54 in W3. Applying the masks produced by $Clumpfind$ to the parameter maps which resulted from our fitting analysis routines (Section \ref{sec:DR}) allows us to extract the distribution of column density, kinetic temperature, etc. for each detected core. In this paper we present the integrated properties of these extracted cores and leave analysis of any spatial variation of these parameters within the cores for subsequent publications. The central positions of each source, measured integrated intensities, fitted linewidths LSR velocities and any relevant matches from the point source catalogue of \citet{Rosolowsky2008} are presented in Tables \ref{tbl:P_clump_positions} and \ref{tbl:W_clump_positions}. Where a source position listed in the catalogues of either \citet{Hatchell2005} or \citet{Moore2007} is within the lowest contour of the ammonia emission as extracted by $Clumpfind$ in this work, we have also listed that association.

\begin{table*}
\begin{center}
\caption{Central positions and matching sources for each mapped source in Perseus.}
\label{tbl:P_clump_positions}
\begin{minipage}{\linewidth}
\begin{center}
\begin{tabular}{lccccccc}
\hline
\hline
{}		& \multicolumn{2}{c}{Central Position}	& W(\nh(1,1)) & dV     & \Vlsr  & {Matching Submm}		& {Matching Source}\\
{Source}	& {R.A.(J2000)}	& {Dec.(J2000)}		& (K km/s)    & (km/s) & (km/s) & {Source$^{1}$}	& {\citep{Rosolowsky2008}}\\
\hline
P01		& 03:25:29.5	& 30:44:47	& 7.92$\pm$0.52 & 0.78  & 4.11	    & 31		    & \\
P02		& 03:25:36.9	& 30:45:23	& 2.95$\pm$0.52 & 0.73  & 4.32	    & 27, 28		    & R10, R12      \\
P03		& 03:25:39.8	& 30:43:53	& 7.82$\pm$0.52 & 0.75  & 4.73	    & 29, Bolo11	    & R15, R14      \\
P04		& 03:25:48.1	& 30:45:25	& 2.47$\pm$0.52 & 0.61  & 4.45	    &			    & \\
P05		& 03:25:50.3	& 30:42:32	& 6.00$\pm$0.52 & 0.59  & 4.28	    & 32		    & R17, R18, R19\\
P06		& 03:25:55.7	& 30:41:02	& 1.86$\pm$0.52 & 0.81  & 3.99	    &			    & R20\\
P07		& 03:26:00.3	& 30:42:07	& 2.12$\pm$0.52 & 0.95  & 4.21	    &			    & \\
P08		& 03:26:00.6	& 30:40:51	& 1.85$\pm$0.52 & ---   & ---	    &			    & \\
P09		& 03:27:31.0	& 30:15:09	& 4.42$\pm$0.51 & 0.42  & 4.87	    &			    & R29\\
P10		& 03:27:35.0	& 30:10:22	& 2.68$\pm$0.51 & 0.52  & 5.02	    &			    & R30\\
P11		& 03:27:36.5	& 30:13:58	& 3.84$\pm$0.51 & 0.81  & 4.55	    & 39		    & R31   \\
P12		& 03:27:37.8	& 30:08:37	& 3.95$\pm$0.51 & 0.53  & 5.28	    &			    & \\
P13		& 03:27:41.3	& 30:12:17	& 8.31$\pm$0.51 & 0.70  & 4.79	    & 35, 36, 40	    & R32, R33, R34\\
P14		& 03:27:45.7	& 30:08:39	& 2.21$\pm$0.51 & 0.73  & 5.30	    &			    & \\
P15		& 03:27:49.7	& 30:11:40	& 6.07$\pm$0.51 & 0.66  & 4.82	    & 37		    & R35   \\
P16		& 03:27:54.6	& 30:09:46	& 2.45$\pm$0.51 & 0.65  & 4.60	    &			    & \\
P17		& 03:27:56.6	& 30:11:15	& 4.47$\pm$0.51 & 0.51  & 4.89	    &			    & \\
P18		& 03:27:56.8	& 30:10:24	& 3.90$\pm$0.51 & 0.60  & 4.64	    &			    & \\
P19		& 03:28:28.6	& 31:10:07	& 2.85$\pm$0.37 & 0.34  & 6.98	    &			    & \\
P20		& 03:28:30.7	& 31:15:14	& 3.15$\pm$0.37 & 0.40  & 7.40	    &			    & \\
P21		& 03:28:31.8	& 31:11:26	& 3.34$\pm$0.37 & 0.42  & 7.02	    & 74		    & R40   \\
P22		& 03:28:33.7	& 31:13:12	& 3.71$\pm$0.37 & 0.47  & 7.13	    & 49		    & R44   \\
P23		& 03:28:33.8	& 31:04:39	& 3.67$\pm$0.37 & 0.46  & 6.57	    & Bolo26		    & R41   \\
P24		& 03:28:35.0	& 31:06:47	& 1.80$\pm$0.37 & 0.44  & 6.67	    & 69		    & R43   \\
P25		& 03:28:39.5	& 31:05:50	& 5.40$\pm$0.37 & 0.49  & 6.87	    & 71		    & R46   \\
P26		& 03:28:44.3	& 31:06:10	& 4.18$\pm$0.37 & 0.59  & 7.11	    & 75		    & R50   \\
P27		& 03:28:49.9	& 31:12:50	& 2.32$\pm$0.37 & 1.28  & 7.16	    &			    & \\
P28		& 03:28:52.6	& 31:14:51	& 7.21$\pm$0.37 & 1.04  & 7.54	    & 44		    & R58   \\
P29		& 03:28:56.5	& 31:13:57	& 6.07$\pm$0.37 & 0.83  & 6.98	    &			    & \\
P30		& 03:29:01.7	& 31:12:23	& 4.48$\pm$0.37 & 0.72  & 6.95	    & 65		    & R65   \\
P31		& 03:29:02.6	& 31:15:26	& 9.58$\pm$0.37 & 1.22  & 7.53	    & 43, 52		    & R67, R68      \\
P32		& 03:29:08.3	& 31:17:22	& 4.95$\pm$0.37 & 0.72  & 8.16	    & 46, 62, Bolo44	    & R72   \\
P33		& 03:29:09.0	& 31:15:21	& 2.05$\pm$0.37 & 1.02  & 7.63	    & 50, 51		    & R70, R73      \\
P34		& 03:29:13.4	& 31:13:20	& 6.32$\pm$0.37 & 0.98  & 7.35	    & 41, 42, 48, 59, 72    & R75, R78, R80, R83\\
P35		& 03:29:17.1	& 31:28:15	& 4.49$\pm$0.36 & 0.40  & 7.34	    & 61		    & R81   \\
P36		& 03:29:17.7	& 31:25:51	& 3.01$\pm$0.36 & 0.55  & 7.11	    & 57		    & R82   \\
P37		& 03:29:19.5	& 31:23:36	& 1.97$\pm$0.36 & 0.59  & 7.35	    & 63, 67		    & R84   \\
P38		& 03:29:20.9	& 31:29:22	& 2.10$\pm$0.36 & ---   & ---	    &			    & \\
P39		& 03:29:22.6	& 31:26:11	& 2.72$\pm$0.36 & 0.35  & 7.17	    &			    & \\
P40		& 03:29:26.0	& 31:28:14	& 3.53$\pm$0.36 & 0.40  & 7.30	    & 64		    & R88   \\
P41		& 03:32:16.1	& 30:49:38	& 5.52$\pm$0.47 & 0.55  & 6.63	    & 76		    & R103  \\
P42		& 03:32:55.3	& 31:21:02	& 3.10$\pm$0.83 & 0.72  & 6.34	    &			    & \\
P43		& 03:32:57.8	& 31:04:10	& 5.79$\pm$0.86 & 0.40  & 6.42	    &			    & \\
P44		& 03:32:58.6	& 31:20:43	& 2.90$\pm$0.83 & 0.70  & 6.37	    &			    & R112\\
P45		& 03:32:59.9	& 31:03:36	& 6.10$\pm$0.86 & 0.40  & 6.45	    &			    & \\
P46		& 03:33:02.2	& 31:20:13	& 2.84$\pm$0.83 & 0.77  & 6.44	    &			    & \\
P47		& 03:33:04.7	& 31:05:01	& 6.93$\pm$0.86 & 0.50  & 6.41	    & 5 		    & R113, R114\\
P48		& 03:33:05.7	& 31:06:28	& 4.32$\pm$0.86 & 0.48  & 6.32	    &			    & R115\\
P49		& 03:33:10.5	& 31:19:47	& 5.17$\pm$0.83 & 0.50  & 6.62	    & 82		    & R118  \\
P50		& 03:33:14.5	& 31:07:12	& 0.28$\pm$0.86 & 0.64  & 6.24	    & 4, 7		    & R119  \\
P51		& 03:33:17.4	& 31:09:20	& 7.72$\pm$0.86 & 0.72  & 6.21	    & 1 		    & R121  \\
P52		& 03:33:21.0	& 31:07:37	& 3.54$\pm$0.86 & 0.74  & 6.45	    & 2 		    & R123  \\
P53		& 03:33:24.3	& 31:05:55	& 4.58$\pm$0.86 & 0.48  & 6.59	    &			    & R124\\
P54		& 03:33:25.0	& 31:19:51	& 2.95$\pm$0.83 & 0.89  & 6.29	    &			    & R125\\
P55		& 03:33:26.3	& 31:06:59	& 3.95$\pm$0.86 & 0.69  & 6.61	    & 10		    & R126  \\
P56		& 03:33:31.0	& 31:20:22	& 3.58$\pm$0.83 & 0.82  & 6.30	    &			    & R127\\
P57		& 03:41:29.8	& 31:57:42	& 1.83$\pm$0.42 & ---   & ---	    &			    & \\
P58		& 03:41:32.9	& 31:59:09	& 1.61$\pm$0.42 & ---   & ---	    &			    & \\
P59		& 03:41:33.5	& 31:58:05	& 1.48$\pm$0.42 & ----  & ---	    &			    & \\
P60		& 03:41:36.4	& 31:54:41	& 1.48$\pm$0.42 & ---   & ---	    &			    & \\
P61		& 03:41:36.5	& 32:00:07	& 1.54$\pm$0.42 & ---   & ---	    &			    & \\
P62		& 03:41:41.1	& 31:57:58	& 3.64$\pm$0.42 & 0.32  & 9.23	    & Bolo92		    & R145  \\
P63		& 03:41:46.8	& 31:57:24	& 3.62$\pm$0.42 & 0.42  & 9.24	    & Bolo94		    & R147  \\
\hline
\end{tabular}\\
\end{center}
\end{minipage}
$^{1}$Sources taken from \citet{Hatchell2007}, incorporating several Bolocam sources from \citet{Enoch2006} labeled as `Boloxx'.
\end{center}
\end{table*}

\addtocounter{table}{-1}

\begin{table*}
\begin{center}
\caption{(cont.) Central positions and matching sources for each mapped source in Perseus.}
\begin{minipage}{\linewidth}
\begin{center}
\begin{tabular}{lccccccc}
\hline
\hline
{}		& \multicolumn{2}{c}{Central Position}	& W(\nh(1,1)) & dV     & \Vlsr  & {Matching Submm}		& {Matching Source}\\
{Source}	& {R.A.(J2000)}	& {Dec.(J2000)}		& (K km/s)    & (km/s) & (km/s) & {Source$^{1}$}		& {\citep{Rosolowsky2008}}\\
\hline
P64		& 03:41:48.0	& 31:54:20	& 1.76$\pm$0.42 & ---   & ---	   &			   & \\
P65		& 03:41:53.2	& 32:00:25	& 1.89$\pm$0.42 & ---    & ---	   &			   & \\
P66		& 03:41:53.4	& 31:54:33	& 1.68$\pm$0.42 & ---    & ---	   &			   & \\
P67		& 03:41:55.3	& 31:59:24	& 1.94$\pm$0.42 & 0.42   & 9.20	   &			   & \\
P68		& 03:41:58.8	& 31:58:54	& 2.36$\pm$0.42 & 0.54   & 9.13	   &			   & R148\\
P69		& 03:42:00.4	& 31:57:47	& 1.73$\pm$0.42 & 0.61   & 9.07	   &			   & \\
P70		& 03:43:39.4	& 32:03:10	& 4.09$\pm$0.48 & 0.80   & 8.50	   & 23 		   & R156  \\
P71		& 03:43:48.0	& 32:03:18	& 5.50$\pm$0.48 & 0.62   & 8.49	   & 15, 24, 26 	   & R157, 158, 159, 160\\
P72		& 03:43:55.9	& 32:00:45	& 4.24$\pm$0.48 & 0.78   & 8.69	   & 12 		   & R161  \\
P73		& 03:43:57.3	& 32:04:03	& 3.76$\pm$0.48 & 0.75   & 8.26	   & 17 		   & R163  \\
P74		& 03:44:01.0	& 32:02:43	& 5.02$\pm$0.48 & 0.62   & 8.47	   & 13, 16, 18, 21	   & R162, 164, 165\\
P75		& 03:44:05.3	& 32:01:34	& 5.73$\pm$0.48 & 0.62   & 8.40	   & 20 		   & R169  \\
P76		& 03:44:31.0	& 32:00:35	& 1.68$\pm$0.45 & 0.66   & 9.00	   &			   & \\
P77		& 03:44:36.9	& 32:00:59	& 2.00$\pm$0.45 & 0.53   & 9.11	   &			   & \\
P78		& 03:44:37.4	& 31:58:42	& 3.00$\pm$0.45 & 0.68   & 9.71	   & 19 		   & R176  \\
P79		& 03:44:39.8	& 31:59:36	& 1.73$\pm$0.45 & 0.90   & 9.55	   &			   & \\
P80		& 03:44:44.1	& 32:01:17	& 2.03$\pm$0.45 & 1.08   & 9.57	   & 14 		   & R178  \\
P81		& 03:44:49.4	& 32:00:40	& 2.45$\pm$0.45 & 0.50   & 8.79	   & 25 		   & R180  \\
P82		& 03:44:56.1	& 32:00:25	& 2.35$\pm$0.45 & 0.53   & 8.88	   &			   & R181\\
P83		& 03:45:01.1	& 32:00:42	& 2.95$\pm$0.45 & 0.40   & 8.93	   &			   & \\
P84		& 03:45:06.3	& 32:00:35	& 2.70$\pm$0.45 & 0.54   & 8.93	   &			   & \\
\hline
\end{tabular}\\
\end{center}
\end{minipage}
$^{1}$Sources taken from \citet{Hatchell2007}, incorporating several Bolocam sources from \citet{Enoch2006} labeled as `Boloxx'.
\end{center}
\end{table*}

\begin{table*}
\begin{center}
\caption{Central positions and matching sources for each mapped source in W3.}
\label{tbl:W_clump_positions}
\begin{minipage}{\linewidth}
\begin{center}
\begin{tabular}{lcccccc}
\hline
\hline
{}		& \multicolumn{2}{c}{Central Position}	& W(\nh(1,1)) & dV     & \Vlsr  & {Matching Submm}	\\
{Source}	& {R.A.(J2000)}	& {Dec.(J2000)}		& (K km/s)    & (km/s) & (km/s) & {Source \citep{Moore2007}}\\
\hline
W01		& 02:20:34.9	& 61:26:43	& 1.26$\pm$0.27		& ---	    &  ---	    & 27 \\
W02		& 02:20:39.9	& 61:09:59	& 0.82$\pm$0.18 	& 0.67      & -48.93	    & 29 \\
W03		& 02:20:44.2	& 61:27:06	& 1.06$\pm$0.27 	& 1.03      & -51.92	    & 30    \\
W04		& 02:20:44.6	& 61:25:58	& 1.02$\pm$0.27 	& ---	    &  ---	    & \\
W05		& 02:20:48.6	& 61:12:24	& 0.72$\pm$0.18 	& ---	    &  ---	    & \\
W06		& 02:20:53.1	& 61:26:58	& 1.56$\pm$0.27 	& 1.75      & -51.05	    & 31    \\
W07		& 02:21:00.9	& 61:26:58	& 1.59$\pm$0.27 	& 1.43      & -51.48	    & 35    \\
W08		& 02:21:04.4	& 61:27:41	& 1.38$\pm$0.27 	& 0.97      & -51.17	    & 38    \\
W09		& 02:21:05.8	& 61:05:58	& 1.12$\pm$0.18 	& 0.78      & -49.86	    & 37    \\
W10		& 02:21:35.8	& 61:05:33	& 1.30$\pm$0.18 	& 0.96      & -49.97	    & 39, 42, 43, 44, 45    \\
W11		& 02:21:53.9	& 61:06:23	& 0.96$\pm$0.18 	& 1.20      & -50.01	    & 47, 49, 50    \\
W12		& 02:22:03.4	& 61:07:18	& 0.66$\pm$0.18 	& ---	    &  ---	    & 51    \\
W13		& 02:22:24.7	& 61:06:04	& 0.70$\pm$0.18 	& 0.82      & -49.99	    & 54    \\
W14		& 02:23:20.6	& 61:12:41	& 1.13$\pm$0.19 	& 0.58      & -50.11	    & 60    \\
W15		& 02:23:27.2	& 61:12:11	& 0.96$\pm$0.19 	& 0.67      & -50.11	    & 61    \\
W16		& 02:24:58.6	& 62:05:07	& 1.73$\pm$0.37 	& 0.97      & -39.68	    & \\
W17		& 02:25:07.3	& 62:05:29	& 1.25$\pm$0.37 	& 1.05      & -39.92	    & 78    \\
W18		& 02:25:25.7	& 61:15:05	& 1.63$\pm$0.21 	& 1.04      & -48.86	    & 85, 86, 91    \\
W19		& 02:25:27.6	& 61:16:25	& 1.33$\pm$0.21 	& ---	    &  ---	    & 87, 92, 98    \\
W20		& 02:25:29.2	& 62:07:23	& 1.30$\pm$0.37 	& ---	    &  ---	    & \\
W21		& 02:25:29.7	& 61:11:01	& 1.10$\pm$0.21 	& 1.10      & -48.73	    & 90    \\
W22		& 02:25:31.1	& 61:13:03	& 1.71$\pm$0.21 	& 1.18      & -48.30	    & 100, 95, 97   \\
W23		& 02:25:32.2	& 62:06:03	& 3.12$\pm$0.37 	& 3.50      & -42.84	    & 109, 99\\
W24		& 02:25:40.4	& 61:10:29	& 1.50$\pm$0.21 	& 1.22      & -47.92	    & \\
W25		& 02:25:40.7	& 61:13:55	& 2.02$\pm$0.21 	& 0.92      & -47.83	    & 103, 105, 107, 111, 112\\
W26		& 02:25:51.4	& 62:06:18	& 1.36$\pm$0.37 	& 2.08      & -38.77	    & \\
W27		& 02:25:52.9	& 62:04:17	& 2.93$\pm$0.37 	& 1.75      & -38.78	    & 119   \\
W28		& 02:26:02.2	& 62:08:43	& 1.97$\pm$0.37 	& 1.41      & -38.42	    & 127   \\
W29		& 02:26:20.8	& 62:04:21	& 1.99$\pm$0.37 	& 2.06      & -39.84	    & 142   \\
W30		& 02:26:26.3	& 62:05:18	& 1.77$\pm$0.37 	& 1.64      & -39.38	    & 148   \\
W31		& 02:26:30.1	& 61:29:38	& 1.20$\pm$0.34 	& 0.85      & -47.08	    & 150   \\
W32		& 02:26:33.3	& 61:26:48	& 1.42$\pm$0.34 	& ---	    &  ---	    & \\
W33		& 02:26:35.6	& 62:06:42	& 1.51$\pm$0.37 	& 1.42      & -39.15	    & 152, 161, 166\\
W34		& 02:26:37.3	& 61:25:46	& 1.55$\pm$0.34 	& ---	    &  ---	    & \\
W35		& 02:26:41.6	& 61:32:49	& 1.31$\pm$0.34 	& 0.94      & -47.88	    & 175 \\
W36		& 02:26:42.1	& 61:25:23	& 1.77$\pm$0.34 	& ---	    &  ---	    & \\
W37		& 02:26:43.1	& 61:33:34	& 1.23$\pm$0.34 	& ---	    &  ---	    & 171   \\
W38		& 02:26:43.4	& 61:31:49	& 1.17$\pm$0.34 	& 1.11      & -47.79	    & 175 \\
W39		& 02:26:44.2	& 61:29:48	& 1.80$\pm$0.34 	& 1.25      & -47.69	    & 172   \\
W40		& 02:26:45.8	& 62:08:41	& 1.26$\pm$0.37 	& 1.41      & -38.68	    & 176, 180       \\
W41		& 02:26:48.2	& 61:25:39	& 1.62$\pm$0.34 	& ---	    &  ---	    & \\
W42		& 02:26:49.8	& 62:10:04	& 1.40$\pm$0.37 	& ---	    &  ---	    & \\
W43		& 02:26:59.3	& 61:54:45	& 1.93$\pm$0.47 	& 0.80      & -45.45	    & 196   \\
W44		& 02:26:59.6	& 61:29:51	& 2.03$\pm$0.34 	& 1.78      & -47.69	    & 195, 214  	     \\
W45		& 02:26:60.0	& 61:53:50	& 1.68$\pm$0.47 	& 1.99      & -45.17	    & 198   \\
W46		& 02:27:02.6	& 61:52:03	& 4.25$\pm$0.47 	& 2.78      & -47.20	    & 213	     \\
W47		& 02:27:11.2	& 61:53:05	& 3.24$\pm$0.47 	& 2.47      & -48.48	    & \\
W48		& 02:28:01.8	& 61:24:34	& 2.29$\pm$0.37 	& 1.84      & -48.05	    & 262, 270, 275, 280, 281, 286\\
W49		& 02:28:01.9	& 61:27:22	& 5.22$\pm$0.37 	& 1.61      & -48.63	    & 266, 268, 284, 288\\
W50		& 02:28:05.3	& 61:29:58	& 4.49$\pm$0.37 	& 1.86      & -47.85	    & 264, 285, 287, 290, 292\\
W51		& 02:28:09.8	& 61:25:41	& 2.84$\pm$0.37 	& 2.11      & -47.49	    & 289, 293\\
W52		& 02:28:15.1	& 61:26:38	& 2.83$\pm$0.37 	& 2.46      & -48.02	    & 291\\
W53		& 02:28:23.1	& 61:31:52	& 2.19$\pm$0.37 	& 1.94      & -47.04	    & 297, 299, 300\\
W54		& 02:28:58.7	& 61:33:29	& 1.11$\pm$0.19 	& 0.81      & -51.46	    & 315\\
\hline
\end{tabular}\\
\end{center}
\end{minipage}
\end{center}
\end{table*}

\begin{table*}
\begin{center}
\caption{Derived physical properties for each mapped source in Perseus.}
\label{tbl:P_Clump_Parameters}
\begin{minipage}{\linewidth}
\begin{center}
\begin{tabular}{lcccc|lcccc}
\hline
\hline
{}	       & \Tk	& Optical	& Filling	&	Column Density		& {}	        & \Tk	& Optical	& Filling	&	Column Density	\\
{Source}       & (K)	& Depth		& Factor	&	(/10$^{13}$ cm$^{-2}$)	& {Source}	& (K)	& Depth		& Factor	&	(/10$^{13}$ cm$^{-2}$)\\
\hline
P01    &       11.30    &	4.13	&	0.70	&	42.87	& P43	&	10.40	 &	 4.54	 &	 0.54	 &	 27.49\\
P02    &       11.37    &	4.01	&	0.81	&	46.44	& P44	&	14.09	 &	 0.82	 &	 0.51	 &	 11.06\\
P03    &       11.40    &	2.85	&	0.60	&	36.05	& P45	&	10.01	 &	 4.36	 &	 0.61	 &	 27.05\\
P04    &       13.47    &	1.64	&	0.22	&	22.02	& P46	&	12.55	 &	 1.48	 &	 0.28	 &	 20.63\\
P05    &       10.19    &	4.14	&	0.62	&	34.65	& P47	&	10.30	 &	 4.57	 &	 0.60	 &	 32.43\\
P06    &       15.69    &	1.21	&	0.21	&	18.16	& P48	&	10.82	 &	 2.66	 &	 0.48	 &	 19.73\\
P07    &       18.56    &	1.75	&	0.07	&	54.21	& P49	&	10.14	 &	 5.15	 &	 0.42	 &	 33.99\\
P08    &       ---      &	---	&	---	&	---	& P50	&	11.63	 &	 4.50	 &	 0.62	 &	 44.07\\
P09    &       11.39    &	3.04	&	0.47	&	20.81	& P51	&	12.22	 &	 2.70	 &	 0.53	 &	 34.22\\
P10    &       12.88    &	1.16	&	0.49	&	10.69	& P52	&	11.18	 &	 4.44	 &	 0.72	 &	 52.61\\
P11    &       13.36    &	1.60	&	0.38	&	25.35	& P53	&	10.52	 &	 3.99	 &	 0.45	 &	 26.56\\
P12    &       12.92    &	1.81	&	0.53	&	15.87	& P54	&	12.38	 &	 1.76	 &	 0.25	 &	 28.44\\
P13    &       12.69    &	3.10	&	0.60	&	35.74	& P55	&	12.41	 &	 1.54	 &	 0.48	 &	 19.14\\
P14    &       19.99    &	0.93	&	0.26	&	21.65	& P56	&	12.88	 &	 1.96	 &	 0.26	 &	 33.21\\
P15    &       12.75    &	2.19	&	0.61	&	22.89	& P57	&	---	 &	 ---	 &	 ---	 &	 ---  \\
P16    &       17.26    &	0.87	&	0.33	&	18.94	& P58	&	---	 &	 ---	 &	 ---	 &	 ---  \\
P17    &       11.84    &	2.29	&	0.64	&	18.33	& P59	&	---	 &	 ---	 &	 ---	 &	 ---  \\
P18    &       13.74    &	1.54	&	0.56	&	15.92	& P60	&	---	 &	 ---	 &	 ---	 &	 ---  \\
P19    &       10.35    &	3.66	&	0.35	&	17.79	& P61	&	---	 &	 ---	 &	 ---	 &	 ---  \\
P20    &       10.84    &	5.49	&	0.37	&	34.49	& P62	&	9.45	 &	 4.78	 &	 0.42	 &	 22.59\\
P21    &       11.58    &	3.50	&	0.32	&	23.10	& P63	&	10.49	 &	 3.35	 &	 0.39	 &	 22.87\\
P22    &       12.00    &	3.37	&	0.34	&	27.37	& P64	&	---	 &	 ---	 &	 ---	 &	 ---  \\
P23    &       11.41    &	3.34	&	0.42	&	25.25	& P65	&	---	 &	 ---	 &	 ---	 &	 ---  \\
P24    &       11.61    &	4.40	&	0.15	&	30.65	& P66	&	---	 &	 ---	 &	 ---	 &	 ---  \\
P25    &       10.45    &	3.62	&	0.53	&	27.23	& P67	&	14.33	 &	 1.97	 &	 0.11	 &	 15.37\\
P26    &       11.82    &	3.76	&	0.41	&	32.52	& P68	&	14.52	 &	 1.00	 &	 0.25	 &	 8.50 \\
P27    &       17.82    &	1.10	&	0.29	&	43.15	& P69	&	15.39	 &	 1.31	 &	 0.24	 &	 9.59 \\
P28    &       14.69    &	1.57	&	0.52	&	38.32	& P70	&	16.52	 &	 1.32	 &	 0.36	 &	 23.58\\
P29    &       14.17    &	1.75	&	0.48	&	28.66	& P71	&	13.49	 &	 2.28	 &	 0.48	 &	 22.87\\
P30    &       13.89    &	1.85	&	0.44	&	23.41	& P72	&	16.38	 &	 1.99	 &	 0.32	 &	 34.25\\
P31    &       16.45    &	1.93	&	0.45	&	55.50	& P73	&	15.18	 &	 1.71	 &	 0.29	 &	 25.46\\
P32    &       14.56    &	2.08	&	0.46	&	30.47	& P74	&	13.25	 &	 2.27	 &	 0.44	 &	 23.89\\
P33    &       13.74    &	1.95	&	0.75	&	42.11	& P75	&	12.16	 &	 2.66	 &	 0.47	 &	 31.10\\
P34    &       13.19    &	1.49	&	0.59	&	30.53	& P76	&	16.88	 &	 0.78	 &	 0.21	 &	 10.81\\
P35    &       12.03    &	3.21	&	0.46	&	22.51	& P77	&	13.59	 &	 1.02	 &	 0.30	 &	 10.76\\
P36    &       12.63    &	2.10	&	0.36	&	21.50	& P78	&	13.23	 &	 1.55	 &	 0.35	 &	 19.04\\
P37    &       14.86    &	1.97	&	0.14	&	24.40	& P79	&	18.86	 &	 0.53	 &	 0.20	 &	 15.41\\
P38    &       ---      &	---	&	---	&	---	& P80	&	15.37	 &	 0.61	 &	 0.24	 &	 17.85\\
P39    &       12.63    &	1.98	&	0.34	&	12.98	& P81	&	12.62	 &	 1.90	 &	 0.28	 &	 17.64\\
P40    &       10.99    &	3.11	&	0.45	&	21.72	& P82	&	12.13	 &	 1.21	 &	 0.43	 &	 9.32 \\
P41    &       12.07    &	2.86	&	0.51	&	31.34	& P83	&	11.76	 &	 2.17	 &	 0.42	 &	 14.60\\
P42    &       17.44    &	0.48	&	0.45	&	11.83	& P84	&	13.21	 &	 2.66	 &	 0.22	 &	 24.50\\
\hline
\end{tabular}\\
\end{center}
\end{minipage}
\end{center}
\end{table*}

\begin{table*}
\begin{center}
\caption{Derived physical properties for each mapped source in W3.}
\label{tbl:W_Clump_Parameters}
\begin{minipage}{\linewidth}
\begin{center}
\begin{tabular}{lcccc|lcccc}
\hline
\hline
{}	       & \Tk	& Optical	& Filling	&	Column Density		& {}	        & \Tk	& Optical	& Filling	&	Column Density	\\
{Source}       & (K)	& Depth		& Factor	&	(/10$^{13}$ cm$^{-2}$)	& {Source}	& (K)	& Depth		& Factor	&	(/10$^{13}$ cm$^{-2}$)\\
\hline
W01    &       ---     &  ---	  &	  ---	  &	  ---	 & W28    &	  21.88   &  0.92    &       0.11    &       46.84\\
W02    &       15.08   &  2.70    &	  0.04    &	  41.00  & W29    &	  22.83   &  0.85    &       0.10    &       67.41\\
W03    &       18.20   &  1.65    &	  0.03    &	  71.82  & W30    &	  20.74   &  0.79    &       0.10    &       45.06\\
W04    &       ---     &  ---	  &	  ---	  &	  ---	 & W31    &	  15.10   &  2.59    &       0.05    &       47.99\\
W05    &       ---     &  ---	  &	  ---	  &	  ---	 & W32    &	  ---	  &  ---     &       ---     &       ---  \\
W06    &       17.71   &  1.56    &	  0.06    &	  84.70  & W33    &	  23.11   &  0.49    &       0.11    &       27.09\\
W07    &       15.83   &  1.02    &	  0.12    &	  33.69  & W34    &	  ---	  &  ---     &       ---     &       ---  \\
W08    &       15.41   &  1.29    &	  0.10    &	  32.16  & W35    &	  16.81   &  0.94    &       0.19    &       21.75\\
W09    &       13.49   &  1.42    &	  0.11    &	  23.40  & W36    &	  ---	  &  ---     &       ---     &       ---  \\
W10    &       14.64   &  1.29    &	  0.14    &	  25.19  & W37    &	  ---	  &  ---     &       ---     &       ---  \\
W11    &       16.12   &  1.11    &	  0.08    &	  31.36  & W38    &	  18.19   &  0.43    &       0.16    &       12.29\\
W12    &       ---     &  ---	  &	  ---	  &	  ---	 & W39    &	  15.11   &  1.05    &       0.14    &       28.84\\
W13    &       17.40   &  0.60    &	  0.13    &	  14.07  & W40    &	  20.90   &  0.40    &       0.10    &       20.04\\
W14    &       12.44   &  3.48    &	  0.08    &	  39.87  & W41    &	  ---	  &  ---     &       ---     &       ---  \\
W15    &       15.10   &  3.23    &	  0.05    &	  37.49  & W42    &	  ---	  &  ---     &       ---     &       ---  \\
W16    &       16.91   &  1.71    &	  0.11    &	  43.33  & W43    &	  17.22   &  2.26    &       0.10    &       43.35\\
W17    &       17.24   &  1.18    &	  0.09    &	  31.82  & W44    &	  15.76   &  1.10    &       0.14    &       40.11\\
W18    &       13.75   &  1.32    &	  0.18    &	  27.68  & W45    &	  18.90   &  0.30    &       0.21    &       18.57\\
W19    &       ---     &  ---	  &	  ---	  &	  ---	 & W46    &	  25.84   &  1.39    &       0.09    &      189.02\\
W20    &       ---     &  ---	  &	  ---	  &	  ---	 & W47    &	  22.12   &  1.37    &       0.09    &      134.67\\
W21    &       17.60   &  0.67    &	  0.13    &	  19.53  & W48    &	  17.47   &  0.94    &       0.14    &       46.81\\
W22    &       13.88   &  1.61    &	  0.14    &	  35.70  & W49    &	  16.57   &  1.27    &       0.23    &       50.05\\
W23    &       35.04   &  0.73    &	  0.08    &	 199.64  & W50    &	  17.97   &  1.13    &       0.19    &       56.89\\
W24    &       15.66   &  0.70    &	  0.20    &	  19.31  & W51    &	  17.43   &  1.05    &       0.14    &       64.44\\
W25    &       12.54   &  1.98    &	  0.17    &	  35.18  & W52    &	  19.05   &  1.13    &       0.10    &       84.35\\
W26    &       34.73   &  0.32    &	  0.08    &	  48.97  & W53    &	  19.67   &  0.78    &       0.11    &       54.48\\
W27    &       21.67   &  1.10    &	  0.11    &	  71.82  & W54    &	  23.76   &  1.18    &       0.10    &       35.21\\
\hline
\end{tabular}\\
\end{center}
\end{minipage}
\end{center}
\end{table*}

\begin{figure*}
\begin{center} 
\includegraphics*[width=0.45\textwidth]{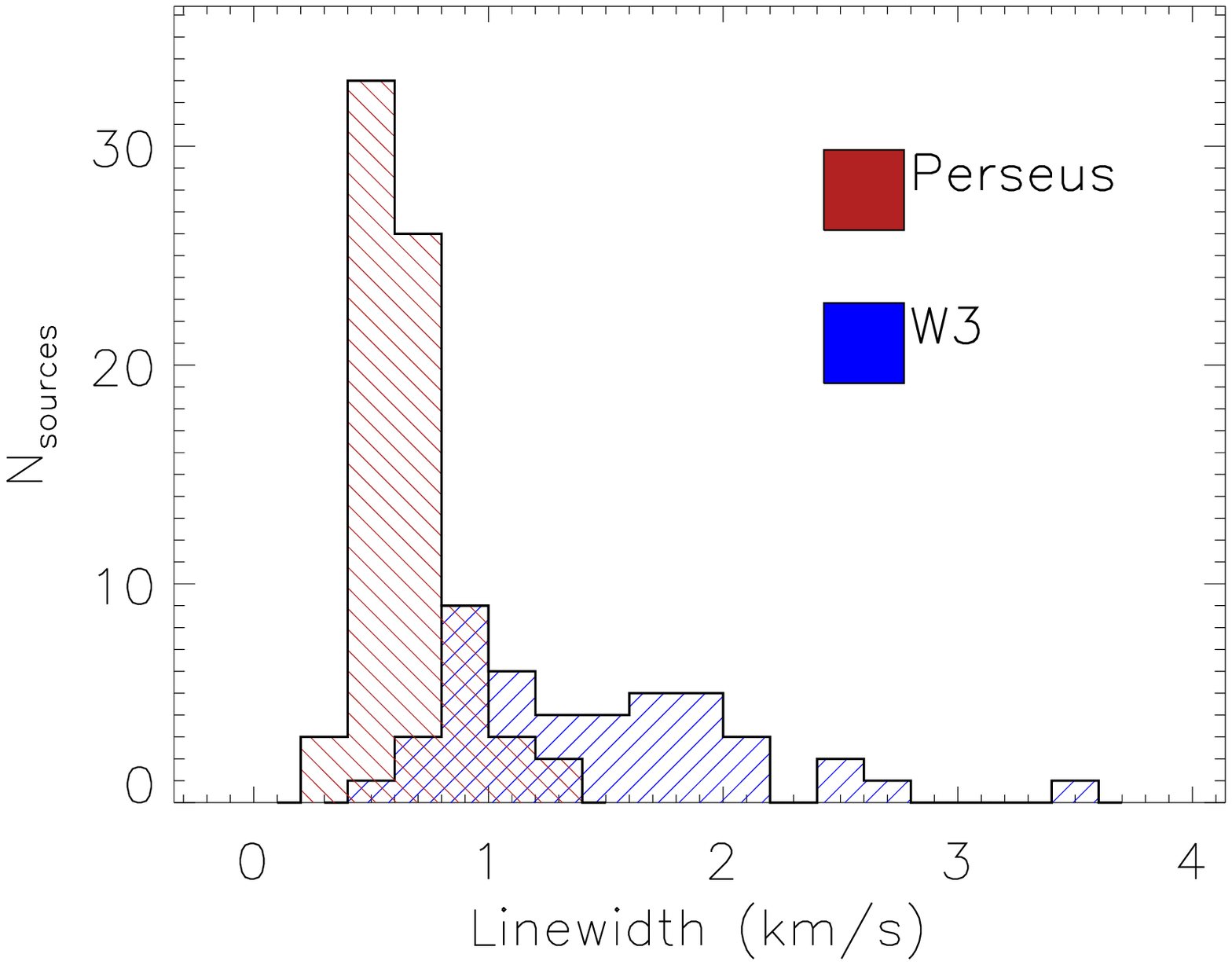}
\includegraphics*[width=0.45\textwidth]{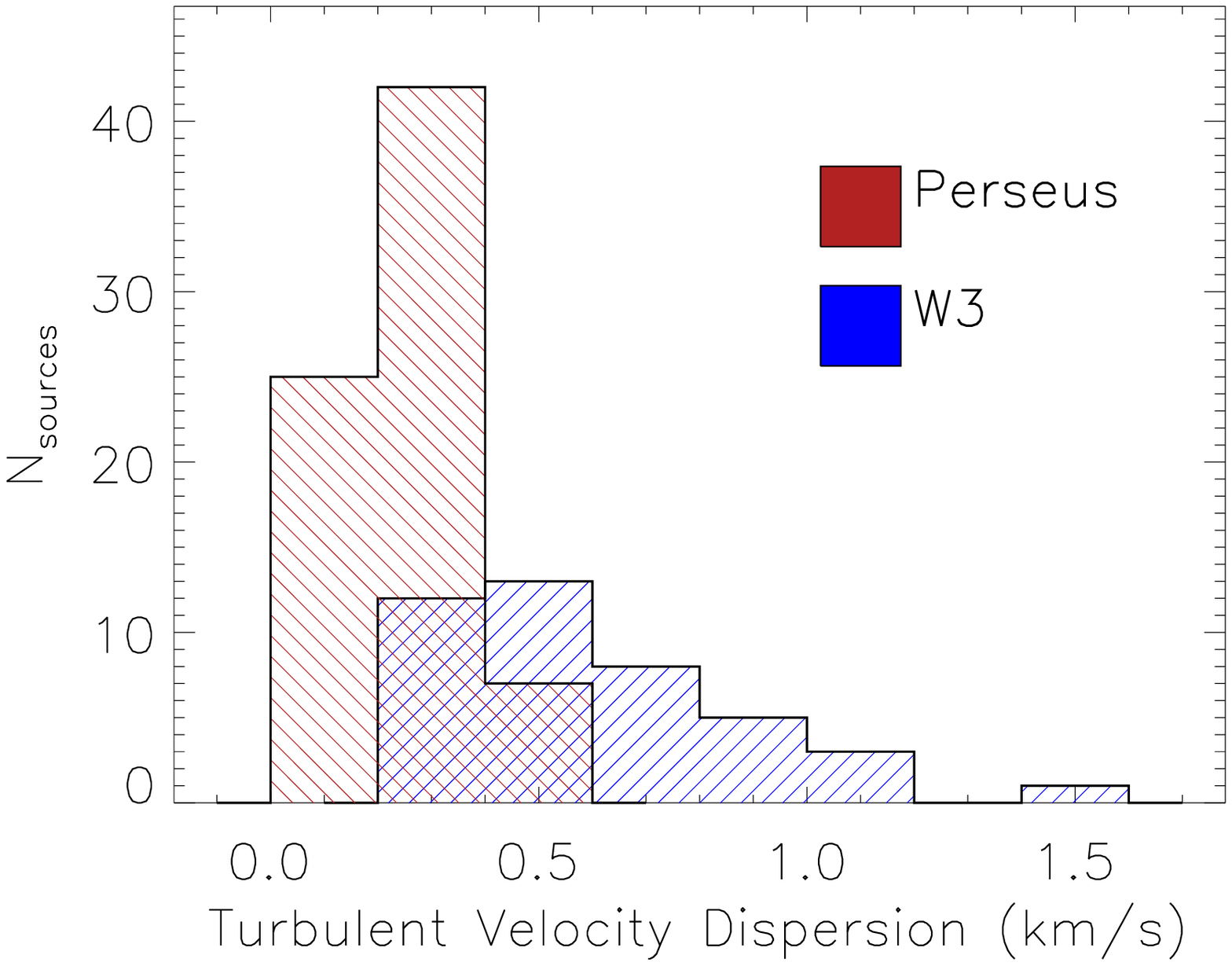}\\
\includegraphics*[width=0.45\textwidth]{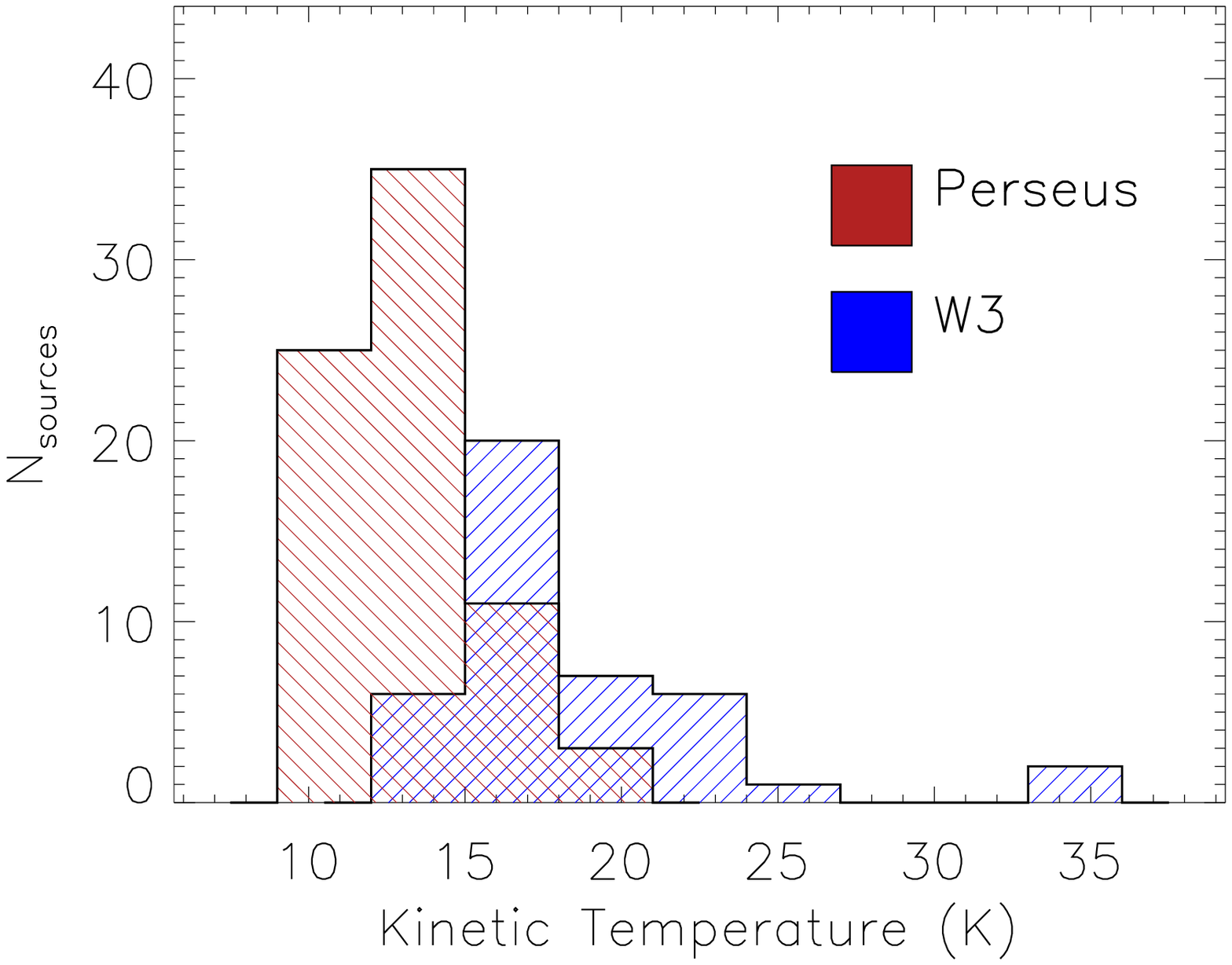}
\includegraphics*[width=0.45\textwidth]{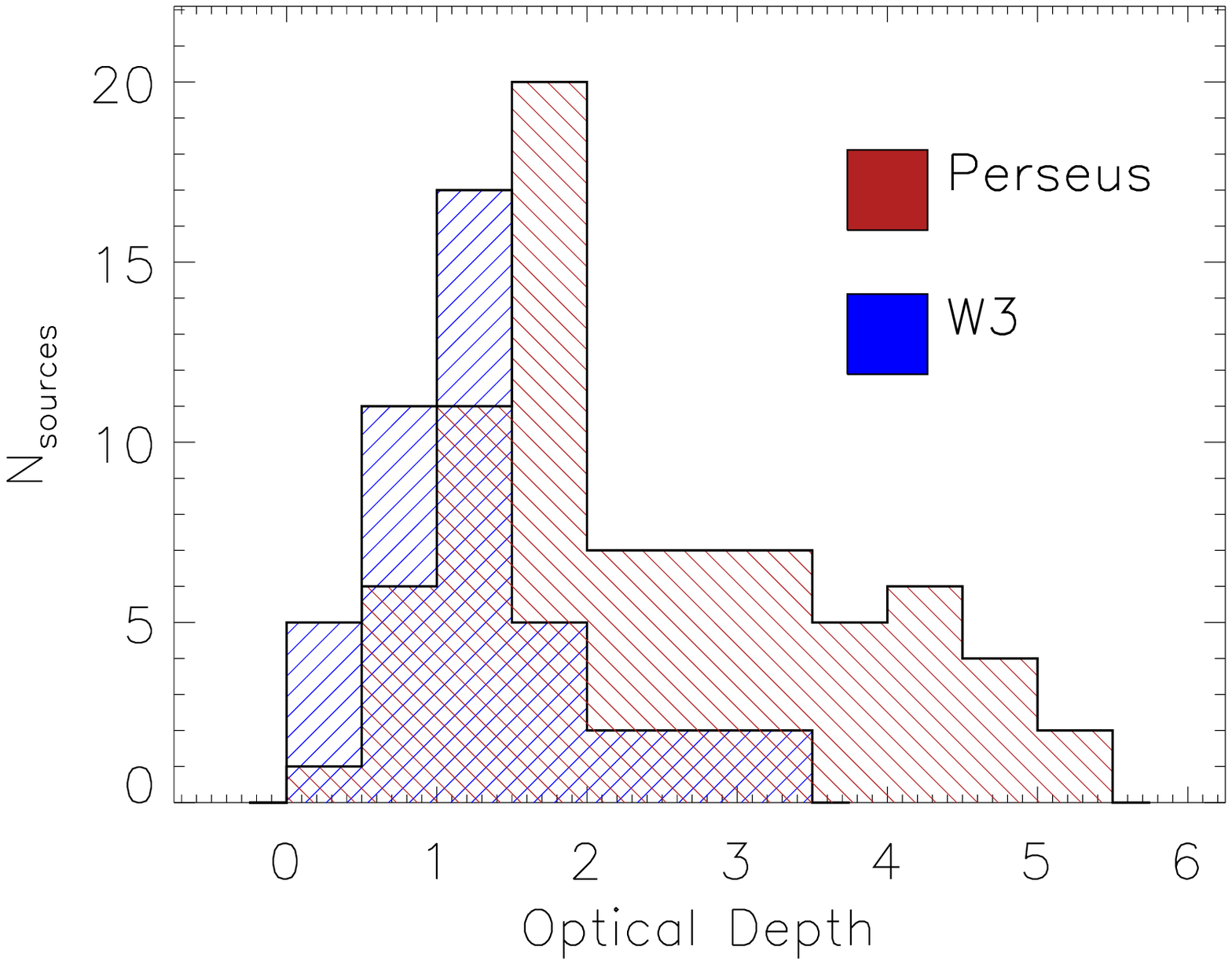}\\
\includegraphics*[width=0.45\textwidth]{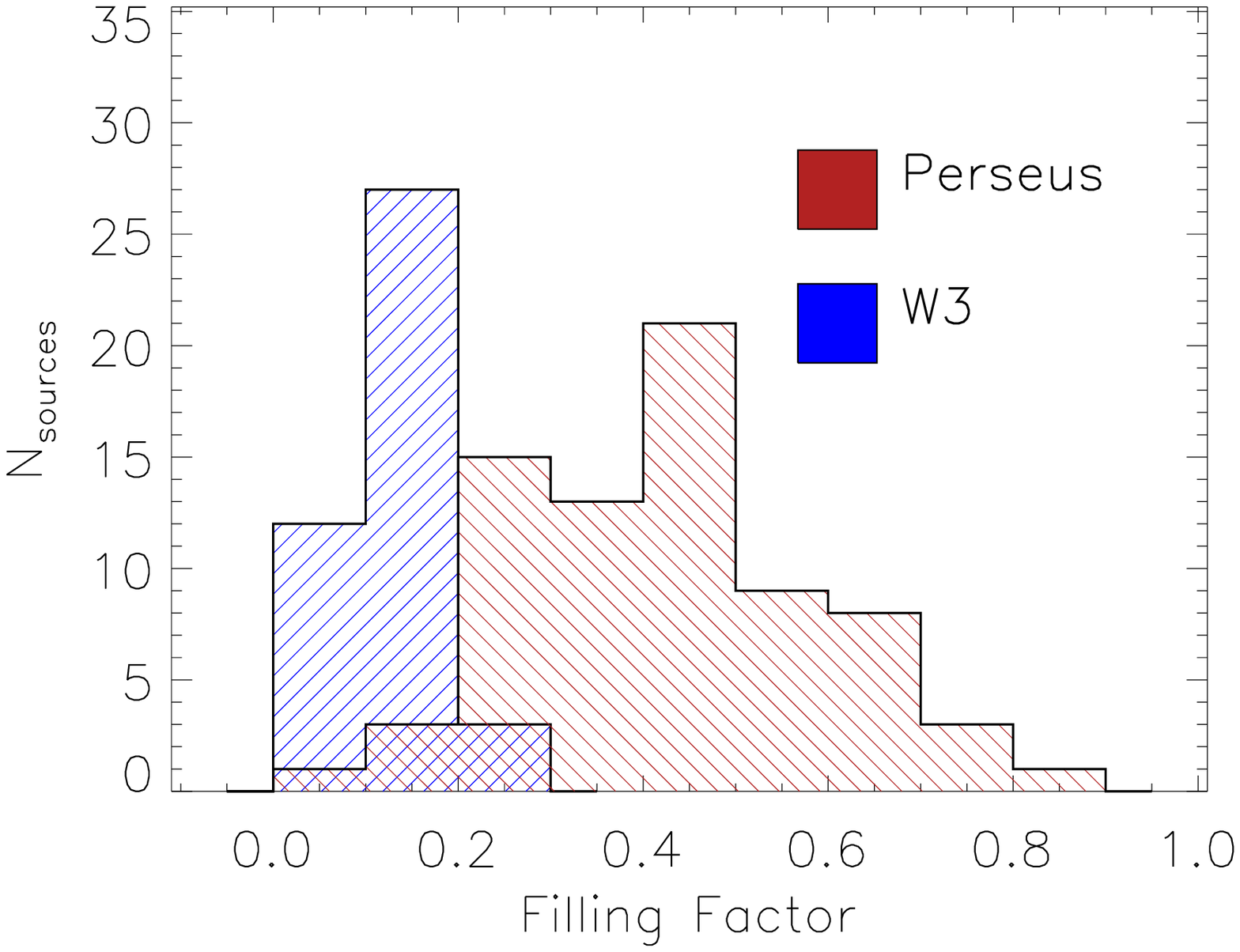}
\includegraphics*[width=0.45\textwidth]{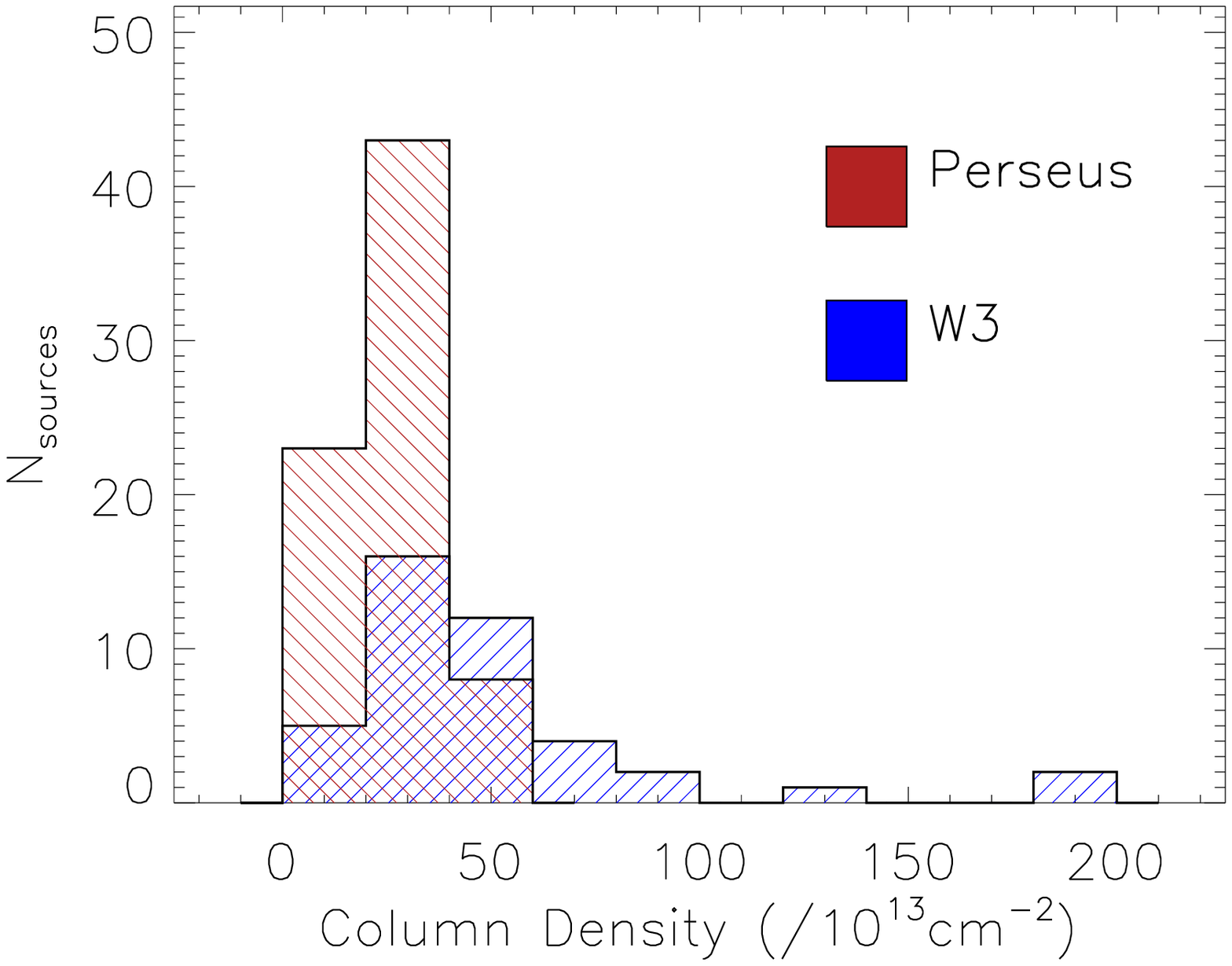}
\caption{Histograms of the physical properties of the Perseus (red) and W3 (blue) regions.}
\label{fig:prop_hists}
\end{center}
\end{figure*}

  Derived physical properties, integrated over the \nh\ source extent, are presented in Tables \ref{tbl:P_Clump_Parameters} and \ref{tbl:W_Clump_Parameters}. Turbulent velocity dispersions of each source were estimated by deconvolving the thermal contributions to the observed linewidths, \mbox{$\mathrm{d}V_{\mathrm{therm}}=\sqrt{8\mathrm{k_{B}}T_\mathrm{k} \mathrm{ln 2/m_{NH_3}}}$}, where \Tk\ is the mean kinetic temperature averaged over the source in question and $\mathrm{m_{NH_3}}$ is the molecular mass of an ammonia molecule (17.03 amu). A statistical summary of the \nh\ source properties is given in Table \ref{tbl:Props_Comparison} and their distributions are presented in Figure \ref{fig:prop_hists}. The properties of the Perseus and W3 samples appear distinct and this is confirmed by Kolmogorov-Smirnov (KS) tests indicating that the physical parameters of the Perseus and W3 samples represent distinct populations with confidence levels of 5$\sigma$, except for column density, which is distinct with a confidence level of 4$\sigma$.

  Despite the significant extent of ammonia emission which is seen in the majority of our sources, it should be noted that filling factors considerably lower than unity are the norm across both the W3 and Perseus regions, with mean values of 0.12 and 0.42 respectively. This likely indicates a significant level of unresolved structure in the objects which we are observing. As noted in Section \ref{sec:DR}, we have made the assumption in our analysis that the gas we are observing is in LTE (i.e. that \Tex\ = \Tk). This has the consequence that filling factors are essentially incorporated into our calculation of the source-averaged column density.

 In general, the Perseus sources have smaller linewidths and turbulent velocity dispersions ($\sigma_{\mathrm{turb}}$), higher optical depths and filling factors than those in W3. These results might be expected for any two given star-forming regions at the relative distances of W3 and Perseus. The observations of Perseus map a physical scale a factor of 7.7 smaller than those of W3, assuming distances to Perseus and W3 of 260 pc and 2 kpc respectively, and it is to be expected that there will be systematic resolution effects upon the measured core properties. For instance, filling factors may be expected to be lower in more distant sources where there is unresolved structure, while linewidths may be larger due to the inclusion of larger physical scales and blending of multiple sources. 

\begin{table*}
\begin{center}
\caption{Summary of physical properties for sources in Perseus and W3.}
\label{tbl:Props_Comparison}
\begin{minipage}{\linewidth}
\begin{center}
\begin{tabular}{lcccccc}
\hline
\hline
{}	       				& \multicolumn{3}{c}{Perseus}	& \multicolumn{3}{c}{W3} \\
{}					& {Mean} & {Median} & {Std. Dev.} & {Mean} & {Median} & {Std. Dev.} \\
\hline
dV (km/s)				& 0.64	& 0.61	& 0.20  & 1.44  &  1.25  &  0.63  \\
$\sigma_{\mathrm{turb}}$ (km/s)		& 0.25	& 0.24	& 0.10  & 0.59  &  0.54  &  0.27  \\
\Tk\ (K)				& 13.14 & 12.69 & 2.31  & 18.50 &  17.43 &  4.85  \\
Optical Depth				& 2.42	& 2.08	& 1.24  & 1.27  &  1.13  &  0.72  \\
Filling Factor				& 0.42	& 0.43	& 0.16  & 0.12  &  0.11  &  0.05  \\
Column Density (/10$^{13}$ cm$^{-2}$)	& 25.78 & 23.89 & 10.69 & 50.07 &  40.11 &  39.92  \\
\hline
\end{tabular}\\
\end{center}
\end{minipage}
\end{center}
\end{table*}

\subsection{Source Associations}
\label{Sec:Source_Associations}
In the Perseus region, 60 protostars or prestellar cores from the catalogue of \citet{Hatchell2007} are covered by our \nh\ observations and all are associated with (1,1) line emission. These are evenly split between individual matches and multiple associations, with 30 submillimetre sources matched one-to-one with a single ammonia clump. A total of 11 ammonia clumps cover the remaining 30 submillimetre sources. There are no cases of a single submillimetre source being associated with more than one ammonia clump in Perseus. We find 43 ammonia clumps that do not have an associated submillimetre source in the catalogue of \citet{Hatchell2007}. Twelve of these are weakly detected and have no (2,2) line detection. Inspection of the position of the remaining detections in the original maps of \citet{Hatchell2005} shows that weak submillimetre emission is often present below the detection limits used in the \citet{Hatchell2007} catalogue. In a few cases there is no apparent submillimetre emission whatsoever, although these are invariably near negative `bowling' associated with nearby bright sources due to the observing method and Fourier-transform-based techniques used in the image reconstruction. 

The ammonia maps of the W3 region cover the positions of 192 of the 316 submillimetre sources in the catalogue of \citet{Moore2007}, 77 of which are detected in the (1,1) line. Compared to Perseus, there is a higher proportion of multiple associations with 25 one-to-one matches and 52 submillimetre sources that are matched with 17 \nh\ clumps. Twelve ammonia clumps were found that were not detected in the catalogue of \citet{Moore2007}. Of these, 8 are weak with no (2,2) detection and 2 are close to the edge of a submillimetre map. The remaining two (W26 and W47) are near to bright submillimetre sources (objects 109 and 213 in the catalogue of \citet{Moore2007}, respectively) and may be associated with these. However, these submillimetre sources are more strongly associated (closer, with more morphological similarity) to two other ammonia clumps (W23 and W46). There is clear separation of the ammonia clump pairs in each case and so these sources have not been amalgamated. There is a single case in our entire set of observations in which two ammonia clumps appear to be clearly associated with a single submillimetre source. The sources W35 and W38 are both associated with the source 175 from the catalogue of \citet{Moore2007}.

\begin{figure*}
\begin{center} 
\includegraphics*[width=0.45\textwidth]{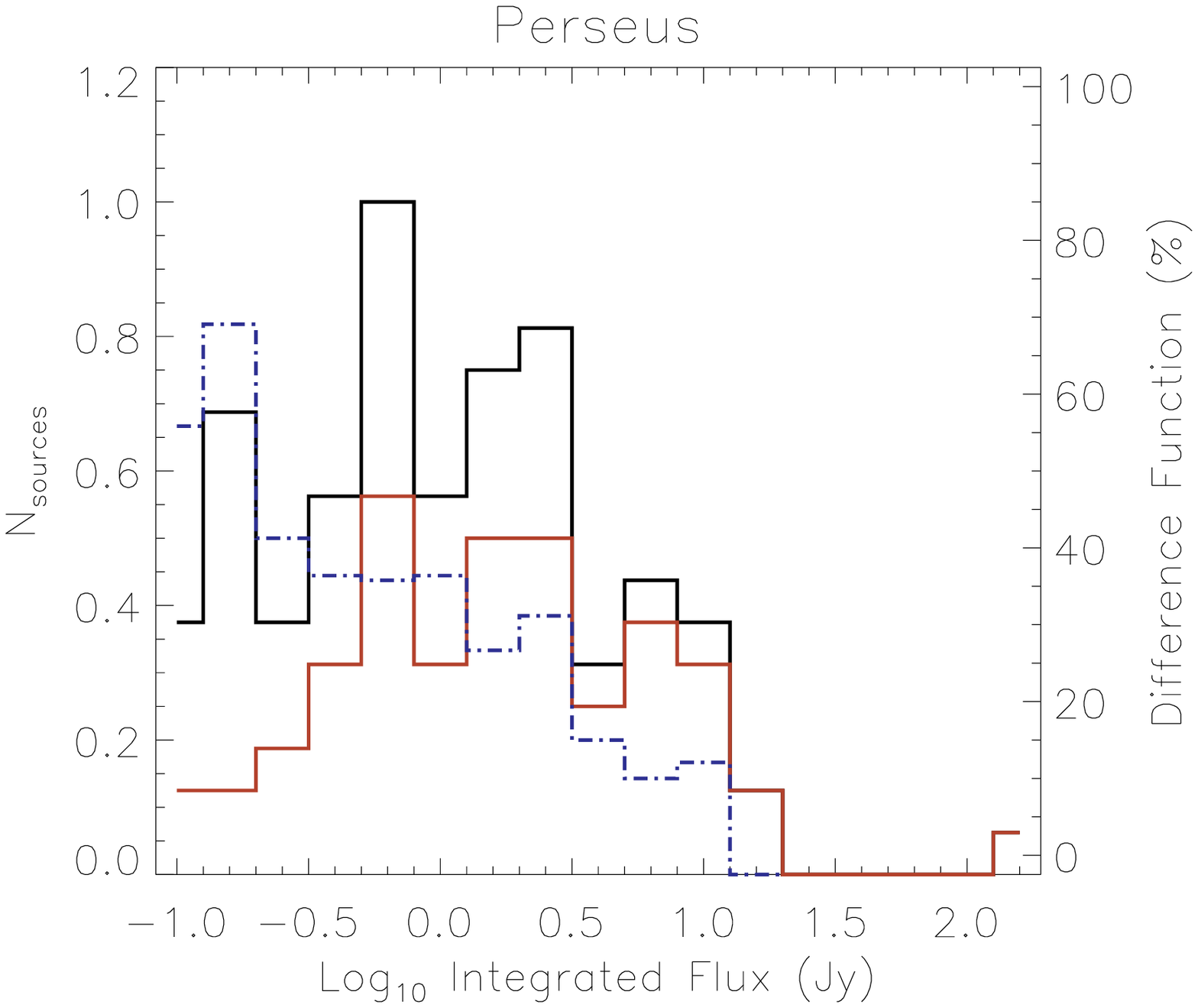}
\includegraphics*[width=0.45\textwidth]{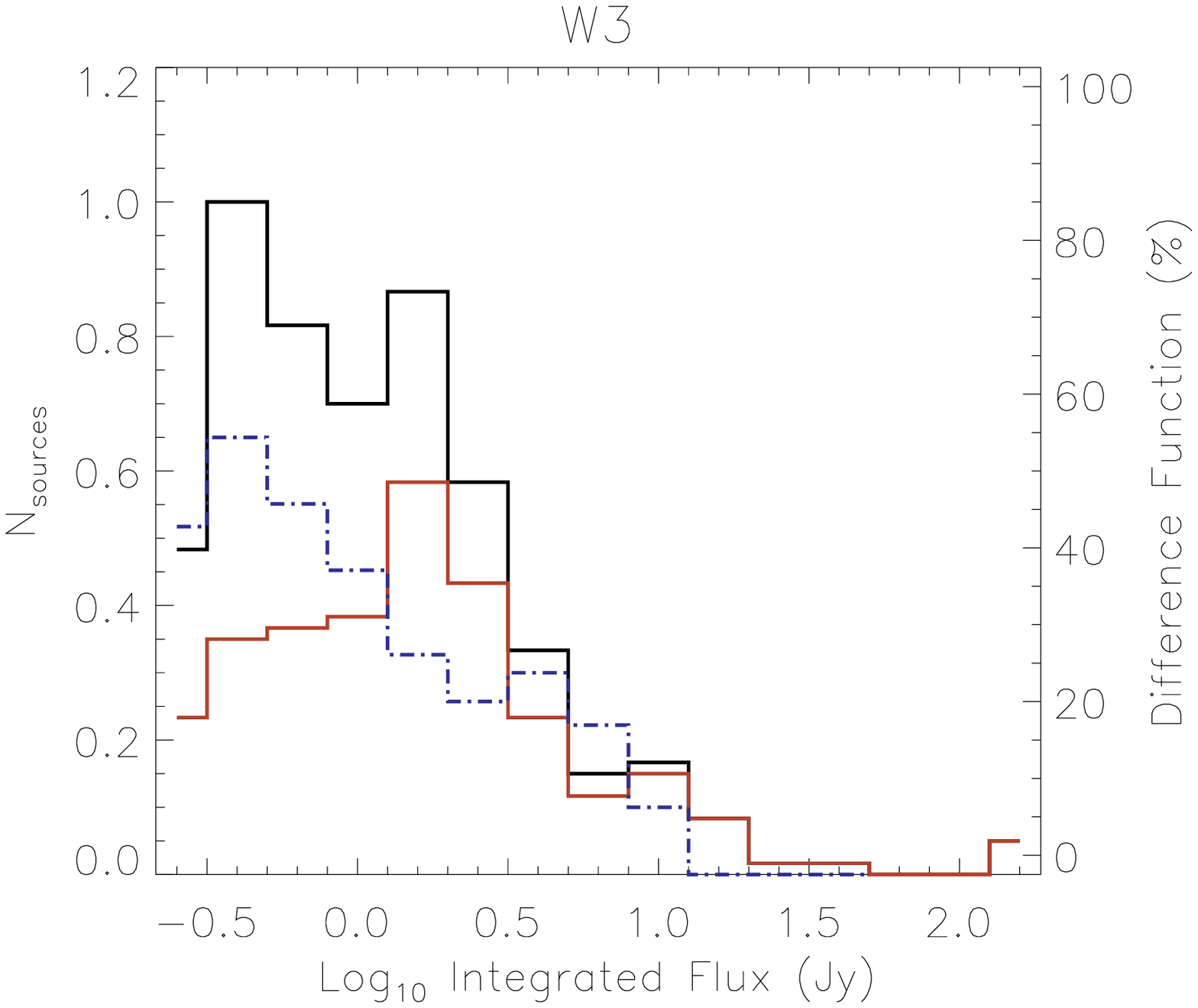}
\caption{Distributions of the submillimetre integrated fluxes of the Perseus (left) and W3 (right). The solid black line shows the distribution of fluxes from the entire catalogue of either \citet{Hatchell2007} for Perseus or \citet{Moore2007} for W3. The solid red line shows the distribution of fluxes of those sources which are covered by our observations and the dashed blue line shows the difference between the two distributions.}
\label{fig:Coverage}
\end{center}
\end{figure*}

In Section \ref{sec:Observations} it was stated that the coverage of sources in both the Perseus and W3 regions was selected to include both clustered and isolated sources, with some bias towards clustered regions. In order to test for selection effects that may affect our results, we have plotted the distributions of integrated submillimetre flux densities for both regions in Figure \ref{fig:Coverage}. It can be seen that a high proportion of sources with low flux densities fall outside our \nh\ observations, but that the populations are virtually fully sampled down to the primary turnovers in both distributions, where the original submillimetre catalogues become incomplete.  Also, the proportion of submillimetre sources covered is very similar (as a function of flux density) in Perseus and W3.  We therefore find no significant difference in the selection of the two samples which might result in artificial differences between them.

\begin{figure*}
\begin{center} 
\includegraphics*[width=0.7\textwidth]{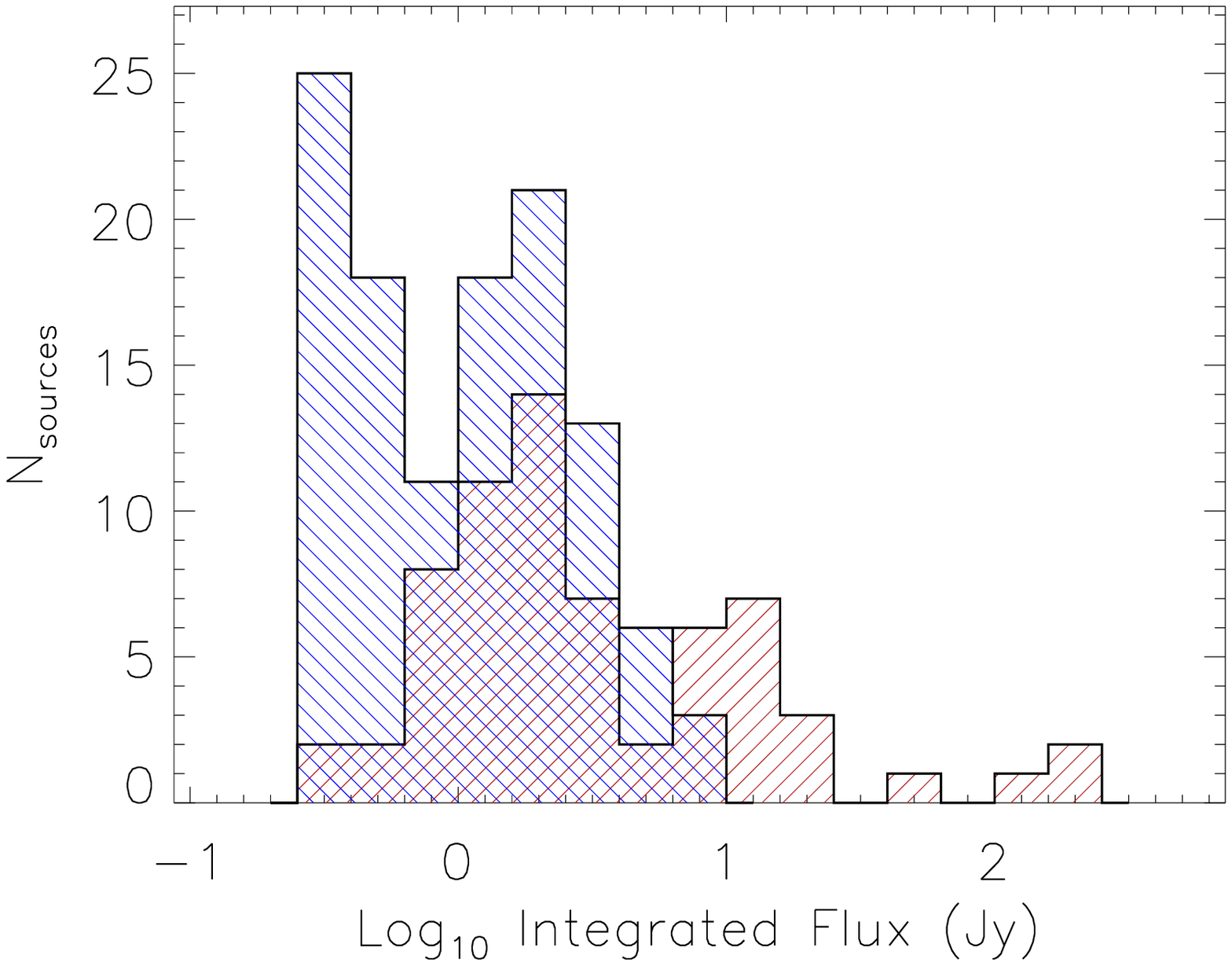}
\caption{Histograms of the (log) integrated fluxes of the sources detected (red) and not-detected (blue) in the W3 region by this survey.}
\label{fig:W3_Dets}
\end{center}
\end{figure*}

A comparison between the integrated submillimetre flux densities of sources detected (red) and not-detected (blue) in ammonia in W3 is shown in Figure \ref{fig:W3_Dets} (no corresponding plot is presented for Perseus as all observed submillimetre sources within that region were detected). The figure shows that, as expected, it is generally the sources that are weaker in submillimetre continuum which are not detected by this survey. However, there is considerable overlap and so the likelihood of detecting a submillimetre source in ammonia is not completely determined by the \nh\ detection limit. 

\begin{figure*}
\begin{center} 
\includegraphics*[width=0.7\textwidth]{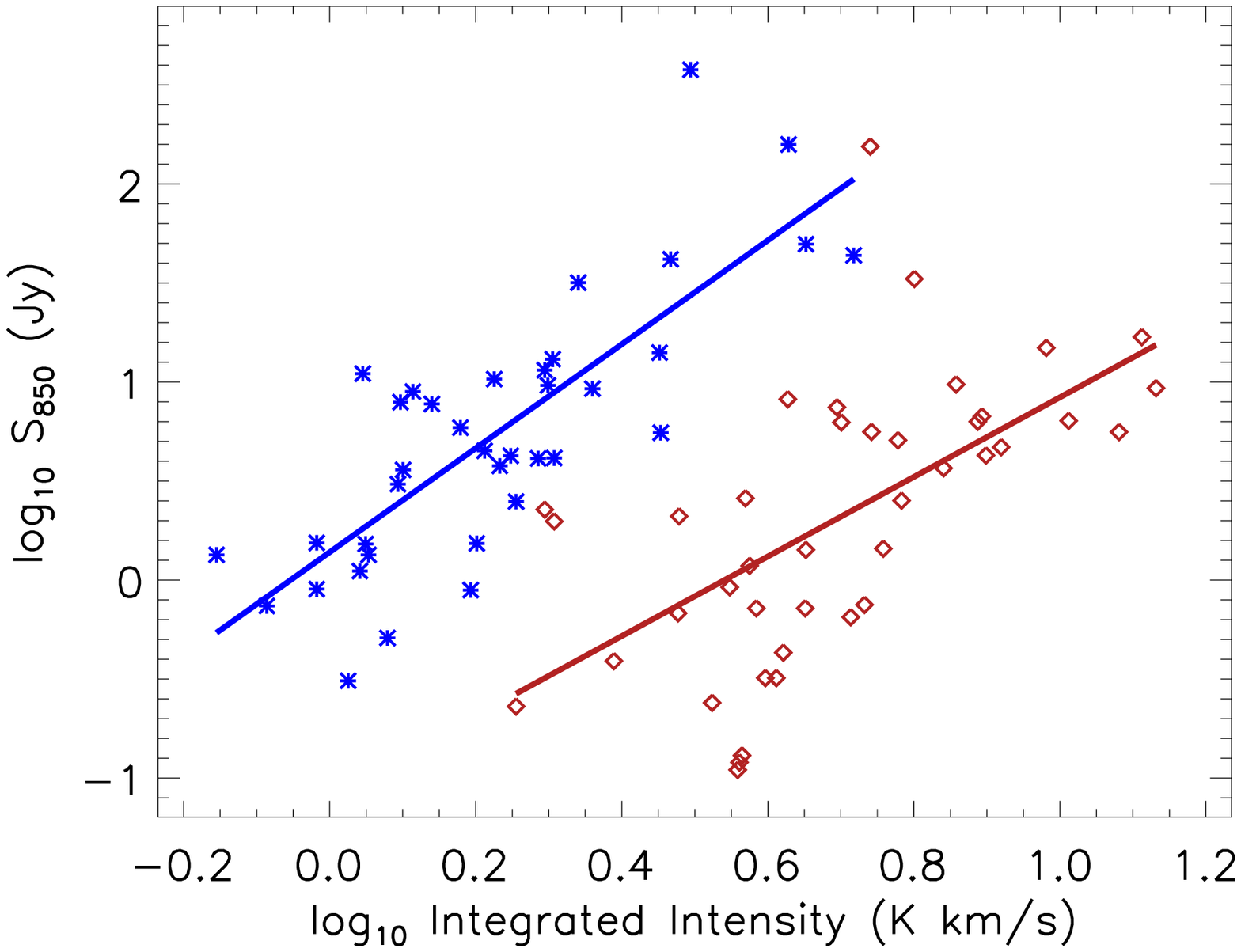}
\caption{Log$_{10}$ integrated 850 \micron\ fluxes vs. \nh\ log$_{10}$ integrated intensities for source associations in Perseus (red diamonds) and W3 (blue stars).}
\label{fig:Int_vs_S850}
\end{center}
\end{figure*}

In Figure \ref{fig:Int_vs_S850} we plot the integrated 850 \micron\ flux density of each source against the integrated \nh\ (1,1) line intensity of the corresponding source. Significant correlations between the two measurements are seen, at $>$5 sigma levels of significance.  The offset between the W3 and Perseus samples in this figure are produced by the higher average temperatures in the W3 sample (Table \ref{tbl:Props_Comparison} and Figure \ref{fig:prop_hists}) which produce higher continuum flux densities and lower \nh\ intensities for a given column density. 

\begin{figure*}
\begin{center} 
\includegraphics*[width=0.7\textwidth]{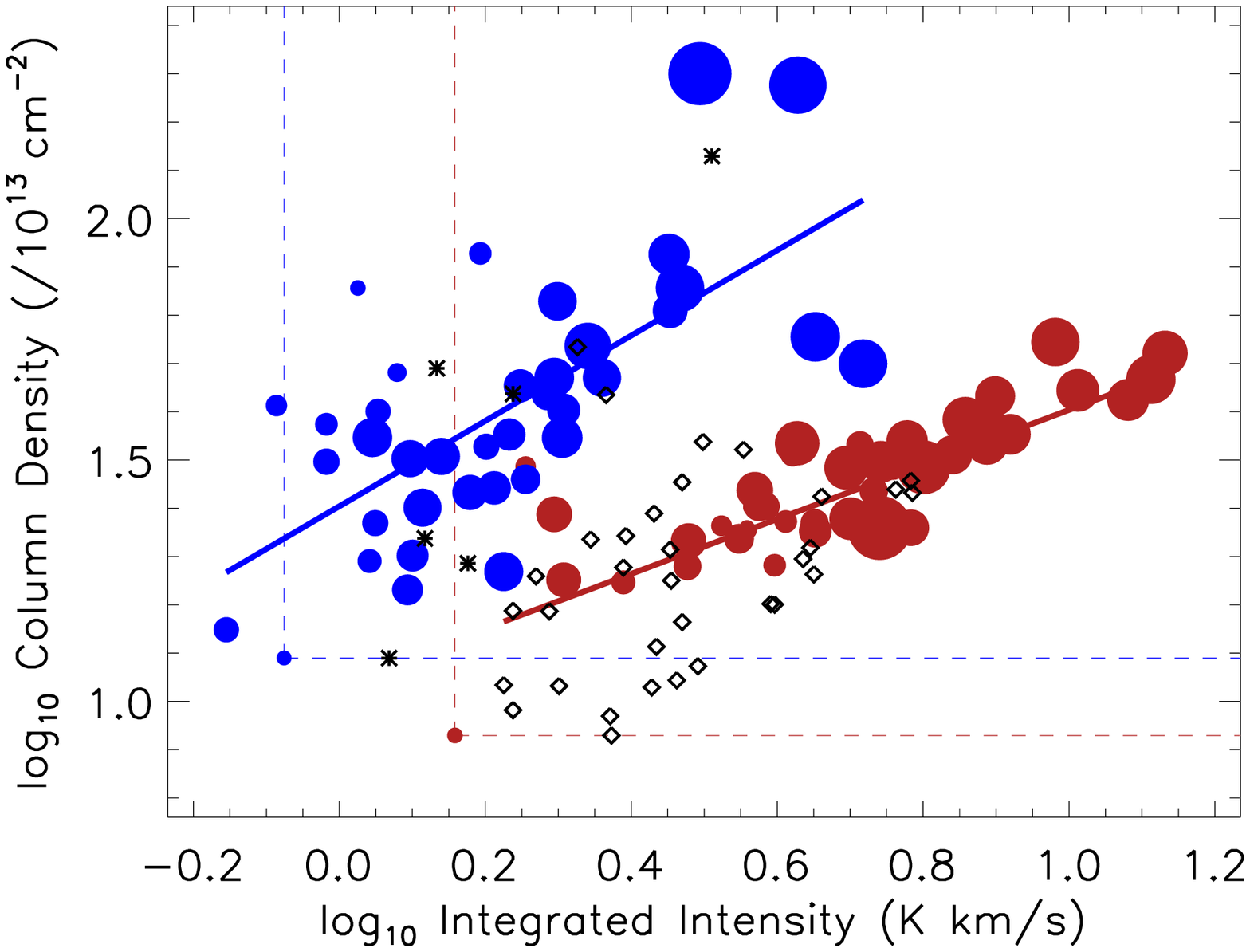}
\caption{The distribution of ammonia column density against integrated intensity on a log-log scale. Circles represent submillimetre associations, the size of the circle representing the log value of the 850 \micron\ integrated flux on a normalised scale. Blue and red circles represent sources in W3 and Perseus respectively, stars (W3) and diamonds (Perseus) show the sources in those regions with no submillimetre association. Lines of best fit to all data points for each region are overlaid. Ammonia sensitivity limits are plotted as dashed lines with the 850 \micron\ completeness estimate represented as a circle at the lower left corner.}
\label{fig:Ncol_vs_Int_w_S850}
\end{center}
\end{figure*}

Figure \ref{fig:Ncol_vs_Int_w_S850} shows the relationship between the column density derived from ammonia and the integrated intensity in the (1,1) line. Where a submillimetre association has been identified, the source is plotted as a circle whose diameter represents the 850 \micron\ integrated flux density. Ammonia detection limits are marked by dashed lines, with small filled circles representing the limiting values of the submm emission, taken from \citet{Hatchell2005} and \citet{Moore2007}. The lower limits for the detection of ammonia integrated intensity represent three times the median r.m.s. values in the integrated intensity images for the relevant region, the value used as the minimum threshold in the $Clumpfind$-based extractions. As there is no unique lower limit on column density measurements, the plotted limit is simply taken as the lowest measured value in the relevant region.

Lines of best fit have been added to Figure \ref{fig:Ncol_vs_Int_w_S850} to emphasise the correlation of column density and integrated intensity for the separate W3 and Perseus samples resulting from the LTE analysis. The offset and slightly differing slopes for the two samples are again due to the higher average temperature of cores in W3 and the scatter results from variations in \Tk\ within each sample. Where there is no associated submillimetre detection, the ammonia source is plotted as a star (W3) or diamond (Perseus). The plot shows that, at least for the Perseus sample, sources that are detected in ammonia with no 850 \micron\ association have the lowest \nh\ integrated intensity and column density. There is considerable overlap between the submillimetre associations and non-associations, but the overall trend of submillimetre associations representing sources with the highest \nh\ intensity and column density is clear. 

Figure \ref{fig:Ncol_vs_Int_w_S850} displays the differing sensitivity limits in what are effectively four observational samples. Our observations of the Perseus and W3 regions have different average sensitivity limits, while there is also a notable difference between the sensitivity of submillimetre observations in W3 and Perseus. These differences explain the number of objects detected at 850 \micron\ by \citet{Moore2007} which are not found in \nh, while all objects detected at 850 \micron\ by \citet{Hatchell2005} are seen in our observations. The relative completeness of these samples is limited by a combination of the limits of our ammonia observations and those of the relevant submillimetre observations. In particular, by submillimetre sensitivity in Perseus and by ammonia sensitivity in W3.

\subsection{The Relationship Between Submillimetre and Ammonia Emission}
\label{sec:SM_NH}

\begin{figure*}
\begin{center} 
\includegraphics*[width=1.0\textwidth]{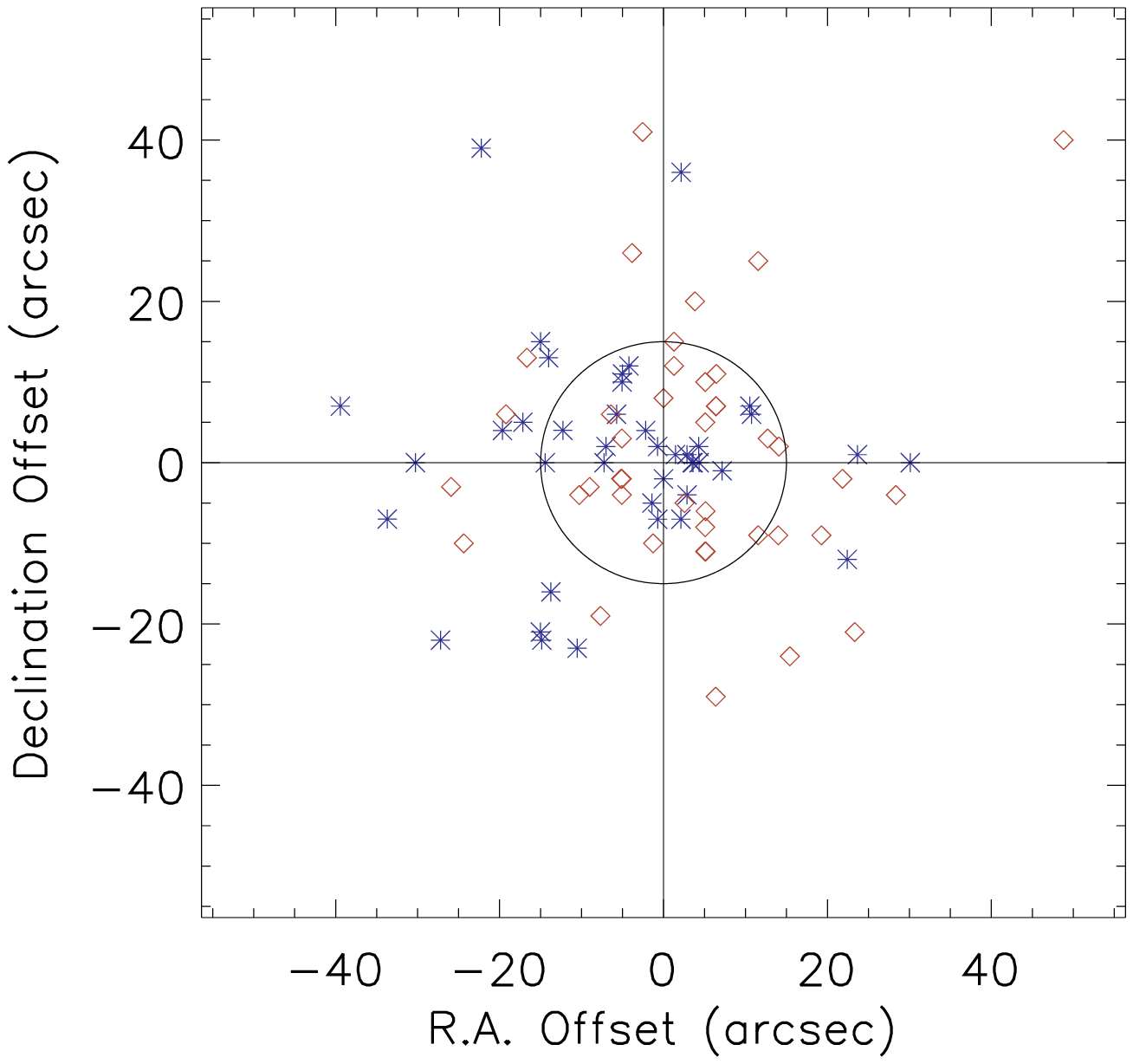}
\caption{Offsets of the peak positions of each submillimetre source in Perseus (red diamonds) and W3 (blue stars) relative to the peak ammonia position centred at 0,0. The circle represents the radius of the GBT beam. Typical pointing errors in our observations are 4\arcsec.}
\label{fig:offsets}
\end{center}
\end{figure*}

\begin{figure*}
\begin{center} 
\includegraphics*[width=0.7\textwidth]{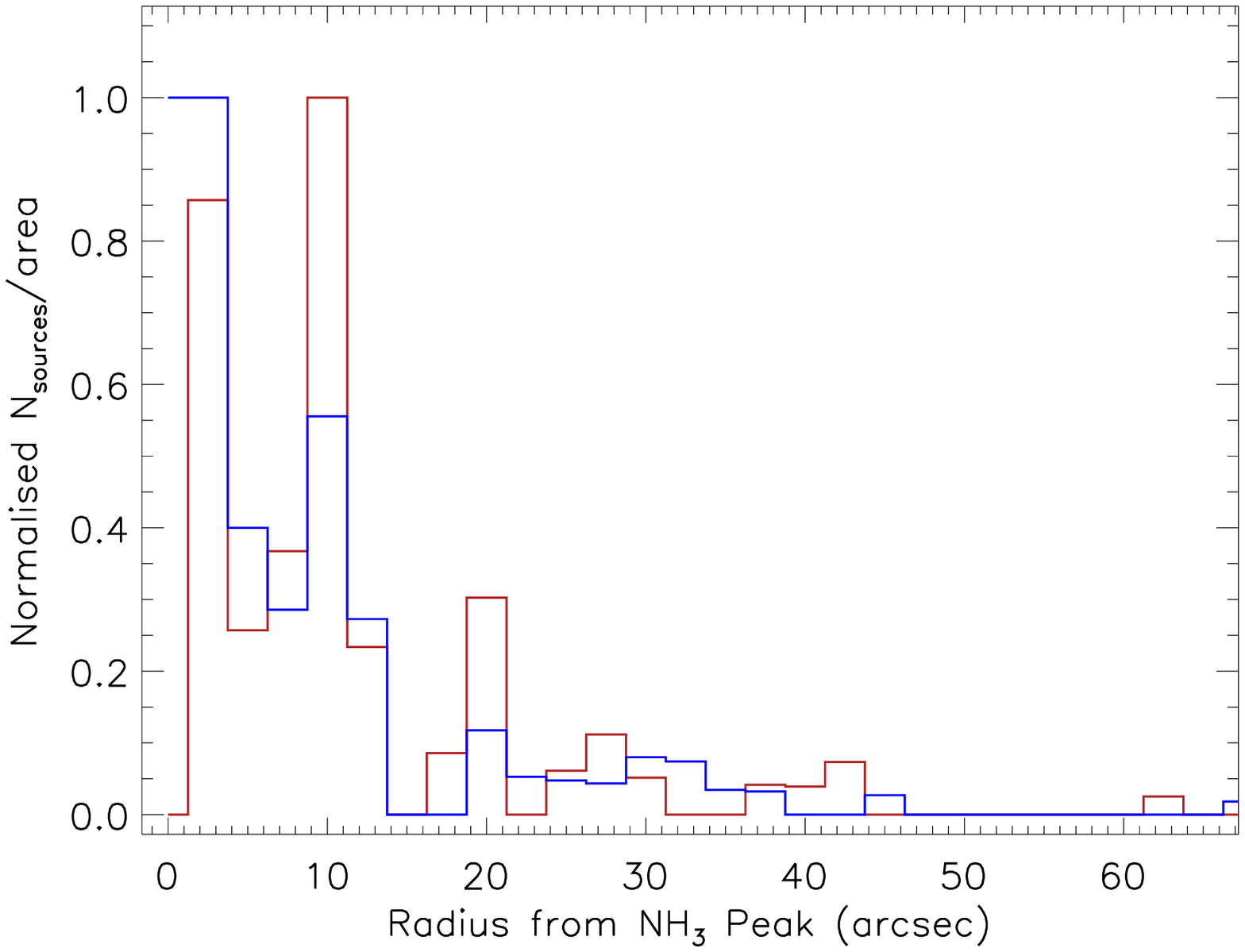}
\caption{The surface density distribution of the offsets of peak positions of submillimetre sources related to the ammonia sources in Perseus (red) and W3 (blue)}
\label{fig:offset_hist}
\end{center}
\end{figure*}

Figure \ref{fig:offsets} shows the positional offsets between the peaks in ammonia integrated intensity and submillimetre continuum for each associated source. In the case of multiple submillimetre source associations, the closest peak is used. While the two tracers share common regions of emission, the peaks may be separated by some distance. The surface density distributions of peak offsets are plotted in Figure \ref{fig:offset_hist}.  While there are some sources with offsets consistent with the pointing errors ($\sim$4\arcsec), there is a significant excess with much larger separations and approximately 40\% are larger than the radius of the GBT beam.  In addition, we see a minimum in the Perseus distribution at zero offset.  These results show that the locations of the emission peaks in the submillimetre continuum are not tightly constrained to those in ammonia.

\begin{figure*}
\begin{center} 
\includegraphics*[width=0.4\textwidth]{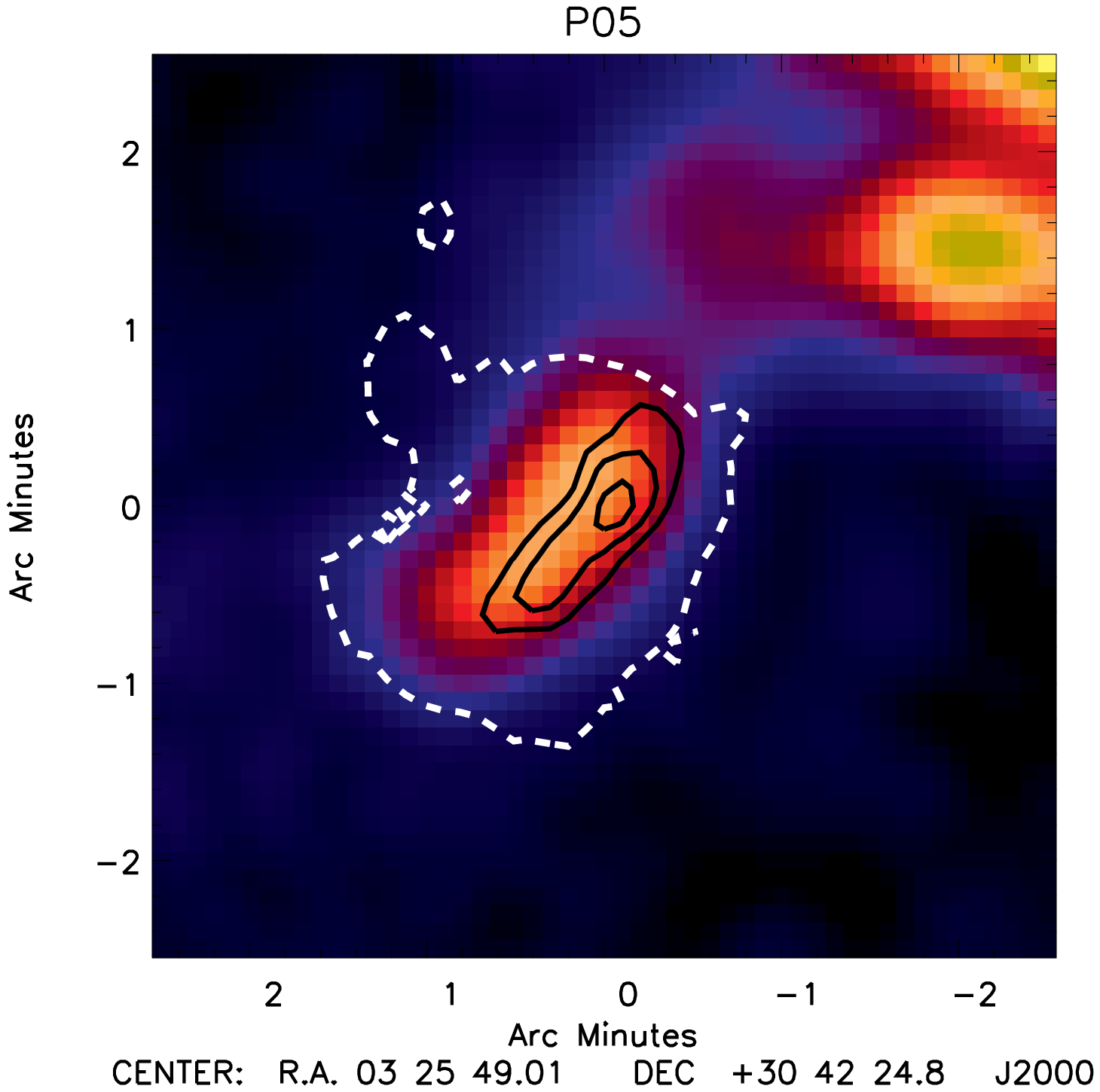}
\includegraphics*[width=0.4\textwidth]{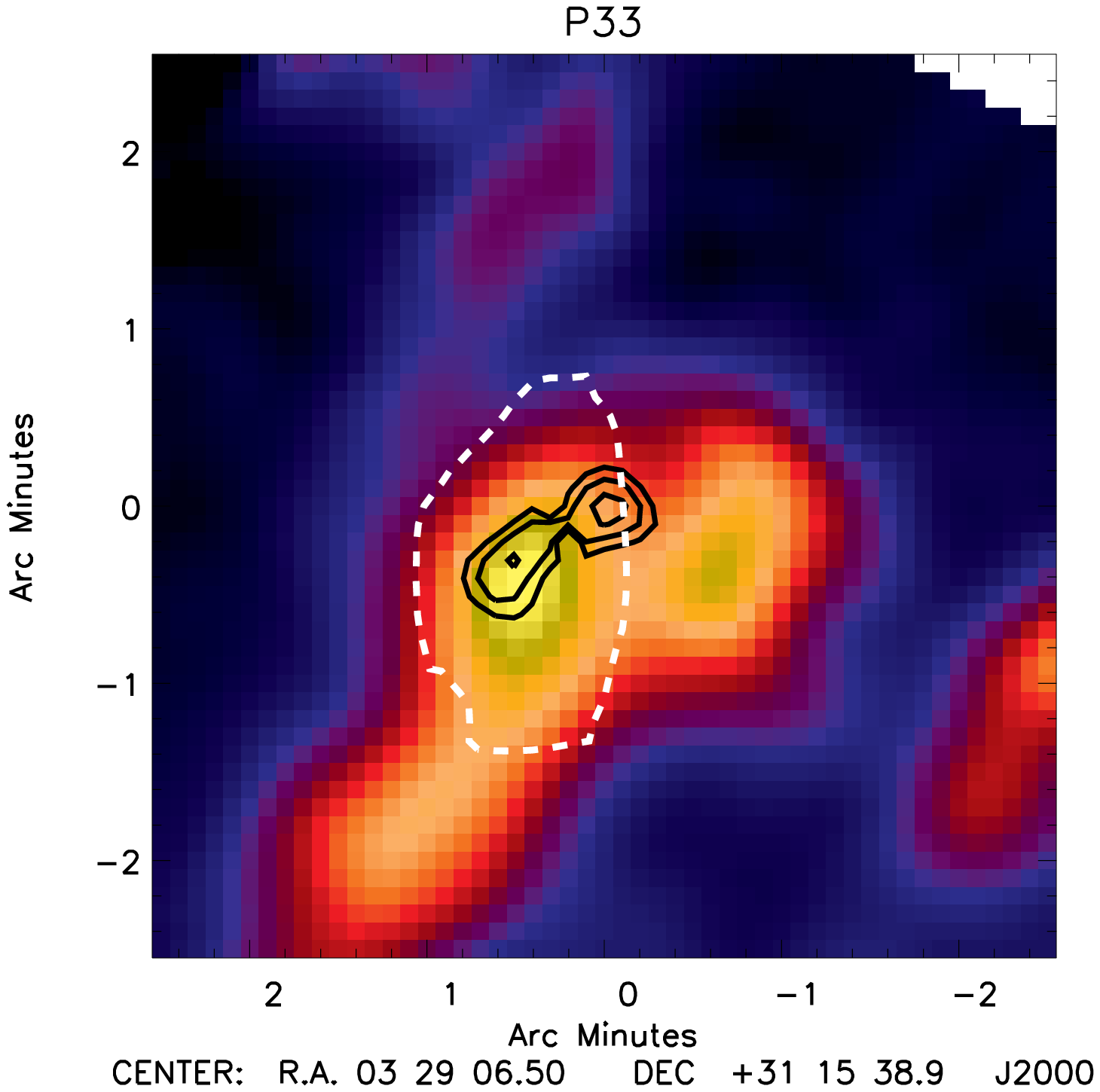}\\
\includegraphics*[width=0.4\textwidth]{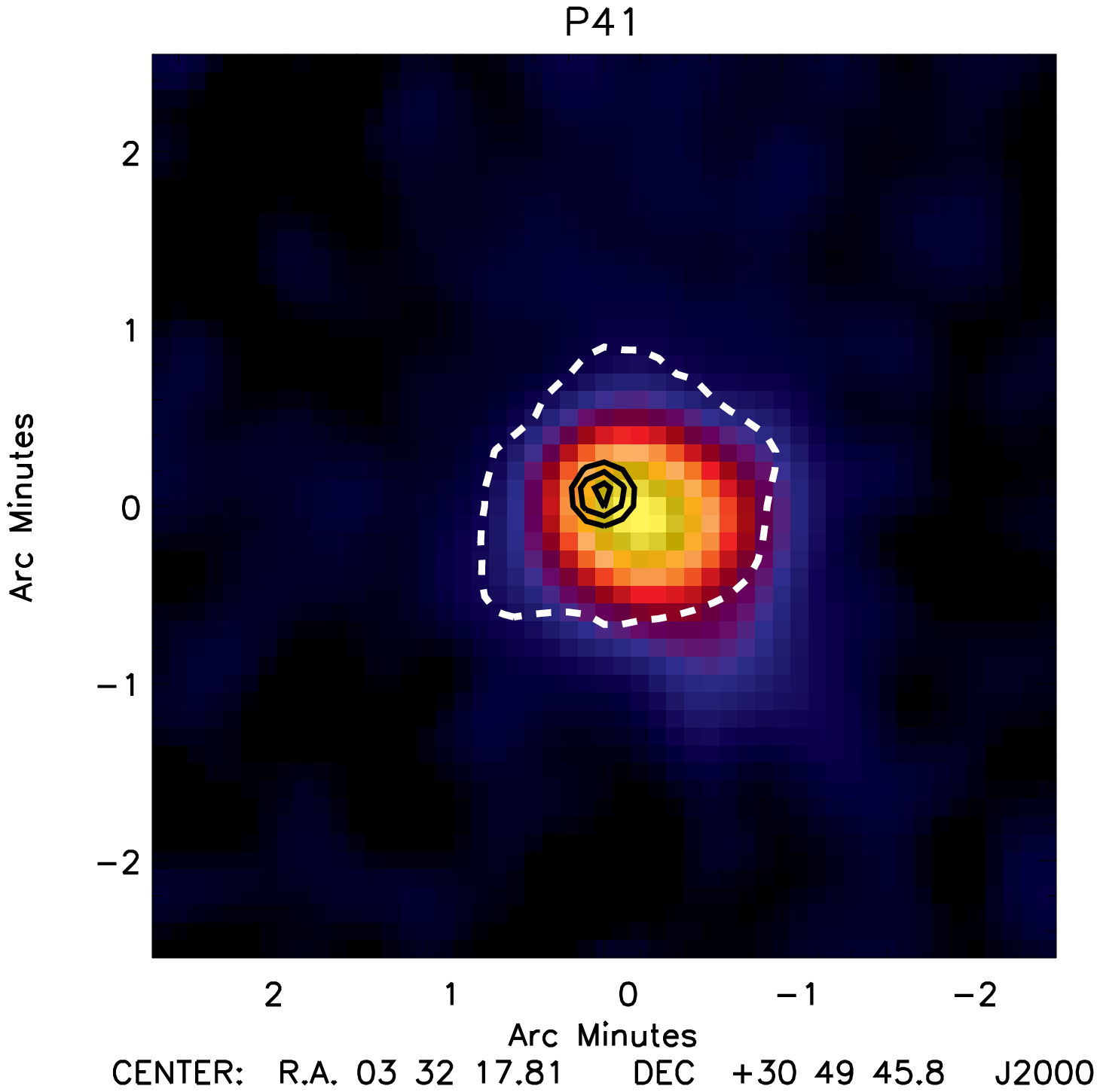}
\includegraphics*[width=0.4\textwidth]{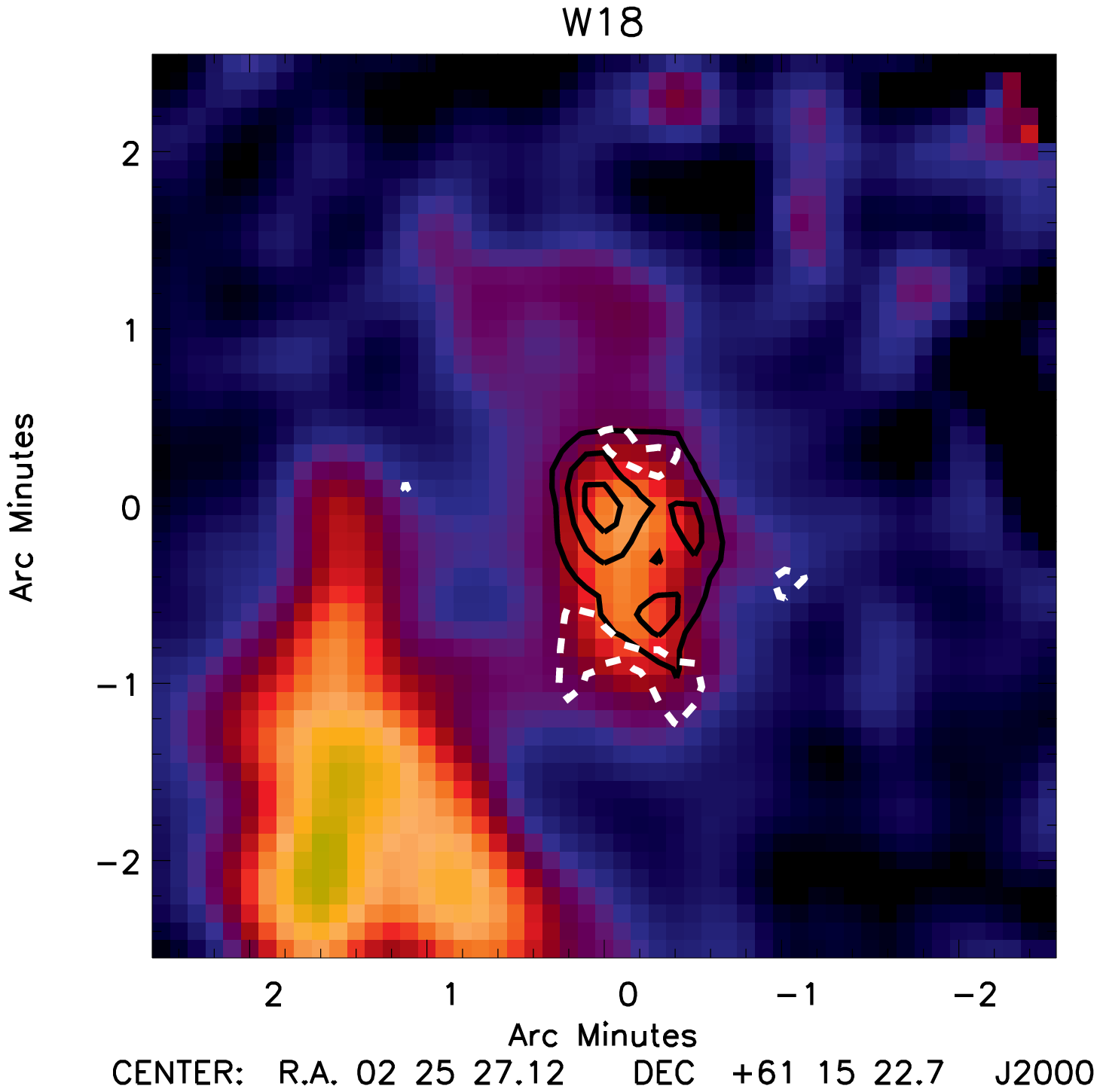}\\
\includegraphics*[width=0.4\textwidth]{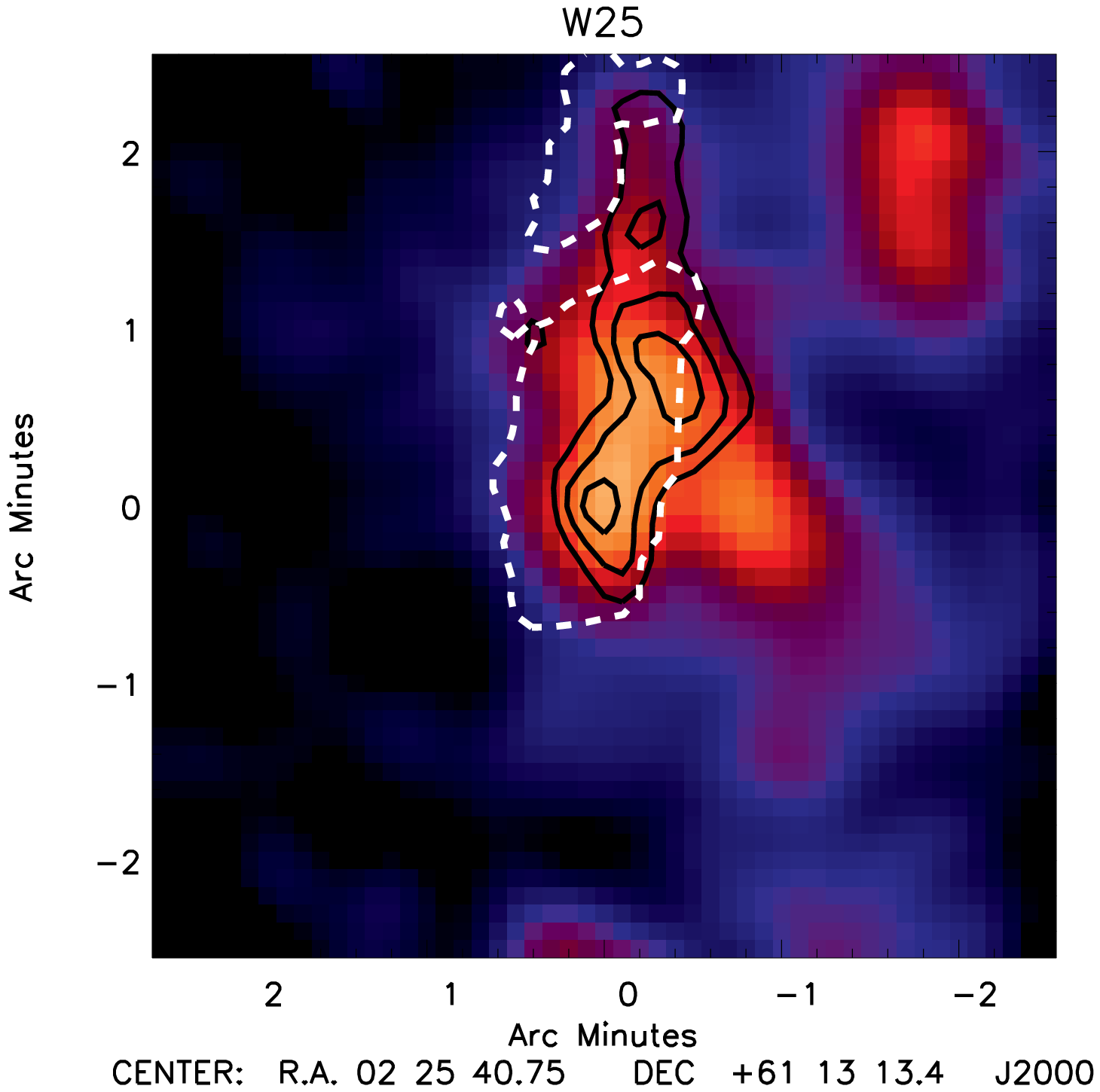}
\includegraphics*[width=0.4\textwidth]{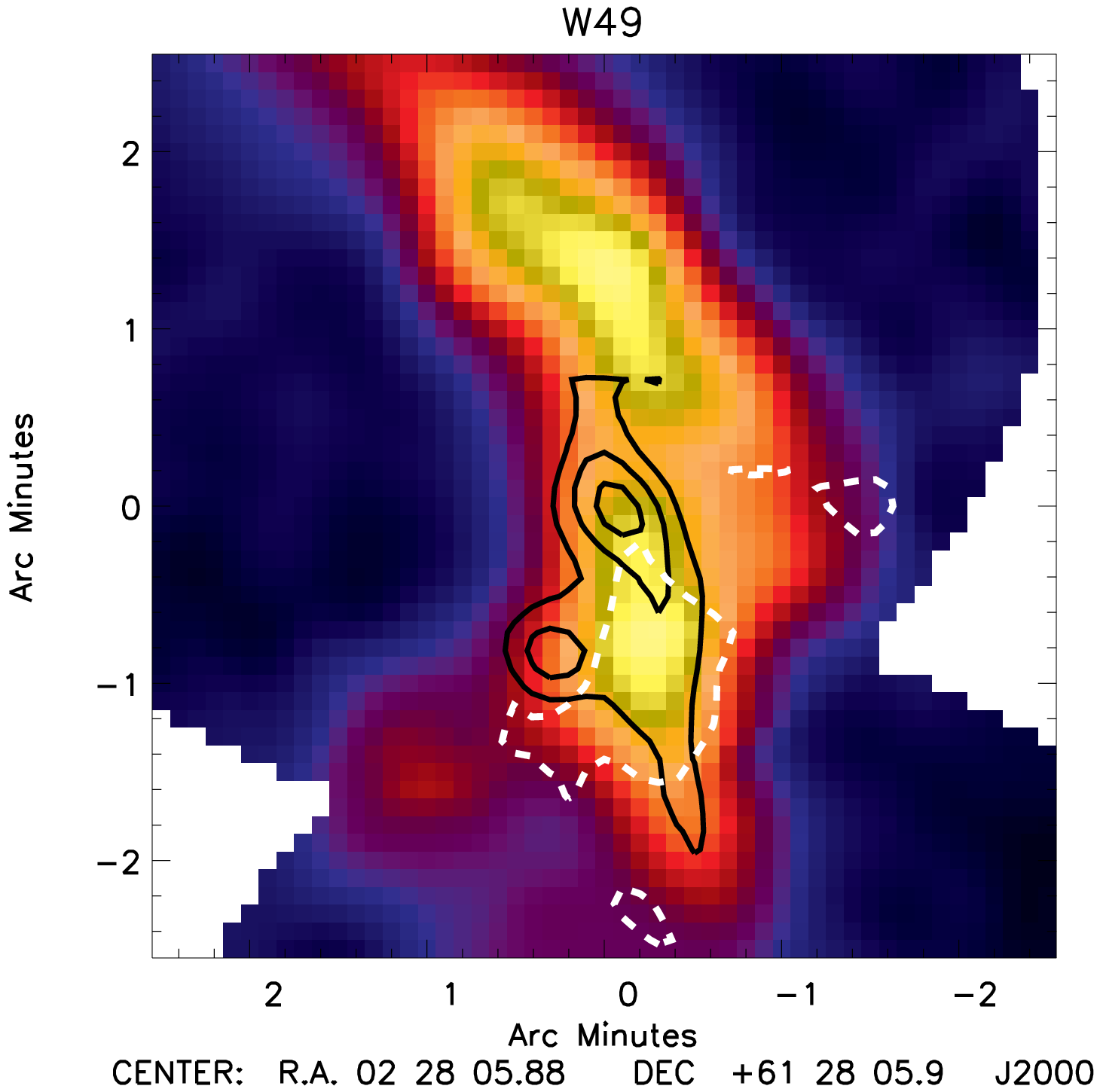}\\
\caption{\nh\ integrated intensity in sources P05, P33 \& P41 and W18, W25 \& W49, overlaid with black contours of submillimetre emission at 50, 70 and 90\% of the peak flux density. A white, broken contour traces the \nh\ column density distribution at a level of 50\% of the local peak for all sources except W18, for which the contour level is 40\%}
\label{fig:SM_NH3_Maps}
\end{center}
\end{figure*}

The relative distribution of the two tracers is shown in Figure \ref{fig:SM_NH3_Maps} for six representative sources (P05, P33 \& P41 and W18, W25 \& W49), with submillimetre continuum contours overlaid upon \nh\ (1,1) line intensity images. A single white, broken contour is overlaid on the images which traces the \nh\ column density distribution at a level representative of the source shape and size (this is the 50\% contour unless explicitly stated otherwise). In the Perseus sources, the ammonia emission is typically significantly more extended than the continuum while in the W3 sources there is less of a disparity. P33 and all the W3 images in Figure \ref{fig:SM_NH3_Maps} show multiple submillimetre cores within the 50\% ammonia emission contour. Equivalent images for all associated sources are presented in online material.

\begin{figure*}
\begin{center} 
\includegraphics*[width=0.45\textwidth]{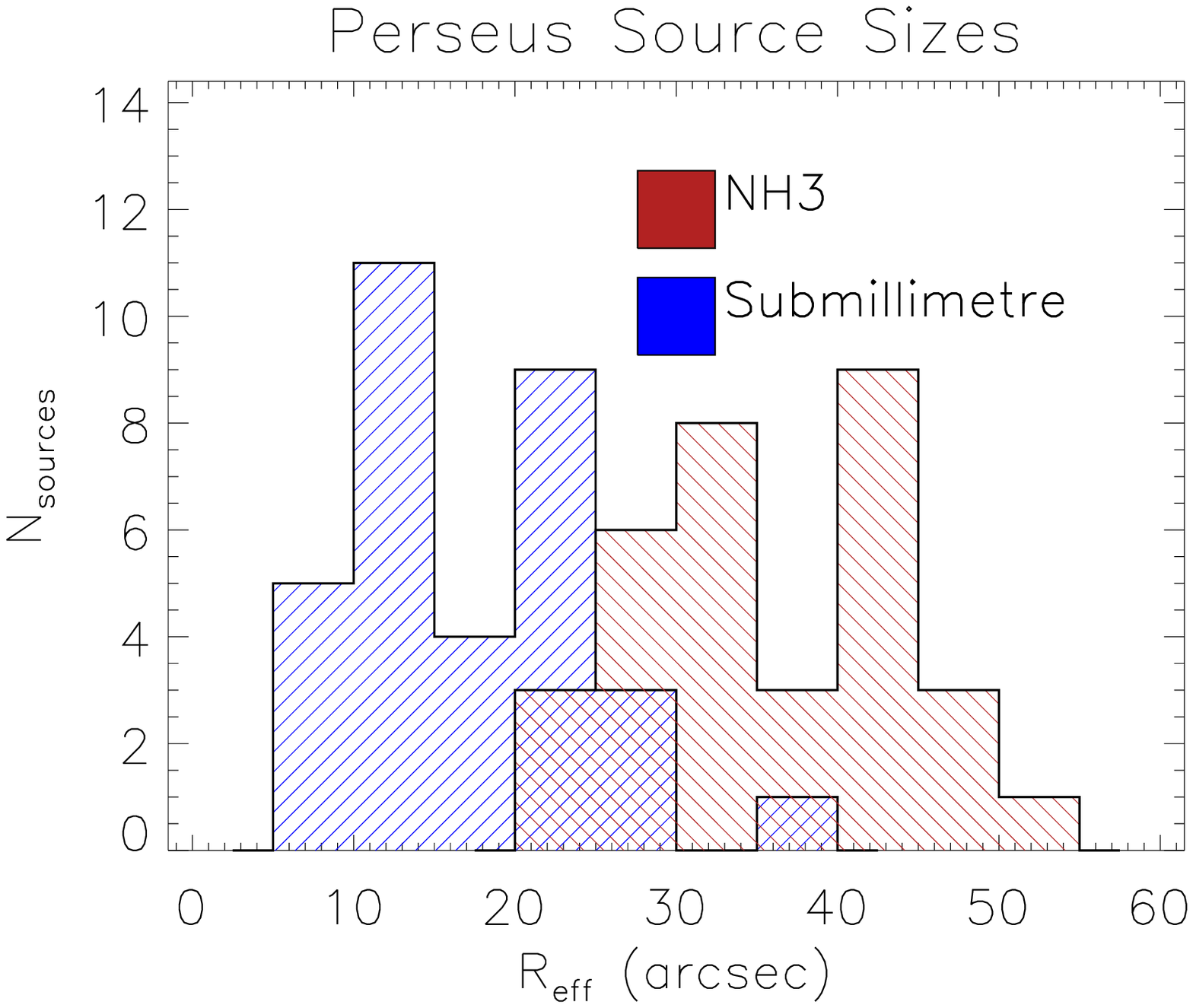}
\includegraphics*[width=0.45\textwidth]{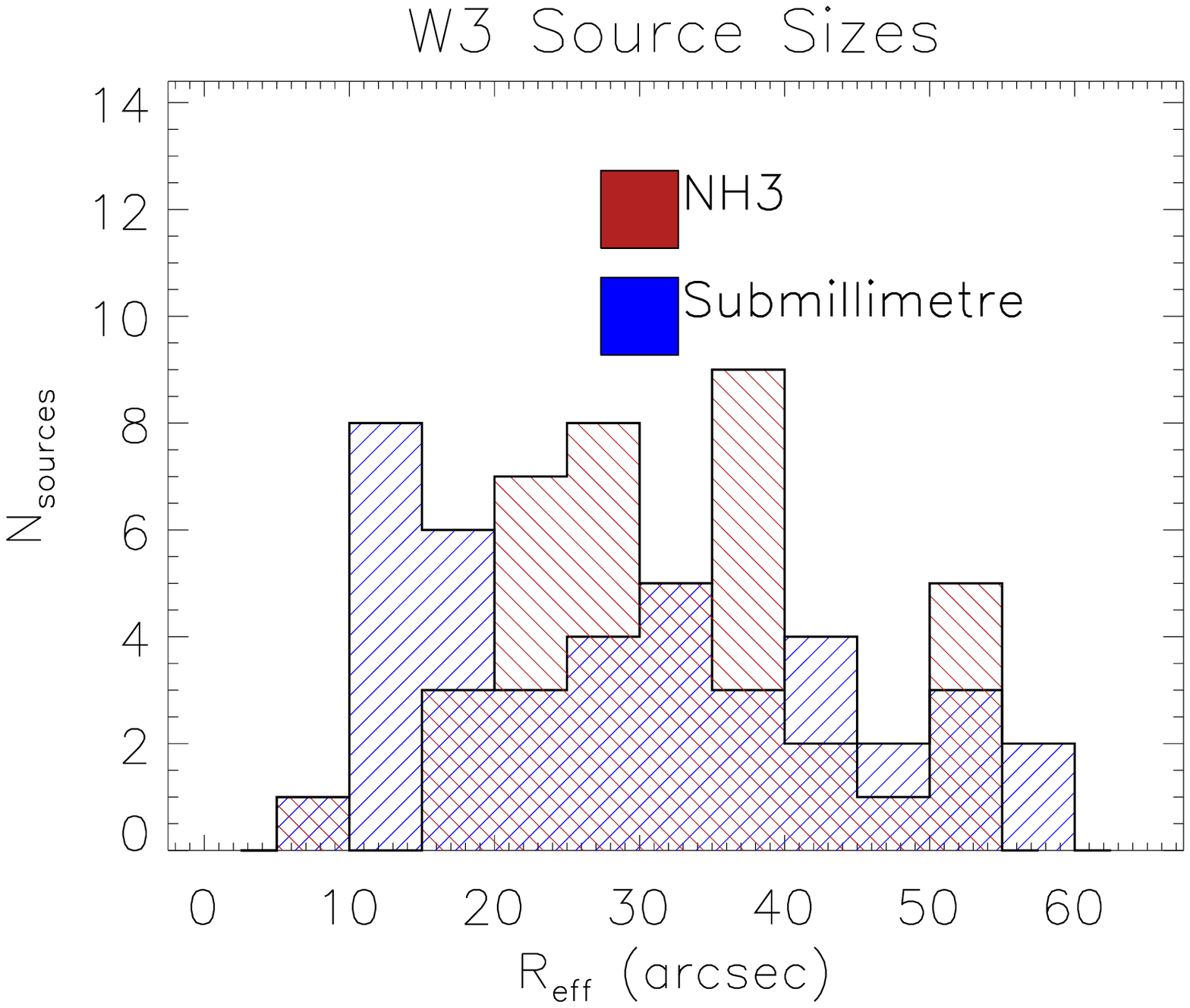}
\includegraphics*[width=0.45\textwidth]{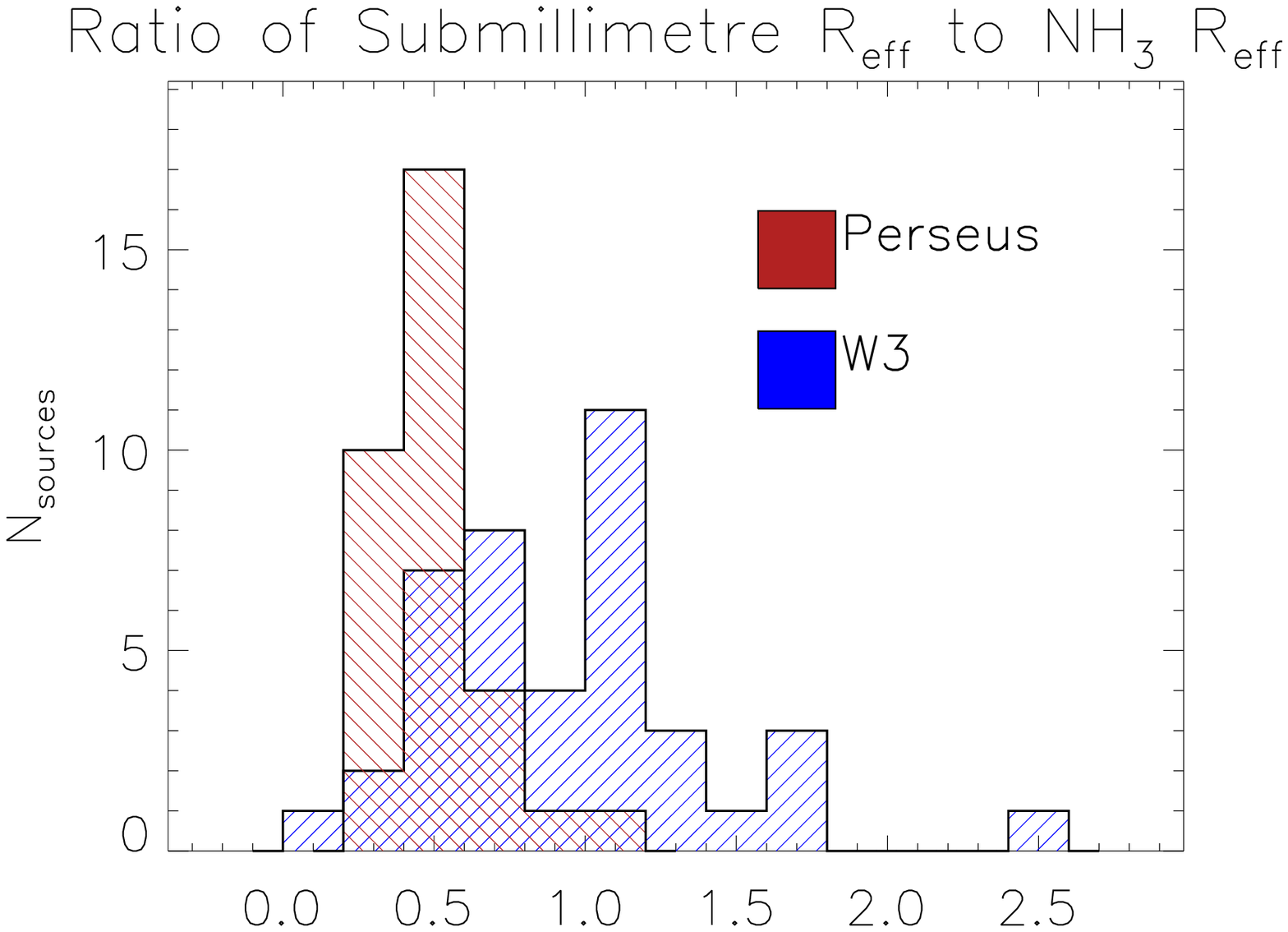}
\caption{The distributions of submillimetre to ammonia source size ratios for Perseus (red) and W3(blue).}
\label{fig:Size_Hists}
\end{center}
\end{figure*}

It has been noted by previous authors (see \citealt{Reid2010} and references therein, also \citealt{Curtis2010}) that different source-extraction techniques may produce significantly different results. It may therefore be important to state that all sources identified here, in both ammonia and submillimetre emission, have been identified using $Clumpfind$. The size of each source was defined by its 50\% surface-brightness contour. Following \citet{Hatchell2007} we define R$_{\mathrm{eff}}$, the effective radius of a source, as $\sqrt{\left(\frac{A}{\pi}\right)}$, where $A$ is the area within the 50\% contour, after deconvolving the beam assuming both a Gaussian source and beam. This deconvolution is only approximate as the assumption of a Gaussian source is somewhat unrealistic. 
R$_{\mathrm{eff}}$ values for all ammonia sources are presented in Tables \ref{tbl:P_source_sizes} and \ref{tbl:W_source_sizes}, along with the sizes of the associated submillimetre sources, estimated in the same way. Histograms of the source-size distributions, comparing the two tracers and the two cloud samples, are shown in Figure \ref{fig:Size_Hists}. Where the emission of an ammonia clump covers more than one submillimetre source, the combined area of all associations is used to calculate R$_{\mathrm{eff}}$ values. Ammonia sizes vary surprisingly little within and between the two regions, with mean deconvolved R$_{\mathrm{eff}}$ measurements of 35.6\arcsec\ and 32.6\arcsec\ for the Perseus and W3 samples, respectively. Despite the similarity in angular size between the regions, at the assumed distances of Perseus and W3 (260 pc and 2 kpc respectively) these correspond to physical radii of 0.05 and 0.31 pc. This large difference in physical scale seen here is discussed in more detail in Section \ref{sec:General_Results}.

 Median angular sizes ($\widetilde{\mathrm{\mathbf{Reff}}}$) agree with the means ($\overline{\mathrm{\mathbf{Reff}}}$) to significantly better than the sample standard deviation in both cases, indicating that each sample is reasonably symmetrical ($\frac{\overline{\mathrm{Reff}} - \widetilde{\mathrm{Reff}}}{\sigma}$ = 0.33 and 0.05 for Perseus and W3 respectively). The submillimetre and \nh\ sizes in W3 are distributed rather similarly whereas, in Perseus, submillimetre radii are significantly smaller, with barely any overlap between the histograms of the two tracers. KS tests of the submillimetre/ammonia R$_{\mathrm{eff}}$ ratios suggest that the populations of W3 and Perseus are significantly different in terms of relative source size. The probability that the distributions are drawn from the same population is measured at 2.3$\times 10^{-7}$.

To test for the effects of distance upon the derived source sizes, we convolved and regridded the Perseus maps (both ammonia and submillimetre) to an equivalent beam and pixel size to place it, effectively, at the distance of W3 (2 kpc) and then re-ran the source-extraction routines and subsequent analyses. In the majority of our KFPA `daisy' maps this smoothing results in only a single source per map. We are thus limited in our analysis by our observing strategy of multiple isolated maps, rather than one, large-scale map.
The source sizes found in the smoothed, regridded Perseus maps are shown in Table \ref{tbl:conv_source_sizes}. It can be seen that the effect of artificially placing the Perseus sources at the distance of W3 decreases the measured size of sources in both submillimetre and ammonia emission by a factor of $\sim$2.

  For a fixed beam size, sources with an intensity profile described by a power law will have a size dependent upon the index of that power law \citep{Young2003}. Therefore, the fact that the ammonia source sizes in both Perseus and W3 are so similar may indicate that similar structures, with power law intensity profiles, are being traced in each region. For example, large clumps/clouds of ammonia which envelop protostellar condensations. However, this scenario is contradicted in two ways. Firstly, the source sizes measured in our smoothed Perseus maps are significantly smaller than those measured in the native resolution maps, indicative of a Gaussian or solid disk intensity profile \citep{Enoch2007}. Secondly, we see significant differences between the sizes of submillimetre sources in W3 and Perseus, inconsistent with the conclusion that sources in W3 are structurally similar to those in Perseus. Our findings are explained by either or both of two conclusions; one, that our sources are not well described by power law intensity profiles. Two, that ammonia and submillimetre emission trace different structures, possibly related to different mass components. 

\begin{figure*}
\begin{center}
\includegraphics*[width=0.7\textwidth]{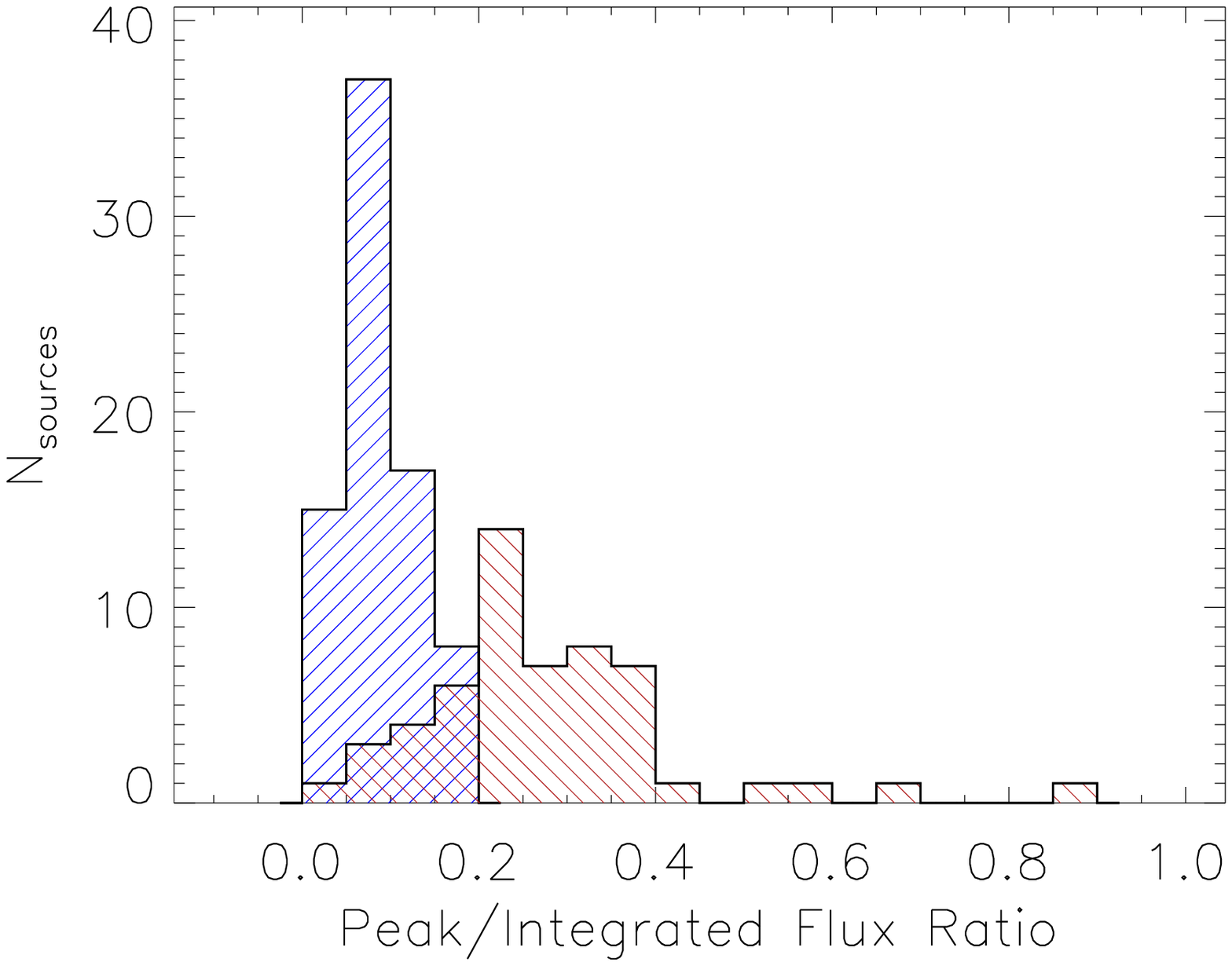}
\caption{The distributions of peak to integrated flux ratios for submillimetre sources associated with our ammonia sources in Perseus (red) and W3(blue).}
\label{fig:Flux_Ratio_Hist}
\end{center}
\end{figure*}

In order to look for differences in source structure in the two star-forming regions, we compared the ratio of peak-to-integrated submillimetre flux densities for each source (hereafter, $\eta^{peak}_{int}$). An object with an unresolved surface-brightness profile should present a $\eta^{peak}_{int}$ value approaching unity. Sources in the W3 sample typically have low values and a narrower distribution than in Perseus (Figure \ref{fig:Flux_Ratio_Hist}). Mean values are 0.09 and 0.28 for W3 and Perseus, respectively. These results, combined with the derived filling factors (mean values of 0.12 and 0.42 in W3 and Perseus, respectively), provide information on the marginally resolved or unresolved structure of our sources. The higher values of $\eta^{peak}_{int}$ in the Perseus sample indicate sources that are more compact than those in W3 and the low filling factors in W3, combined with low values of $\eta^{peak}_{int}$, indicate that these objects are extended but `lumpy', with multiple sources within a beam. This explanation is consistent with the observation that each ammonia source in Perseus is typically associated with fewer submillimetre sources than in W3.

A more thorough examination of source structure, incorporating intensity profiles and the variation of ammonia and submillimetre emission in and around our sources will be more thoroughly examined in a following work (Morgan et al. $in$ $prep$).

\begin{table}
\begin{center}
\caption{Source sizes in Perseus.}
\label{tbl:P_source_sizes}
\begin{minipage}{\linewidth}
\begin{center}
\begin{tabular}{lccccc}
\hline
\hline
{Perseus}	& \multicolumn{2}{c}{\nh\ R$_{\mathrm{eff}}$ }	& \multicolumn{2}{c}{Submillimetre R$_{\mathrm{eff}}$} 	& \underline{Submillimetre R$_{\mathrm{eff}}$}	\\
{Source}	& {(\arcsec)}	&	 {(pc)}			& {(\arcsec)}	&	{(pc)}				& {\nh\ R$_{\mathrm{eff}}$ }			\\
\hline
P01		& 27.4	&	0.035	    & 20.8 	&	0.026	& 0.76 \\
P03		& 42.1	&	0.053	    & 9.5  	&	0.012	& 0.23 \\
P05		& 40.7	&	0.051	    & 26.4 	&	0.033	& 0.65 \\
P11		& 29.4	&	0.037	    & 13.5 	&	0.017	& 0.46 \\
P13		& 45.4	&	0.057	    & 23.5 	&	0.030	& 0.52 \\
P15		& 47.1	&	0.059	    & 20.3 	&	0.026	& 0.43 \\
P22		& 31.8	&	0.040	    & 11.0 	&	0.014	& 0.35 \\
P25		& 41.5	&	0.052	    & 17.4 	&	0.022	& 0.42 \\
P26		& 31.5	&	0.040	    & 12.6 	&	0.016	& 0.40 \\
P28		& 29.2	&	0.037	    & 8.7  	&	0.011	& 0.30 \\
P30		& 32.0	&	0.040	    & 9.7  	&	0.012	& 0.30 \\
P31		& 41.3	&	0.052	    & 18.9 	&	0.024	& 0.46 \\
P32		& 40.8	&	0.051	    & 20.2 	&	0.025	& 0.50 \\
P33		& 42.4	&	0.053	    & 21.1 	&	0.027	& 0.50 \\
P34		& 47.0	&	0.059	    & 29.4 	&	0.037	& 0.63 \\
P35		& 41.9	&	0.053	    & 12.9 	&	0.016	& 0.31 \\
P36		& 38.1	&	0.048	    & 16.9 	&	0.021	& 0.44 \\
P37		& 23.5	&	0.030	    & 26.0 	&	0.033	& 1.11 \\
P40		& 27.1	&	0.034	    & 12.8 	&	0.016	& 0.47 \\
P41		& 27.6	&	0.035	    & 8.0  	&	0.010	& 0.29 \\
P47		& 38.8	&	0.049	    & 37.1 	&	0.047	& 0.96 \\
P49		& 31.6	&	0.040	    & 14.3 	&	0.018	& 0.45 \\
P50		& 42.1	&	0.053	    & 23.4 	&	0.029	& 0.56 \\
P51		& 30.4	&	0.038	    & 10.0 	&	0.013	& 0.33 \\
P52		& 37.0	&	0.047	    & 13.3 	&	0.017	& 0.36 \\
P70		& 31.0	&	0.039	    & 10.2 	&	0.013	& 0.33 \\
P71		& 50.7	&	0.064	    & 24.0 	&	0.030	& 0.47 \\
P72		& 29.6	&	0.037	    & 11.4 	&	0.014	& 0.39 \\
P73		& 32.7	&	0.041	    & 16.9 	&	0.021	& 0.52 \\
P74		& 43.0	&	0.054	    & 21.0 	&	0.026	& 0.49 \\
P75		& 33.1	&	0.042	    & 22.9 	&	0.029	& 0.69 \\
P78		& 24.5	&	0.031	    & 13.0 	&	0.016	& 0.53 \\
P80		& 23.0	&	0.029	    & 9.9  	&	0.012	& 0.43 \\
\hline
Mean		& 35.6	&	0.045	    & 17.2 	&	0.022	& 0.49 \\
Median		& 33.1	&	0.042	    & 16.9 	&	0.021	& 0.46 \\
St.Dev.		& 7.5 	&	0.010	    & 6.9  	&	0.009	& 0.19 \\
\hline
\end{tabular}\\
\end{center}
\end{minipage}
\end{center}
\end{table}

\begin{table}
\begin{center}
\caption{Source sizes in W3.}
\label{tbl:W_source_sizes}
\begin{minipage}{\linewidth}
\begin{center}
\begin{tabular}{lccccc}
\hline
\hline
{W3}	& \multicolumn{2}{c}{\nh\ R$_{\mathrm{eff}}$ }	& \multicolumn{2}{c}{Submillimetre R$_{\mathrm{eff}}$} 	& \underline{Submillimetre R$_{\mathrm{eff}}$}	\\
{Source}	& {(\arcsec)}	&	 {(pc)}			& {(\arcsec)}	&	{(pc)}				& {\nh\ R$_{\mathrm{eff}}$ }			\\
\hline
W01		& 26.3  &	0.26	    & 10.5	&	0.10	& 0.40 \\
W02		& 28.6  &	0.28	    & 30.0	&	0.29	& 1.05 \\
W03		& 29.6  &	0.29	    & 19.6	&	0.19	& 0.66 \\
W06		& 32.7  &	0.32	    & 22.7	&	0.22	& 0.69 \\
W07		& 31.5  &	0.31	    & 17.7	&	0.17	& 0.56 \\
W08		& 35.7  &	0.35	    & 26.2	&	0.25	& 0.73 \\
W09		& 24.7  &	0.24	    & 17.4	&	0.17	& 0.70 \\
W10		& 27.8  &	0.27	    & 46.4	&	0.45	& 1.67 \\
W11		& 35.7  &	0.35	    & 31.4	&	0.30	& 0.88 \\
W12		& 22.3  &	0.22	    & 20.9	&	0.20	& 0.94 \\
W13		& 25.2  &	0.24	    & 13.8	&	0.13	& 0.55 \\
W14		& 19.2  &	0.19	    & 19.4	&	0.19	& 1.01 \\
W15		& 21.7  &	0.21	    & 13.9	&	0.13	& 0.64 \\
W17		& 18.3  &	0.18	    & 46.6	&	0.45	& 2.55 \\
W18		& 37.6  &	0.36	    & 35.1	&	0.34	& 0.93 \\
W19		& 34.6  &	0.34	    & 44.0	&	0.43	& 1.27 \\
W21		& 41.0  &	0.40	    & 21.0	&	0.20	& 0.51 \\
W22		& 36.9  &	0.36	    & 38.8	&	0.38	& 1.05 \\
W23		& 27.6  &	0.27	    & 29.2	&	0.28	& 1.06 \\
W25		& 45.8  &	0.44	    & 52.2	&	0.51	& 1.14 \\
W27		& 22.8  &	0.22	    & 12.6	&	0.12	& 0.55 \\
W28		& 23.8  &	0.23	    & 26.5	&	0.26	& 1.11 \\
W29		& 25.4  &	0.25	    & 31.0	&	0.30	& 1.22 \\
W30		& 37.4  &	0.36	    & 26.0	&	0.25	& 0.70 \\
W31		& 27.4  &	0.27	    & 13.3	&	0.13	& 0.49 \\
W33		& 52.6  &	0.51	    & 41.8	&	0.41	& 0.79 \\
W35/W38		& 30.6	&	0.30	    & 52.1	&	0.51	& 1.70 \\
W37		& 16.4  &	0.16	    & 18.4	&	0.18	& 1.12 \\
W39		& 36.6  &	0.35	    & 13.6	&	0.13	& 0.37 \\
W40		& 36.6  &	0.35	    & 40.2	&	0.39	& 1.10 \\
W43		& 9.8	&	0.10	    & 16.6	&	0.16	& 1.69 \\
W44		& 42.2  &	0.41	    & 30.0	&	0.29	& 0.71 \\
W45		& 21.7  &	0.21	    & 33.1	&	0.32	& 1.53 \\
W46		& 36.9  &	0.36	    & 11.4	&	0.11	& 0.31 \\
W48		& 54.7  &	0.53	    & 53.2	&	0.52	& 0.97 \\
W49		& 51.7  &	0.50	    & 55.2	&	0.54	& 1.07 \\
W50		& 54.8  &	0.53	    & 58.2	&	0.56	& 1.06 \\
W51		& 32.0  &	0.31	    & 44.4	&	0.43	& 1.39 \\
W52		& 37.9  &	0.37	    & 38.4	&	0.37	& 1.01 \\
W53		& 53.5  &	0.52	    & 7.6	&	0.07	& 0.14 \\
W54		& 23.5  &	0.23	    & 12.6	&	0.12	& 0.54 \\
\hline
Mean		& 32.5  &	0.31	& 29.1	&	0.28 	& 0.94 \\
Median		& 31.5  &	0.31	& 26.5  &	0.26	& 0.94 \\
St.Dev.		& 10.9  &	0.11	& 14.4  &	0.14	& 0.46 \\
\hline
\end{tabular}\\
\end{center}
\end{minipage}
\end{center}
\end{table}

\begin{table}
\begin{center}
\caption{Convolved and regridded source sizes in Perseus}
\label{tbl:conv_source_sizes}
\begin{minipage}{\linewidth}
\begin{center}
\begin{tabular}{lccccc}
\hline
\hline
{Perseus}	& \multicolumn{2}{c}{\nh\ R$_{\mathrm{eff}}$ }	& \multicolumn{2}{c}{Submillimetre R$_{\mathrm{eff}}$} 	& \underline{Submillimetre R$_{\mathrm{eff}}$}	\\
{Source}	& {(\arcsec)}	&	 {(pc)}			& {(\arcsec)}	&	{(pc)}				& {\nh\ R$_{\mathrm{eff}}$ }			\\
\hline
Map01		& 11.9	&	0.12	& 7.7  &    0.07    & 0.65 \\
Map02		& 22.8	&	0.22	& 10.3 &    0.10    & 0.45 \\
Map07		& 18.7	&	0.18	& 7.8  &    0.08    & 0.42 \\
Map08		& 23.1	&	0.22	& 6.0  &    0.06    & 0.26 \\
Map09		& 18.0	&	0.17	& 16.1 &    0.16    & 0.89 \\
Map10		& 10.4	&	0.10	& 8.9  &    0.09    & 0.86 \\
Map11		& 14.9	&	0.14	& 8.5  &    0.08    & 0.57 \\
Map12		& 18.3	&	0.18	& 10.5 &    0.10    & 0.57 \\
Map13		& 14.5	&	0.14	& 8.6  &    0.08    & 0.59 \\
P\_b\_1		& 16.0	&	0.16	& 9.6  &    0.09    & 0.60 \\
P\_b\_2		& 20.7	&	0.20	& 8.8  &    0.09    & 0.43 \\
\hline
Mean		& 17.2  &	0.17	& 9.4  &    0.09    & 0.57 \\
Median		& 18.0  &	0.17	& 8.8  &    0.09    & 0.57 \\
St.Dev		&  4.1  &	0.04	& 2.6  &    0.03    & 0.19 \\
\hline
\end{tabular}\\
\end{center}
\end{minipage}
\end{center}
\end{table}

\subsection{The Fractional Abundance of Ammonia}
\label{sec:Abundance}
  The fractional abundance of ammonia in protostellar clouds is a poorly defined quantity with estimates in specific regions varying from 7 $\times$ 10$^{-10}$ \citep{DiFrancesco2002} to $\sim$1 $\times$ 10$^{-8}$ \citep{Tafalla2004,Friesen2009}. More recent values averaged over wide-field samples and Galactic plane surveys favour higher abundances with averages of $\sim$4 $\times$ 10$^{-8}$ \citep{Dunham2011} and 1.2 $\times$ 10$^{-7}$ \citep{Wienen2012} found in single-pointing ammonia observations of the positions of peak surface brightness of sources detected in the thermal dust continuum at 1.1 mm (the Bolocam Galactic Plane Survey, \citealt{Aguirre2011}) and 870 \micron\ (the Apex Telescope Large Area Survey of the Galaxy, ATLASGAL, \citealt{Schuller2009}), respectively. The abundance has also been shown to be dependent upon the internal structure of cores in some case studies \citep{Tafalla2004}. The apparent variation of ammonia abundance, both within cores and from source-to-source, means that the ammonia-derived gas mass is not a good estimator of total core mass, since it requires independent knowledge of the NH3 abundance.

  We have calculated the fractional abundance of ammonia for each source, both as an integrated property as well as within each source individually. The imperfect correlation of ammonia and submillimetre column density, as noted above, means that fractional abundances may only be determined for certain portions of many cores.

  The abundance was calculated through the simple ratio of \nh\ column density to H$_2$ column density measured from submillimetre maps, where the latter were aligned and regridded to match the ammonia maps using a nearest neighbour interpolation. The H$_2$ column density is determined through the equation
\begin{equation}
\label{eq:SM_Ncol}
N(H_2) = S_{\nu}/\left[\Omega_m \mu m_H \kappa_{\nu} B_{\nu}(T_d)\right]
\end{equation}
where $S_{\nu}$ represents the 850 \micron\ flux density (evaluated per pixel), $\Omega_m$ is the main beam solid angle, $\mu$ is the mean molecular weight (taken as 2.29), $m_H$ is the mass of a hydrogen atom and $\kappa_{\nu}$ is the dust opacity per unit mass at 850 \micron, taken here and elsewhere as 0.01 cm$^2$g$^{-1}$, in agreement with the OH5 model of \citet{Ossenkopf1994}. $B_{\nu}(T_d)$ represents the Planck function evaluated at a dust temperature $T_d$. We have taken $T_d$ to be equal to the integrated \nh\ kinetic temperature over our sources. We are constrained in this choice mostly by a lack of options, a possible alternative being to simply select a single average dust temperature for all sources, a solution followed by many authors (e.g. \citealp{Johnstone2010,Moore2007,Mitchell2001}). We have used the approximation that $T_d \sim$ \Tk\ as, firstly, we observe little variation of \Tk\ over any given source (Morgan et al. $in$ $prep$). Secondly, the kinetic temperature of ammonia in star-forming sources has been found to be roughly equivalent to the dust temperature (see, e.g. \citealt{Morgan2010}). These findings indicate that there is likely less error in the $T_d \sim$ \Tk\ assumption than in the use of a single temperature for all sources.

\begin{figure*}
\begin{center} 
\includegraphics*[width=0.7\textwidth]{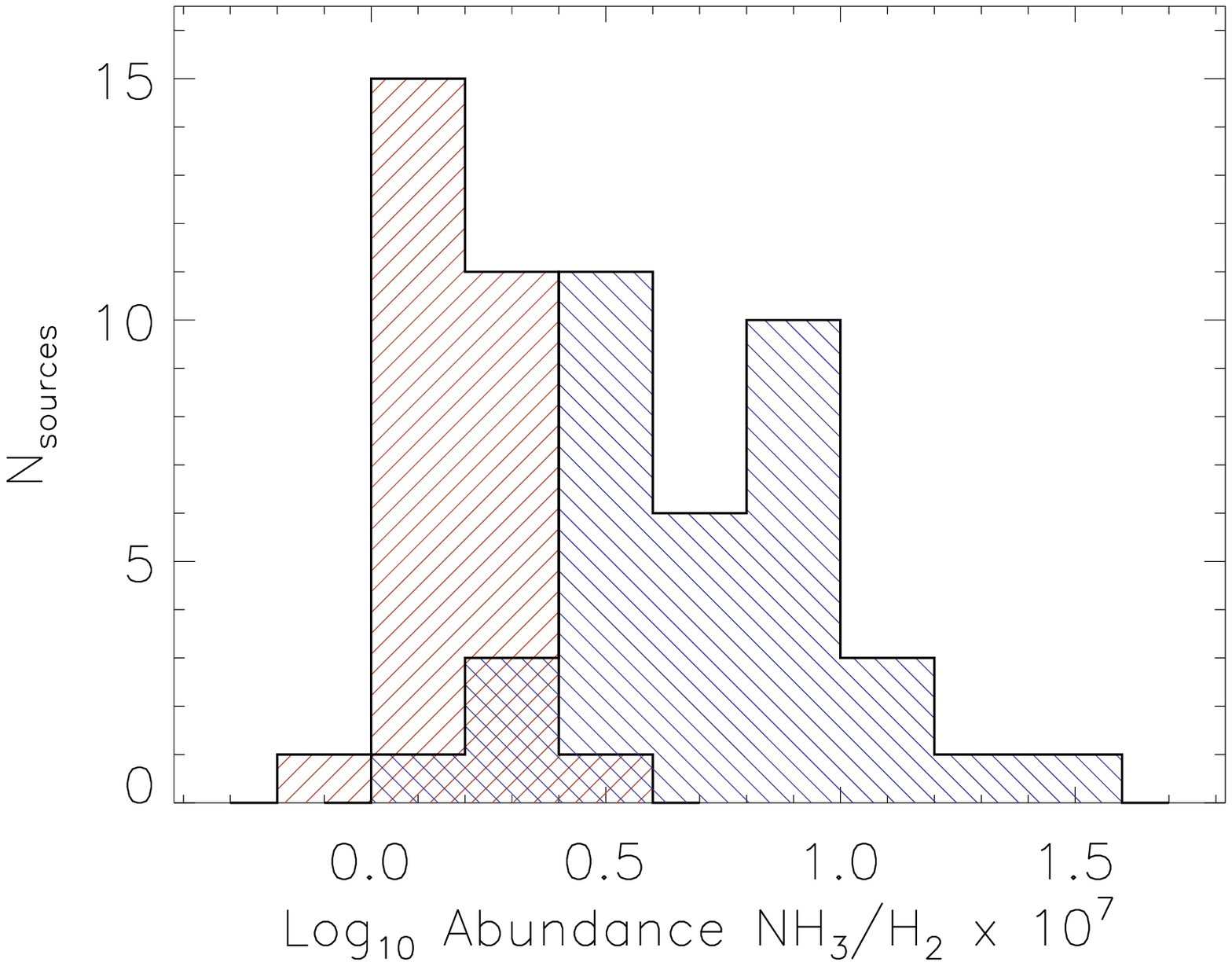}
\caption{\nh\ fractional abundances found in Perseus (red) and W3 (blue) cores.}
\label{fig:Abun_Hist}
\end{center}
\end{figure*}

 It should be noted that we have used the temperature-averaging method described in \citet{Morgan2012} to extend our maps of ammonia column density for maximal coverage of the submillimetre emission. We find that fractional abundances of ammonia, averaged over each source, range from 7.9 $\times$ 10$^{-8}$ to 2.7 $\times$ 10$^{-7}$ in our Perseus sample (mean = 1.6 $\times$ 10$^{-7}$) and from 1.1 $\times$ 10$^{-7}$ to 3.6 $\times$ 10$^{-6}$ in W3 (mean = 6.2 $\times$ 10$^{-7}$). These values are likely to be prone to a high degree of uncertainty due to e.g. the $\sim$50\% uncertainty in the estimated dust opacity, $\kappa_{\nu}$ \citep{Shirley2011}. We therefore estimate the uncertainty in abundance values at 50\%, although we recognise the arbitrary nature of this estimate and the possibility that uncertainties may in fact be even higher. The distributions of abundance values for both Perseus and W3 cores are shown in Figure \ref{fig:Abun_Hist}. Despite the large uncertainty in individual abundance values, there is a large difference in the spread of the abundance distributions. A KS test of the distributions returns a 100\% probability that the samples are drawn from separate populations.
 The standard deviation of abundance values in Perseus is typically 29\% of the mean while, in W3, it is 98\%. In addition, the variation $within$ individual cores is seen to be significantly higher in W3 than in Perseus with average standard deviations of 77\% and 39\% of the individual mean value respectively.

\begin{figure*}
\begin{center} 
\includegraphics*[width=0.7\textwidth]{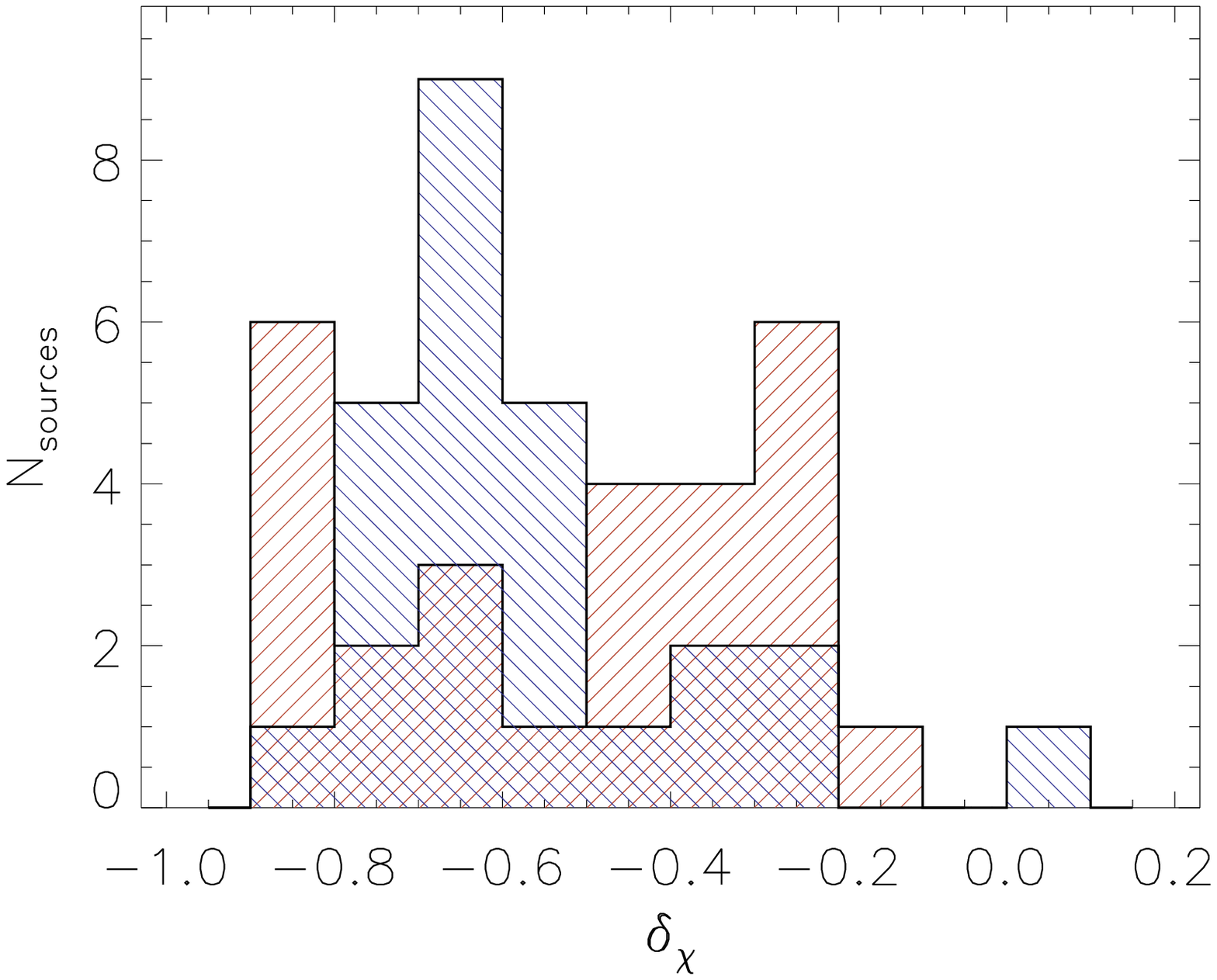}
\caption{Histogram plot of the difference in the fractional abundance at the submillimetre peak position compared to the mean found in Perseus (red) and W3 (blue) cores.}
\label{fig:Peak_Abun_Hist}
\end{center}
\end{figure*}

  It is common practice to estimate the fractional abundance of ammonia in star-forming regions through single-pointing observations directed at previously identified peaks of submillimetre emission (e.g. \citealt{Wienen2012,Urquhart2011,Dunham2011,Rosolowsky2008}). In order to test the assumption that measurements of ammonia abundance made at the position of dust emission peaks are representative of a given source, we compare the abundance found at the submillimetre peak position, $\chi_{sm}$, with the average value over the entire source, $\overline{\chi}$.  We define the quantity $\delta_{\chi} = \left[\chi_{sm}-\overline{\chi}\right]/\overline{\chi}$.
  Figure \ref{fig:Peak_Abun_Hist} shows the variation of $\delta_{\chi}$, the value of ammonia abundance found at the position of peak submillimetre emission is significantly lower than the mean in all but one case and is not representative of the region as a whole. On average, the Perseus and W3 samples show similar values of $\delta_{\chi}$ with mean differences of -52\% and -58\% respectively, i.e. values of \nh\ abundance, at the position of peak submillimetre emission, are approximately a factor of two lower than the average for a given source. However, Figure \ref{fig:Peak_Abun_Hist} shows that the distributions of $\delta_{\chi}$ are skewed with the median in Perseus ($\widetilde{\delta_{\chi}} = -48\%$) higher than the mean. Conversely, the median in W3 ($\widetilde{\delta_{\chi}} = -65\%$) is lower than the mean. These results echo those of \citet{Friesen2009} in which they find that dust clumps in the region of Ophiuchus, as identified by 850 \micron\ submillimetre emission, are associated with minima in the fractional abundance of ammonia.
  The possibility that this effect might be caused by the submillimetre peaks being coincident with dust temperature enhancements unresolved by the GBT beam was checked for by smoothing the submillimetre data to the resolution of the ammonia maps, resulting in similar results to the unsmoothed map analysis. Also, there is typically no evidence within our ammonia observations for kinetic temperature increases towards the peaks of submillimetre emission. Given the typical moderate-to-low optical depths of ammonia (1,1) line emission it seems unlikely that submillimetre emission is tracing temperature peaks that are unseen in our ammonia observations. However, it should be noted that the lack of correlation between submillimetre peak positions and ammonia temperature does not take into account any variation of temperature within the clouds along the line-of-sight. If the submillimetre emission really is tracing higher temperatures towards embedded cores, then it is clear that these higher temperatures are not traced by the ammonia emission. It is also possible that gas and dust temperatures have become decoupled at scales below that of our observations, although this is thought to be unlikely in what are probably the highest density regions of the cores, where log (n) $\sim$ 5-6 \citep{Myers1992}.

  It is particularly noteworthy that the median values of $\delta_{\chi}$ are comparable to the standard deviations found for fractional abundance within individual cores, indicating that the practice of taking the position of peak submillimetre emission as representative of an entire source when measuring fractional abundance will lead to a bias greater than 1$\sigma$ in a large proportion of cases.

\subsection{Virial Ratios}
\label{sec:VR}
\begin{figure*}
\begin{center}
\includegraphics*[width=0.7\textwidth]{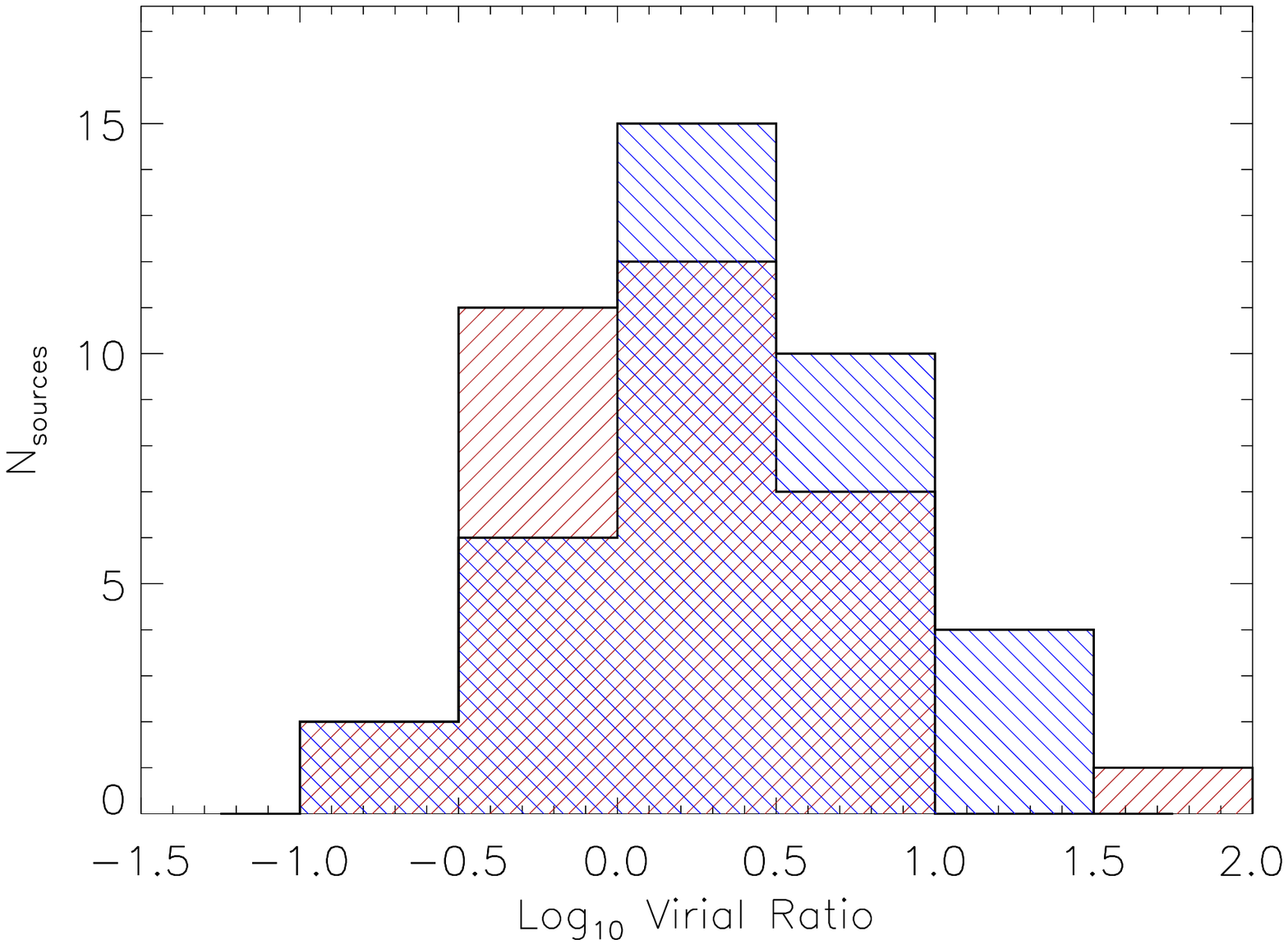}
\caption{Number distribution plot of the log virial ratios found in the Perseus (red) and W3 (blue) regions.}
\label{fig:VR_Hist}
\end{center}
\end{figure*}

In order to further examine the nature of the observed sources in W3 in comparison to those in Perseus, we calculated the virial ratios of each source for which the necessary parameters had been obtained. Assuming spherical geometry and constant density we use the formulation
\begin{equation}
M_v = \frac{5 R_c\ \sigma_{V_{tot}}^2}{\mathrm{G}}
\end{equation}
to determine the virial mass of each core, where $\mathrm{G}$ is the gravitational constant and $R_c$ is the submillimetre core radius as presented in Tables \ref{tbl:P_source_sizes} and \ref{tbl:W_source_sizes}. $\sigma_{V_{tot}}$ is the 1-D velocity dispersion integrated over the core in question, calculated by substituting the thermal contribution to the measured line width from ammonia with that assuming a mean molecular weight of 2.29 (i.e. $\sigma_{V_{tot}}^2$ = $\sigma_{V_{obs}}^2$-k$_\mathrm{B}T_\mathrm{k} \left[ \frac{1}{\mathrm{M_{NH_3}}} - \frac{1}{2.29 \mathrm{M_{H}}}\right]$). Having determined the virial mass of each core, we then calculated the virial ratio ($M_{\mathrm{observed}}/M_v$) using the observed mass, determined using the submillimetre fluxes published by \citet{Moore2007} and \citet{Hatchell2007} and the kinetic temperatures determined from our ammonia observations, the resulting virial ratio values are plotted as a histogram in Figure \ref{fig:VR_Hist}.

This approach requires the use of a mixture of \nh\ and submillimetre continuum data. The differing resolutions between these two types of observation in this study may introduce a bias in our values as larger physical regions are expected to exhibit greater turbulent linewidths \citep{Larson1981}. As our observations of sources in W3 necessarily include greater amounts of material, turbulent linewidths are expected to be larger, thus increasing virial masses - some further discussion of this problem is presented in Section \ref{sec:General_Results}. Further, while molecular line data are is necessary in order to determine velocity dispersion, submillimetre emission is our best tracer of core masses and radii, with the necessary assumption that they trace the same mass component. Since the column density match between \nh\ and submm continuum is not good in detail, this assumption is far from ideal and may introduce systematic biases or significant scatter into the results.  The alternative would be to use \nh-based masses and radii but this is likely to introduce even larger errors.  In particular, core mass estimates from \nh\ data require an assumed \nh\ abundance, average values of which may be uncertain by orders of magnitude, as we have seen.   Abundances in individual sources are more tightly constrained but are originally derived from submillimetre column densities.  We are therefore obliged to use submillimetre continuum masses since they are more fundamental and so the submillimetre-traced core radii are also appropriate.

  The results in Figures \ref{fig:VR_Hist} show significant and similar scatter in the virial ratios of our sources in both the Perseus and W3 regions. This scatter may be due in part to the uncertainties in determining virial ratio values. Estimates for submillimetre-derived mass require estimates of $\kappa_{\nu}$, the frequency-dependent dust opacity per unit mass. This is highly uncertain with estimates varying from 0.01 cm$^2$g$^{-1}$ \citep{Moore2007} to 0.02 cm$^2$g$^{-1}$ \citep{Thompson2004a}, even in similar regions. In addition to this, a significant uncertainty in mass ($\sim$ $\sqrt{2}$ \citealp{Morgan2008}) can be attributed to temperature estimates through the Planck function. Therefore, we estimate uncertainties in virial ratio to be as large as 50\%. However, even this uncertainty cannot account for the scatter seen in virial ratios. This scatter is likely to be real, reflecting the physical state of the cores.

  Median virial ratios for each region are 1.50 and 2.40 for Perseus and W3 respectively. Mean values are significantly higher at 3.50 and 4.50 which indicates a skew in both populations with tendencies to values above unity. The difference between the mean values is not large in comparison to the standard error at 1.46 and 0.90 for Perseus and W3 respectively. A KS test also shows no significant probability of the virial ratio values being drawn from separate populations at the 1 sigma level.

In the previous section, we described the effects on measured source sizes of convolving the Perseus data to lower resolution, in order to simulate observing it at the distance of W3. We have also examined the effect of this artifically reduced resolution on the apparent virial ratio for the eleven resulting convolved cores whose effective radii are listed in Table \ref{tbl:conv_source_sizes}. The results show that the effect of reduced resolution, and so of greater distance, is to increase the apparent virial ratio of sources. In this case, the aggregated Perseus region sample has a significantly higher median virial ratio (4.2) than the native resolution Perseus sample. The explanation of this shift appears to be that the increase in mass included within the larger effective beam is greater than the corresponding increase in velocity dispersion.

\section{Discussion}
\label{sec:Discussion}
\subsection{General Results}
\label{sec:General_Results}
We find reasonable agreement between our results for ammonia sources in Perseus and those of \citet{Rosolowsky2008} who completed a single-pointing survey of ammonia observations, mainly toward submillimetre peaks in the Perseus region. A comparison of integrated intensity measurements at the positions of sources listed in \citet{Rosolowsky2008} show that they are statistically indistinguishable. The median of the absolute difference between integrated intensity measurements of all matching sources is 13\%, consistent with expected uncertainties in our observational techniques.

 In terms of the \nh-derived source parameters, the average values and distributions show significant differences between the Perseus and W3 regions. In general terms these differences may be largely accounted for by the difference in distance between the two regions. Lower filling factors and optical depths are expected for regions at a greater distance due to the greater level of unresolved structure within our beam. Also, if core emission is extended, this may account for the differences seen in observed linewidth, with a greater amount of material being captured by the beam in W3 compared to Perseus, consistent with more sources being contained by the beam and/or a greater velocity dispersion associated with a greater scale of emission, as per the scaling relation of \citet{Larson1981}. The radii covered by the GBT beam correspond to 0.05 pc at the distance of Perseus and 0.31 pc at the distance of W3. This difference in physical scale means that there may be a distinct difference in the type of structure being traced in Perseus as compared to W3. However, it should be noted that the mean effective angular radii of our sources are approximately equal to our beam diameter in both Perseus and W3 (see Section \ref{sec:SM_NH}), with R$_{\mathrm{eff}}$ measurements of 35.6 and 32.5\arcsec\ respectively. This similarity in the angular scale of the objects in our observations indicates that we may be tracing objects following power law density profiles. For a power law density profile in a given object the R$_{\mathrm{eff}}$/FWHM ratio stays approximately constant between observations of varying resolution \citep{Young2003}, while for Gaussian-like or solid-disk density profiles this ratio will decrease approximately linearly. If the objects we are observing are hierarchical structures following a power law density distribution (such as the commonly cited singular isothermal sphere model, which has a radial profile following $\rho \propto r^{-2}$ at all radii \citealt{Shu1977}), then the determined size of our objects may primarily depend upon the resolution and sensitivity of our observations. This dependence of source size upon the spatial distribution of matter, i.e. the density profile, may then directly relate to the variation we see in source size, as discussed in Section \ref{sec:SM_NH}. The slope of the radial density profile may be important in moderately resolved sources \citep{Terebey1993} and source size may equally be an indicator of the age of a source or the total source mass \citep{Young2003}. However, as a full analysis of core density distributions and other structure parameters is beyond the scope of the current work, this will be discussed in a forthcoming publication (Morgan et al, $in$ $prep$). A final point to be noted here is that convolution of the physical parameter maps derived from the Perseus data to a resolution comparable to that of W3 indicates that differences in distance are not able to explain the difference in physical parameters completely, resulting clump sizes are significantly smaller than native resolution observations, indicating that there is some real difference in the dense structures in the two regions, with larger clusters of objects in W3 than in Perseus. Higher average filling factors in the Perseus sample indicate that we are resolving the structure within Perseus to a finer level than within W3.

  There are clear correlations between submillimetre and \nh\ detections (Section \ref{sec:SM_NH}), although correlations in brightnesses of the two tracers are imperfect. The reasons for this are not clear; being reliably optically thin, submillimetre continuum emission has uncomplicated radiative transfer and thus is often considered a definitive tracer of column density. Line emission, conversely, has relatively complex radiative transfer and excitation processes requiring knowledge of or assumptions about optical depth and the relative population of energy levels.

  83\% of the \nh\ sources in our combined W3 and Perseus catalogue, which have an associated submillimetre source, are found to contain at least one 22\micron\ source from the WISE catalogue \citep{Wright2010} within their 50\% flux contour. These mid-IR associations likely indicate the presence of embedded protostars within the larger regions traced by ammonia and submillimetre continuum emission. However, it should be noted that the majority are offset from the position of peak submillimetre emission, with only 40\% of the WISE associations falling within a JCMT beam size (14\arcsec) of the measured position of the submillimetre peak.

  The value of the fractional abundance of ammonia has been found to have a large range in the literature, with extreme values ranging over four orders of magnitude. Recent values over large self-consistent samples have considerable scatter within samples (see \citealt{Wienen2012,Dunham2011}) and, although there are well-established models which predict ammonia column density as a function of temperature (see \citealt{Morgan2012}), so far there is no well-established theory relating ammonia abundance with any physical parameter (e.g. density or temperature). Fractional abundance ratios found here by comparing integrated \nh\ and submillimetre-derived H$_2$ column densities are in a comparitively narrow range, within a range of a factor of two of the mean for sources in W3 and only 30\% in Perseus. These values are higher by a factor between two and ten than those generally found in the literature at $\mathbf{N_{\mathrm{NH_3}}/N_{\mathrm{H_2}}}$ = $\chi \sim$ 10$^{-7}$ (c.f. $\chi \sim$10$^{-8}$, \citealp{Tafalla2004,Friesen2009}), though recent results based over large survey samples and with comparable resolution to our observations \citep{Wienen2012,Dunham2011} are in better agreement. We find abundance ratios of ammonia to be considerably higher in W3 than in Perseus with a large degree of confidence, indicating that there is true variation of this metric from region to region. Given this, as well as the variation found within cores and the suggestion of an anti-correlation of submillimetre column density with ammonia abundance it is clear that determining \nh\ abundances using single-point data is not advisable. Also, when adopting abundance values to calculate total gas masses from NH3 data, we must take account of the size scale under examination.

\subsection{Association of Gas and Dust Emission}
\label{sec:gdass}
  In general, submillimetre sources are found to be associated with \nh\ emission. Submillimetre source peaks are found within \nh\ source boundaries and the detection rate of \nh\ associated with submillimetre cores is high. However, the correlation between the two tracers is not overly strong, there are many ($\sim$40\%) sources for which peaks in \nh\ and submillimetre emission do not coincide (c.f. the study of \citealt{Friesen2009} of the Ophiuchus region) and the overall contours of emission are poorly correlated in a significant number of sources. In addition, an anti-correlation exists between peaks of the column density measured through submillimetre emission and the fractional abundance of ammonia (see Section \ref{sec:Abundance}) with the anti-correlation stronger on average in our W3 sample than for sources in Perseus. Relative abundances at the position of maximum column density are consistently lower than the average over the whole core for most of our sources, implying that the \nh\ traced column distributions are flatter than the submm ones, even when smoothed to a matching resolution. The relationship of fractional abundance of ammonia to submillimetre column density is complex as, not only may the dust opacity factor ($\kappa_{\nu}$) inherent in the derivation of submillimetre column density vary from region-to-region, it may actually vary considerably within regions.

  The number of sources which have a weak correlation between their position of emission peaks for submillimetre continuum and ammonia emission has a possible explanation in terms of temperature gradients. Derived column densities derived from submillimetre and ammonia emission depend in opposite ways upon the assumed temperature, with a direct dependence in the ammonia case and an inverse relationship for the continuum (for a fixed intensity). Column density measurements made from submillimetre observations are dependent upon temperature estimates through the Planck function and are therefore quite strongly subject to errors in the assumed T. However, measurements of the error in column density compared to errors in T indicate that the relationship is linear for temperatures around the mean found in our observations (13.14 K and 18.50 K for Perseus and W3 respectively), with a 20\% error in T translating to 29\% and 20\% errors in column density for Perseus and W3, respectively.
   
  The lack of detailed correlation (and particularly the dips seen in ammonia fractional abundance towards submillimetre peaks) may be explained by undetected peaks in the temperature distribution. While a full analysis of the spatial variation of various physical parameters measured from our ammonia observations (temperature, linewidth, etc.) is left for a separate study, kinetic temperature distributions generally show minima at the submillimetre continuum emission peaks, rather than maxima.  Any undetected local temperature rises must therefore be negligible on the scales probed by the current data. Also, the 22 \micron\ WISE sources which we observe in and around our sources, which we might expect to trace internally heated protostellar cores, show considerable positional scatter around both the submillimetre and ammonia \Tk\ peaks. This indicates that, while the presence of the WISE source is likely to indicate a peaked temperature distribution, this is not traced by either our ammonia observations or the thermal emission at the scale at which we can resolve that emission. We suggest that the 22 \micron\ sources are likely tracing more evolved star formation than the submillimetre and/or ammonia emission.

   Our results imply that \nh\ emission and the submillimetre continuum do not reliably trace the same mass component in detail, suggesting variations in excitation conditions (i.e. temperature or density) on scales smaller than the beam or deviations from LTE within cores. Explicitly, an inequality of dust and gas temperatures invalidates the assumption of thermal coupling between the gas and dust, while mis-assigned temperatures create disproportionate errors in gas/dust associations. This is likely to be particularly relevant when determining fractional abundances of ammonia. In the case that densities in the observed regions are lower than the critical density of ammonia ($\sim$4 $\times$ 10$^3$ cm$^{-3}$, \citealt{Maret2009}) and \nh\ is sub-thermally excited then we may expect \nh\ emission to be fainter than predicted in a LTE analysis and so low density sources may be responsible for the submillimetre sources we find with no corresponding ammonia detection.

There are multiple chemical effects which potentially have a bearing on our study. Firstly, the \nh\ molecule is, like all molecules, subject to depletion in dense, cold gas by freezing out of the gas phase onto the surfaces of dust grains. This has the effect of significantly altering the abundance in a somewhat unpredictable way, though nitrogen-bearing molecules such as \nh\ are usually predicted by models to be slow to freeze out, even at high densities \citep{Charnley1997,Bergin1997}. Secondly, the presence (or absence) of CO in star-forming cores has a significant effect on less abundant species \citep{Lee2004}, including ammonia due to the interaction between freezeout and evaporation of the molecules themselves and CO. The competition between `freeze-out' and CO evaporation in an internally heated core is significant when determining the abundance of N$_{2}$H$^{+}$ and \nh. However, these are largely confined to a small radius (see \citealt{Lee2004}). Recent models which invoke episodic accretion provide an estimate of the zone of CO evaporation and ammonia destruction \citep{Kim2012}. Using spectral energy distribution determinations for protostellar cores in Perseus \citep{Dunham2013} we find a median extinction corrected bolometric luminosity of 1.20 \lsun\ (with a range of 0.05 to 37 \lsun ). From figure 3 of \citet{Kim2012} we can use these values to estimate the radius at which cores in Perseus reach the evaporation temperature of CO ($\sim$25 K, \citealt{Lee2004}) - this is approximately 500-600 AU for a 1.2 \lsun\ core, increasing to $\sim$2000 AU for the maximum luminosity of 37 \lsun\ which was determined. These radii are considerably below the resolution of our ammonia observations with a GBT FWHM beam size of $\sim$8300 AU at the distance of the Perseus molecular cloud.
 
While there are physical reasons relating to temperature gradients and protostellar chemistry why we may not observe strong correlations between ammonia and submillimetre continuum emission peaks and why there is a prevalent anti-correlation of the fractional abundance of ammonia toward submillimetre emission peak, these are hard to reconcile with the observations that we have presented, potentially due to the physical scale which these observations are sampling.

\subsection{Virial Ratio}
\label{sec:VR_Discussion}
In Section \ref{sec:VR} we described the determination of virial ratios for each region. The distribution of virial ratios in the two samples is similar, though with a lower median in Perseus than in W3. There are a significant number of sources with values of virial ratios indicating unbound objects, these may represent cores which have formed stars and started to be disrupted, or where the bound part of the core has collapsed and disappeared from the gas and dust tracers but its gravitational potential energy remains.

Differences seen in virial ratio values between the two regions in our sample may be understood in terms of a structural difference. The spatial resolution of our ammonia observations (0.04 pc and 0.29 pc for Perseus and W3, respectively) is such that we may well be selecting sources which will result in the formation of multiple stars and, while we have generally described our sources as `cores' (the precursors of single stars or small binary systems), the presence of multiple 22 \micron\ sources indicates that they might be more correctly referred to as `clumps' (overdense regions out of which stellar clusters form - \citealt{McKee2007}).Further, analysis of our ammonia maps shows several sources that exhibit filamentary structure (e.g. P35 and W11). This is likely to have stronger repercussions for our analysis of the W3 region than for Perseus, given the larger spatial scale we are examining in W3. We have attempted to account for this through the convolution of our Perseus observations to mimic the distance of the W3 region, resulting in higher virial ratios for our convolved sources than even the native resolution W3 observations. This implies that the value of virial ratio found for a given source is a function of spatial resolution. A more physical interpretation is that star-forming clumps are bound on large scales but cores are pressure-confined on smaller scales although these possibilities are not mutually exclusive.

We have suggested several times in this work that an inherent difference in the physical structure of the Perseus and W3 regions may be responsible for some of the observed differences we have described (see Section \ref{sec:General_Results}). \citet{Myers2013} presents a model in which low-mass cores draw mass from inner filament gas in their formation, while higher-mass cores draw from the more extended filament gas. Assuming that our observations are of an appropriate scale in both regions, this model would provide some explanation for our results in that, for Perseus, virial ratios are based upon linewidths taken from the more extended gas which we do not expect to form part of the final stellar product. In W3, this more extended material is part of the bound structure, possibly resulting in higher virial ratio values (though this might be expected to be offset by higher velocity dispersions). For this interpretation to be correct, we expect that the submillimetre observations are tracing the compact `core' structure in both regions, while the ammonia traces the more extended, filamentary material.

\section{Summary and Conclusions}
\label{sec:Conclusion}
This study was designed to assess the way in which observable parameters associated with gas and dust emission are related, a further goal was to compare those results between a region of high-mass star formation and a region of low-mass star formation. In terms of source brightness, our observations of ammonia in the Perseus and W3 star-forming regions are consistent with previous observations, with differences explainable by measurement error. 

There are clear differences in the measured physical parameters of dense cores in the Perseus and W3 samples. Most of these differences can be attributed to resolution effects introduced by the considerably larger distance of the W3 cloud.  For example, lower filling factors in W3 indicate additional structure within the beam and larger linewidths are expected from larger scale emission. Several parameter differences are not so easily explained this way, however.  Source sizes in the Perseus data are significantly smaller than in W3, even when convolved to compensate for distance, indicating real differences in source structure between the two regions, with individual cores more likely to be traced within Perseus and larger clusters of objects or filaments being traced in W3.

A general correlation of submillimetre and ammonia emission has led to many studies using the two tracers in combination to derive, e.g. \nh\ fractional abundances and virial ratios.  However, we have shown that the correlation is imperfect in detail, with significant scatter in the positions of peak surface brightness and an anti-correlation of submillimetre-traced column density and differential \nh\ abundance within individual cores. This shows that a full understanding of the relationship of ammonia to the cold dust component traced by submillimetre emission is yet to be attained and that mapping of \nh\ emission is preferable to single-pointing measurements for determination of integrated source physical parameters.

  Although a comparison of ammonia sources to submillimetre emission reveals that sensitivity limitations may be responsible for certain sources being detected in one tracer but not the other, the lack of detailed correlation between the emission features of \nh\ and submillimetre continuum may also indicate variations in \nh\ abundance or excitation conditions within star-forming cores. For example, we see that ammonia temperature dips towards the position of peak submillimetre emission, indicating that the two tracers are not necessarily thermally coupled and that ammonia may well be cooled through efficient cooling in regions of high density. While ongoing star formation is indicated in the general environment of our catalogue through the presence of 22 \micron\ IR sources, these are only coincident with peaks of submillimetre emission in a minority. Overall, our source catalogue then appears to consist primarily of prestellar or early protostellar cores which are likely still undergoing infall.
  The anti-correlation between \nh\ abundance and submillimetre-traced column density within cores varies in significance and slope between the Perseus and W3 regions, with a stronger correlation and steeper slope in Perseus. Depletion models of ammonia do not describe a situation easily applied to our observations and this phenomenon requires further study.

Virial ratios were found to be higher in the W3 sources than for sources in Perseus.  However, we also find that, by artificially placing Perseus cores at an equivalent distance to W3 by degrading the resolution, the virial ratios are increased.  These results may indicate a selection effect, but may also indicate a difference in the underlying structure and physics of the two star-forming regions, dependent upon their relative mass.\\

\noindent \textit{Acknowledgments}\\
The authors would like to thank Neal Evans and Erik Rosolowsky for insightful and thorough comments on this work which have considerably improved its quality.
The authors acknowledge the data analysis facilities provided by the Starlink Project under continual development by the JAC. In addition, the following Starlink packages have been used: Kappa, Cupid, Gaia, Convert and Coco.
We would like to thank the helpful staff of the Green Bank Telescope in the collection of data used in this paper related to the project GBT10C\_024. LKM is supported by a STFC postdoctoral grant (ST/G001847/1) and DJE is supported by a STFC PhD studentship. This research would not have been possible without the SIMBAD astronomical database service operated at CDS, Strasbourg, France and the NASA Astrophysics Data System Bibliographic Services.

\appendix
\section{Online Material}
\begin{figure*}
\begin{center}
\includegraphics*[width=0.4\textwidth]{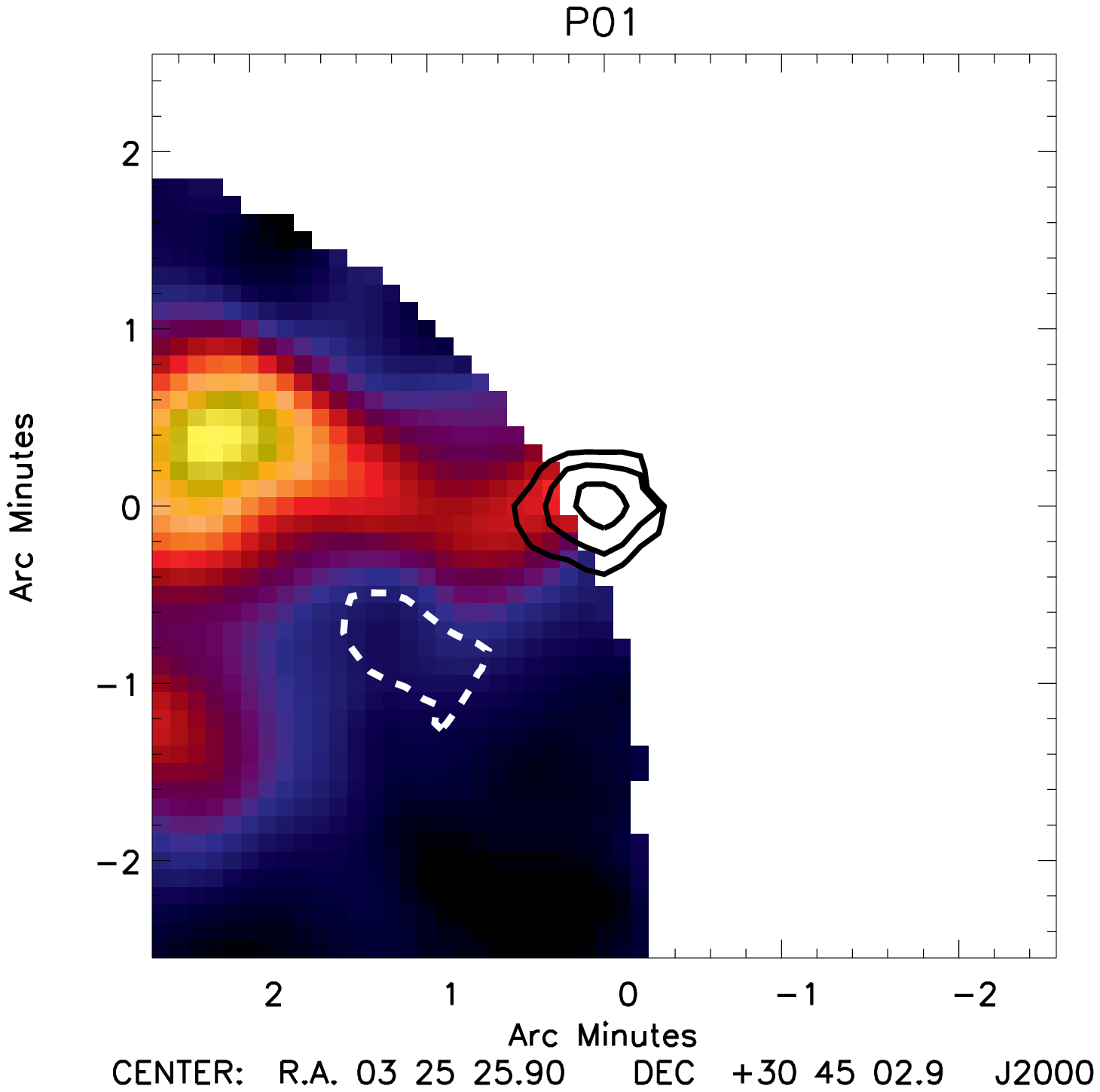}
\includegraphics*[width=0.4\textwidth]{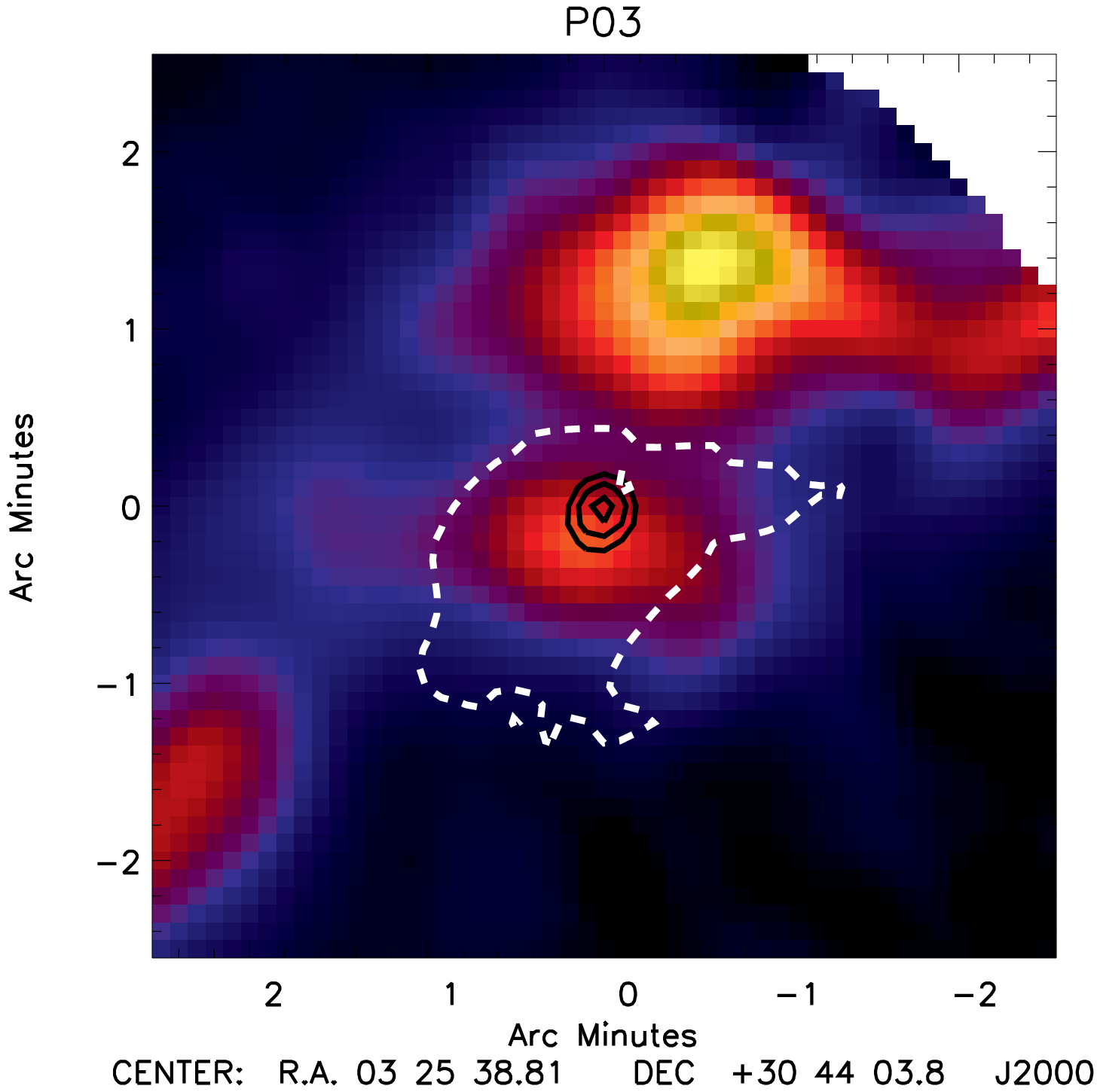}\\
\includegraphics*[width=0.4\textwidth]{P_05_SM_NH3_Conv_Image.eps}
\includegraphics*[width=0.4\textwidth]{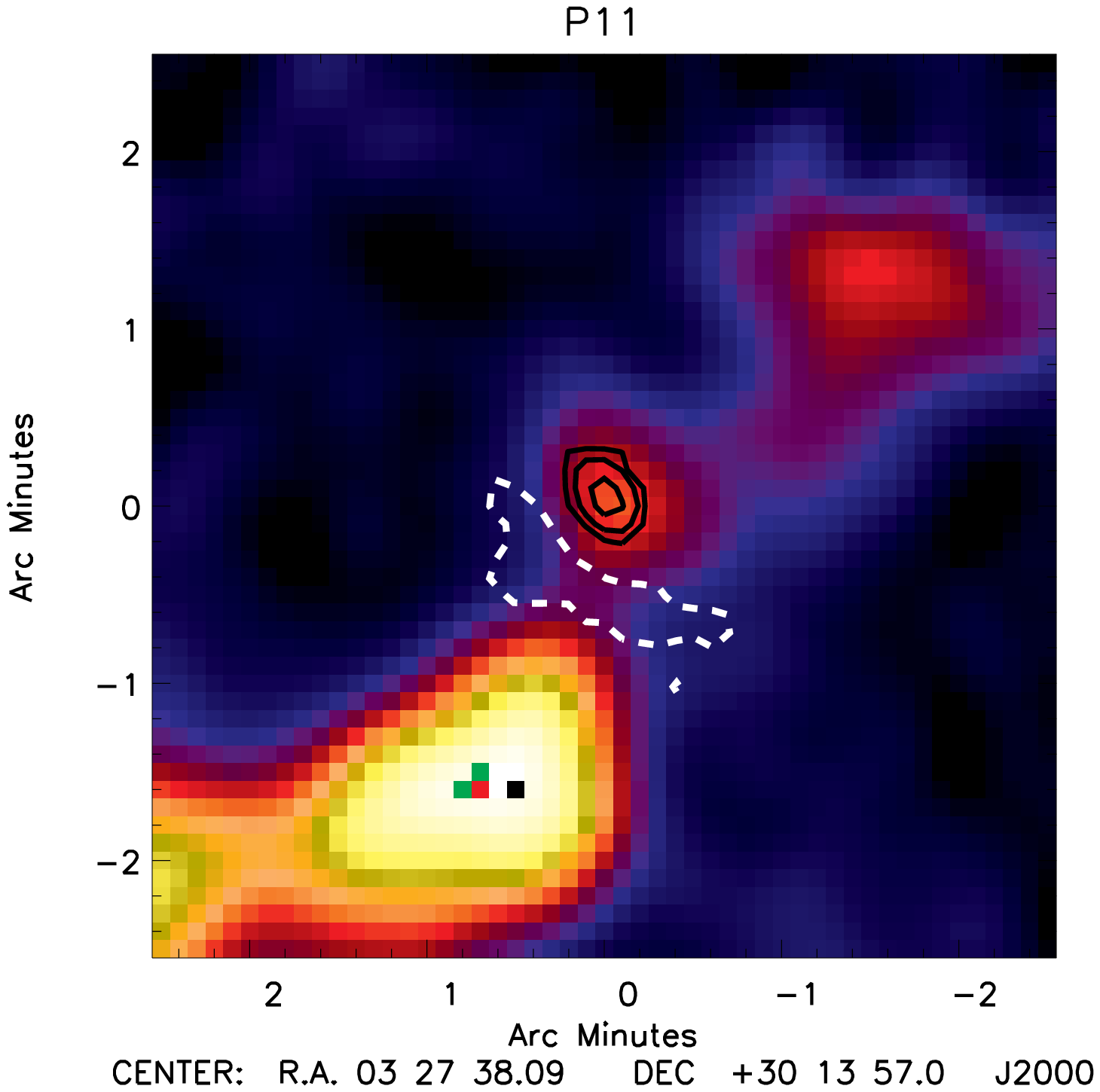}\\
\includegraphics*[width=0.4\textwidth]{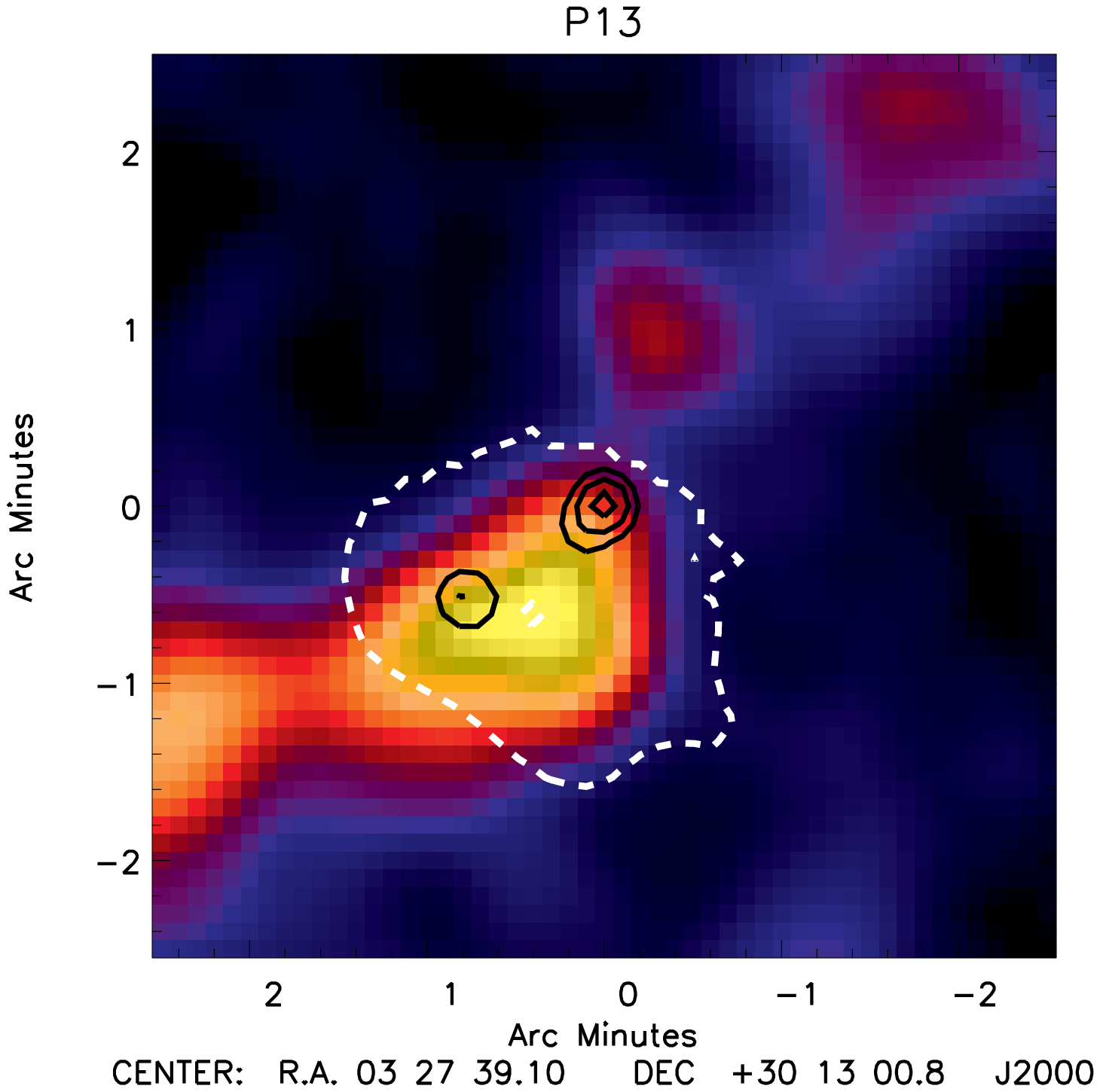}
\includegraphics*[width=0.4\textwidth]{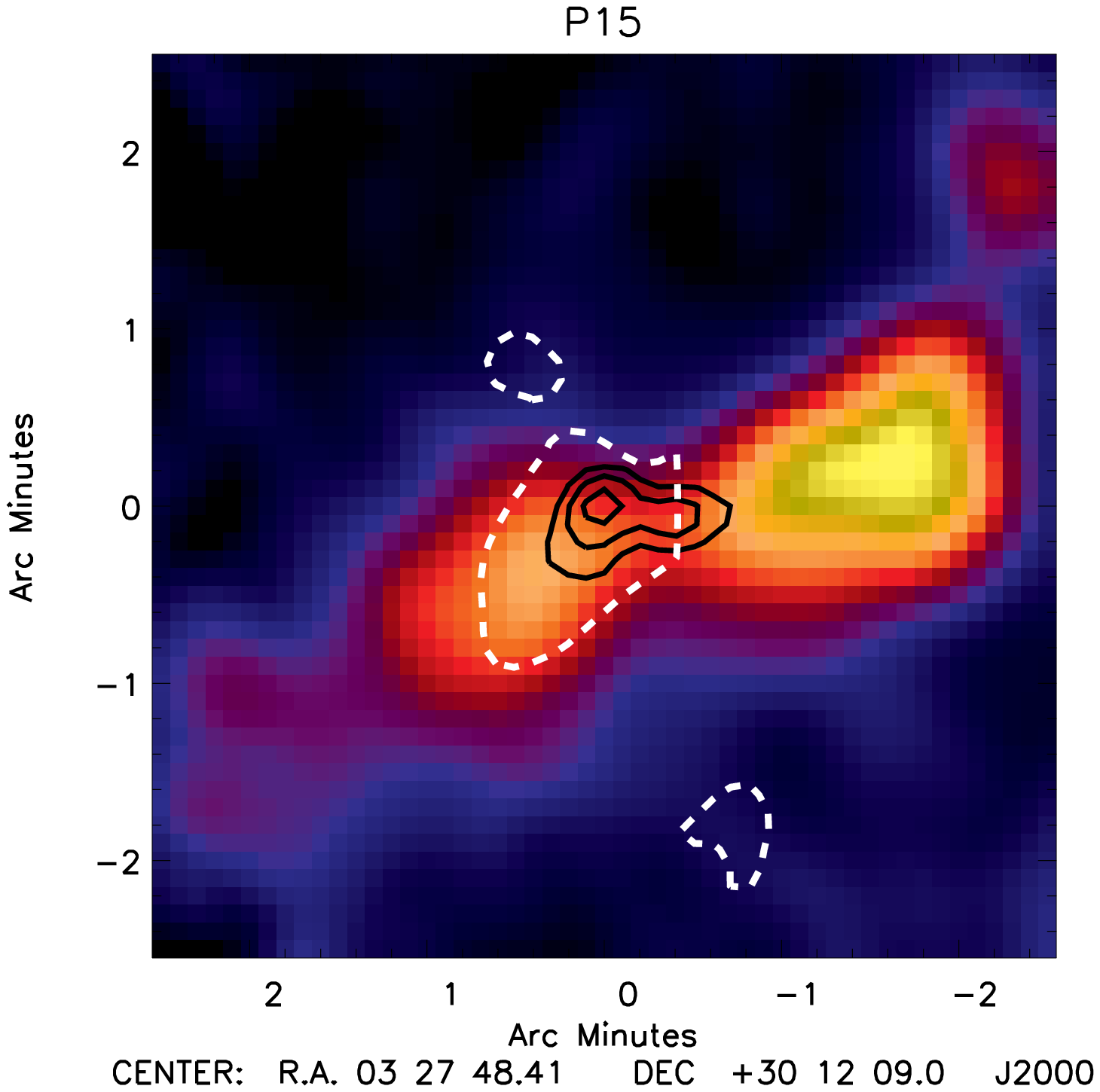}\\
\caption{\nh\ integrated intensity in sources P01, P03, P05, P11, P13 \& P15, overlaid with black contours of submillimetre emission at 50, 70 and 90\% of the peak flux density. A white, broken contour traces the 50\% level of the \nh\ column density distribution.}
\label{fig:All_Sources}
\end{center}
\end{figure*}
\addtocounter{figure}{-1}

\begin{figure*}
\begin{center} 
\includegraphics*[width=0.4\textwidth]{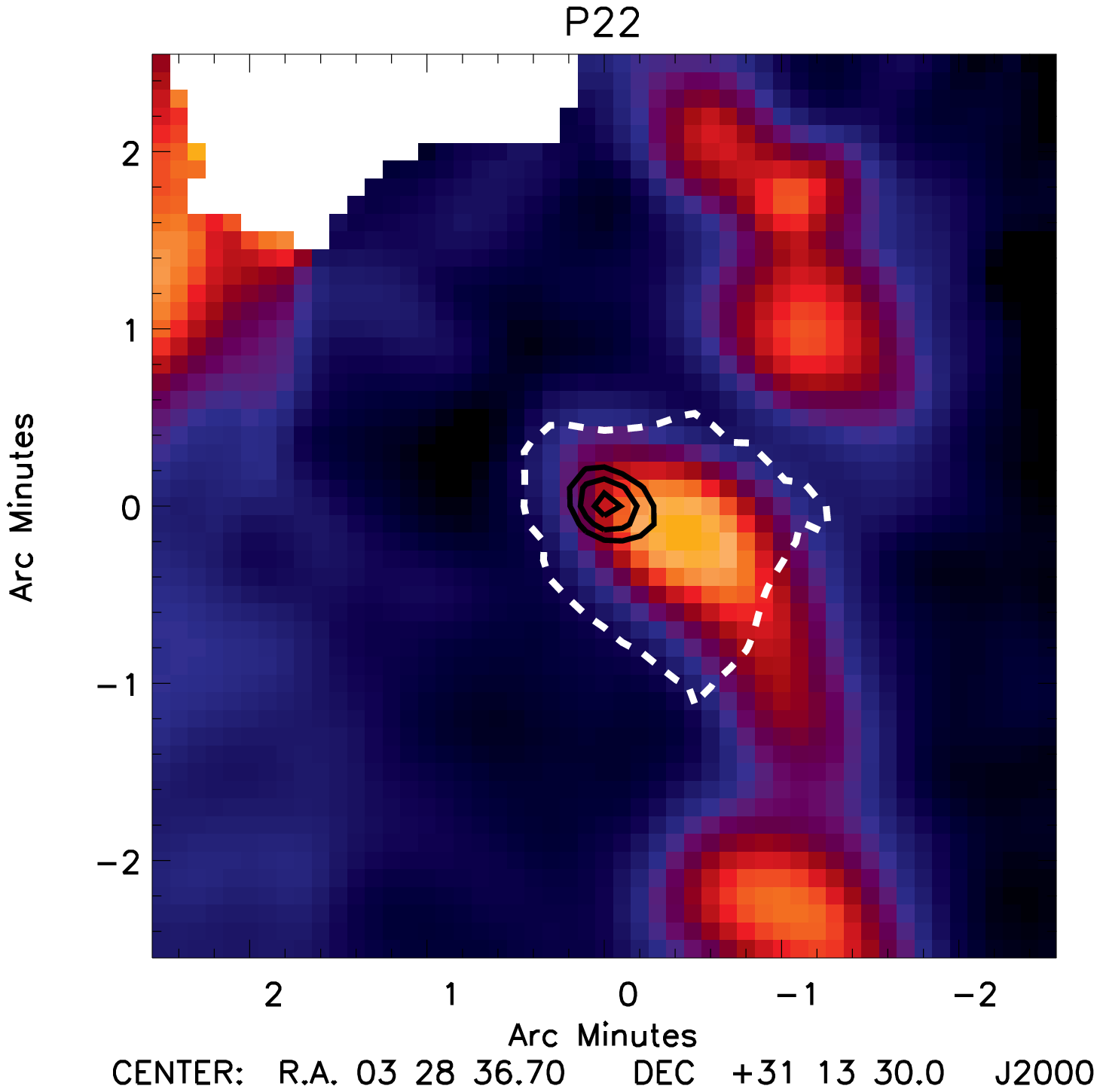}
\includegraphics*[width=0.4\textwidth]{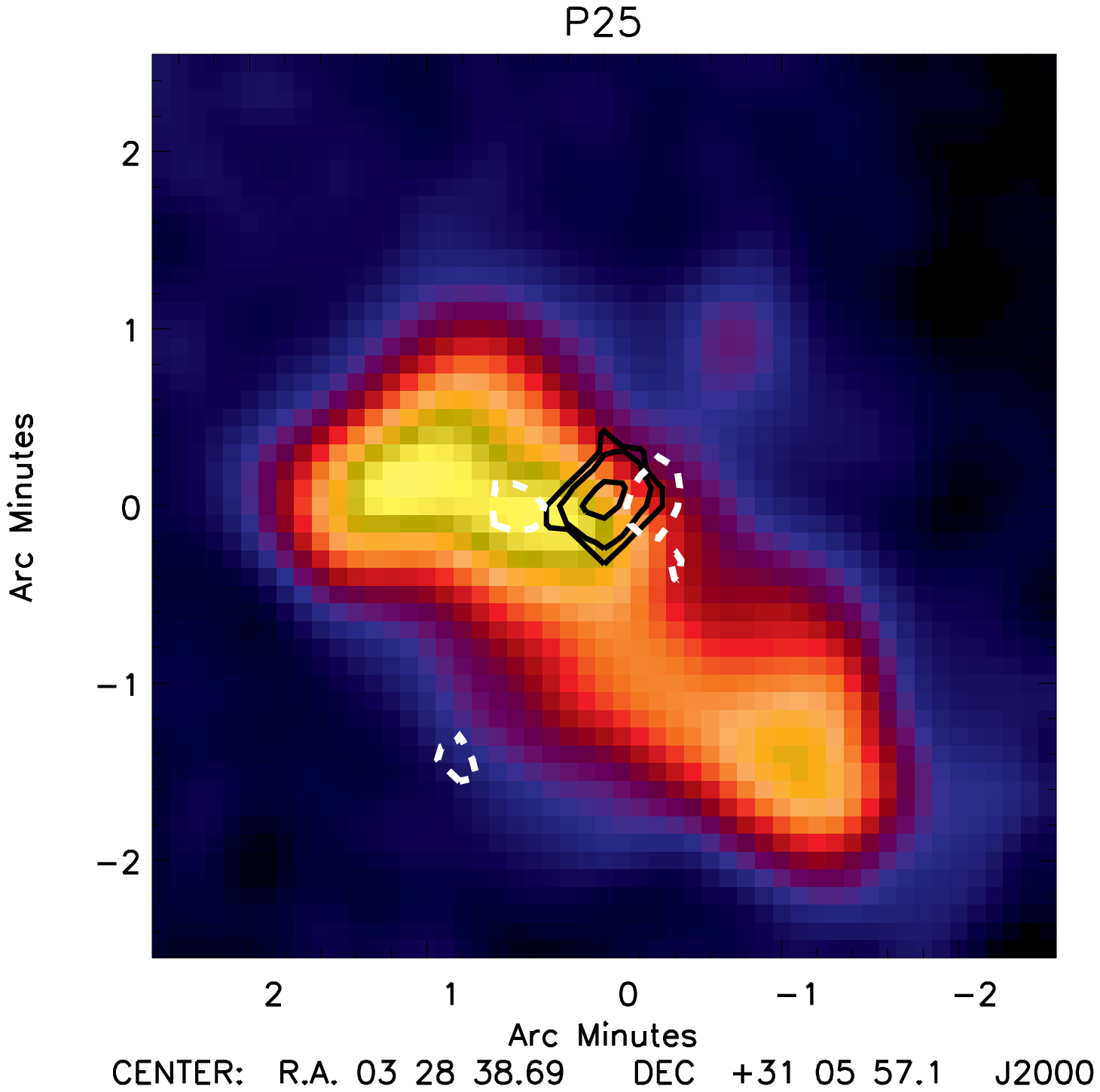}\\
\includegraphics*[width=0.4\textwidth]{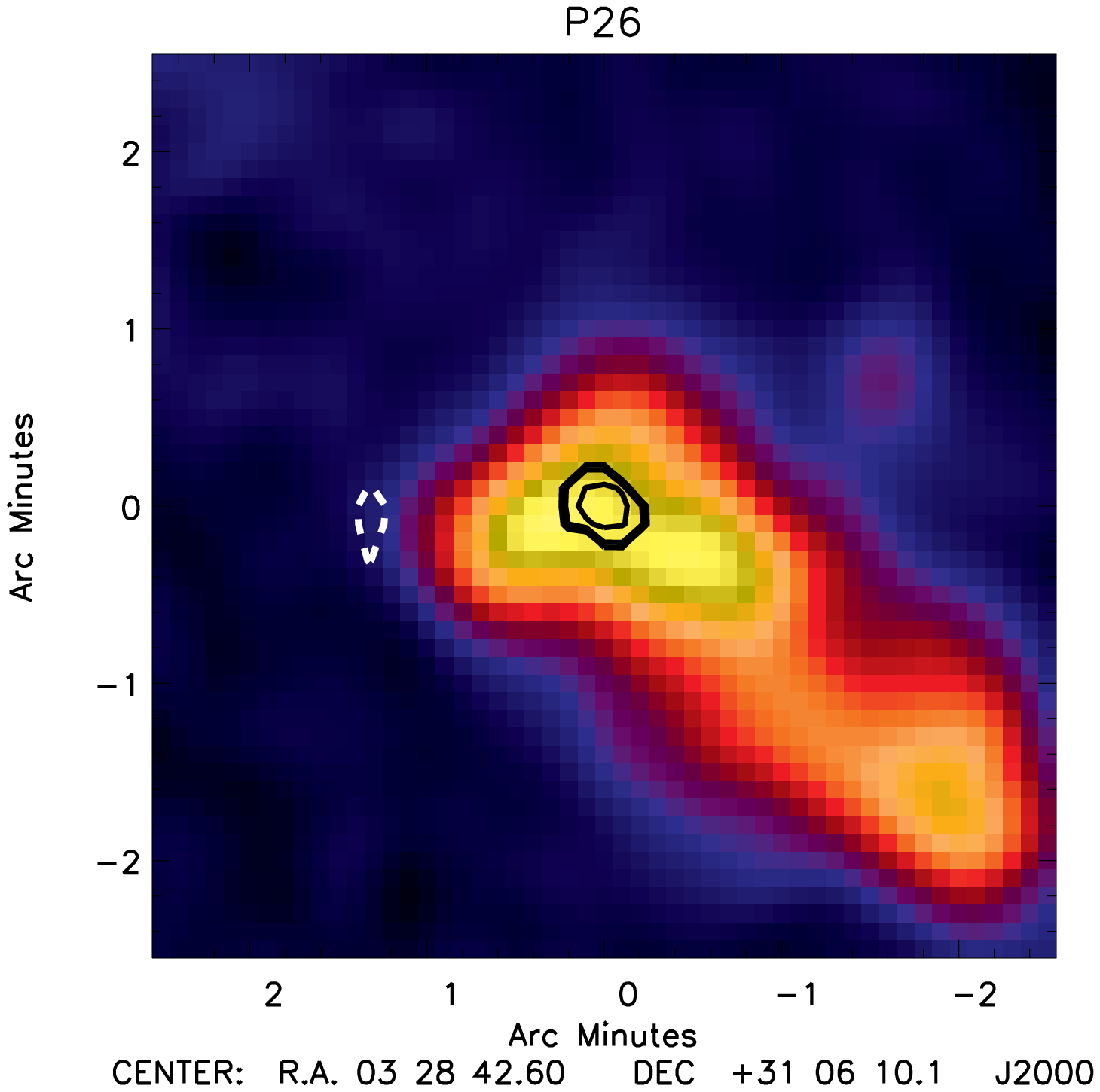}
\includegraphics*[width=0.4\textwidth]{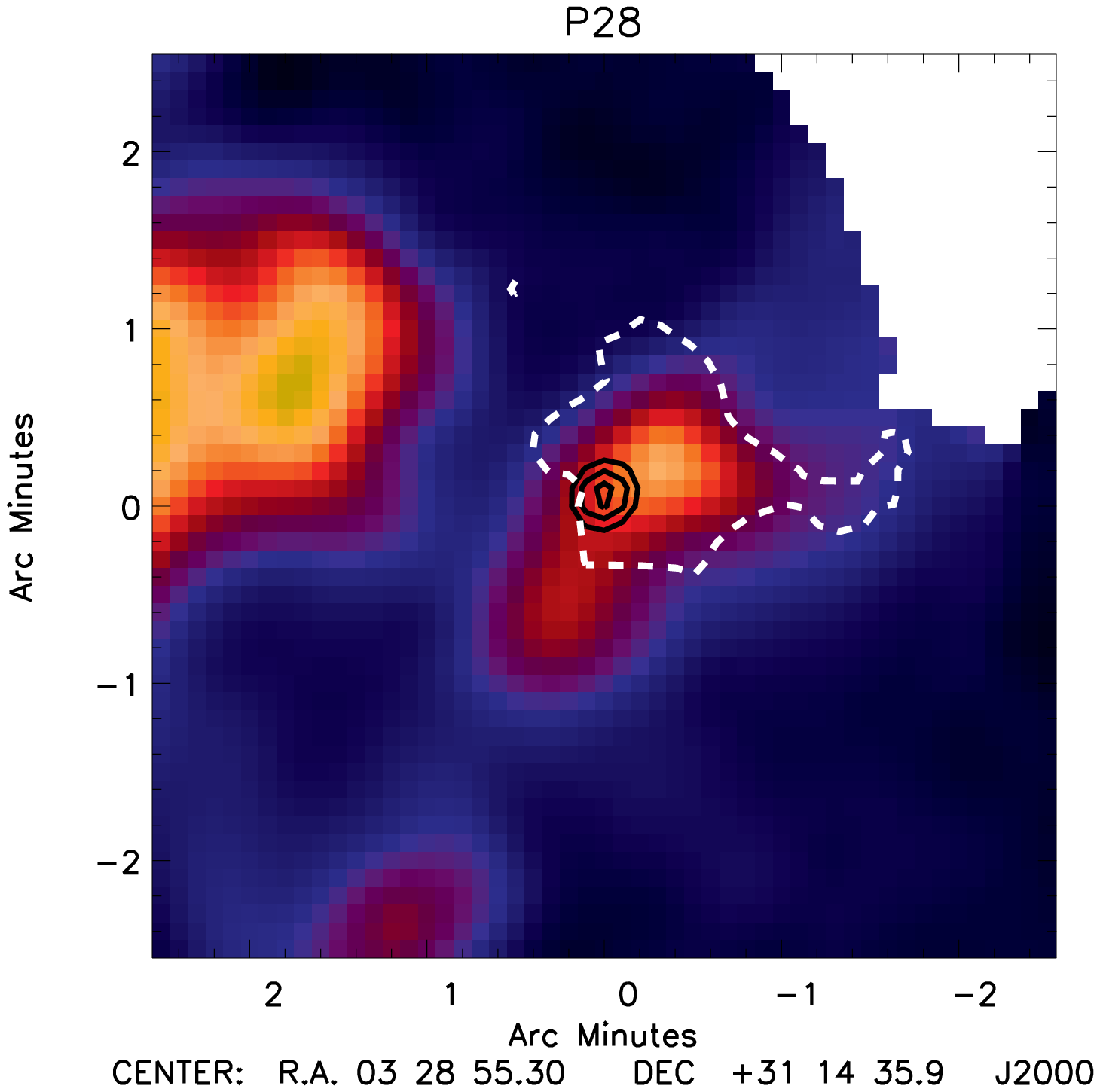}\\
\includegraphics*[width=0.4\textwidth]{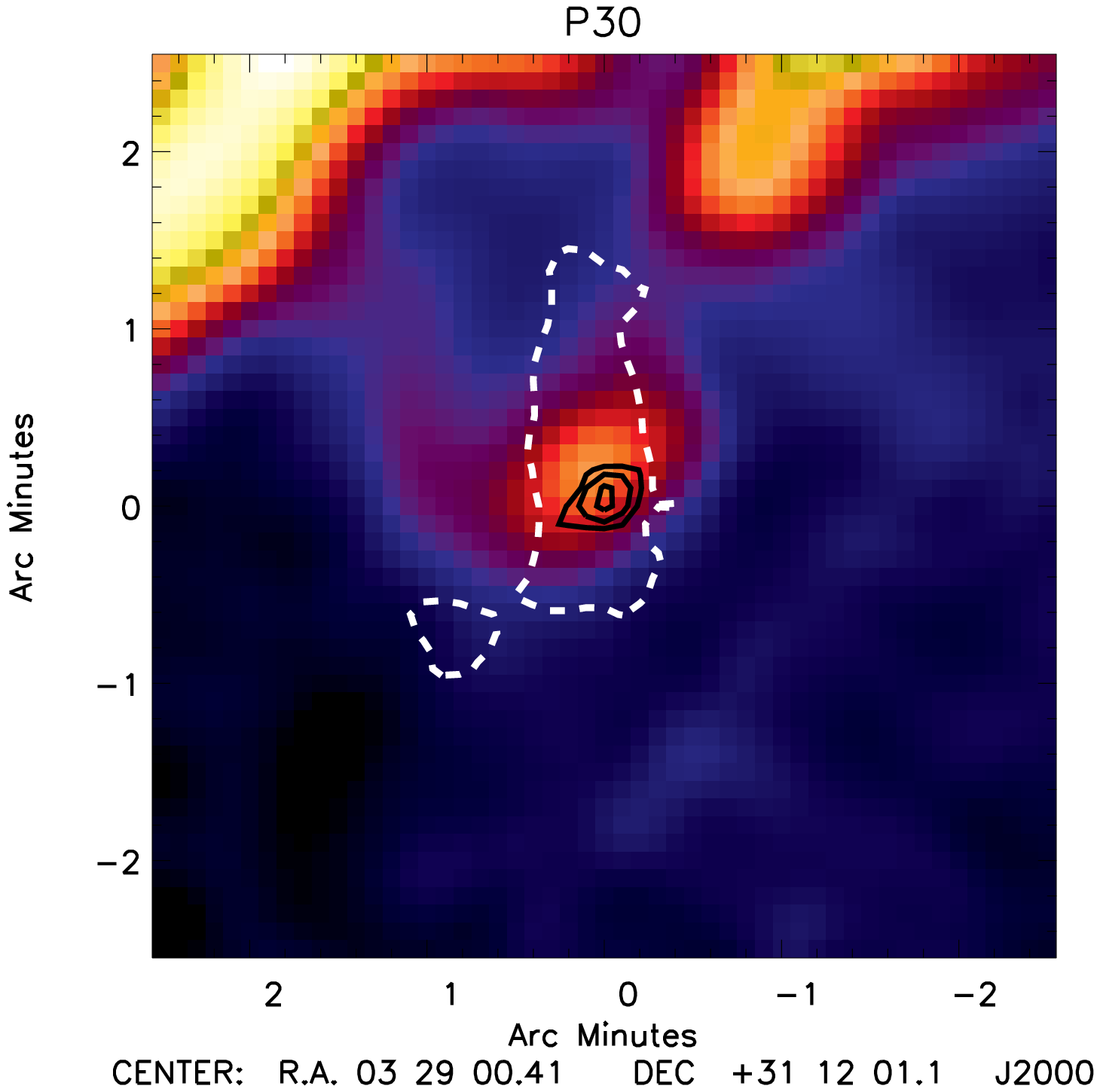}
\includegraphics*[width=0.4\textwidth]{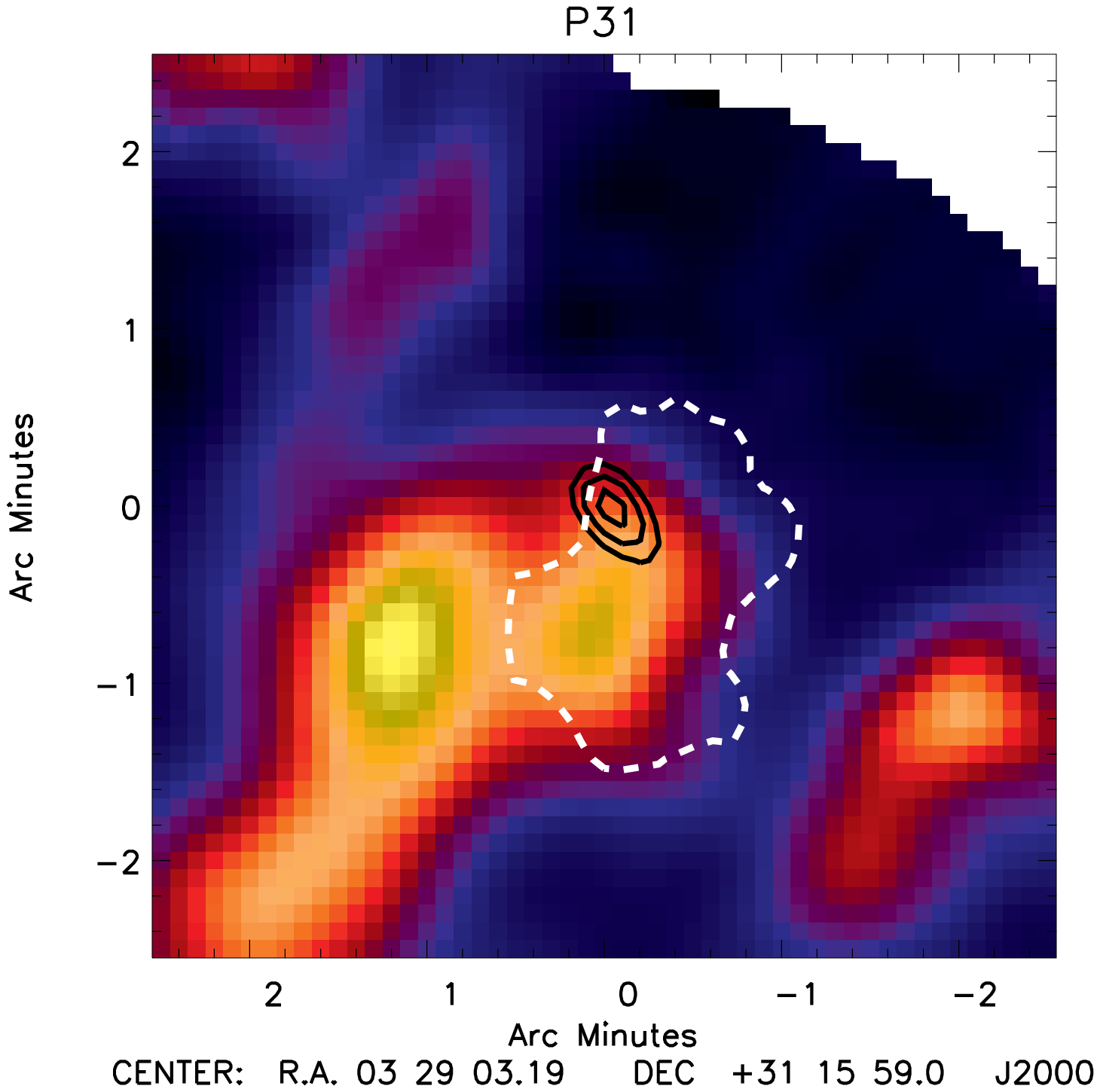}\\
\caption{\nh\ integrated intensity in sources P22, P25, P26, P28, P30 \& P31, overlaid with black contours of submillimetre emission at 50, 70 and 90\% of the peak flux density. A white, broken contour traces the 50\% level of the \nh\ column density distribution.}
\end{center}
\end{figure*}
\addtocounter{figure}{-1}

\begin{figure*}
\begin{center} 
\includegraphics*[width=0.4\textwidth]{P_33_SM_NH3_Conv_Image.eps}
\includegraphics*[width=0.4\textwidth]{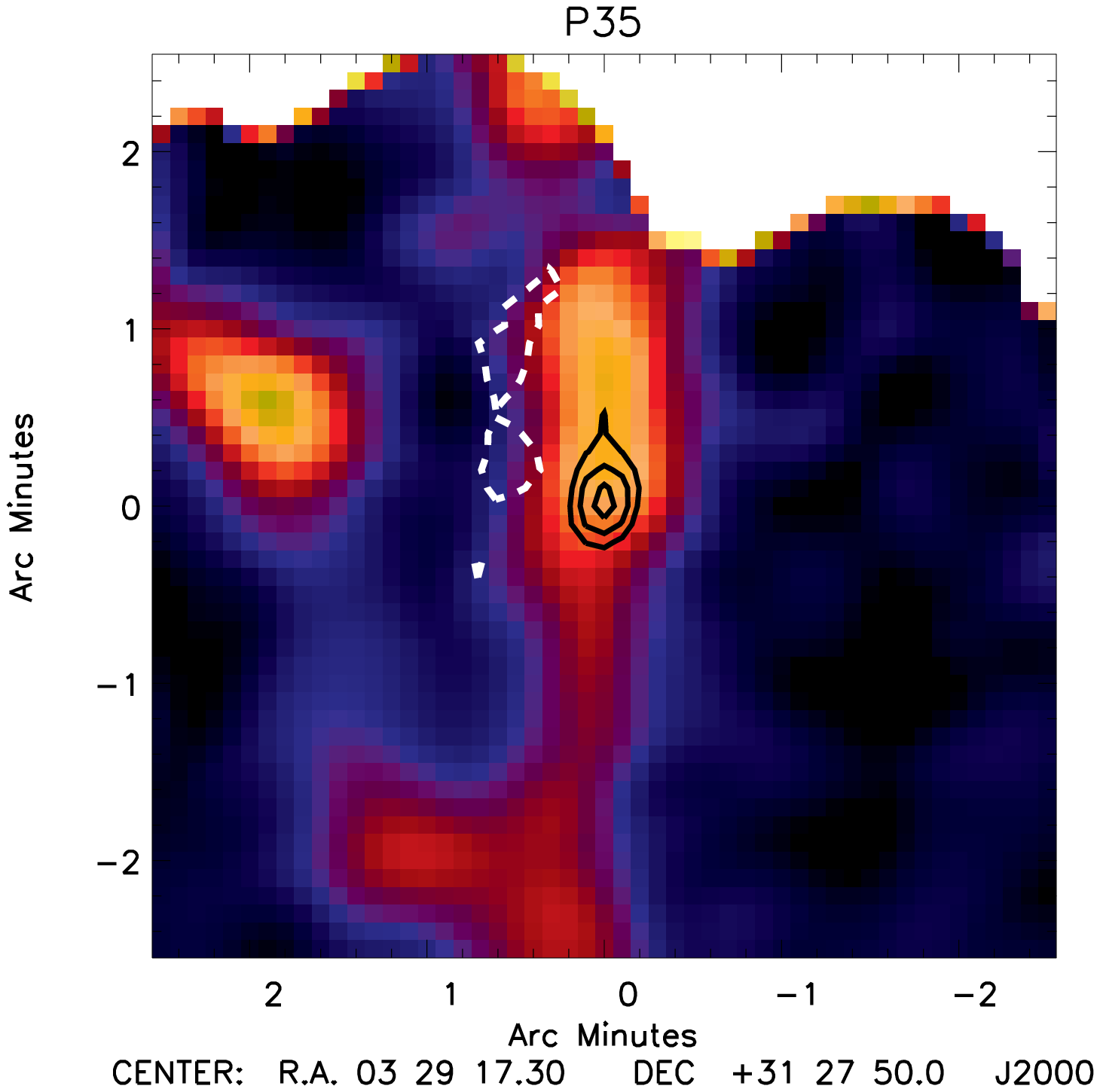}\\
\includegraphics*[width=0.4\textwidth]{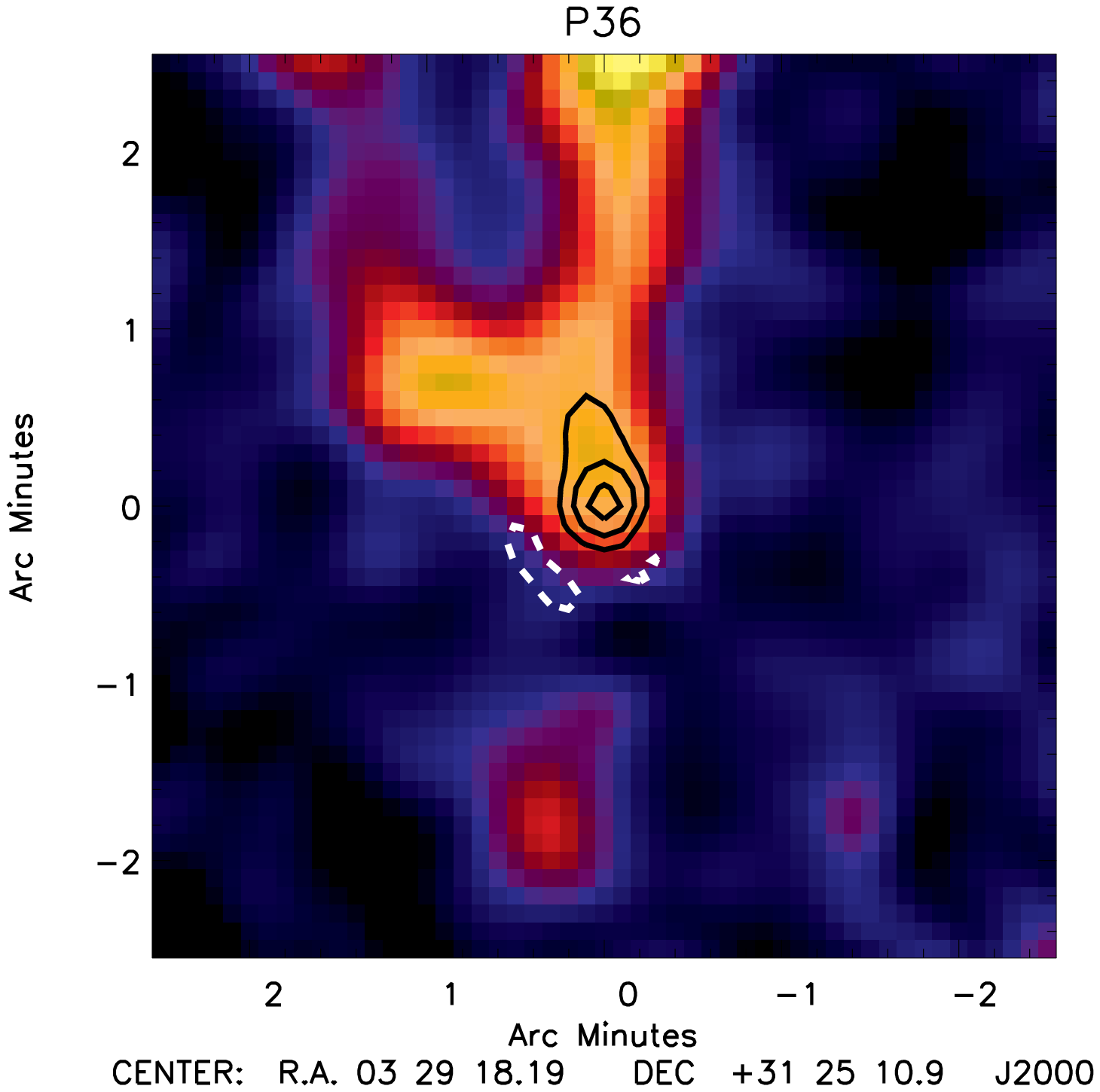}
\includegraphics*[width=0.4\textwidth]{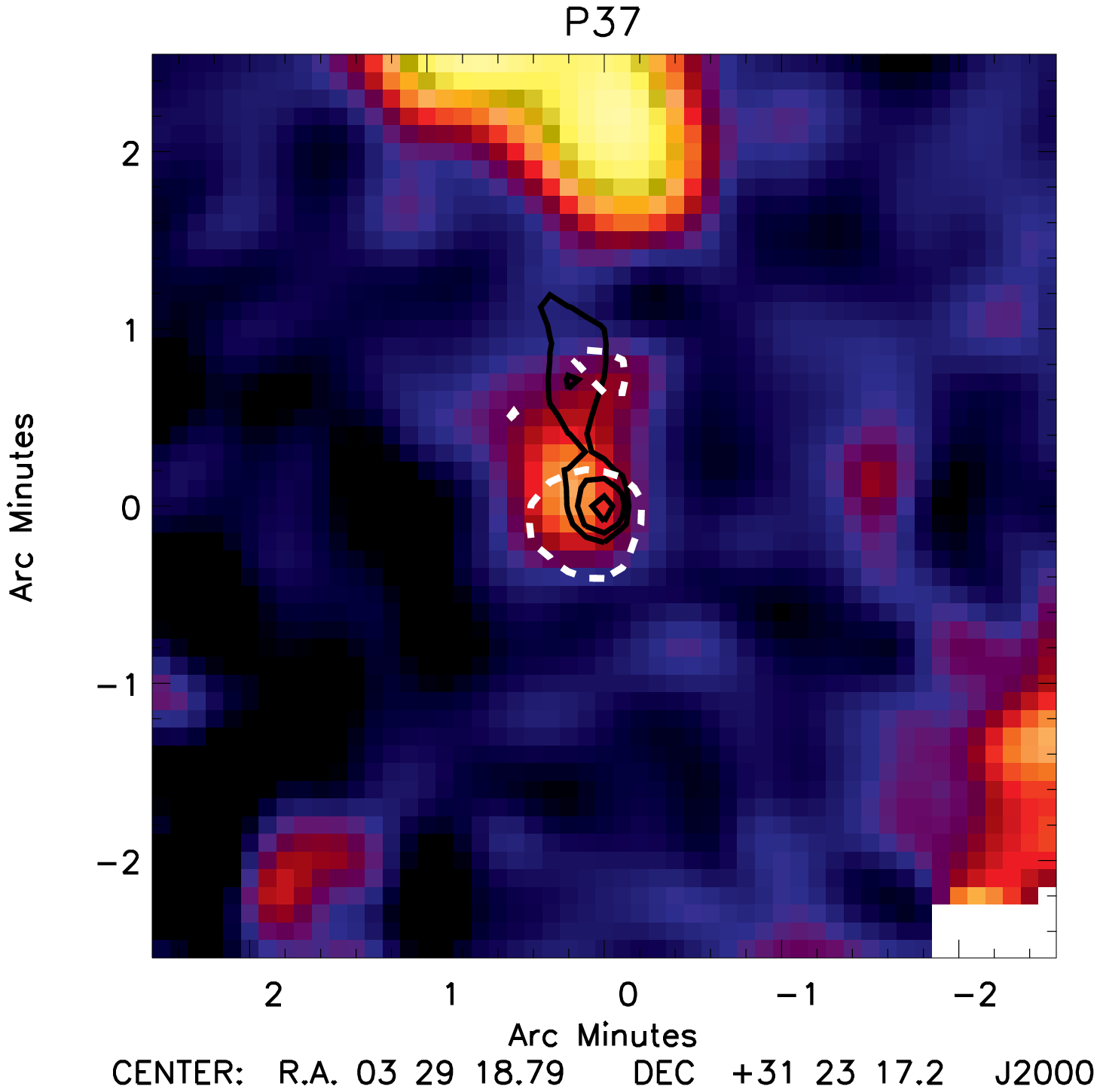}\\
\includegraphics*[width=0.4\textwidth]{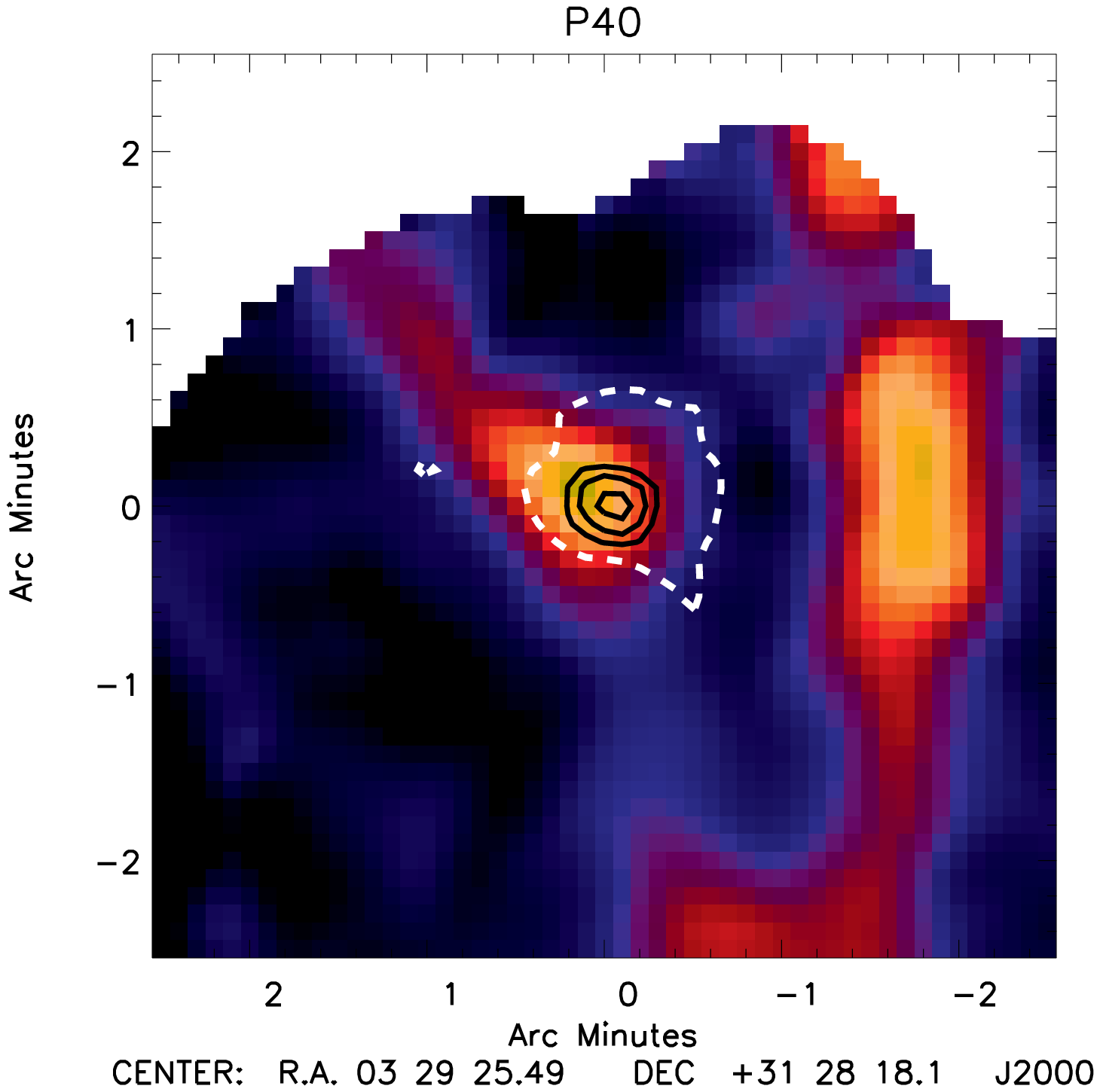}
\includegraphics*[width=0.4\textwidth]{P_41_SM_NH3_Conv_Image.eps}\\
\caption{\nh\ integrated intensity in sources P33, P35, P36, P37, P40 \& P41, overlaid with black contours of submillimetre emission at 50, 70 and 90\% of the peak flux density. A white, broken contour traces the 50\% level of the \nh\ column density distribution.}
\end{center}
\end{figure*}
\addtocounter{figure}{-1}

\begin{figure*}
\begin{center} 
\includegraphics*[width=0.4\textwidth]{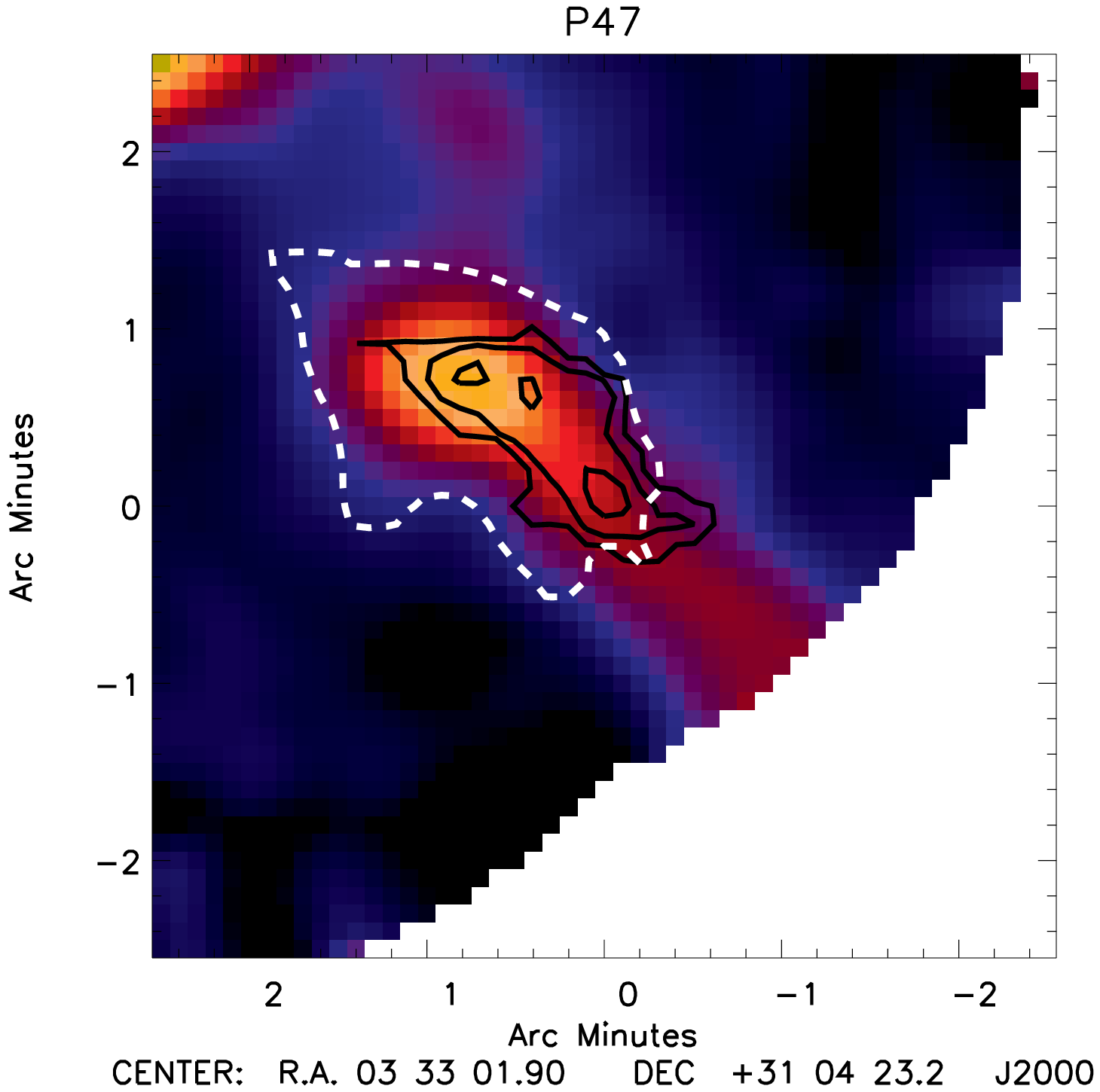}
\includegraphics*[width=0.4\textwidth]{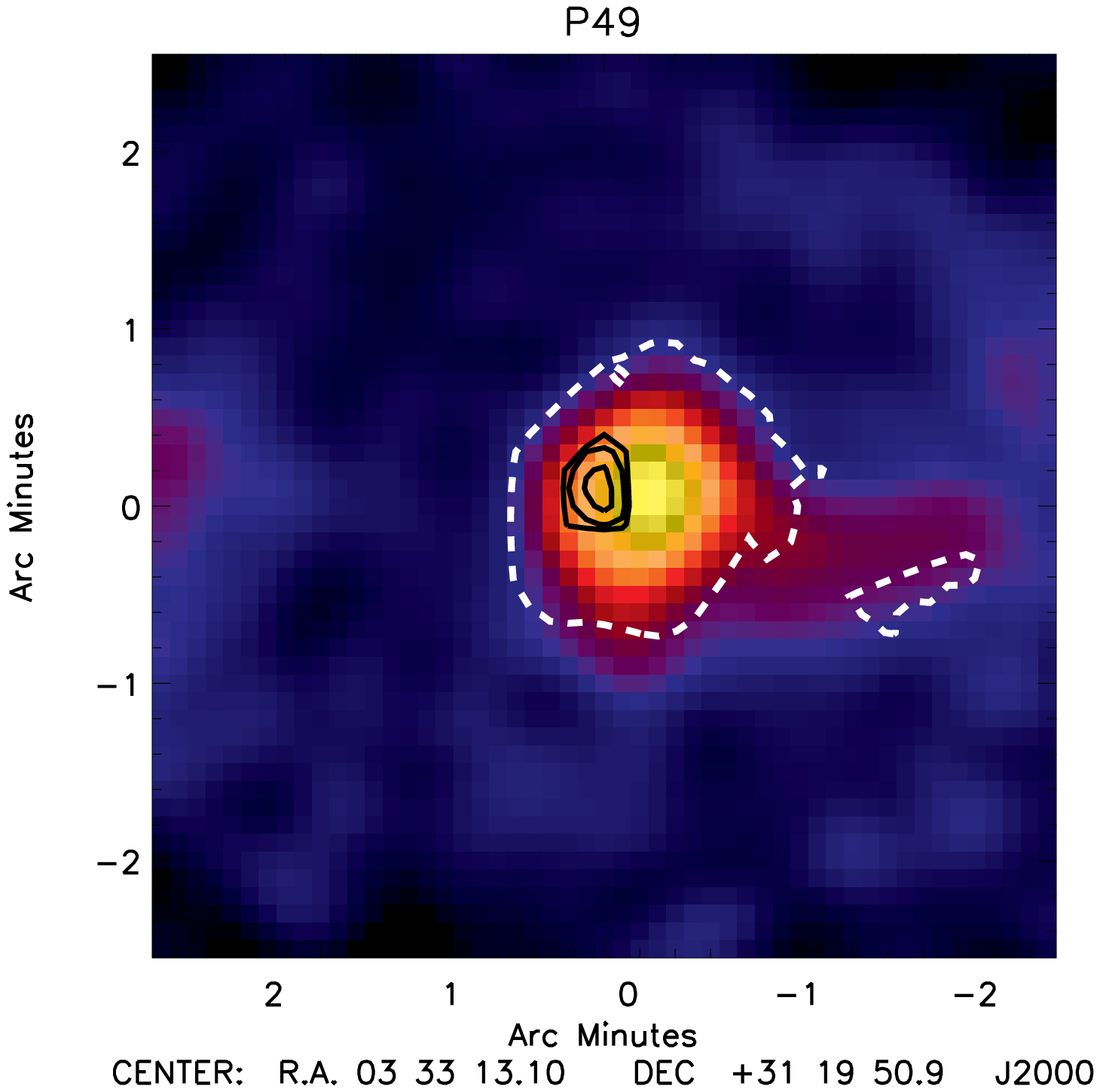}\\
\includegraphics*[width=0.4\textwidth]{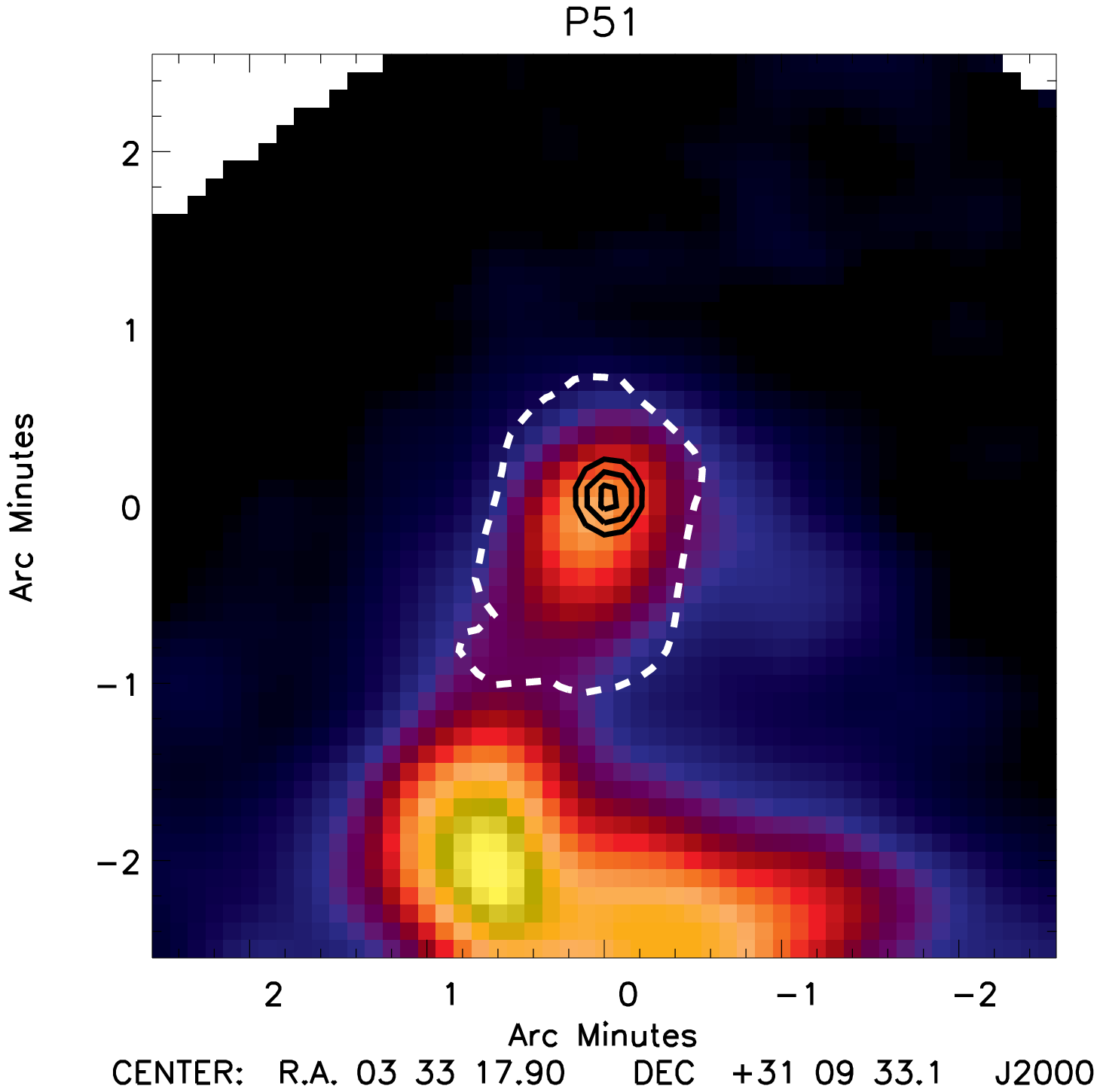}
\includegraphics*[width=0.4\textwidth]{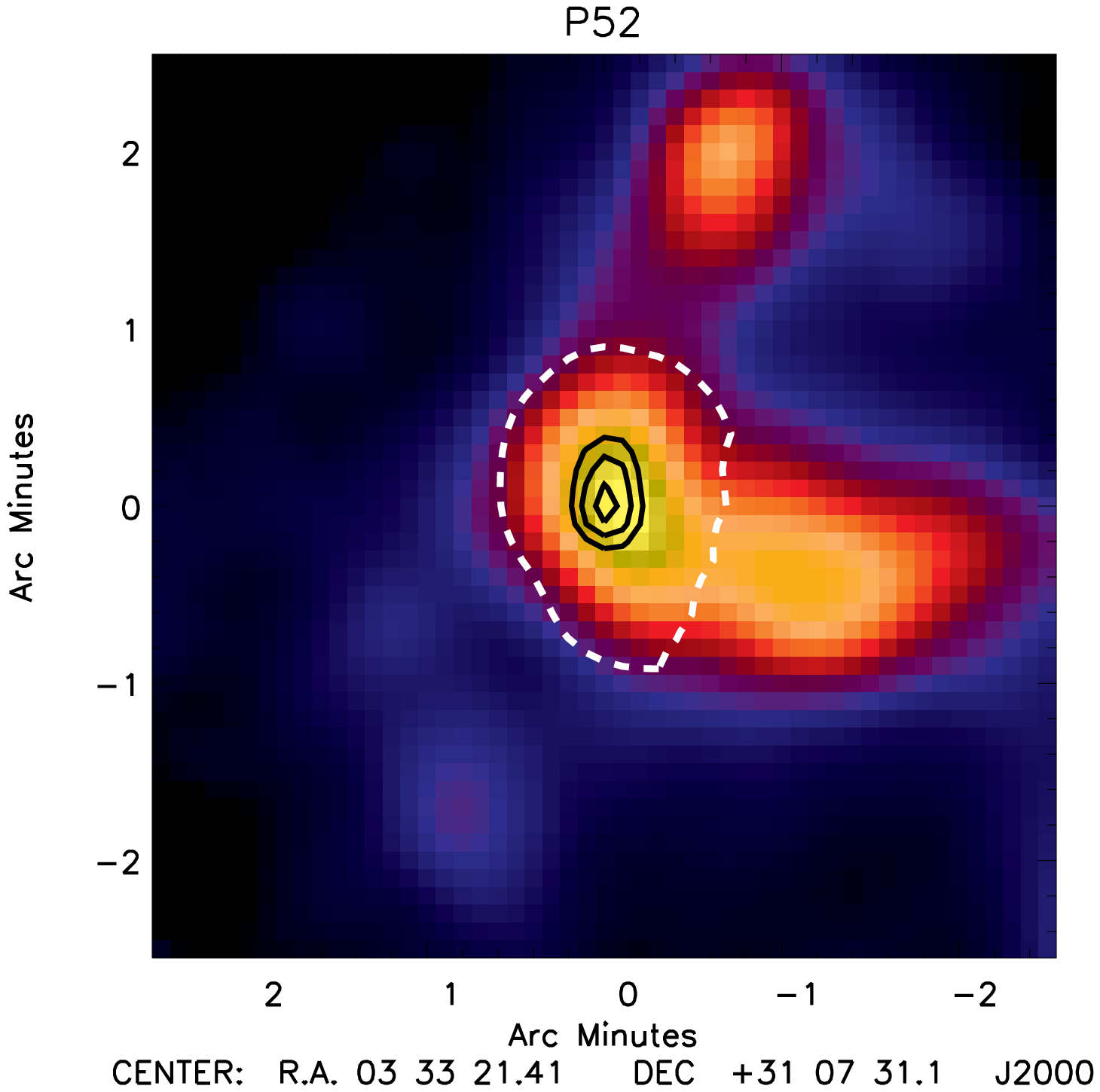}\\
\includegraphics*[width=0.4\textwidth]{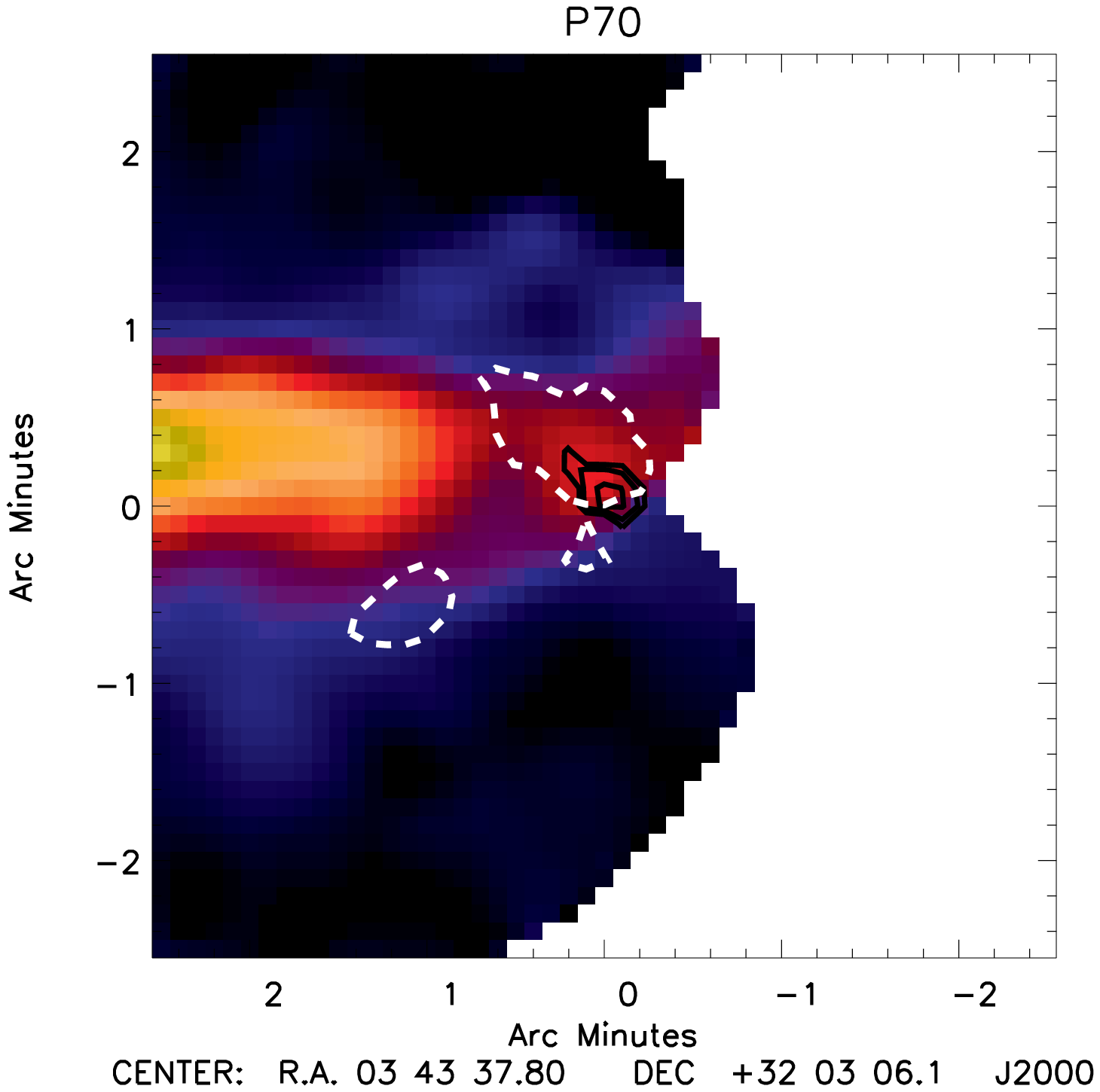}
\includegraphics*[width=0.4\textwidth]{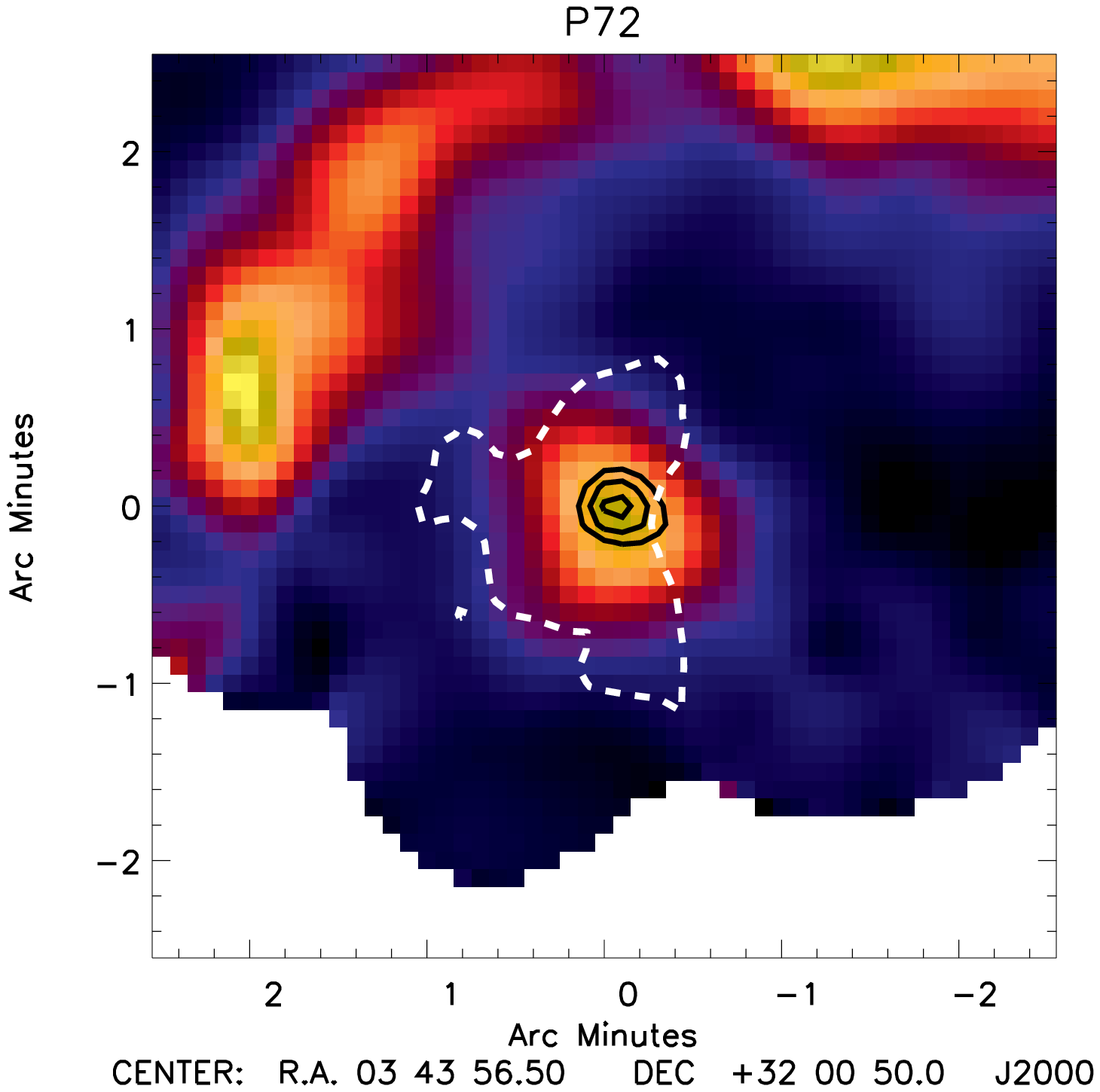}\\
\caption{\nh\ integrated intensity in sources P47, P49, P51, P52, P70 \& P72, overlaid with black contours of submillimetre emission at 50, 70 and 90\% of the peak flux density. A white, broken contour traces the 50\% level of the \nh\ column density distribution.}
\end{center}
\end{figure*}
\addtocounter{figure}{-1}

\begin{figure*}
\begin{center} 
\includegraphics*[width=0.4\textwidth]{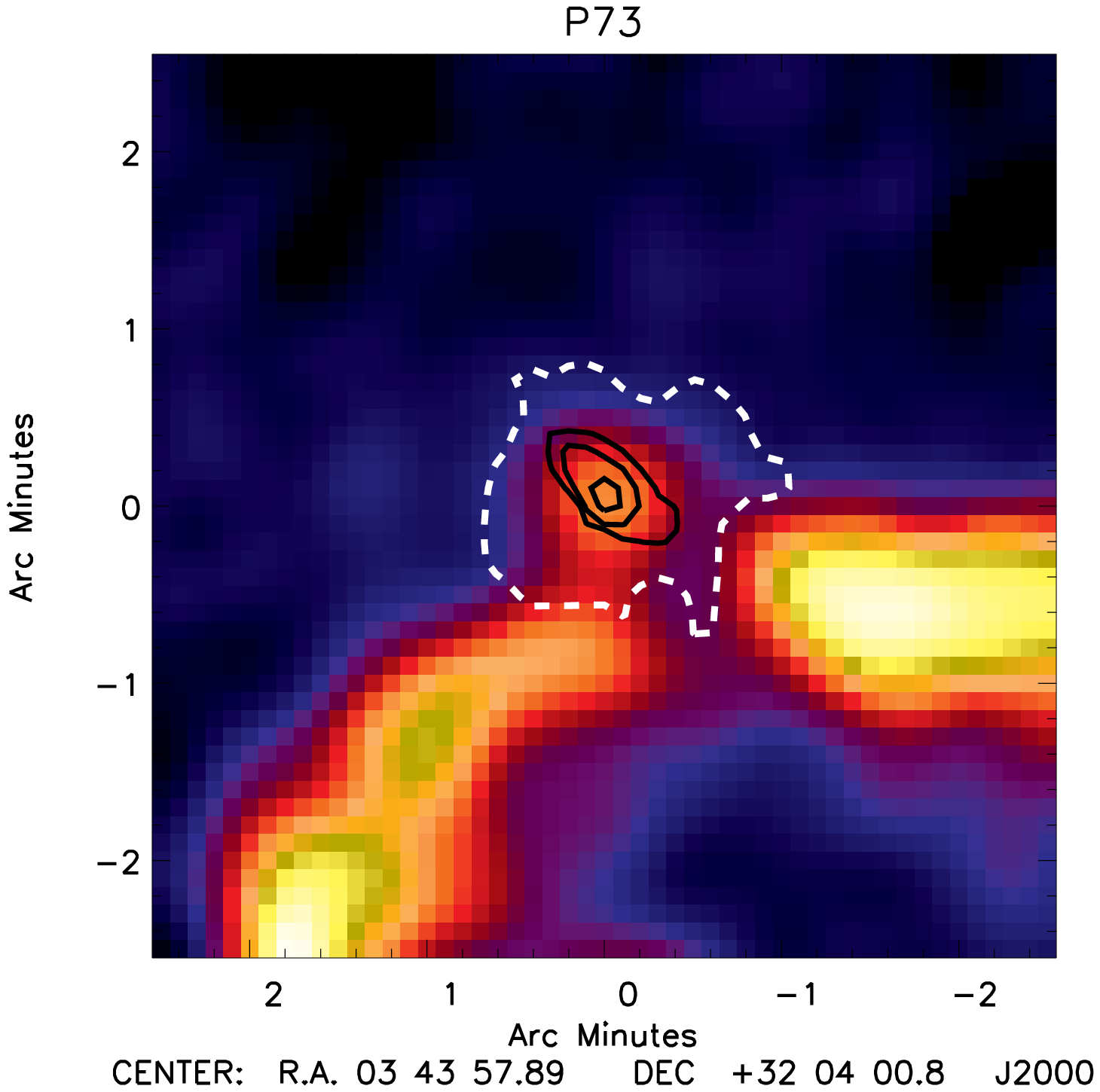}
\includegraphics*[width=0.4\textwidth]{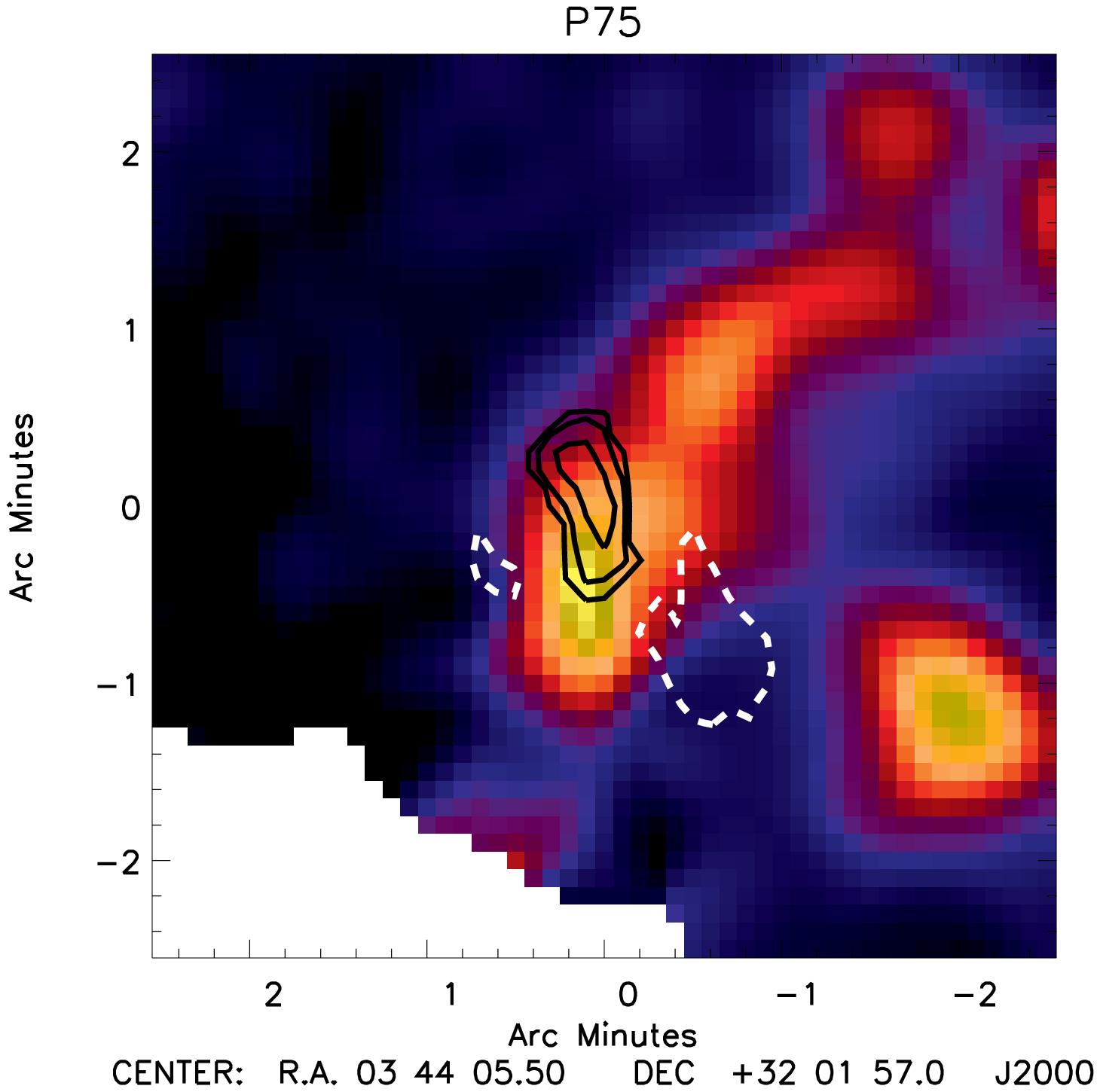}\\
\includegraphics*[width=0.4\textwidth]{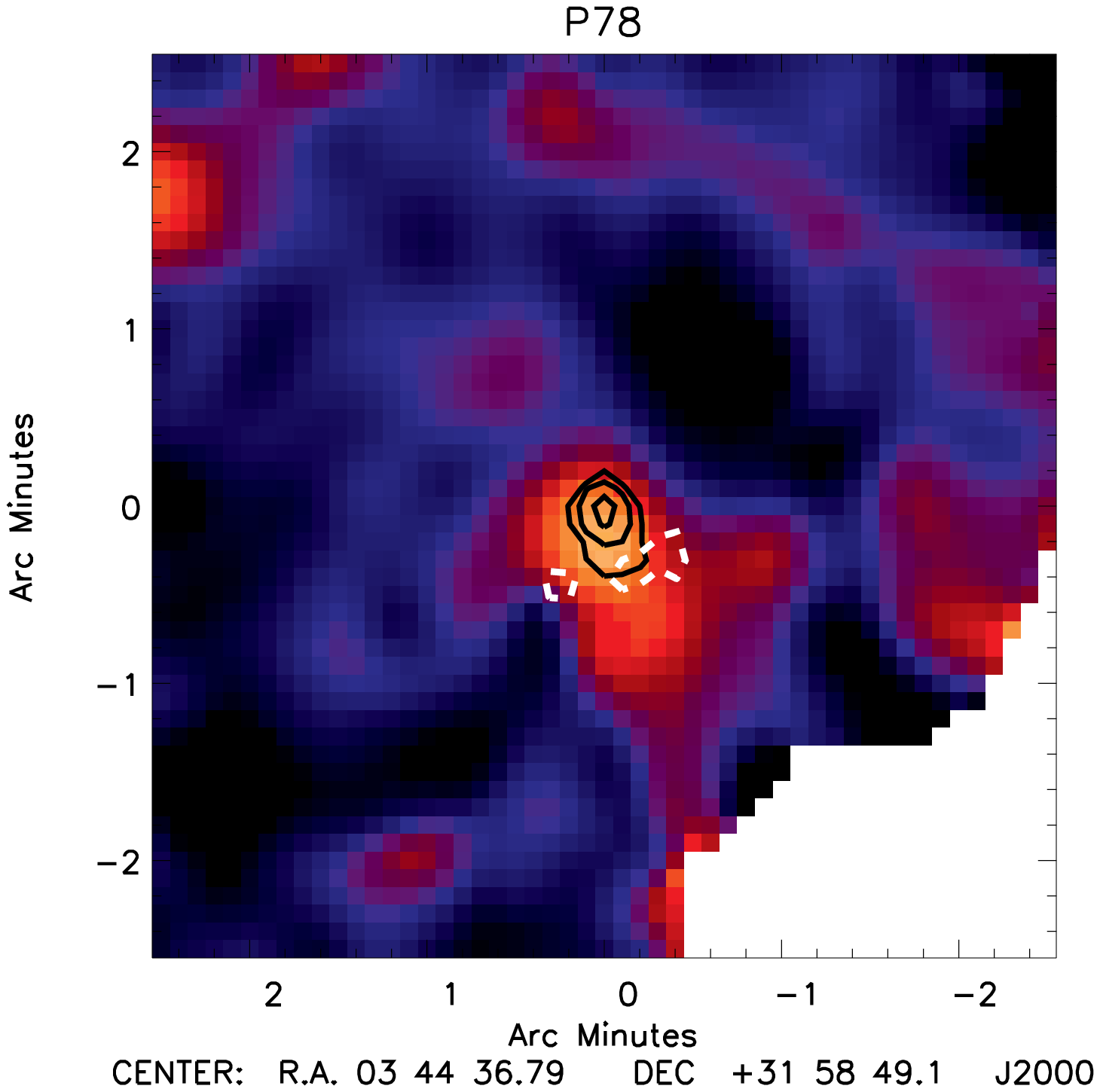}
\includegraphics*[width=0.4\textwidth]{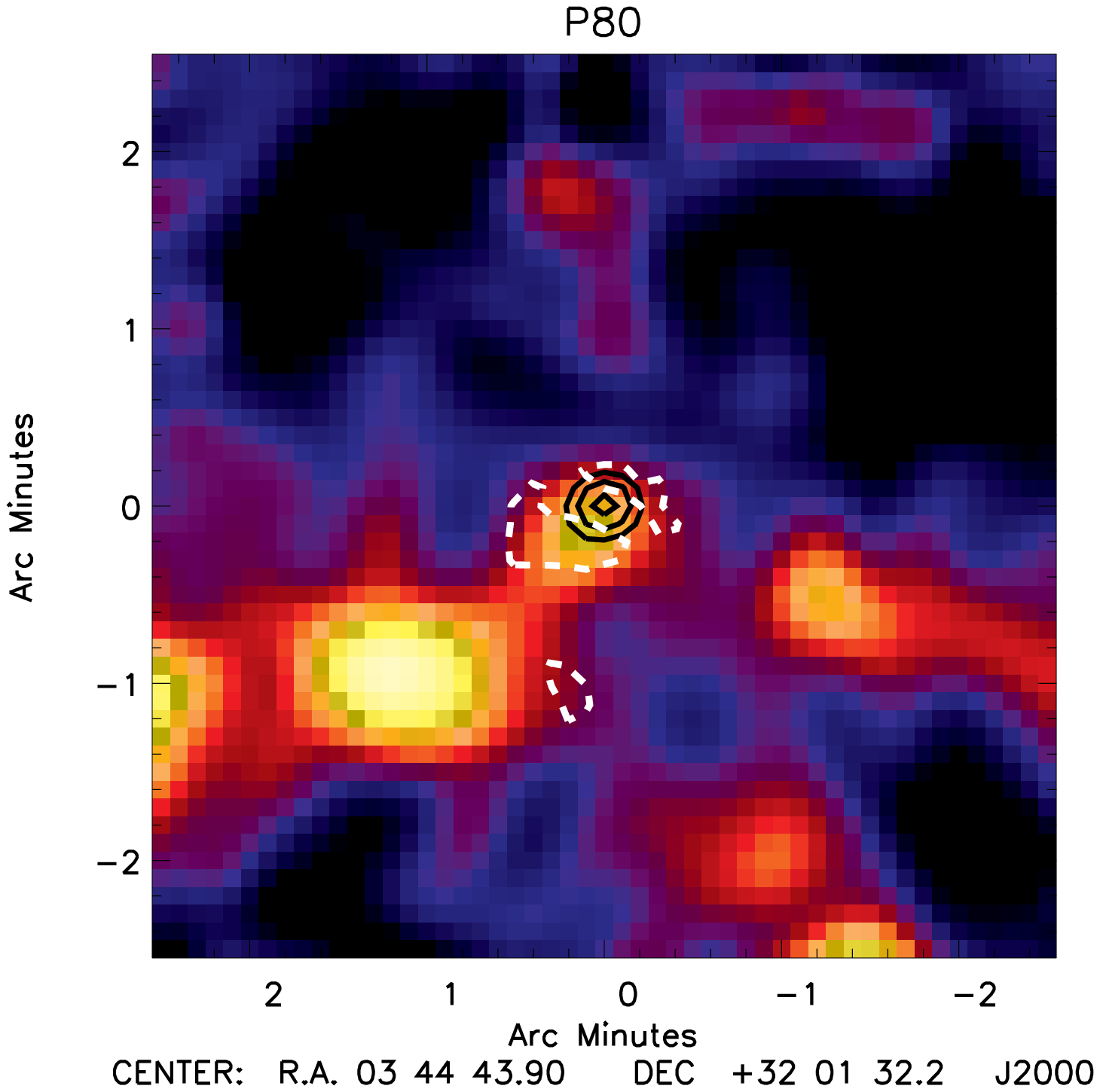}\\
\caption{\nh\ integrated intensity in sources P73, P75, P78 \& P80, overlaid with black contours of submillimetre emission at 50, 70 and 90\% of the peak flux density. A white, broken contour traces the 50\% level of the \nh\ column density distribution.}
\end{center}
\end{figure*}
\addtocounter{figure}{-1}

\begin{figure*}
\begin{center}
\includegraphics*[width=0.4\textwidth]{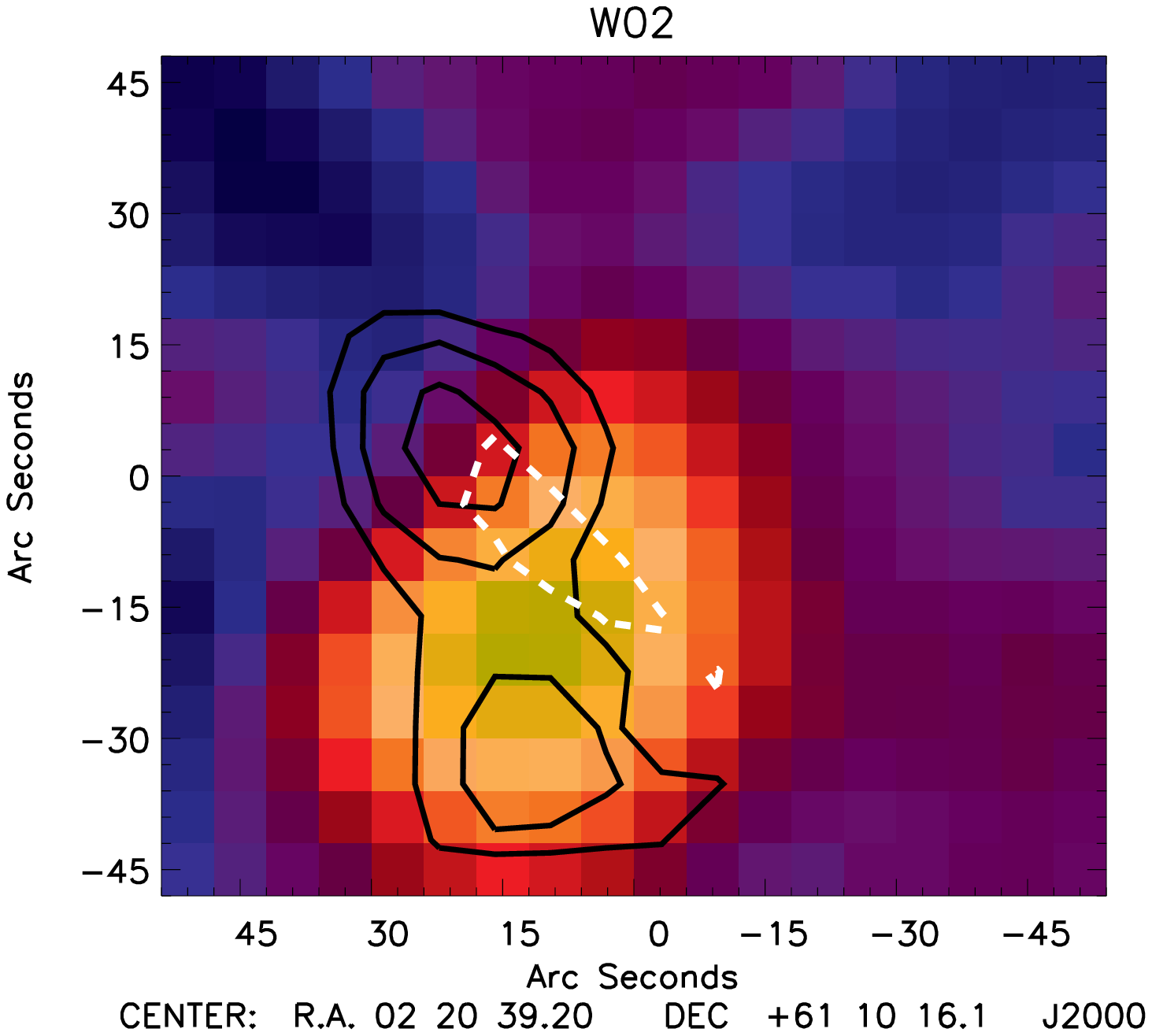}
\includegraphics*[width=0.4\textwidth]{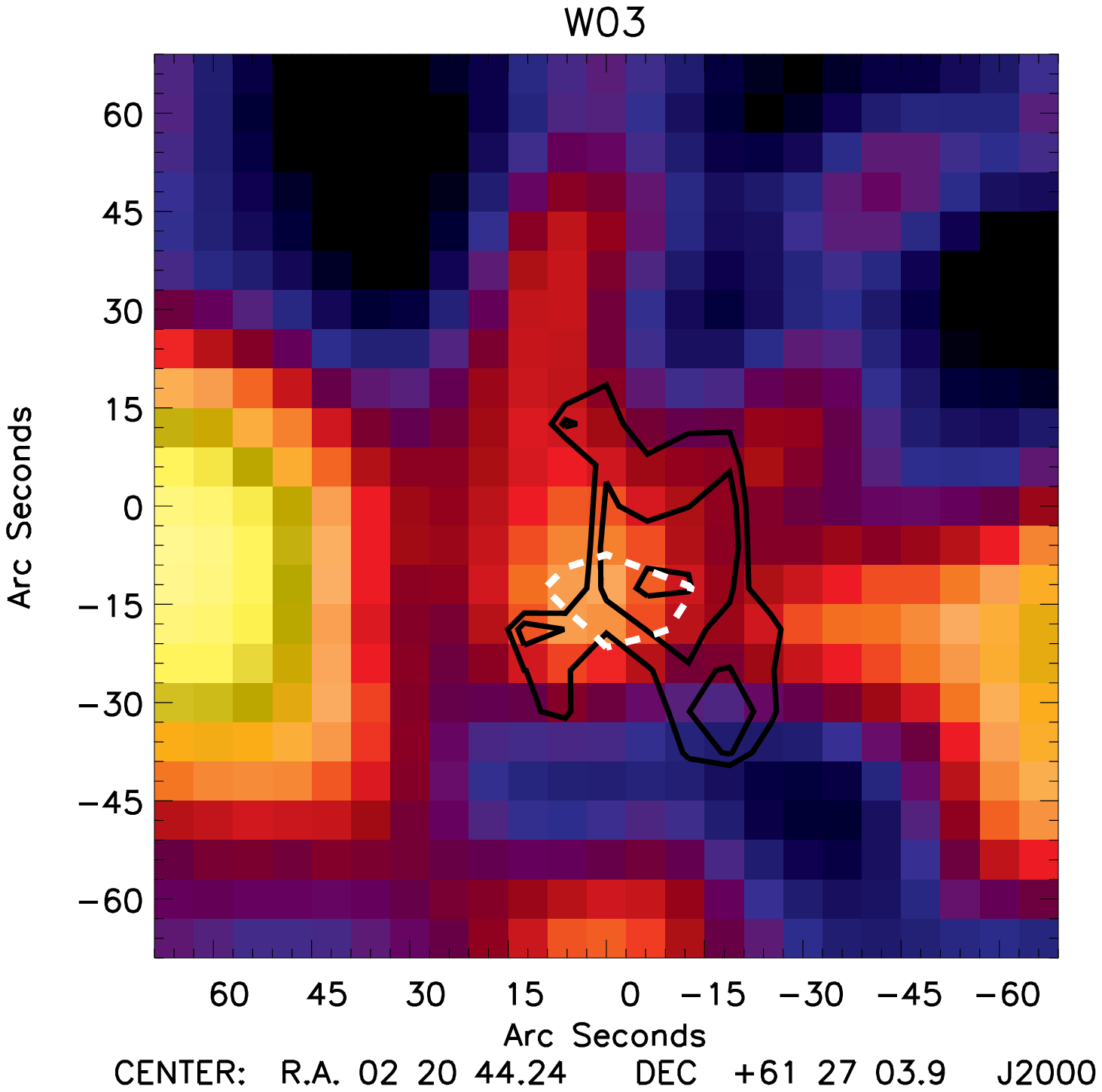}\\
\includegraphics*[width=0.4\textwidth]{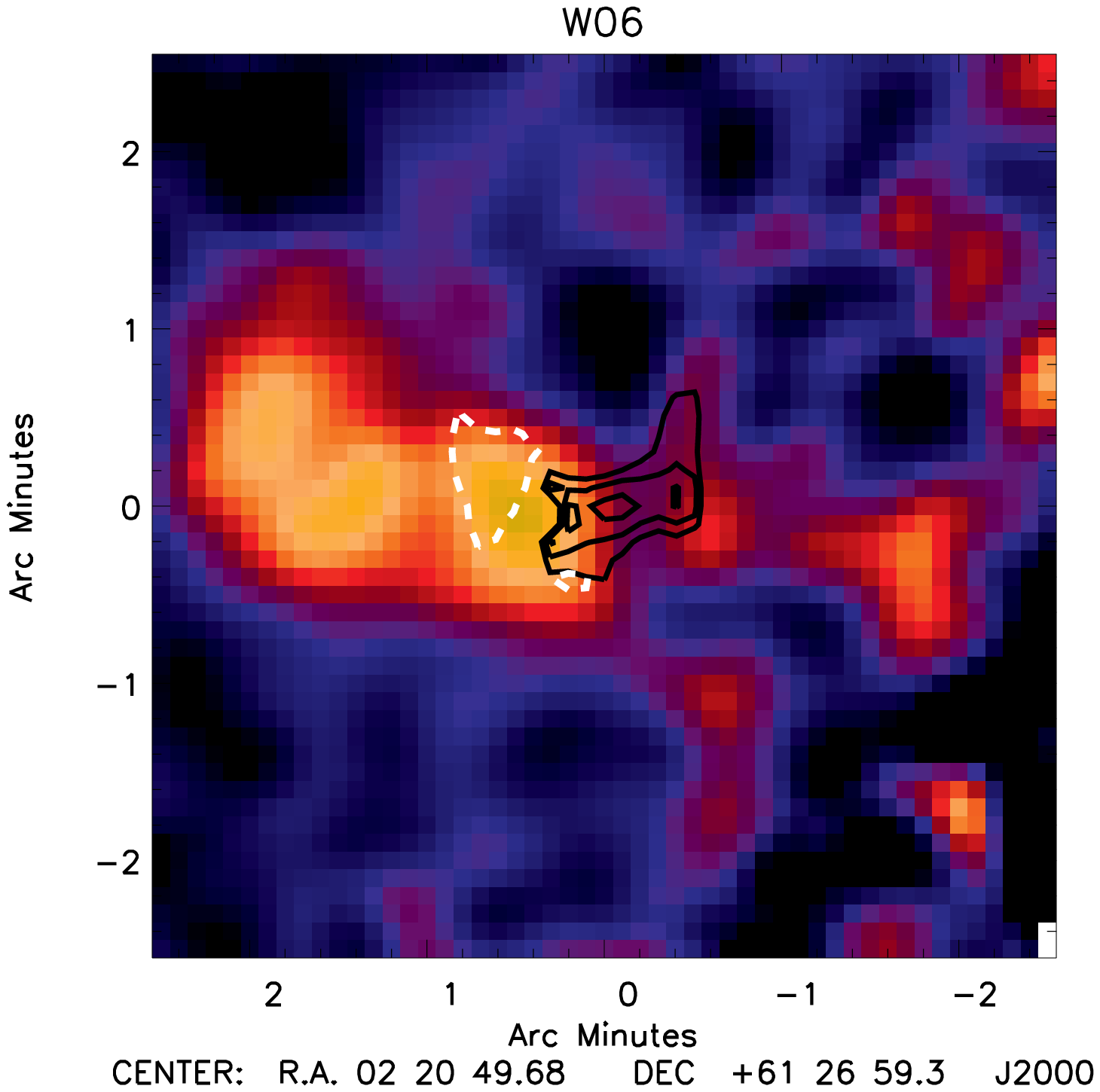}
\includegraphics*[width=0.4\textwidth]{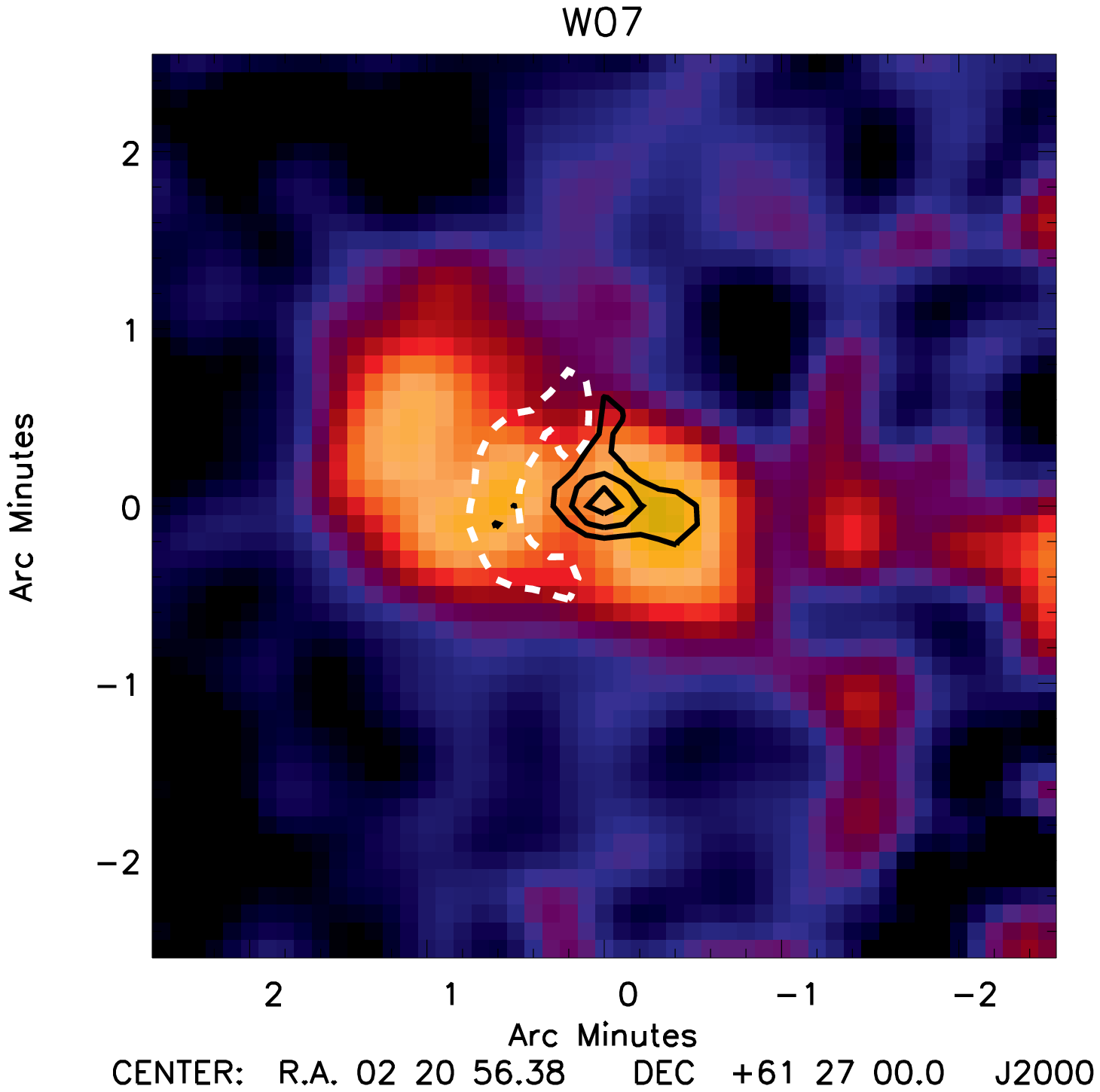}\\
\includegraphics*[width=0.4\textwidth]{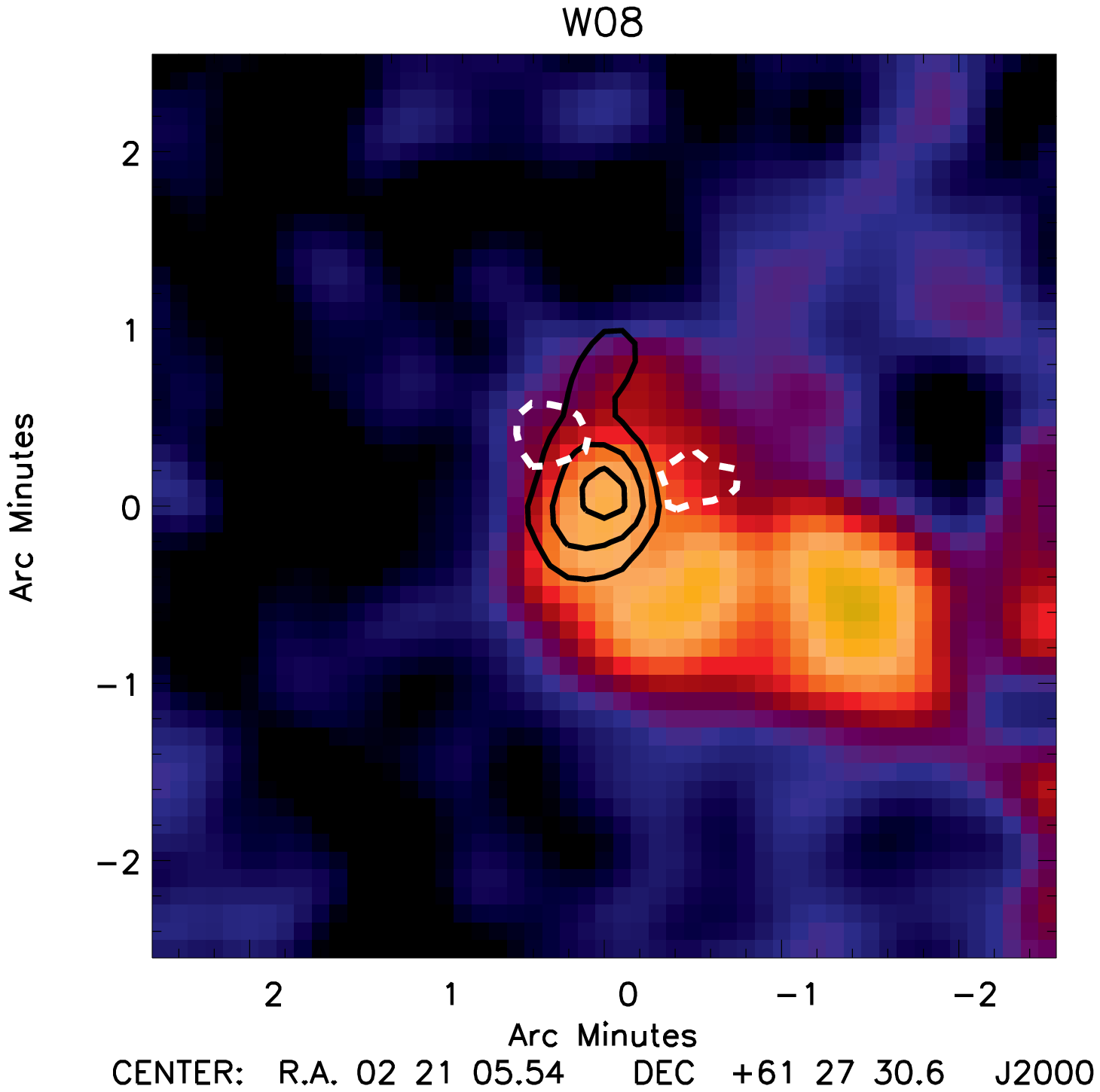}
\includegraphics*[width=0.4\textwidth]{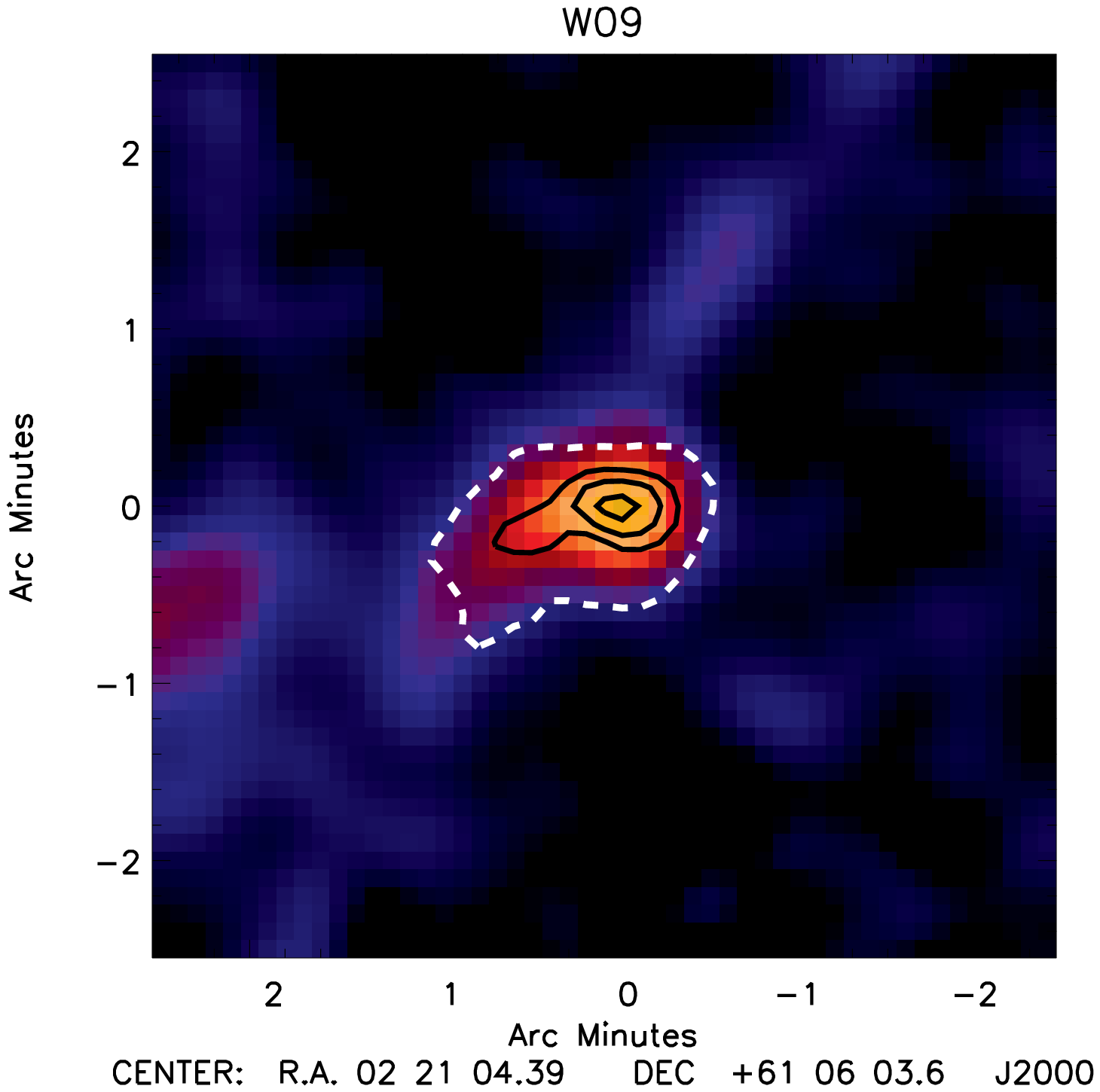}\\
\caption{\nh\ integrated intensity in sources W02, W03, W06, W07, W08 \& W09 overlaid with black contours of submillimetre emission at 50, 70 and 90\% of the peak flux density.  A white, broken contour traces the \nh\ column density distribution at a level of 50\% of the local peak for all sources except W02 and W07, for which the contour levels are 70\% and 40\%, respectively.}
\end{center}
\end{figure*}
\addtocounter{figure}{-1}

\begin{figure*}
\begin{center} 
\includegraphics*[width=0.4\textwidth]{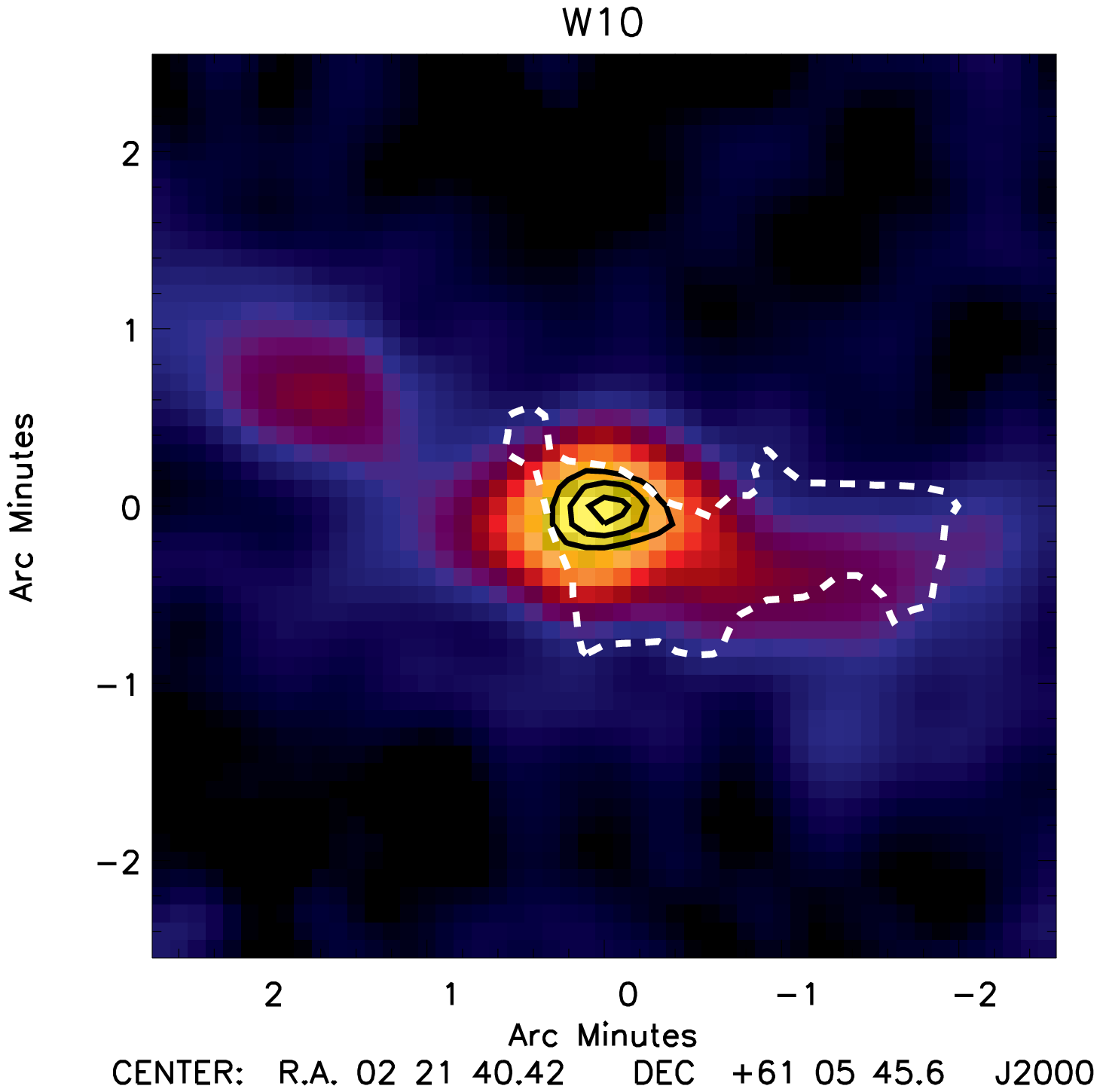}
\includegraphics*[width=0.4\textwidth]{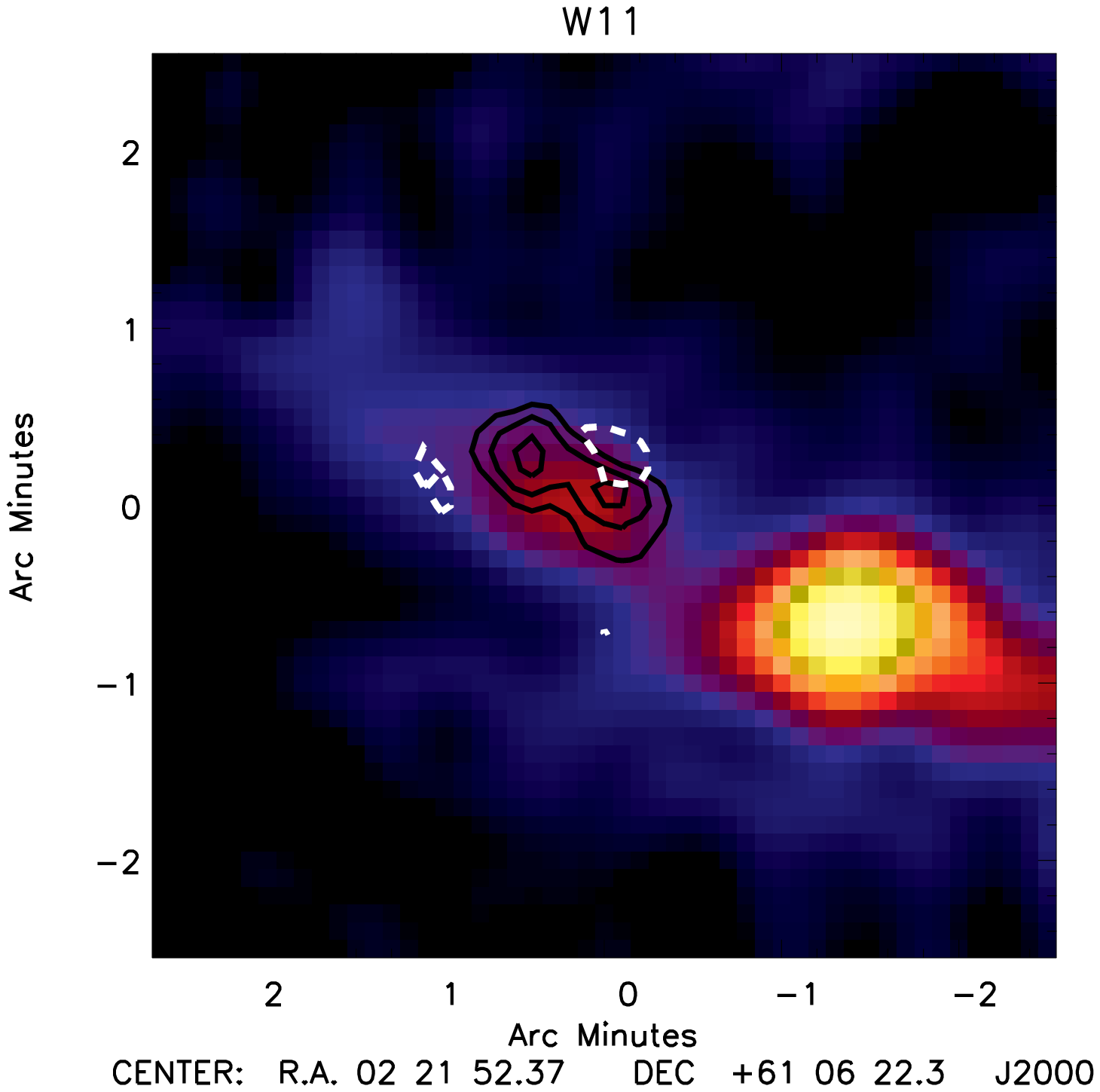}\\
\includegraphics*[width=0.4\textwidth]{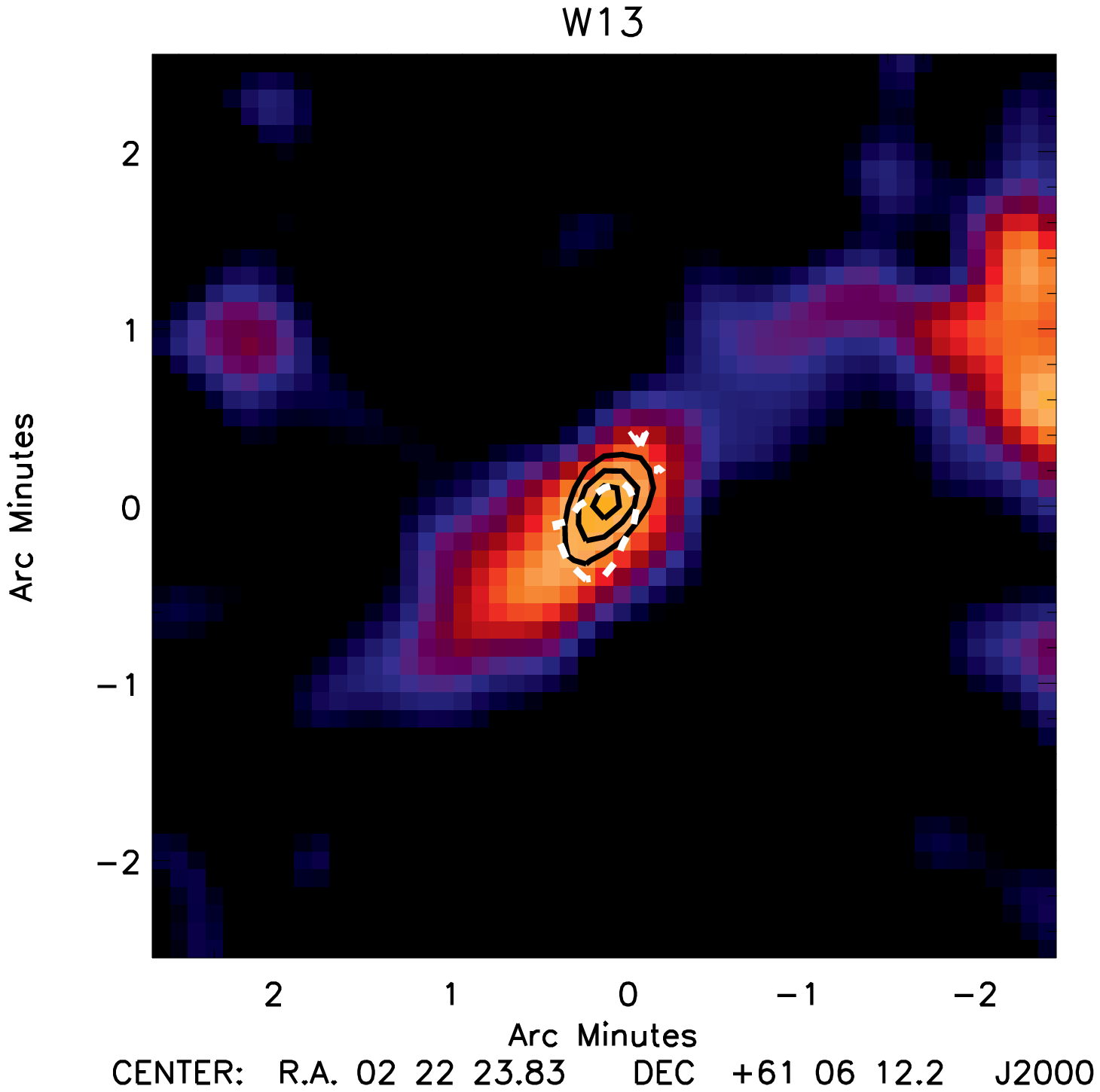}
\includegraphics*[width=0.4\textwidth]{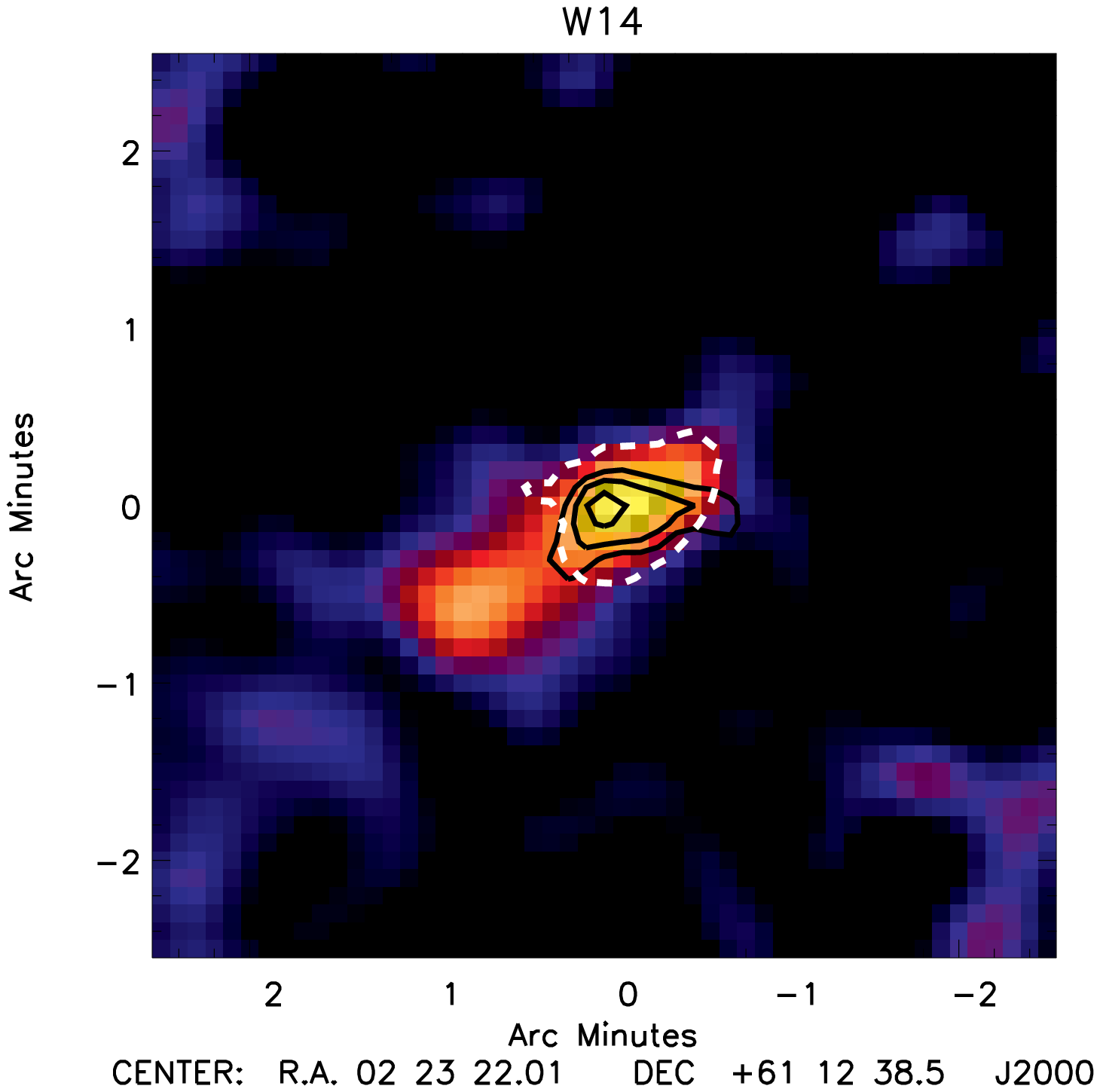}\\
\includegraphics*[width=0.4\textwidth]{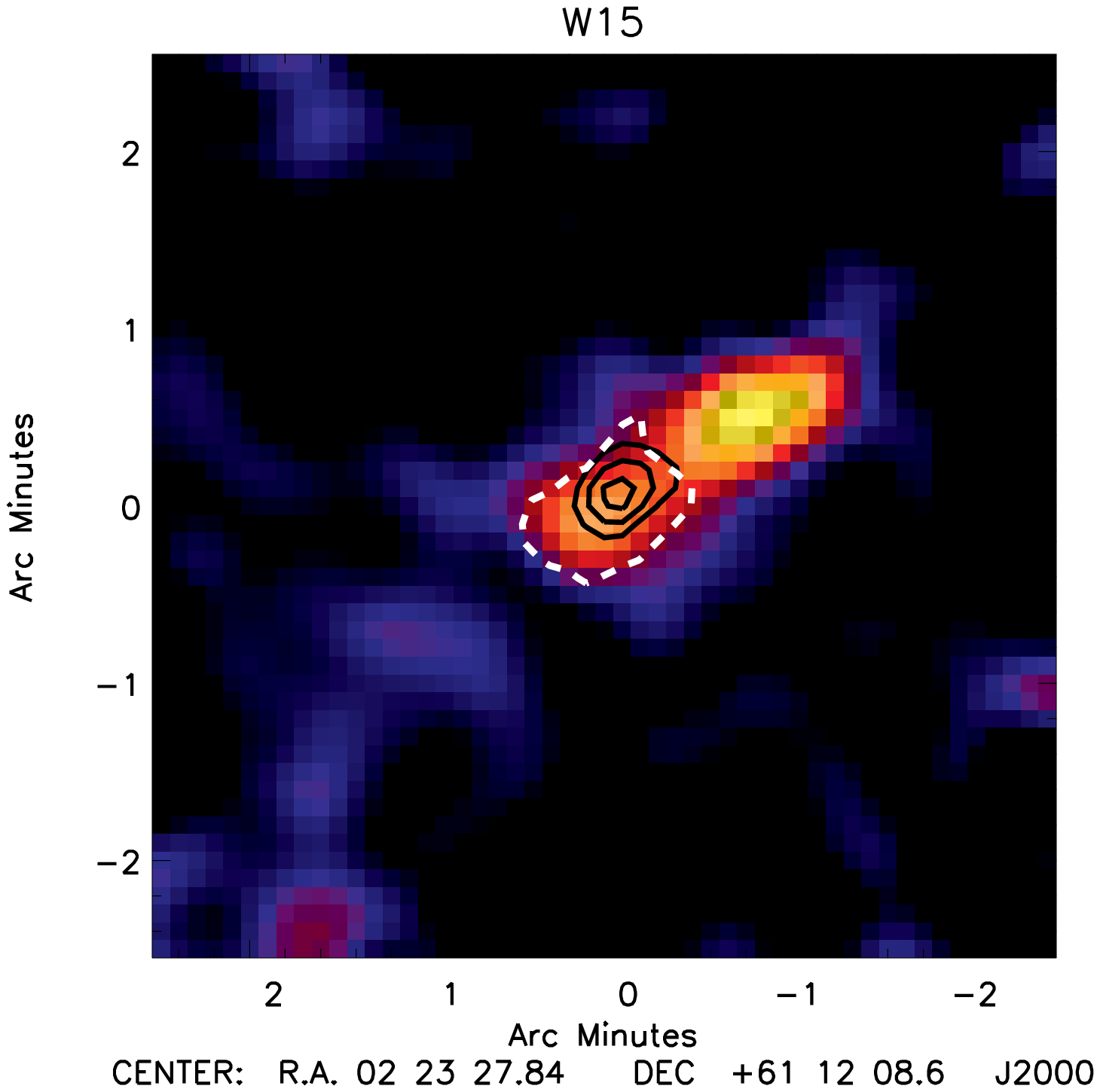}
\includegraphics*[width=0.4\textwidth]{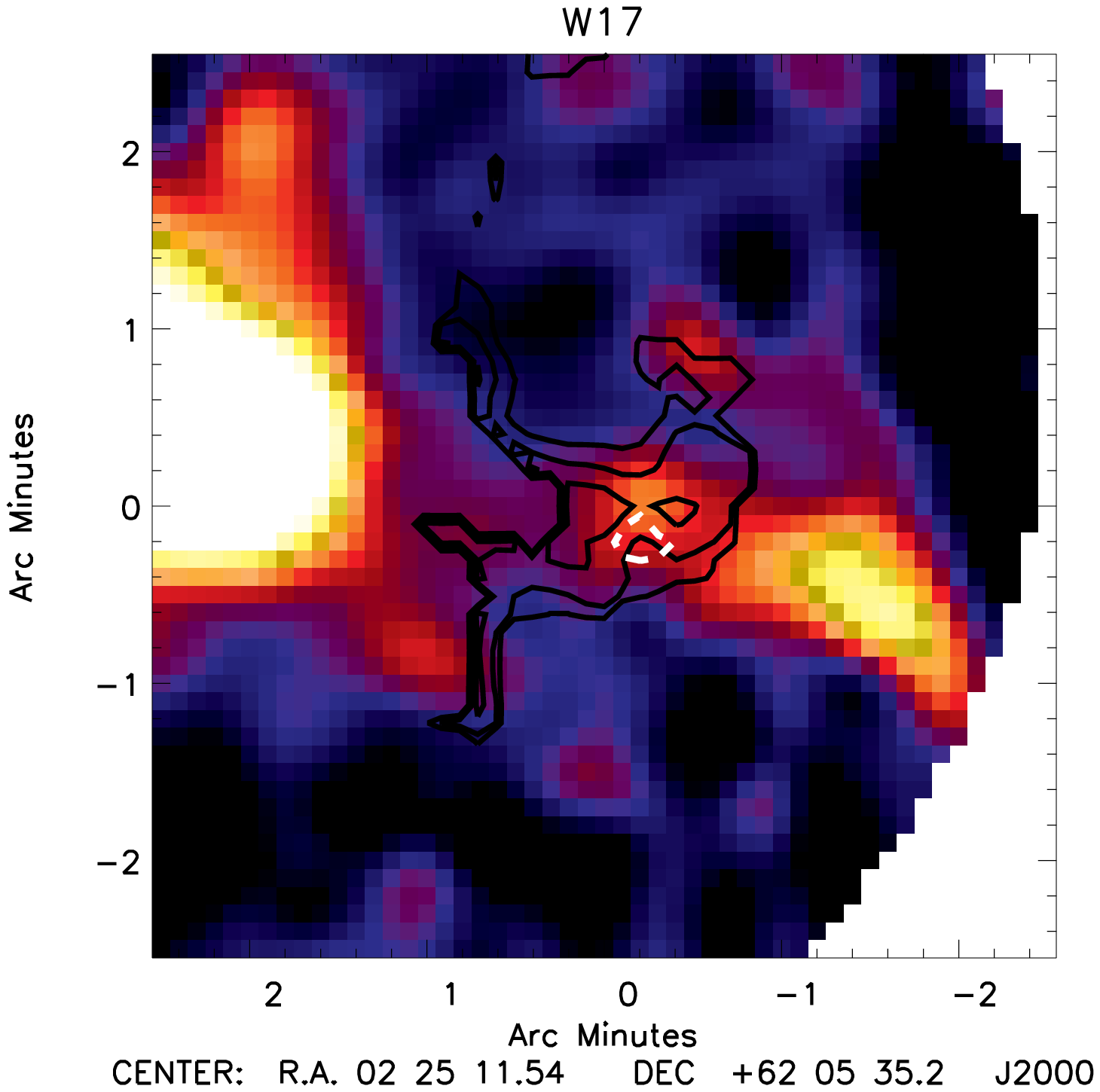}\\
\caption{\nh\ integrated intensity in sources W10, W11, W13, W14, W15 \& W17 overlaid with black contours of submillimetre emission at 50, 70 and 90\% of the peak flux density. A white, broken contour traces the 50\% level of the \nh\ column density distribution.}
\end{center}
\end{figure*}
\addtocounter{figure}{-1}

\begin{figure*}
\begin{center} 
\includegraphics*[width=0.4\textwidth]{W_18_SM_NH3_Conv_Image.eps}
\includegraphics*[width=0.4\textwidth]{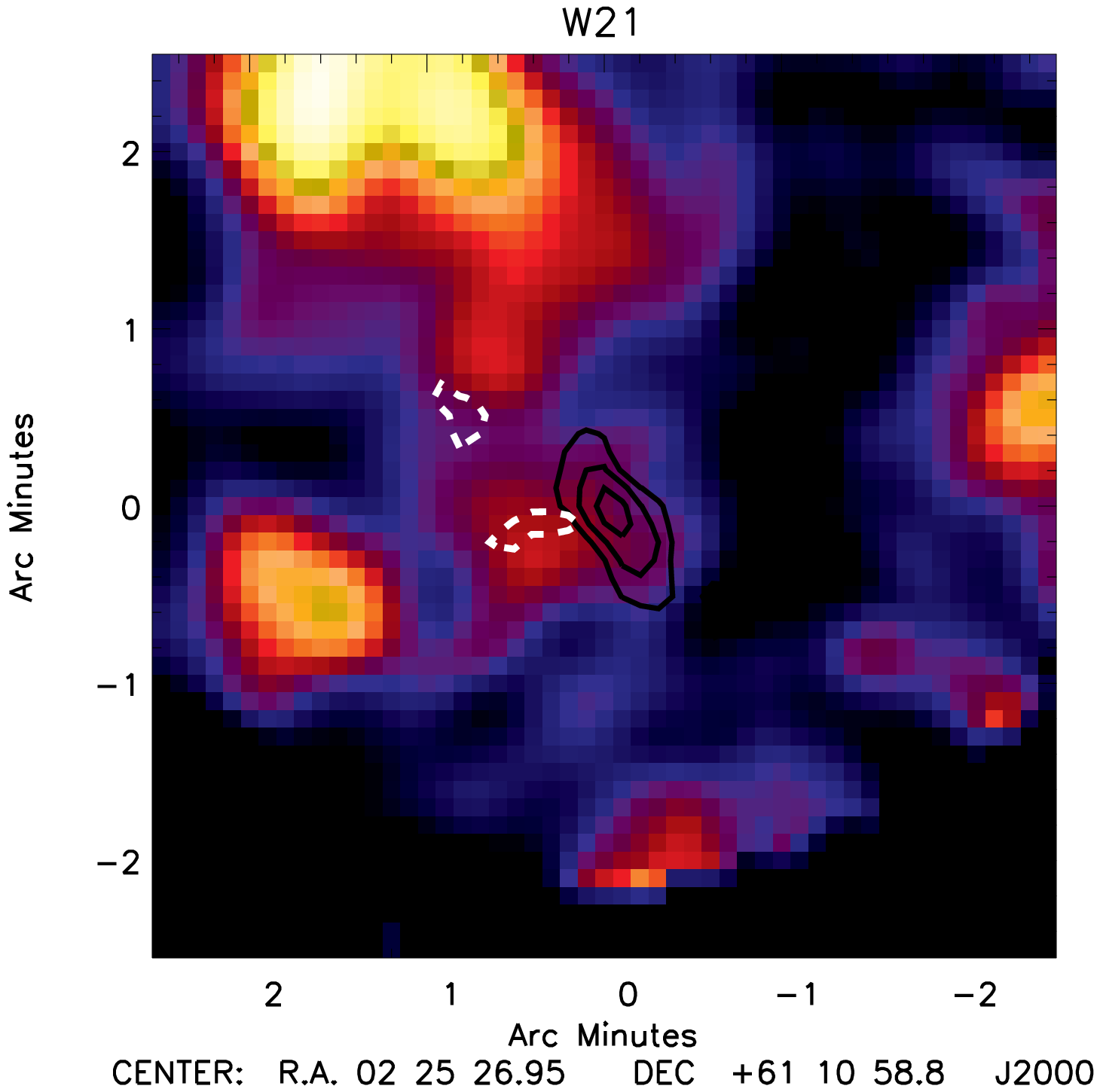}\\
\includegraphics*[width=0.4\textwidth]{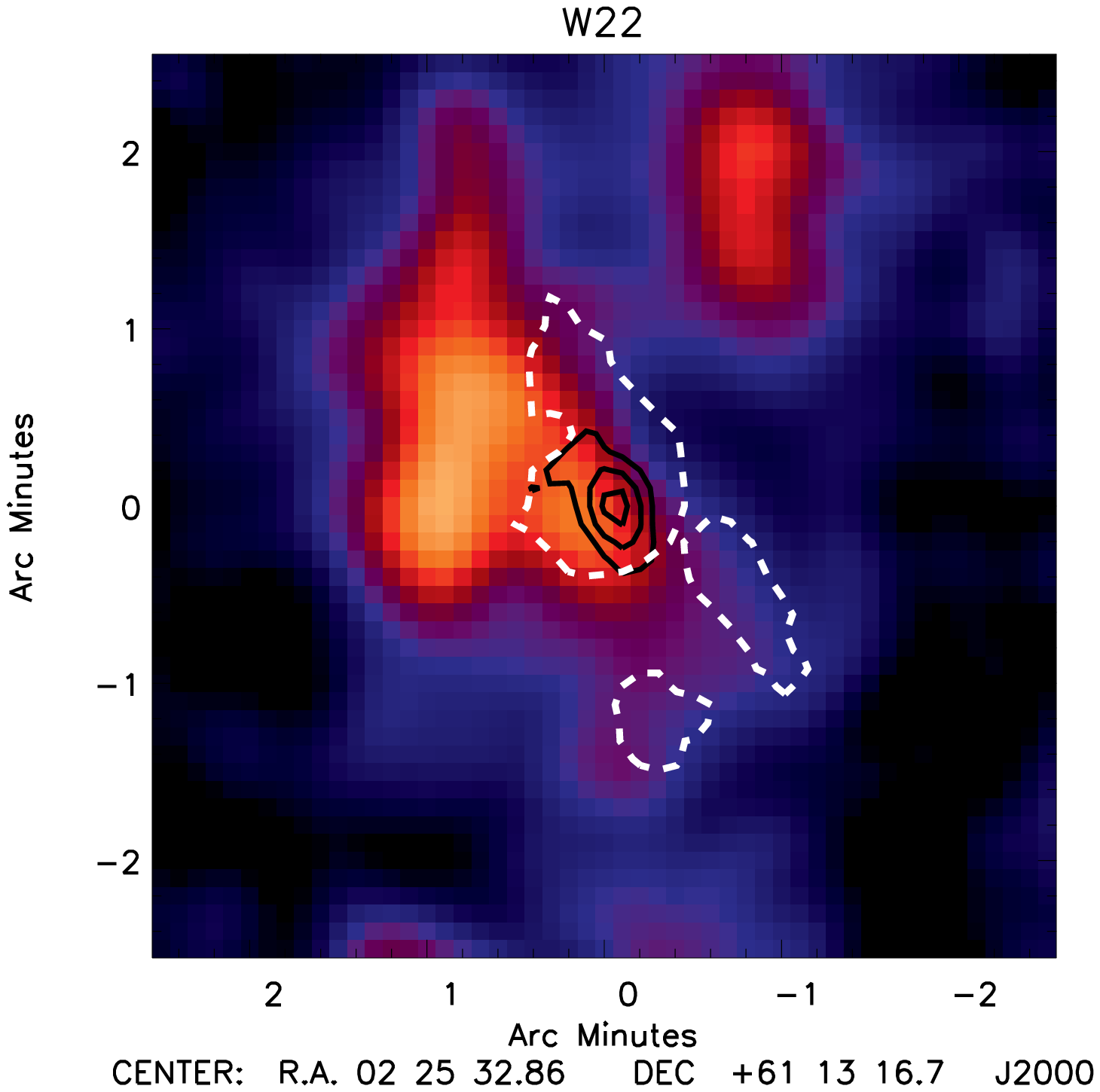}
\includegraphics*[width=0.4\textwidth]{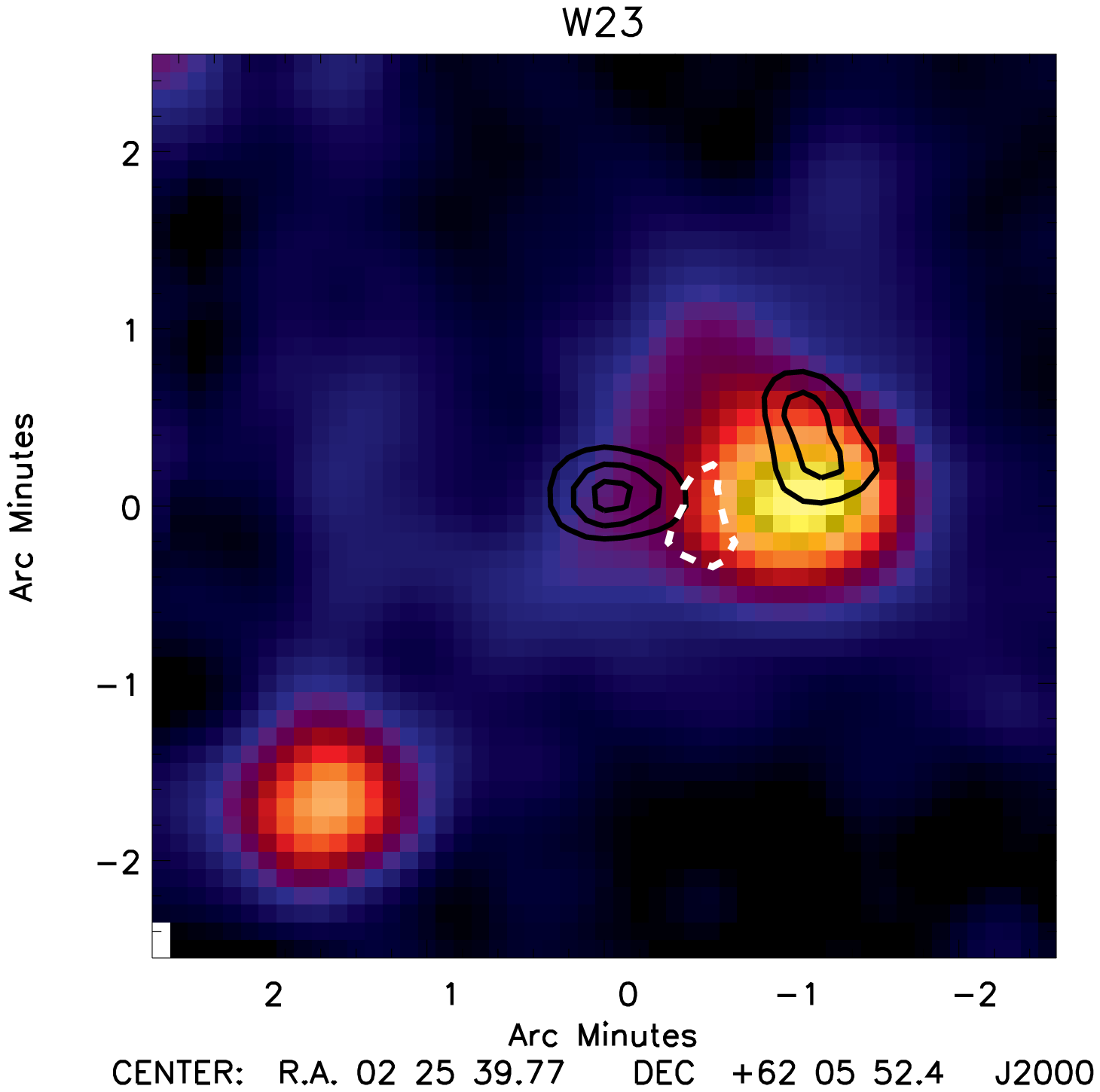}\\
\includegraphics*[width=0.4\textwidth]{W_25_SM_NH3_Conv_Image.eps}
\includegraphics*[width=0.4\textwidth]{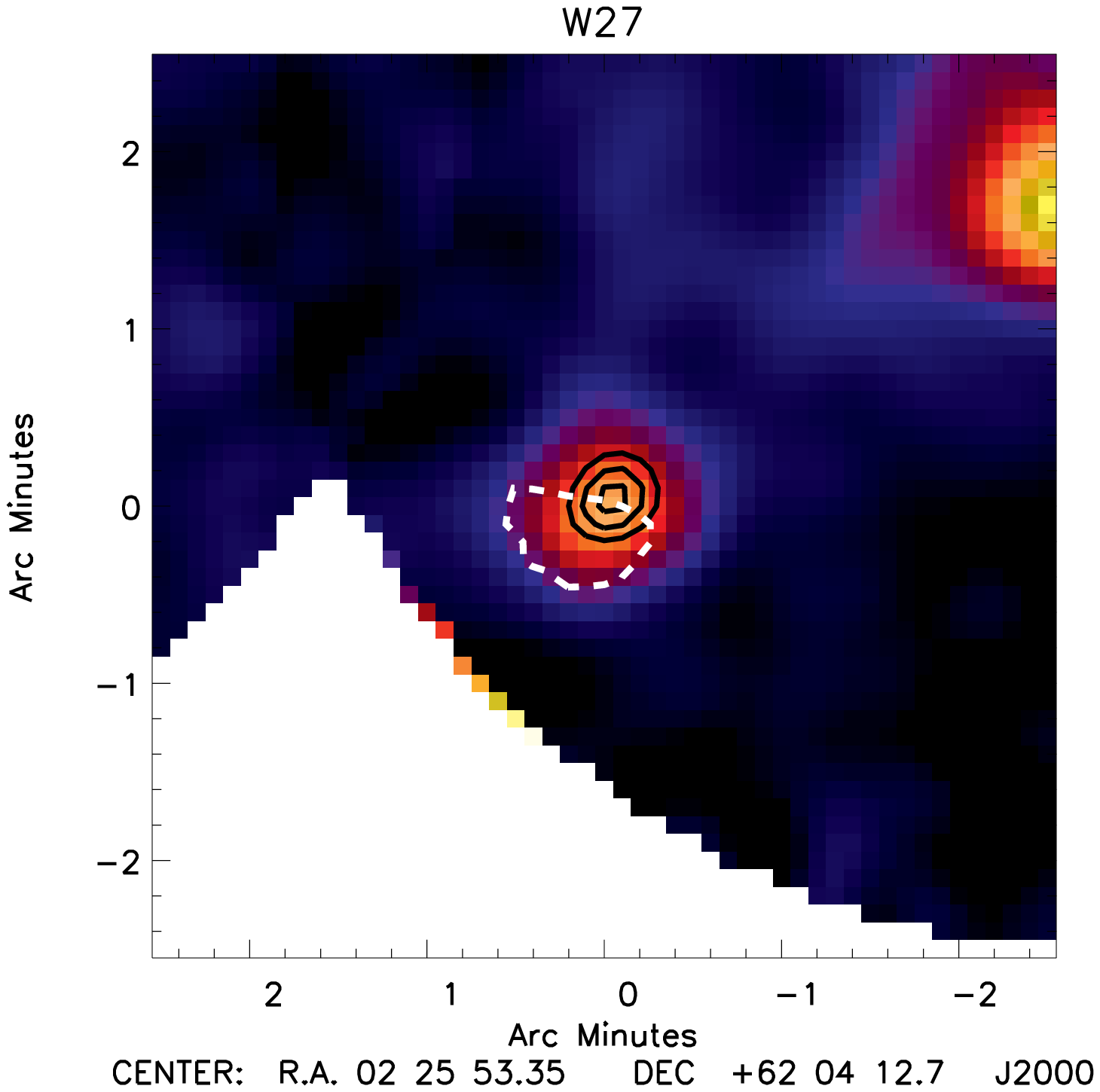}\\
\caption{\nh\ integrated intensity in sources W18, W21, W22, W23, W25 \& W27 overlaid with black contours of submillimetre emission at 50, 70 and 90\% of the peak flux density.  A white, broken contour traces the \nh\ column density distribution at a level of 50\% of the local peak for all sources except W18 and W22, for which the contour levels are 40\% and 20\%, respectively.}
\end{center}
\end{figure*}
\addtocounter{figure}{-1}

\begin{figure*}
\begin{center} 
\includegraphics*[width=0.4\textwidth]{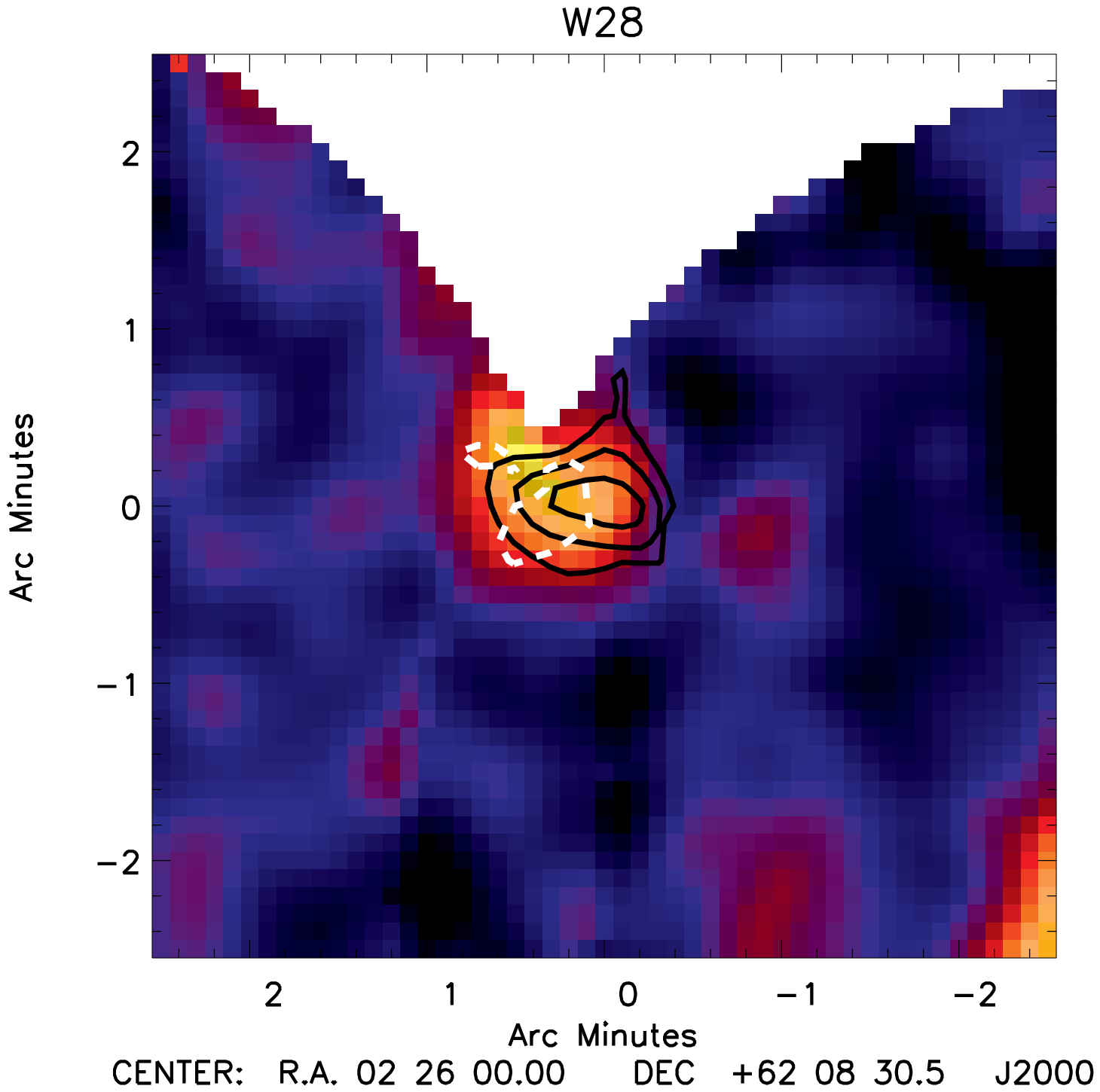}
\includegraphics*[width=0.4\textwidth]{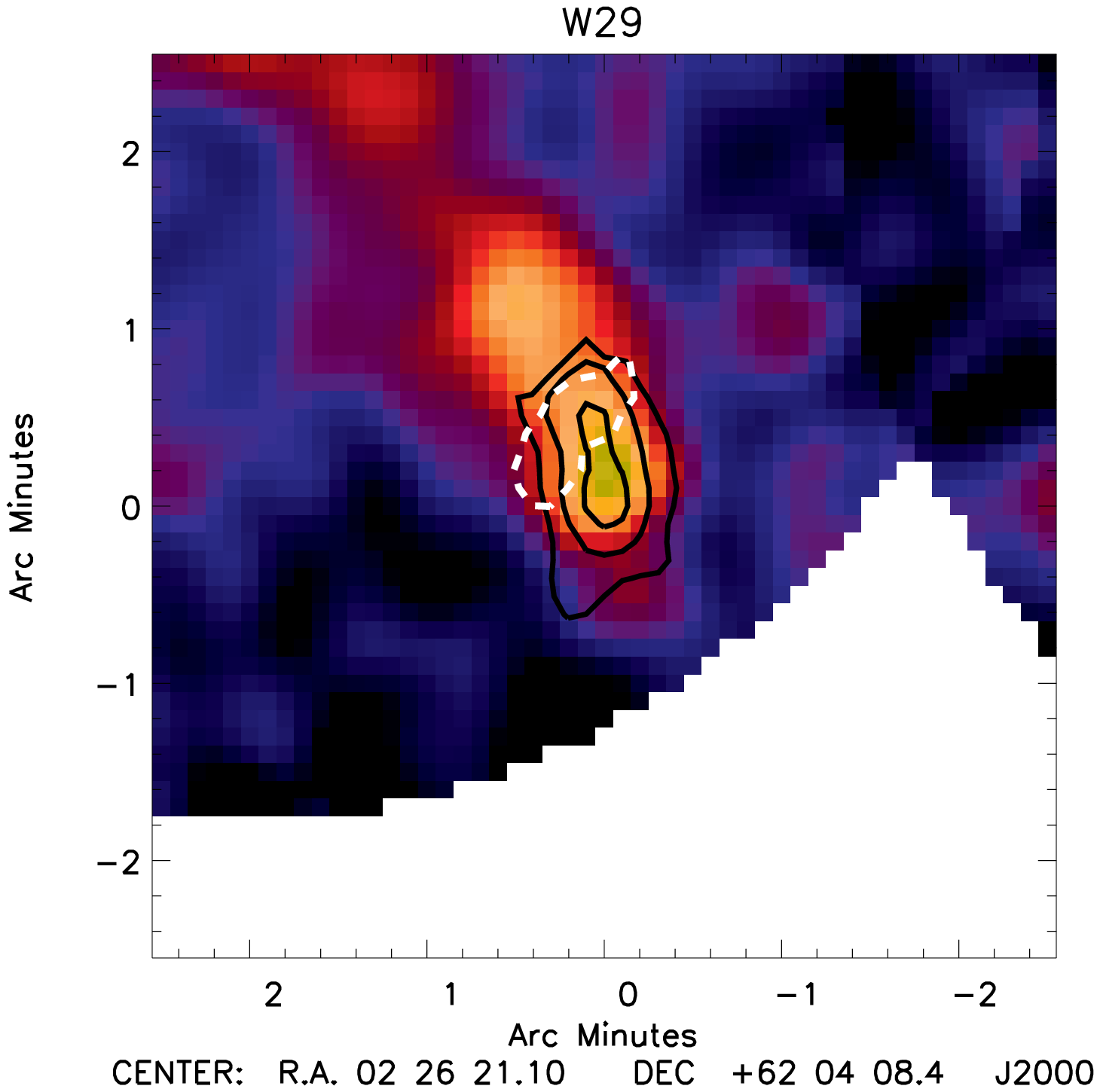}\\
\includegraphics*[width=0.4\textwidth]{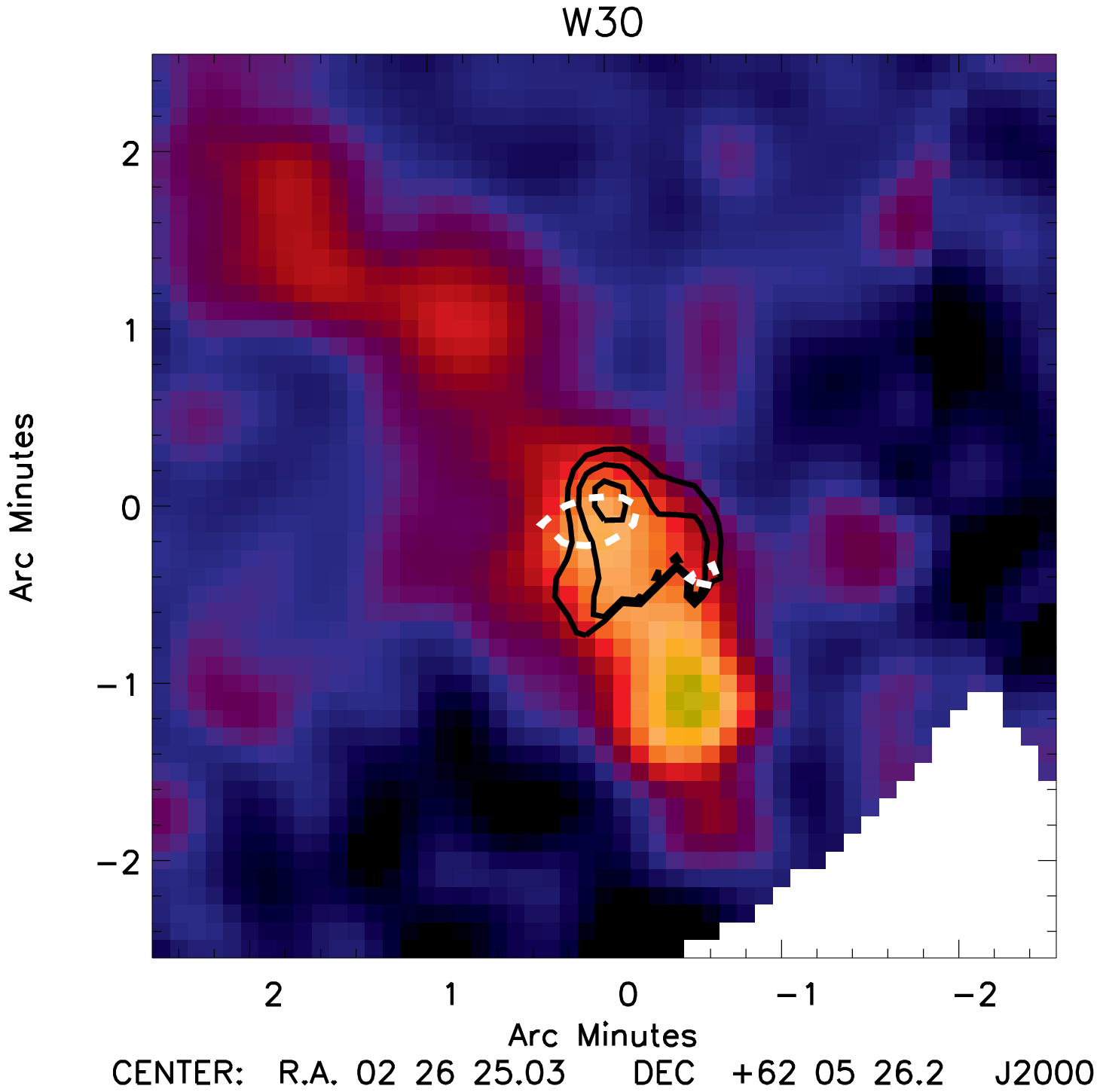}
\includegraphics*[width=0.4\textwidth]{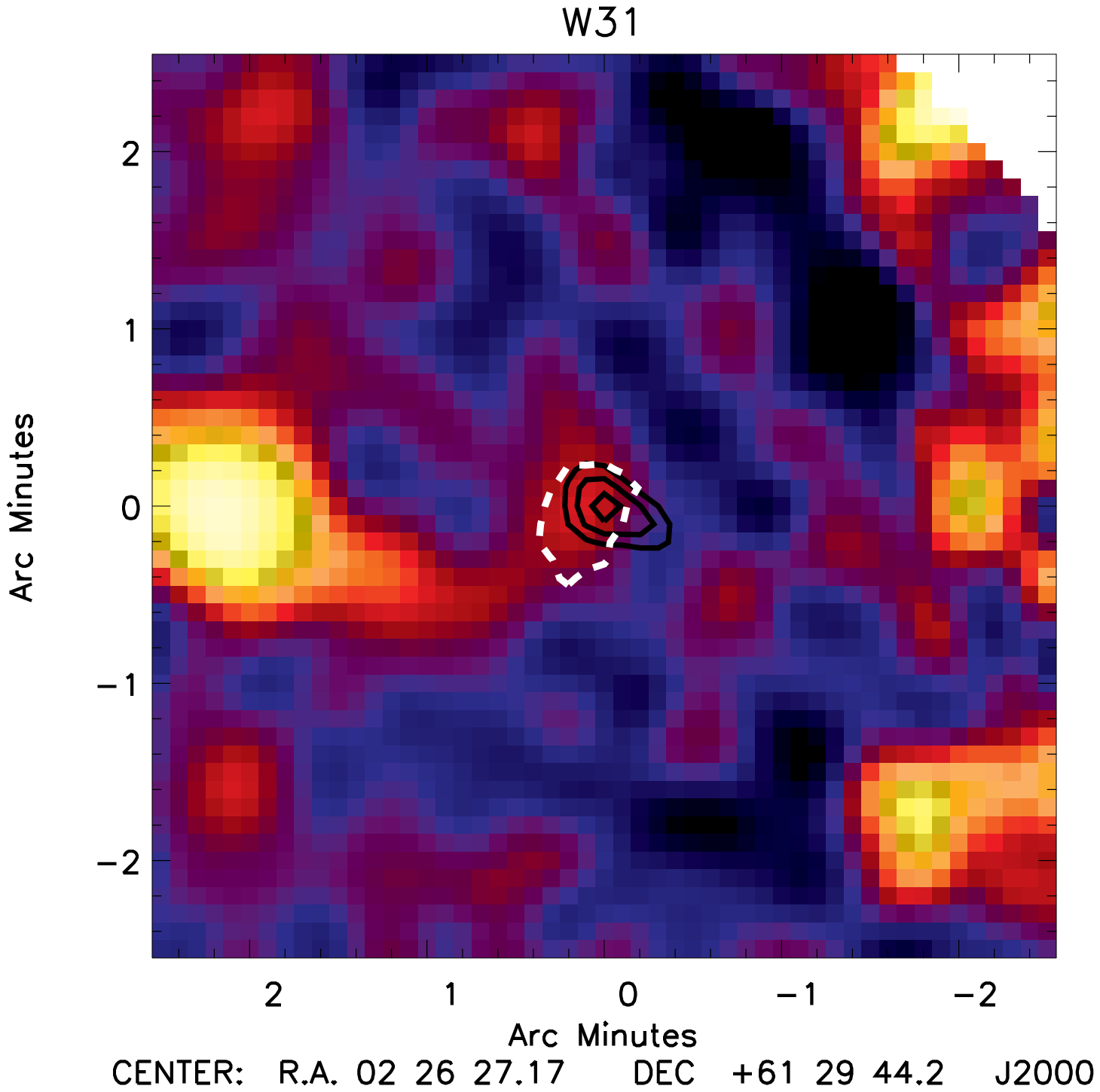}\\
\includegraphics*[width=0.4\textwidth]{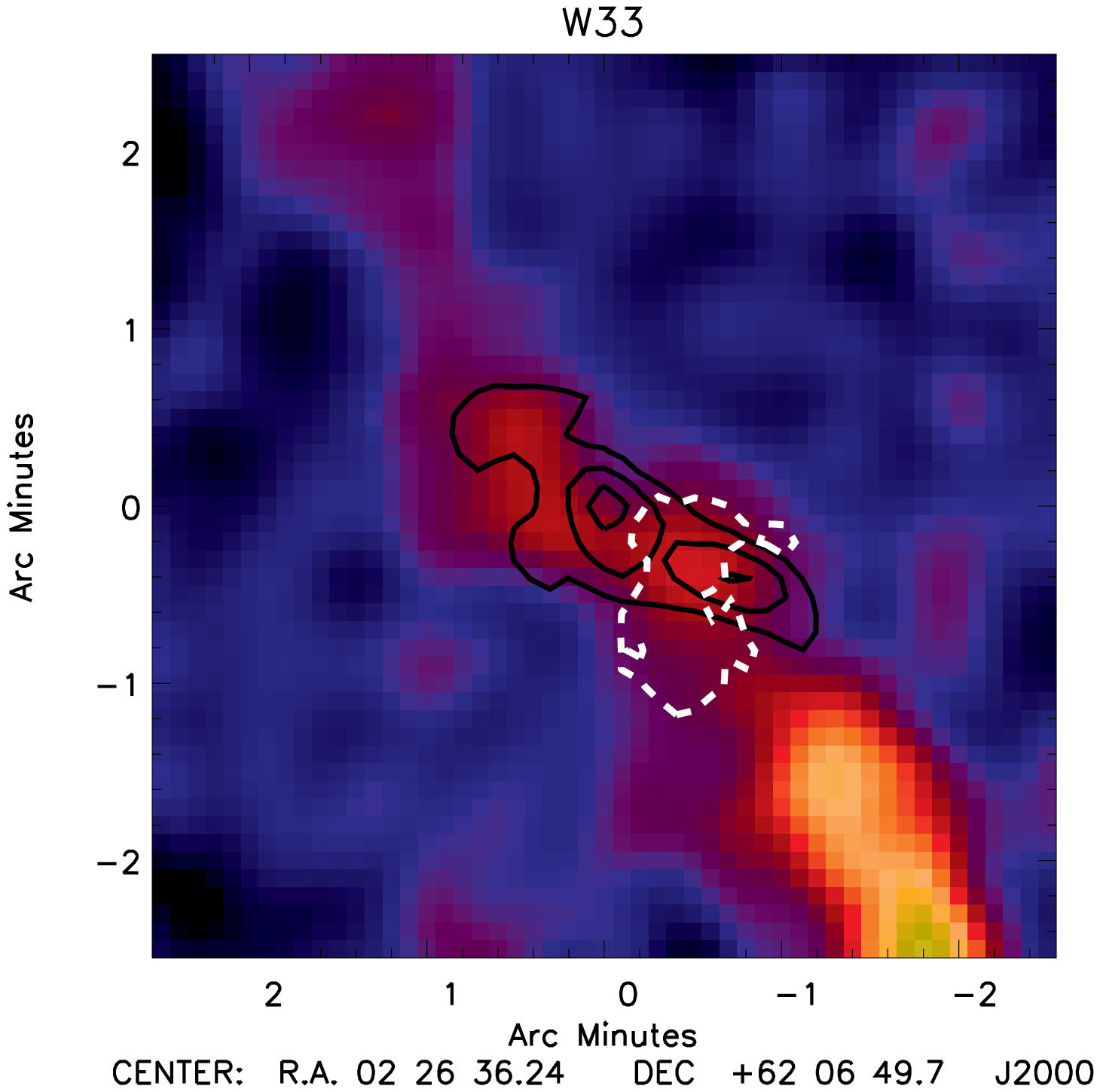}
\includegraphics*[width=0.4\textwidth]{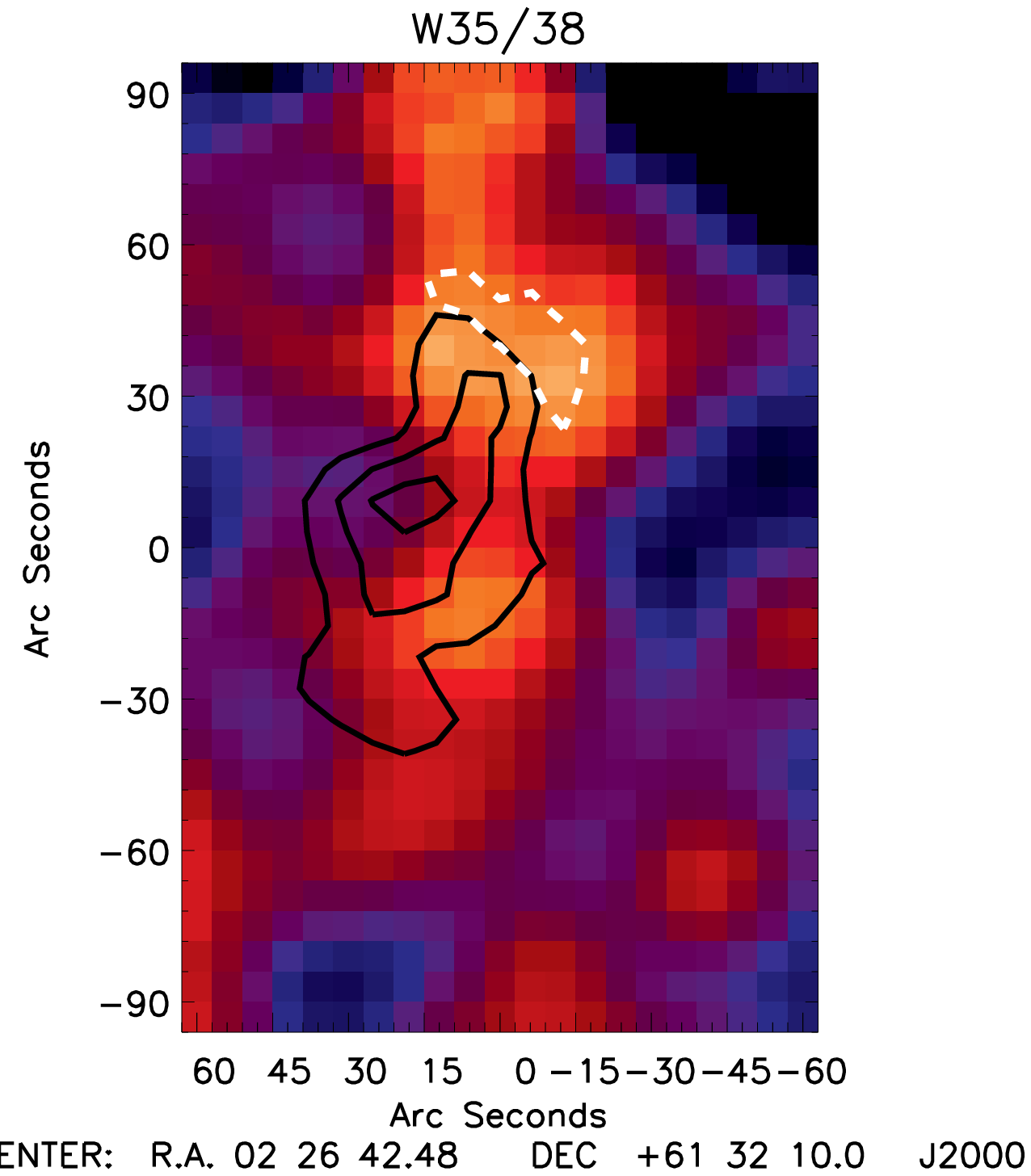}\\
\caption{\nh\ integrated intensity in sources W28, W29, W30, W31, W33 \& W3538 overlaid with black contours of submillimetre emission at 50, 70 and 90\% of the peak flux density. A white, broken contour traces the \nh\ column density distribution at a level of 50\% of the local peak for the sources W30 and W31. For W28, W29, W33 and W3538 the contour is at 30\%}
\end{center}
\end{figure*}
\addtocounter{figure}{-1}

\begin{figure*}
\begin{center} 
\includegraphics*[width=0.4\textwidth]{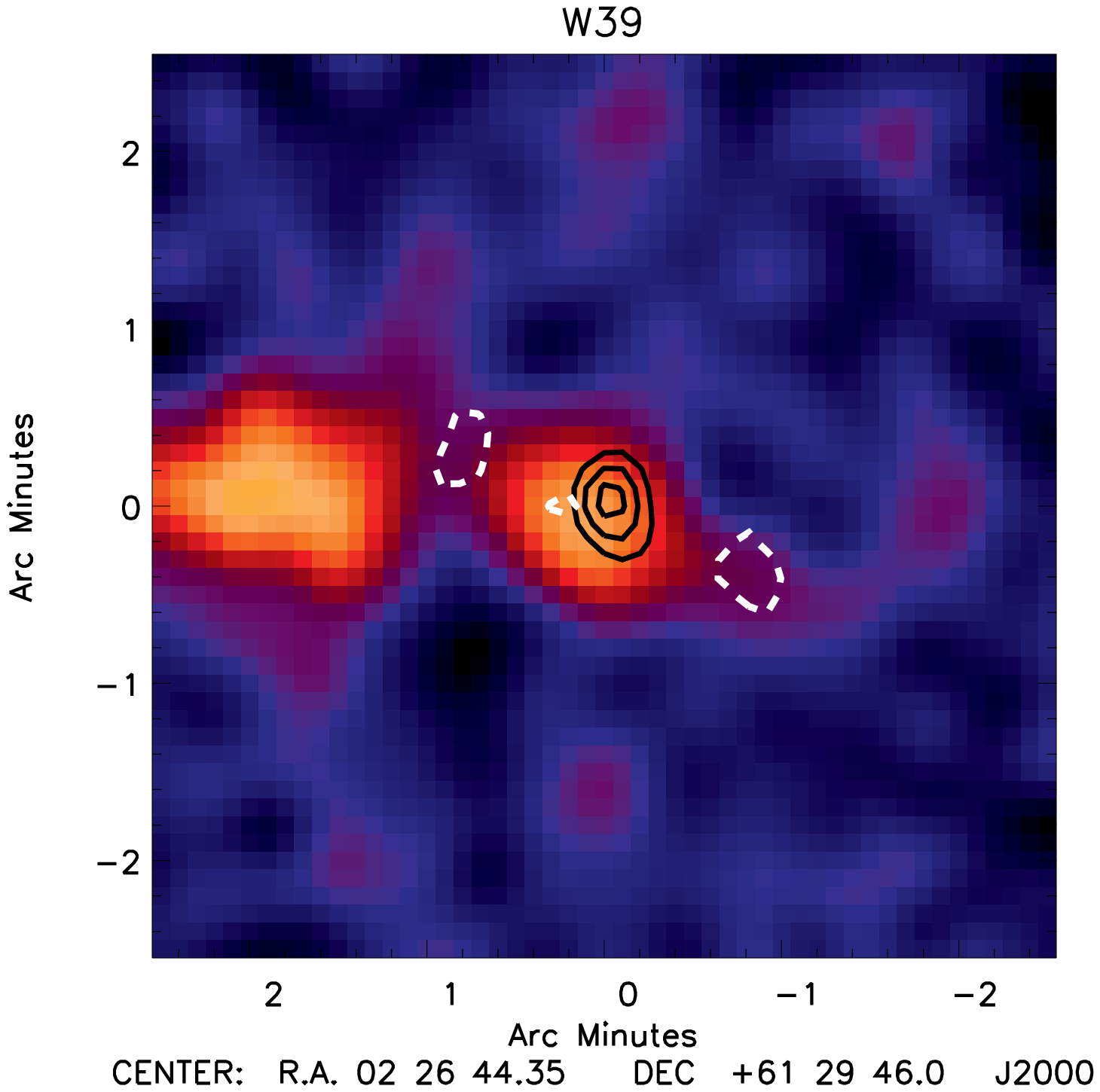}
\includegraphics*[width=0.4\textwidth]{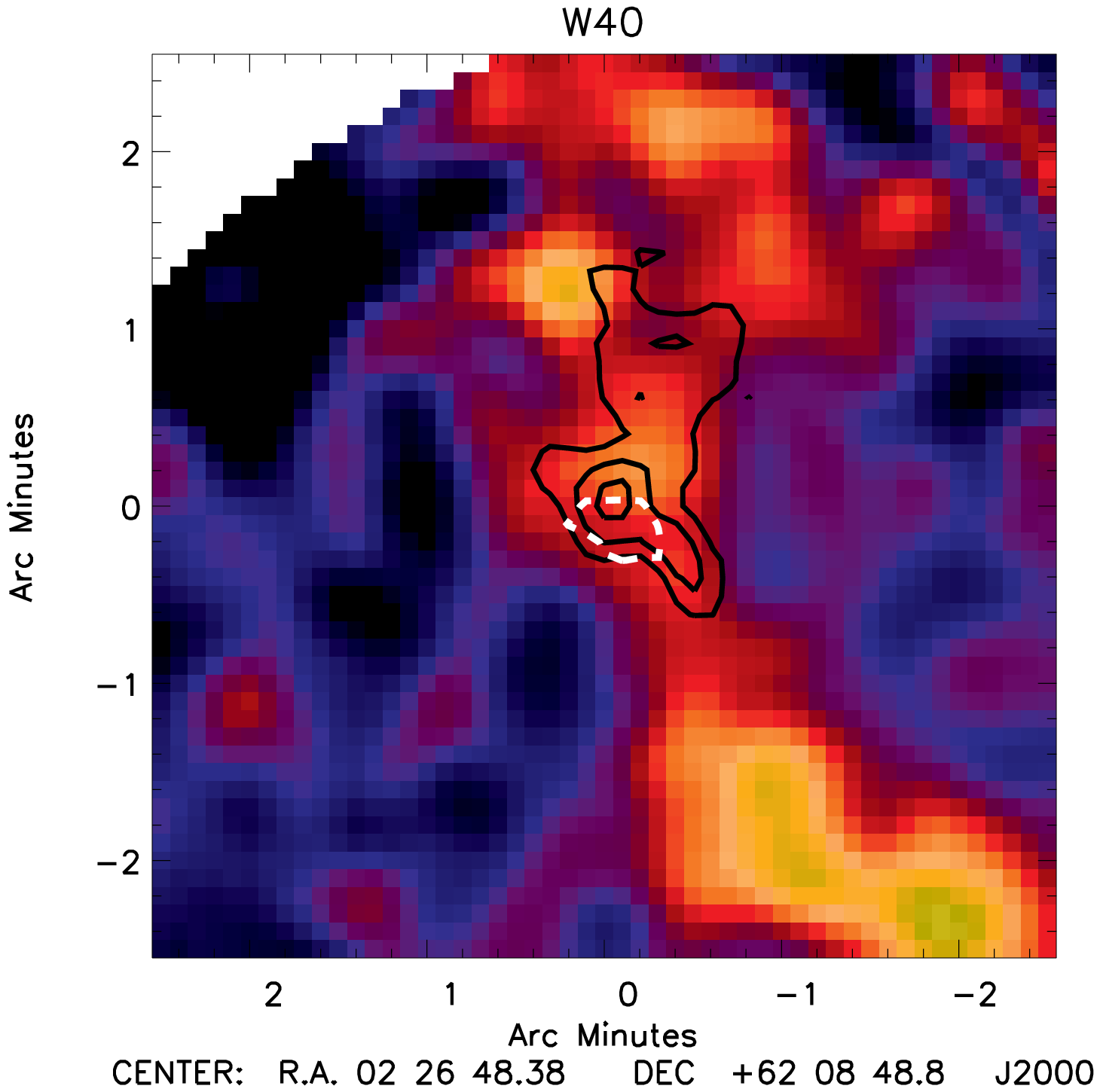}\\
\includegraphics*[width=0.4\textwidth]{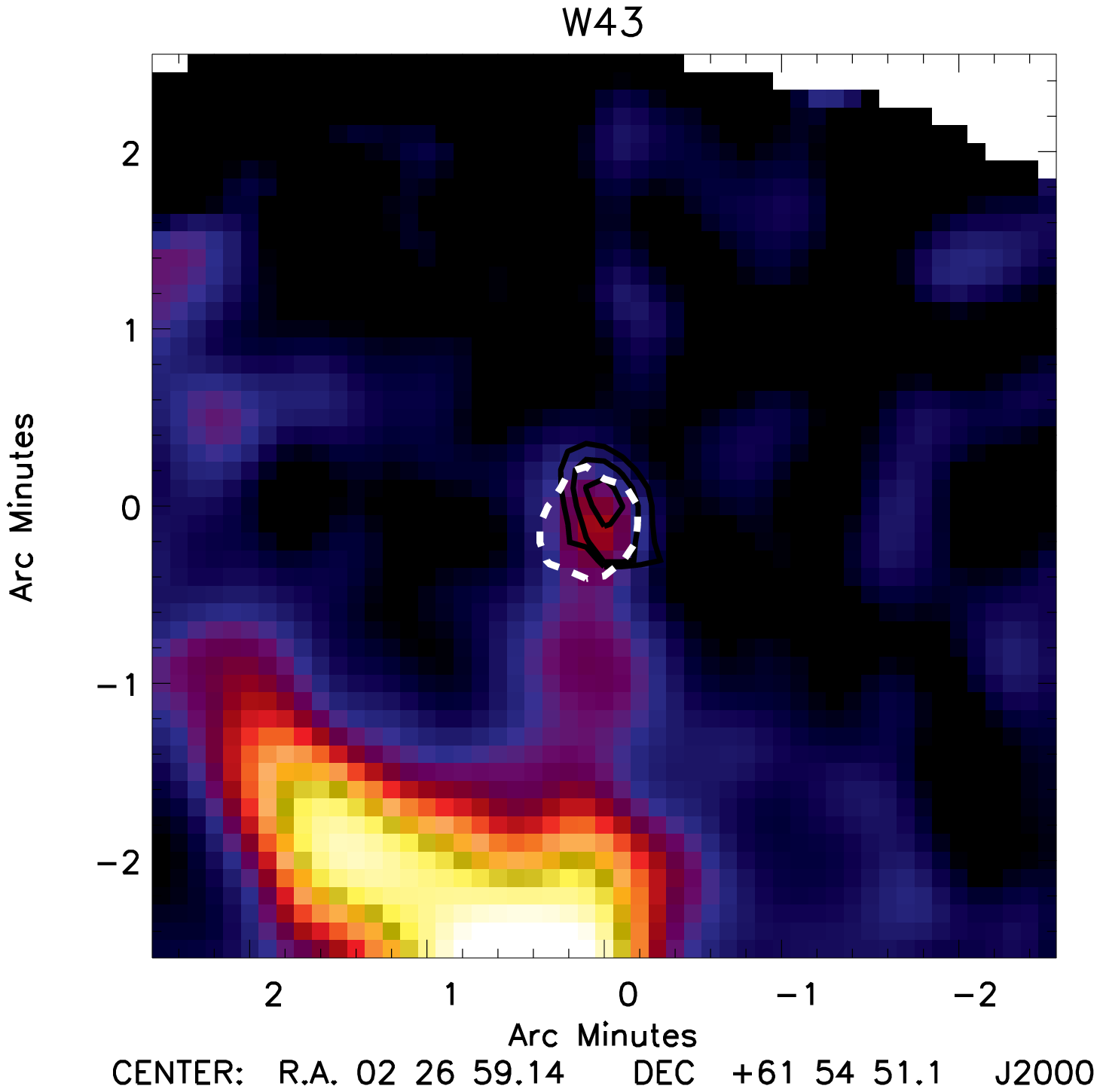}
\includegraphics*[width=0.4\textwidth]{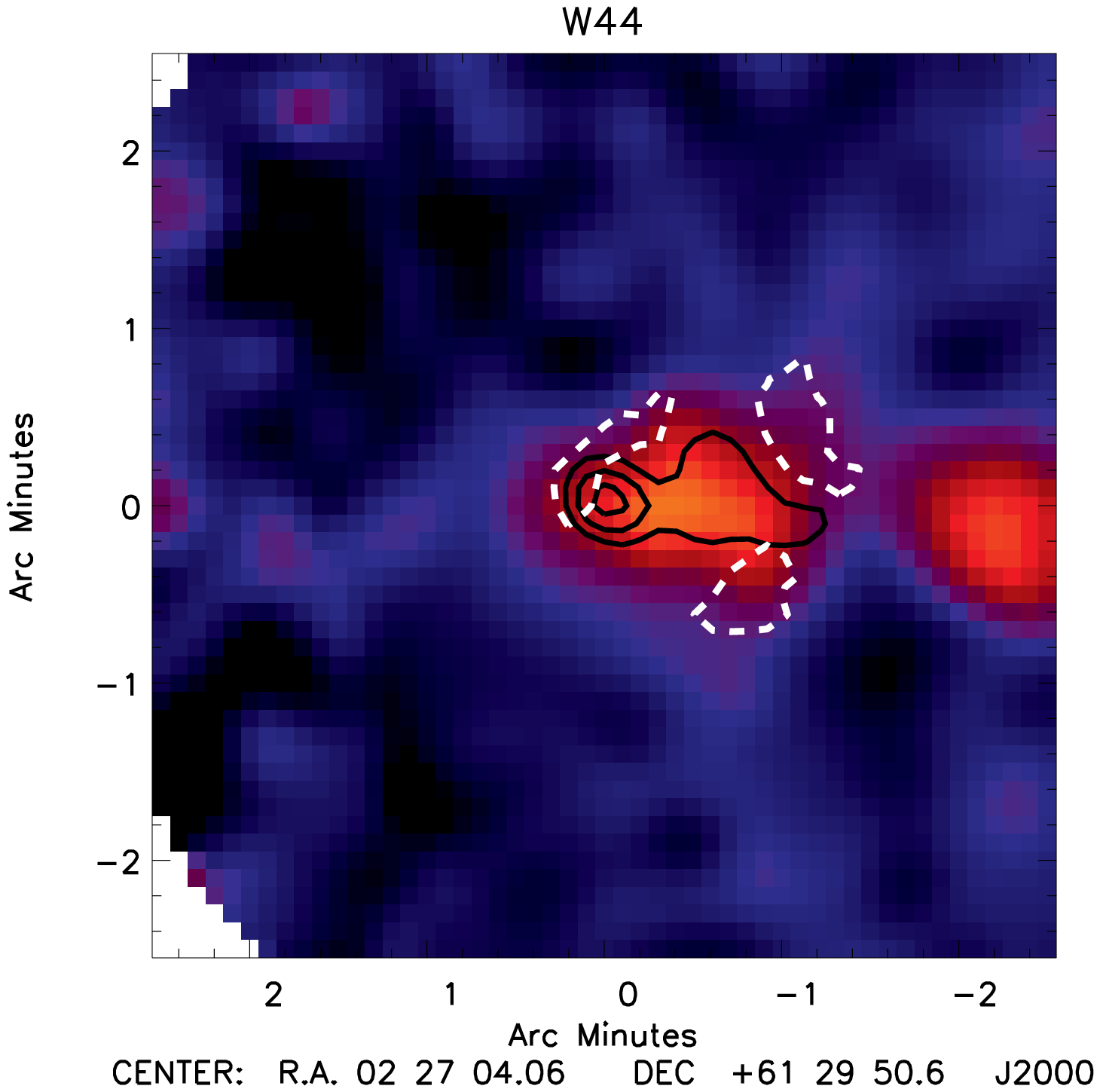}\\
\includegraphics*[width=0.4\textwidth]{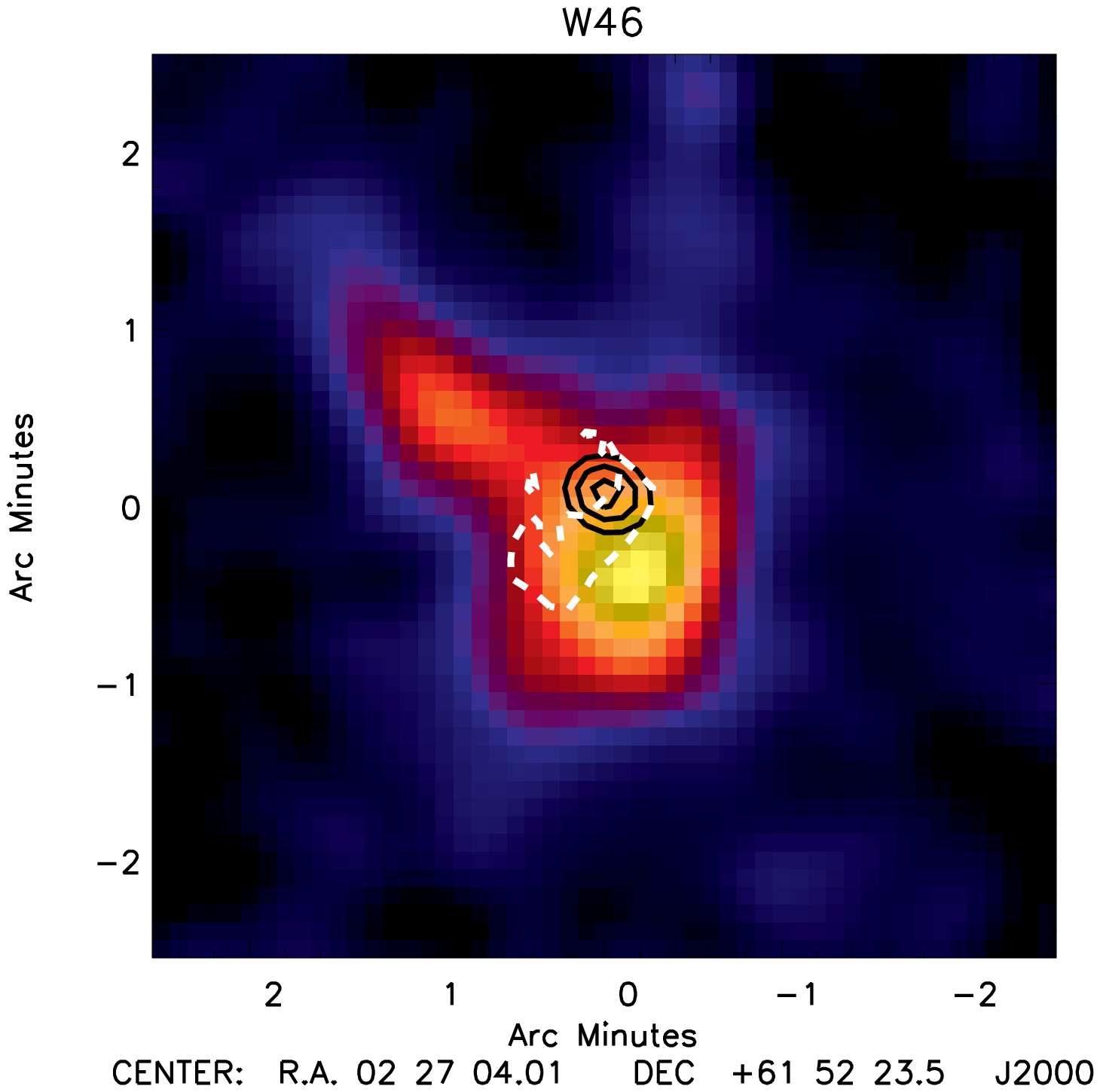}
\includegraphics*[width=0.4\textwidth]{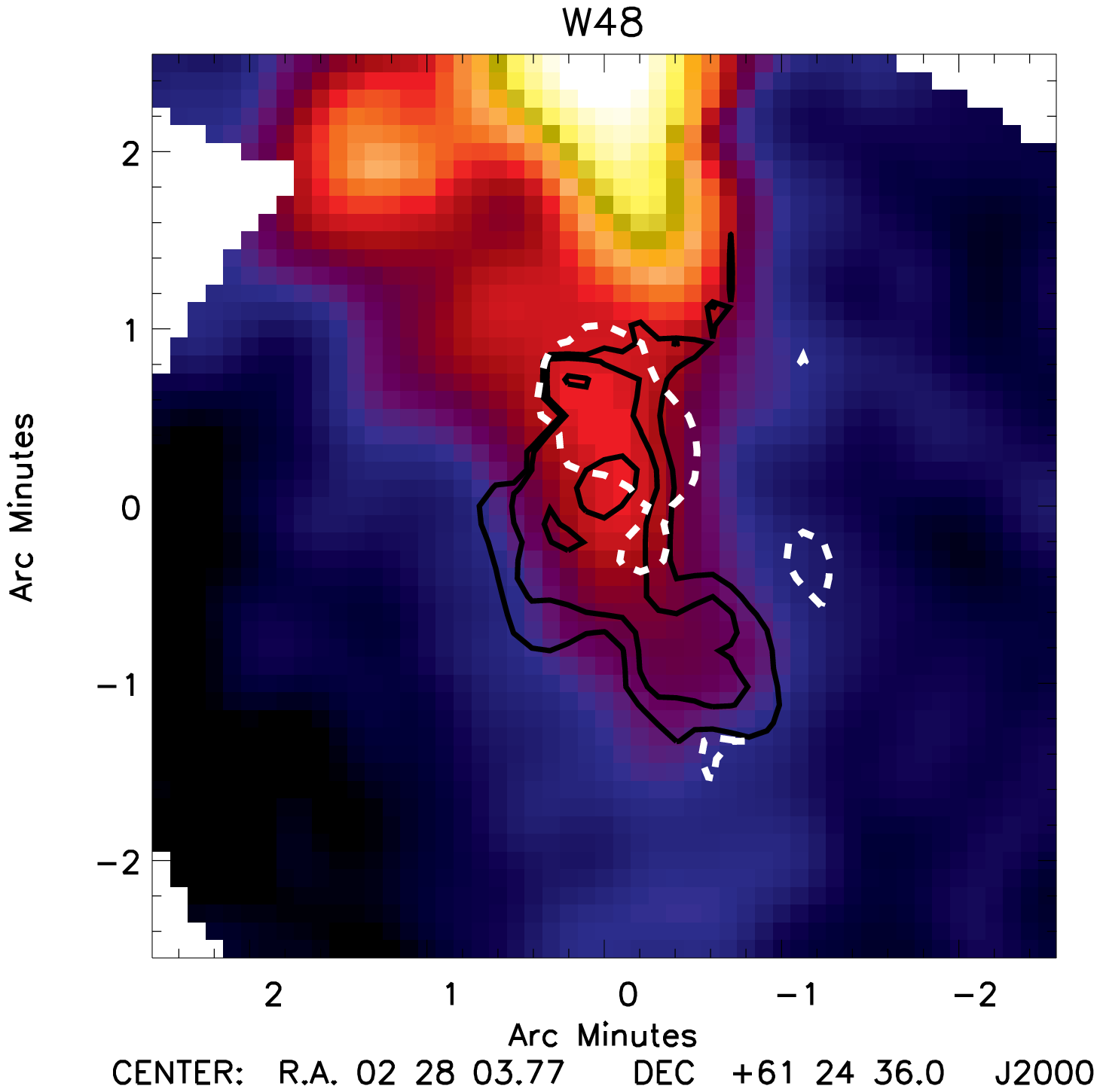}\\
\caption{\nh\ integrated intensity in sources W39, W40, W43, W44, W46 \& 48 overlaid with black contours of submillimetre emission at 50, 70 and 90\% of the peak flux density. A white, broken contour traces the \nh\ column density distribution at a level of 50\% of the local peak for the sources W40, W45 and W46. For W39, W43 and W44 the contour is at 30\%}
\end{center}
\end{figure*}
\addtocounter{figure}{-1}

\begin{figure*}
\begin{center} 
\includegraphics*[width=0.4\textwidth]{W_49_SM_NH3_Conv_Image.eps}
\includegraphics*[width=0.4\textwidth]{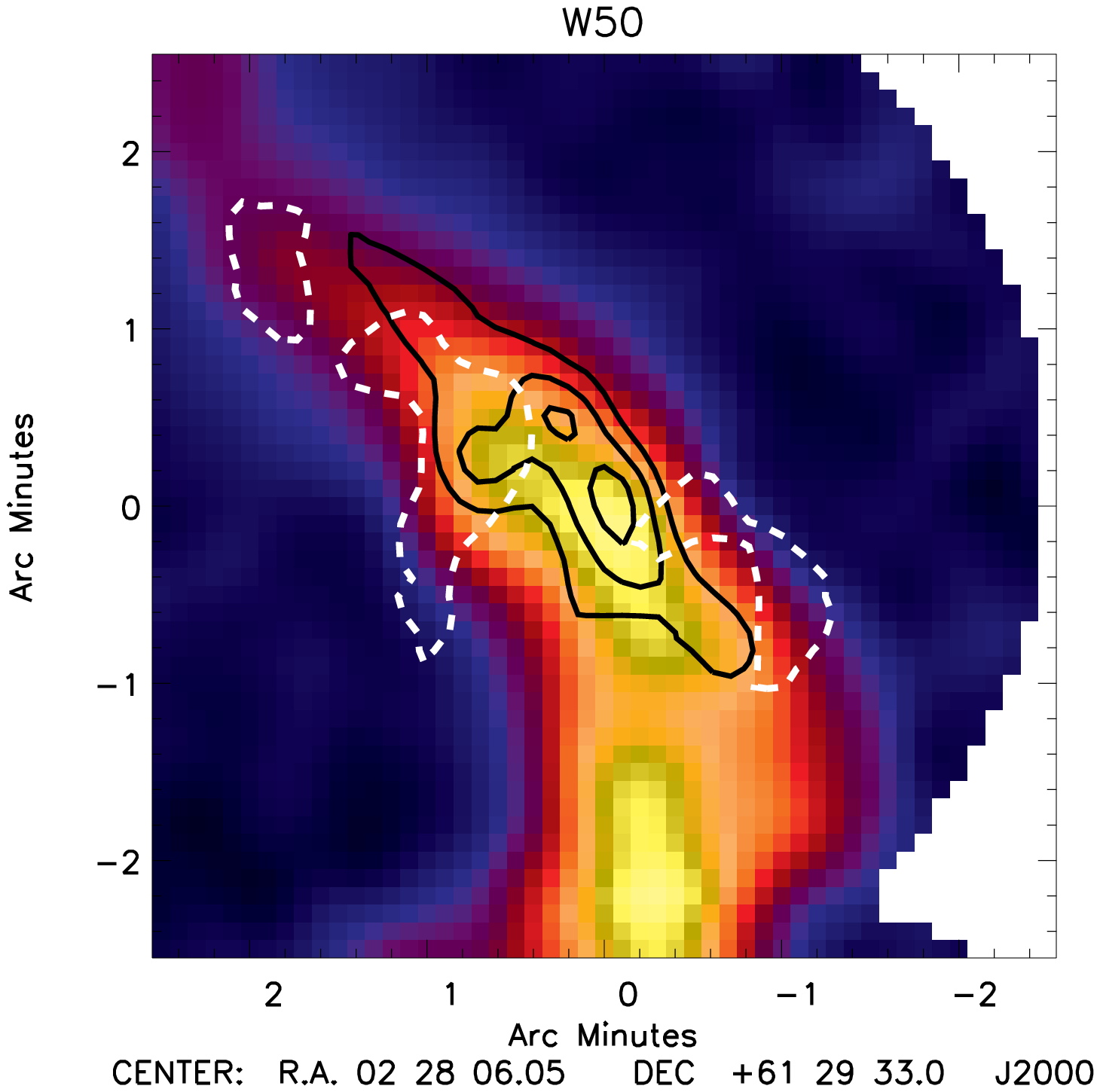}\\
\includegraphics*[width=0.4\textwidth]{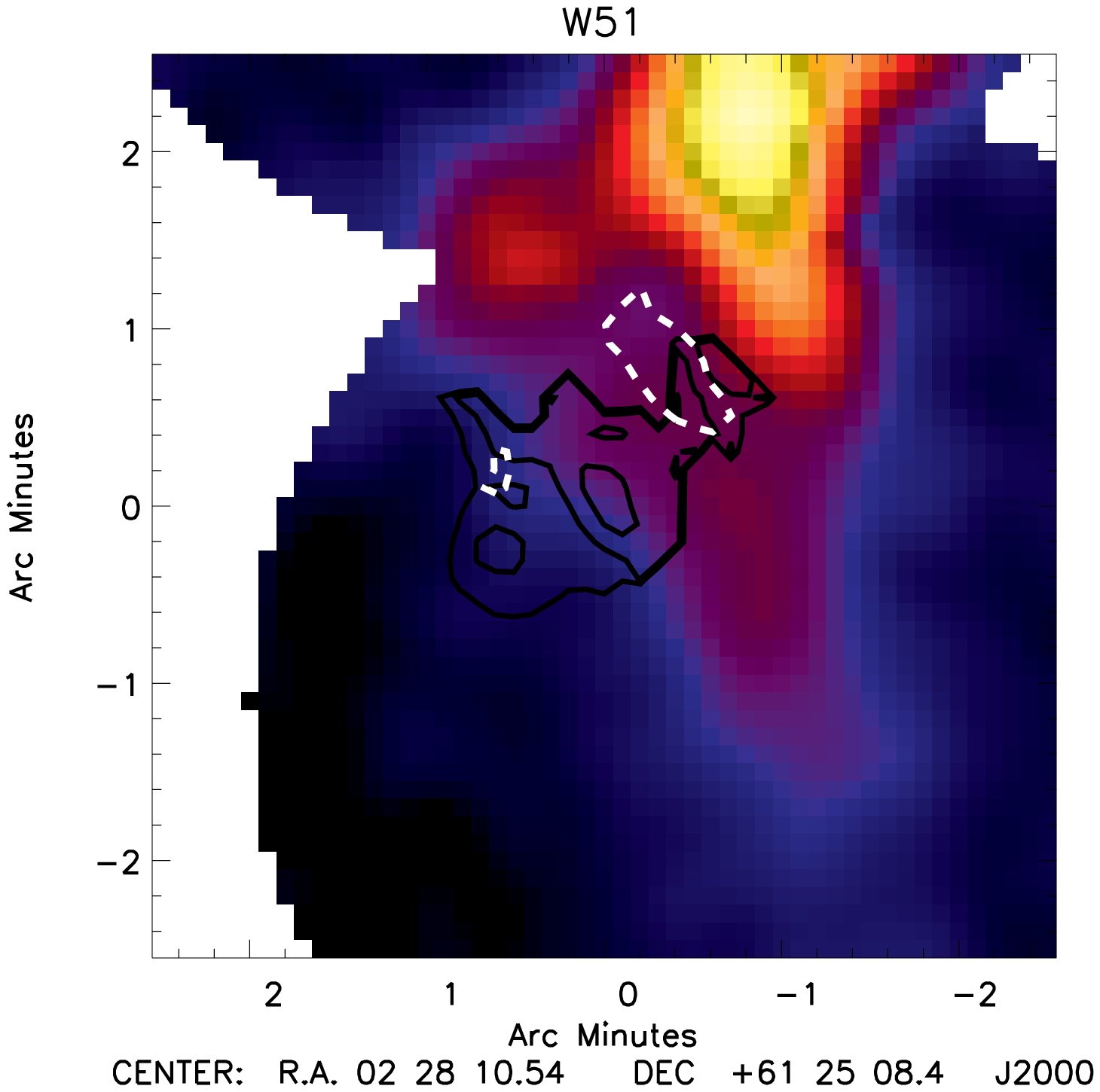}
\includegraphics*[width=0.4\textwidth]{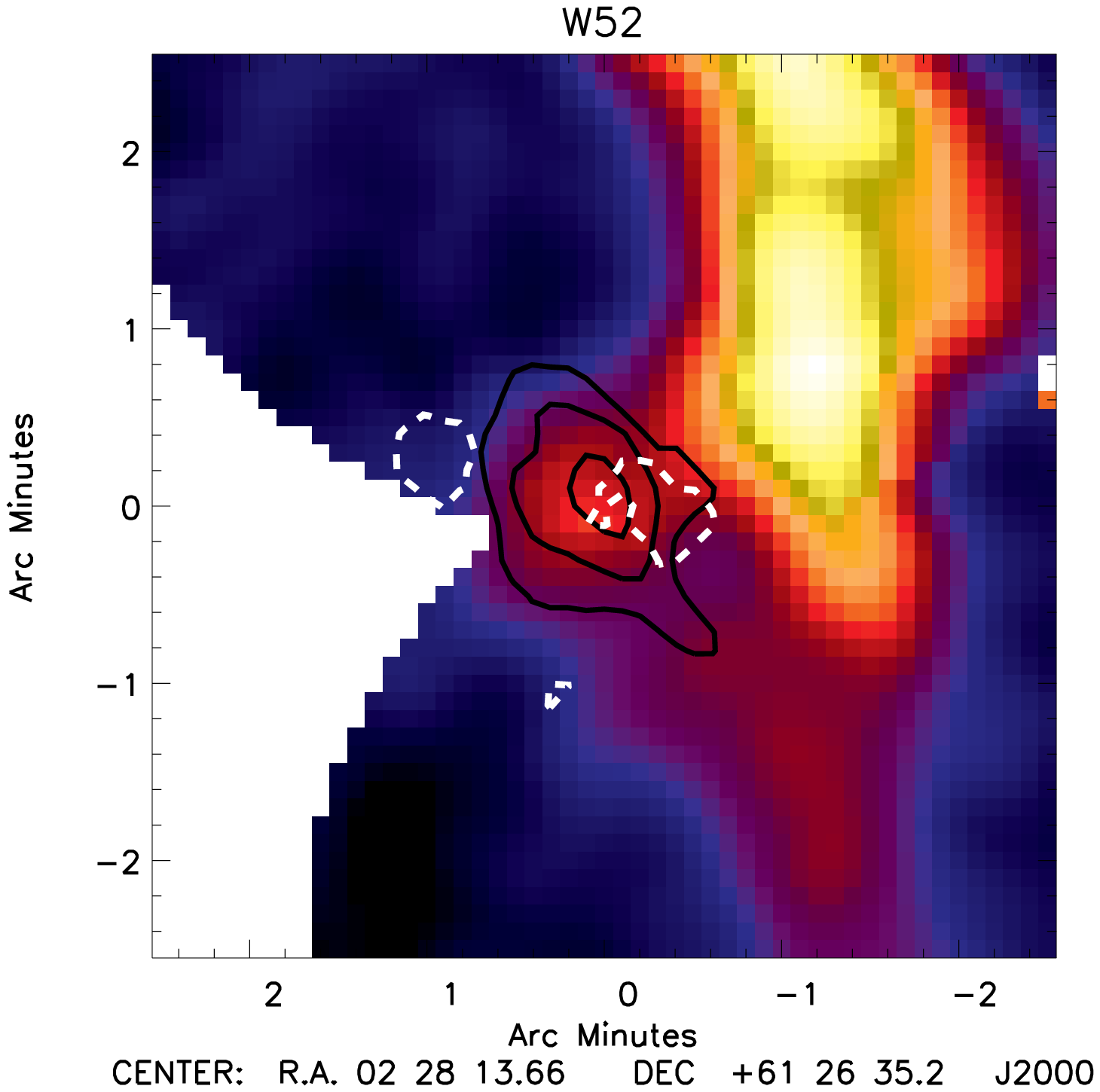}\\
\includegraphics*[width=0.4\textwidth]{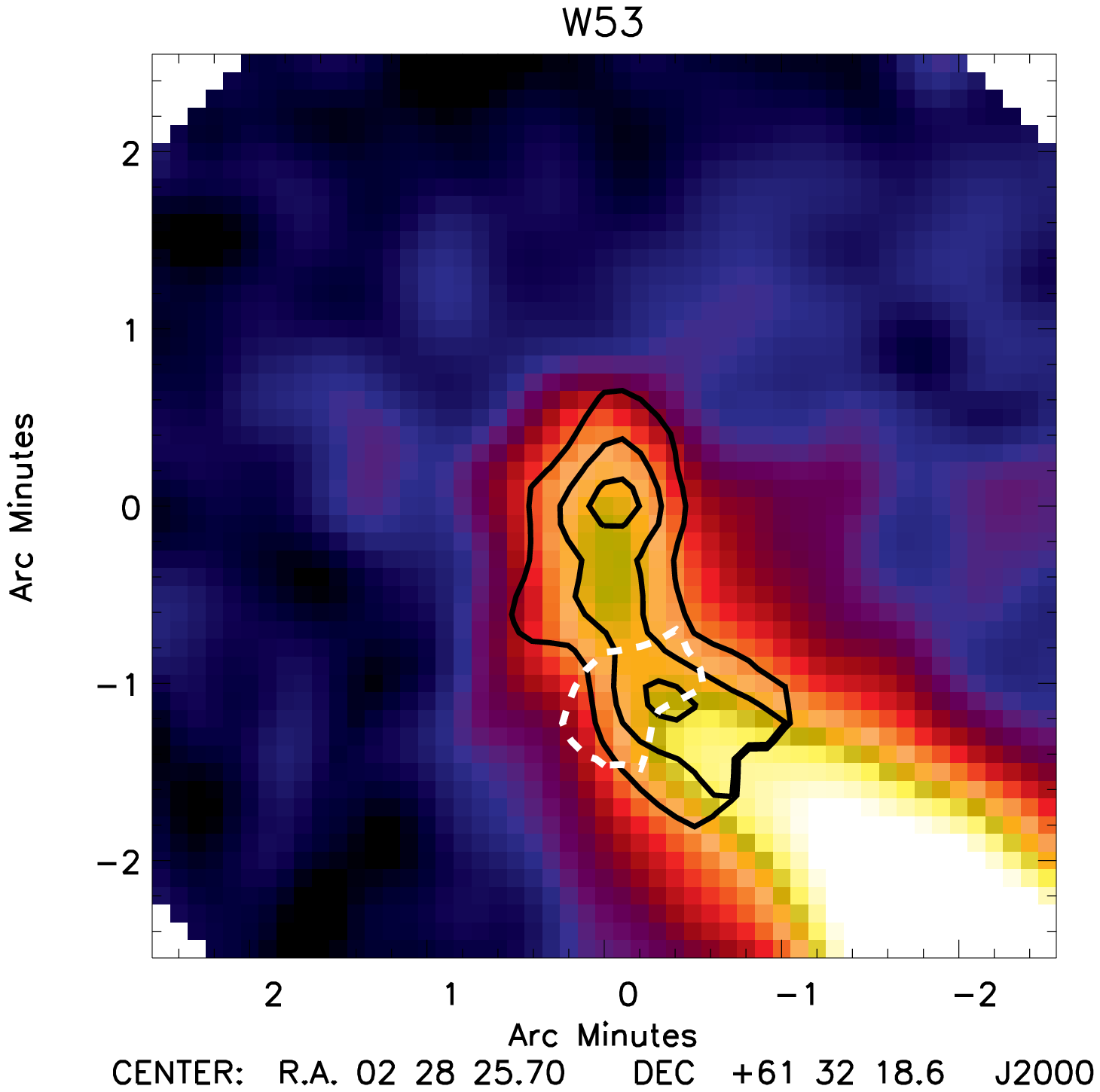}
\includegraphics*[width=0.4\textwidth]{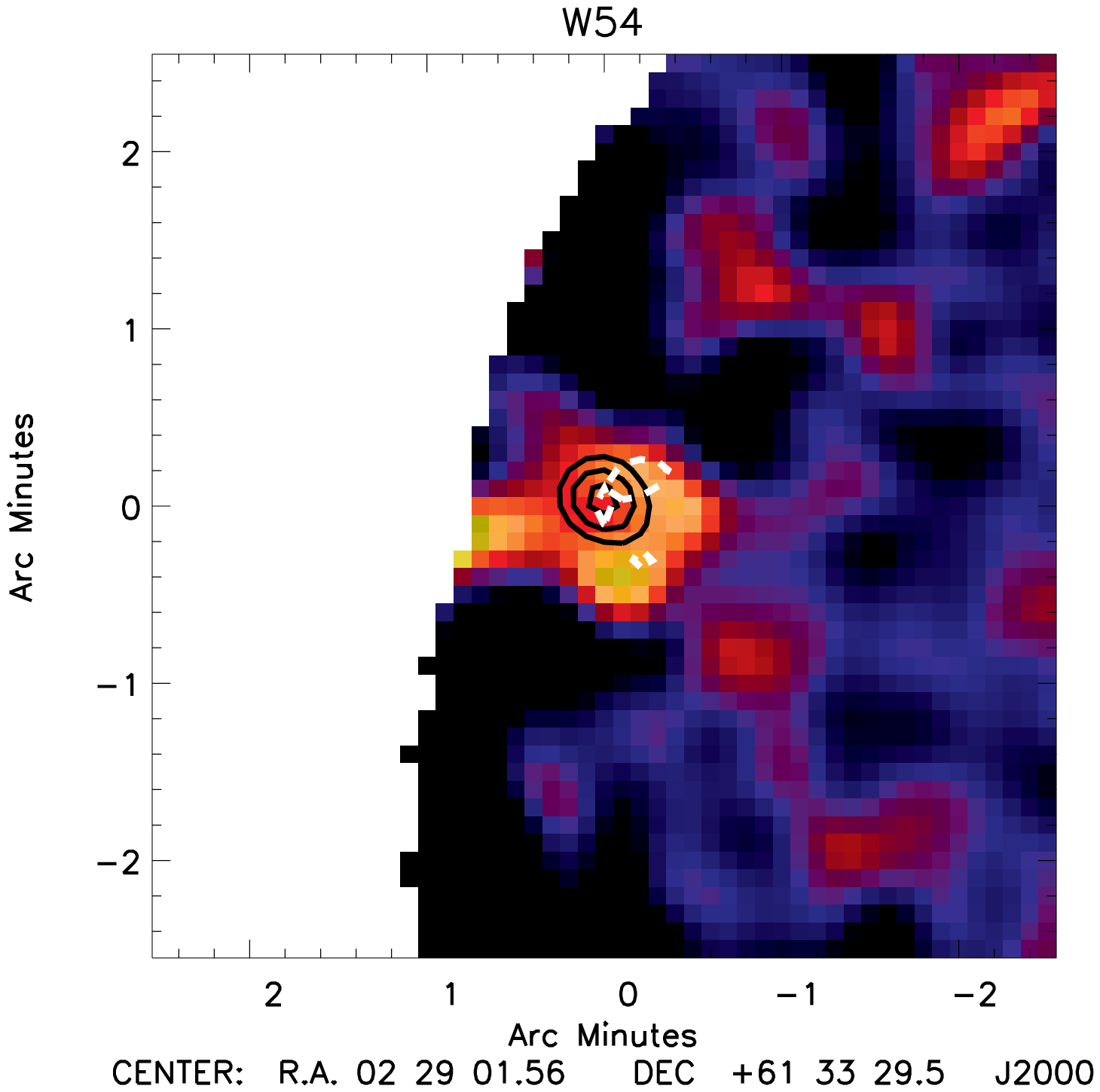}\\
\caption{\nh\ integrated intensity in sources W49, W50, W51, W52, W53 \& W54 overlaid with black contours of submillimetre emission at 50, 70 and 90\% of the peak flux density. A white, broken contour traces the \nh\ column density distribution at a level of 50\% of the local peak for all sources except W50, for which the contour level is at 30\%}
\end{center}
\end{figure*}
\addtocounter{figure}{-1}

\bibliography{References}

\bibliographystyle{mn2e_long_author}
\label{lastpage}

\end{document}